\documentclass[final,3p,times]{elsarticle}

\usepackage[version=4]{mhchem}
\usepackage{amsmath}
\usepackage{url}
\usepackage{graphicx}
\usepackage{epstopdf, epsfig}
\usepackage{amsfonts}
\usepackage{amssymb}
\usepackage{float}
\usepackage{xcolor}
\usepackage{color}

\usepackage[draft]{todonotes}

\usepackage{bm}
\usepackage[acronym]{glossaries}
\usepackage{soul}
\usepackage{comment}

\newcommand{\vv}{{\bm{v}}}
\newcommand{\cc}{{\bm{v}}}

\newcommand{\CC}{\bm{C}}
\newcommand{\rr}{{\bm{r}}}
\newcommand{\jj}{{\bm{j}}}
\newcommand{\FF}{{\bm{F}}}
\newcommand{\gggg}{{\bm{g}}}
\newcommand{\GG}{{\bm{G}}}

\newcommand{\uu}{\bm{u}}
\newcommand{\qq}{\bm{q}}

\newcommand{\xx}{\bm{x}}

\newcommand{\ee}{\bm{e}}
\newcommand{\bnabla}{{\bm{\nabla}}}
\newcommand{\blambda}{{\bm{\lambda}}}
\newcommand{\bONE}{\bm{1}}

\newcommand{\notrace}[1]{\left\langle{#1}\right\rangle}

\usepackage{tikz}
\usetikzlibrary{shapes,arrows,decorations.pathreplacing}
\definecolor{mycolor1}{rgb}{1.00000,0.00000,0.00000}%
\definecolor{mycolor2}{rgb}{1.00000,0.85710,0.00000}%
\definecolor{mycolor3}{rgb}{0.28570,1.00000,0.00000}%
\definecolor{mycolor4}{rgb}{0.00000,1.00000,0.57140}%
\definecolor{mycolor5}{rgb}{0.00000,0.57140,1.00000}%
\definecolor{mycolor6}{rgb}{0.28570,0.00000,1.00000}%
\definecolor{mycolor7}{rgb}{1.00000,0.00000,0.85710}%
\newcommand{\lineone}{\raisebox{2pt}{\tikz{\draw[-,mycolor1,solid,line width = 0.7pt](0,0) -- (5mm,0);}}}
\newcommand{\linetwo}{\raisebox{2pt}{\tikz{\draw[-,mycolor2,solid,line width = 0.7pt](0,0) -- (5mm,0);}}}
\newcommand{\linethree}{\raisebox{2pt}{\tikz{\draw[-,mycolor3,solid,line width = 0.7pt](0,0) -- (5mm,0);}}}
\newcommand{\linefour}{\raisebox{2pt}{\tikz{\draw[-,mycolor4,solid,line width = 0.7pt](0,0) -- (5mm,0);}}}
\newcommand{\linefive}{\raisebox{2pt}{\tikz{\draw[-,mycolor5,solid,line width = 0.7pt](0,0) -- (5mm,0);}}}
\newcommand{\linesix}{\raisebox{2pt}{\tikz{\draw[-,mycolor6,solid,line width = 0.7pt](0,0) -- (5mm,0);}}}
\newcommand{\lineseven}{\raisebox{2pt}{\tikz{\draw[-,mycolor7,solid,line width = 0.7pt](0,0) -- (5mm,0);}}}

\journal{Physics reports}

\begin{document}

\begin{frontmatter}



\title{Lattice Boltzmann for non-ideal fluids: Fundamentals and Practice}

\author[inst1]{S.A. Hosseini}
\affiliation[inst1]{organization={Department of Mechanical and Process Engineering},
            addressline={ETH Zurich}, 
            city={Zurich},
            postcode={8092}, 
            country={Switzerland}}
\author[inst1]{I.V. Karlin\corref{cor1}}
\ead{ikarlin@ethz.ch}
\cortext[cor1]{Corresponding author}
\begin{abstract}
This contribution presents a comprehensive overview of of lattice Boltzmann models for non-ideal fluids, covering both theoretical concepts at both kinetic and macroscopic levels and more practical discussion of numerical nature. In that context, elements of kinetic theory of ideal gases are presented and discussed at length. Then a detailed discussion of the lattice Boltzmann method for ideal gases from discretization to Galilean invariance issues and different collision models along with their effect on stability and consistency at the hydrodynamic level is presented. Extension to non-ideal fluids is then introduced in the context of the kinetic theory of gases along with the corresponding thermodynamics at the macroscopic level, i.e. the van der Waals fluid, followed by an overview of different lattice Boltzmann based models for non-ideal fluids. After an in-depth discussion of different well-known issues and artifacts and corresponding solutions, the article finishes with a brief discussion on most recent applications of such models and extensions proposed in the literature towards non-isothermal and multi-component flows.
\end{abstract}




\end{frontmatter}

\tableofcontents
\pagebreak
\section{Introduction}
Non-ideal fluids are omnipresent in science and technology. From micro-droplets coalescing in clouds, to solidification or melting of alloys and diesel droplets evaporation and subsequent combustion, all involve multiple interacting phases and moving interfaces. This ubiquity fueled wide efforts focused on the development of predictive mathematical models and numerical tools for multi-phase flows.
While significant attention  has been focused on sharp interface methods requiring  efficient tracking of the evolving and deforming interfaces, and imposing jump conditions  \cite{sethian_level_2003,scardovelli_direct_1999,popinet_numerical_2018,prosperetti_computational_2009}, the ever-growing range of temperatures and pressures involved in typical systems of interest is making thermodynamic consistency of the computational models at interfaces essential. Dramatically different thermodynamic regimes are encountered in diesel engines during the compression phase, in aeronautical engines during take-off while most rocket engines operate in trans- and super-critical regimes, where the interface thickness becomes comparable to the flow scales. Nucleation and cavitation are yet another example, where the sharp interface limit does not hold and modifications to the classical nucleation theory \cite{debenedetti_metastable_1997},
related to curvature-dependence of the surface tension, are required.
In such cases, an accurate account of non-ideality of the fluid, including a finite interface thickness, is crucial for predictive numerical simulations of the flow physics.
At a macorscopic level, a primer example for thermodynamics of non-ideal fluids is the second-gradient theory, first introduced by van der Waals for single-component fluids \citep{van_der_waals_thermodynamische_1894},
leading to the Navier-Stokes equations supplemented with the Korteweg stress tensor \citep{korteweg_sur_1901}, and is a starting point for numerical methods known as diffuse interface approach \cite{anderson_diffuse-interface_1998}.
On the other hand, extension of the Boltzmann equation to dense gases within the Enskog hard-sphere collision model \citep{enskog_warmeleitung_1921} and Vlasov mean-field approximation \citep{vlasov_many-particle_1961} provides a kinetic-theory basis for dynamics of non-ideal fluid \citep{chapman_mathematical_1939}.\\
The lattice Boltzmann method (LBM), a discrete solver for the discrete Boltzmann equation ( discrete in phase space) is a numerical tool targeting the hydrodynamic balance equations (initially in the incompressible limit) that has experienced noticeable growth in both popularity and ability to incorporate complex physics. Since the pioneering work of \cite{shan_lattice_1993}, the LBM gained popularity as a viable numerical tool targeting the hydrodynamic regime of non-ideal fluids. Different models for non-ideal fluid dynamics have been developed since and used for a wide variety of applications involving complex physics and geometries. The rapid growth, constant evolution of non-ideal fluid models within the context of the LBM and very large number of yearly publications in that area point to the need for a comprehensive document listing and discussing fundamentals, pitfalls, best practices and most recent developments. A couple of publications have previously taken upon this task in previous years. While well-written and covering a number of topics related to implementation and numerics they do not treat of fundamentals and do not provide in-depth discussions .\\
In the present contribution we aim to provide a comprehensive overview of the LBM for non-ideal fluids simulations. Given the relatively wide area of topics covered in the present document and depth of discussions, sections have been organised in a manner allowing readers to go directly to topics they are interested in without having to go through the document in its entirety. As such readers interested in an introduction to the kinetic theory can read section \ref{sec:el_kinetic_theory} while those looking for a more in-depth discussion can move on to section \ref{sec:frm_bolt_to_hydro}. Those with an interest only for the LBM can directly go to section \ref{sec:LBM_intro}. To give interested readers all necessary background to analyze and understand challenges and shortcomings of different models we start with a detailed introductory discussion covering different aspects of the kinetic theory of ideal gases. This is followed by a section covering all fundamental aspects of the lattice Boltzmann method for ideal gases covering discretization in phase-space physical space and time, incorporation of body forces, an overview of different collision operators and corresponding numerical and physical properties. Extension of the lattice Boltzmann method to non-ideal fluids is then discussed in details in section \ref{sec:non_ideal} where different models are reviewed and a comprehensive discussion covering most important challenges is provided. Some of the most interesting and recent achievement in terms of application with LBMs for non-ideal fluids are then reviewed and discussed in section \ref{sec:applications}. This is followed by a section discussing interesting extensions of the iso-thermal non-ideal fluid solvers to compressible and multi-component fluids in section \ref{sec:comp_and_multcomp}.
\section{Elements of kinetic theory\label{sec:el_kinetic_theory}}

\subsection{The Boltzmann transport equation}
Following Boltzmann, we consider a gas consisting of a large number of identical particles of mass $m=1$. 
The state of the gas is described by the distribution function,
\begin{equation}
	f(\bm{v}, \bm{x}, t),
\end{equation}
with the interpretation that the number of particles $dN_1$ in a differential volume of phase space centered at $(\bm{x}, \bm{v})$ at time $t$ is
\begin{equation}
	dN=f(\bm{v},\bm{x}, t)d^3\bm{v} d^3\bm{x}.
\end{equation}
We further assume that the gas is moderately dilute, that is, the particles mostly fly freely, and experience encounters (collisions) from time to time.
Collisions with a participation of more than two particles at a time can be neglected, so we consider only binary collisions.
More specifically, think of hard spheres of a  diameter $d$. Then if $N$ is the number of such spheres in a container of a volume $V$, we consider the case $N\to\infty$ ("many particles"), $V\to\infty$ ("big container") and $d\to 0$ ("interaction range is short"), so that with both these limits we have $N/V\sim const$ (the average density of particles is finite). Furthermore,
$Nd^3/V\to 0$ (total volume occupied by the particles is negligible to the volume of the container), but $Nd^2\sim const$ (total cross-section area is fixed, so that molecules will be able to "see" each other and collide). This is called the Boltzmann--Grad limit and is used to rigorously justify the domain of validity of the Boltzmann equation.


With the above assumptions we can state that there will be two mechanisms which contribute to changing the distribution function in time and space: the free flight (or a modification thereof if long-range forces are present), and the binary collisions. Thus, we write,
\begin{equation}
	\frac{\partial f}{\partial t} +\cc\cdot\frac{\partial f}{\partial \xx}
	+\frac{\FF}{m}\cdot \frac{\partial f}{\partial \cc}=\mathcal{J}_{\rm B}.
\end{equation}
Here on the left we have the free flight operator ($\cc\cdot\partial/\partial \xx$) and we have also included an effect of action of large-scale forces such as gravity, $(\FF/m)\cdot\partial/\partial \cc$. These two terms signify the net change of the number of particles in the element of the phase space volume $d^3\cc d^3\xx$ centered at $(\cc,\xx)$ due to flying in and out of particles and the change of their velocities through an acceleration caused by external forces.

The term on the right hand side is called the collision integral, and which takes into account the change of the number of particles in the elementary volume $d^3\cc d^3\xx$ through "kicking out" a particle with the velocity about $\cc$ when it collides with some other particle and alters the velocity, and through "kicking in" a particle with the velocity $\cc$ which is produced in a collision of two particles with some different velocities.
Now we shall evaluate the effect of the binary collisions according to Boltzmann.

\subsubsection{Binary collision: Kinematics}
We need first to consider a purely mechanical part of the derivation, namely the collision of two particles. In this section, we shall neglect the long-range forces. 
Collision is vaguely understood as a relatively sharp change of the trajectories of the particles approaching each other from a distance.
The range of the force which causes the sharp change of their trajectories is assumed short so that the particles approach each other along straight trajectories (it is best to think of hard spheres again); their encounter happens almost instantly, after which the molecules fly away from each other along the changed straight trajectories.

We shall consider two molecules labeled "$1$" and "$2$", their velocities before the collision in the lab frame are $\cc_1$ and $\cc_2$, whereas after the collision they are denoted by primes, $\cc_1'$ and $\cc_2'$, respectively. Furthermore, we introduce the center-of-mass velocity $\GG$ and the relative velocity $\gggg_{21}$, and similarly, after the collision,
$\GG'$ and $\gggg'_{21}$,
\begin{align}
	\GG&=\frac{\cc_1+\cc_2}{2},\\
	\gggg_{21}&=\cc_2-\cc_1,\\
	\GG'&=\frac{\cc'_1+\cc'_2}{2},\\
	\gggg_{21}'&=\cc'_2-\cc'_1.
\end{align}
The unit vector in the direction of the relative velocity $\gggg_{21}$ will be denoted $\ee$; correspondingly, $\ee'$ for the post-collision relative velocity $\gggg'_{21}$,
\begin{align}
	\ee&=\frac{\cc_2-\cc_1}{|\cc_2-\cc_1|}=\frac{\gggg_{21}}{g_{21}},\\
	\ee'&=\frac{\cc'_2-\cc'_1}{|\cc'_2-\cc'_1|}=\frac{\gggg'_{21}}{g'_{21}}.
\end{align}
Thus, before and after the collision, the velocities of the molecules in the lab frame are, respectively,
\begin{align}
	\cc_1&=\GG-\frac{1}{2}g_{21}\ee,\\
	\cc_2&=\GG+\frac{1}{2}g_{21}\ee,\\
	\cc_1'&=\GG'-\frac{1}{2}g_{21}'\ee',\\
	\cc_2&=\GG'+\frac{1}{2}g_{21}'\ee'.
\end{align}
Collision are further assumed to be elastic so that the momentum of the pair of molecules and their energy is conserved as the result of the collision (particle's mass $m=1$),
\begin{align}
	\cc_1+\cc_2&=\cc_1'+\cc_2',\label{eq:binarymomentum}\\
	|\cc_1|^2+|\cc_2|^2&=|\cc_1'|^2+|\cc_2'|^2.\label{eq:binaryenergy}
\end{align}
These are $3+1=4$ equations (three components of momentum conservation and one energy conservation) for $6$ unknowns (three components of $\cc'$ and three components of $\cc_1'$).
From the first equation, we infer that the center-of-mass momentum is conserved in the collision. From the second we get that the magnitude of the relative velocity does not change in the collision,
\begin{align}
	\GG'&=\GG,\\
	g'_{21}&=g_{21}.
\end{align}
The post-collision velocities in the lab frame are thus a two-parametric family is parameterized by a vector $\ee'$ belonging to a unit sphere, $|\ee'|^2=1$:
\begin{align}
	\cc'_1&=\frac{\cc_1+\cc_2}{2}-\frac{|\cc_2-\cc_1|}{2}\ee',\label{eq:cprime}\\
	\cc_2'&=\frac{\cc_1+\cc_2}{2}+\frac{|\cc_2-\cc_1|}{2}\ee'.\label{eq:coneprime}
\end{align}
From these relation we see that the vector $|\ee'|$ is a rotation of the relative velocity as the result of the collision.
Deflection angle $\theta$ is the angle between $\ee$ and $\ee'$,
\begin{equation}
	\cos\theta=\ee\cdot\ee'.
\end{equation}

\subsubsection{Differential scattering cross-section}

We now shall move on, still in a pure mechanics mode, to considering a number of encounters. 
This is a scattering problem, familiar from classical mechanics.
We shall further assume that molecules are interacting through a potential which depends only on the absolute distance between them, $U(\rr_1,\rr_2)=U(|\rr_2-\rr_1|)$ (central forces).
In such a case the trajectories of both molecules are confined to a plane, which is orthogonal to the angular momentum (the latter is conserved for the central force interaction).

The analysis is best done in the center-of-mass (CM) frame (co-moving with the center-of-mass velocity $\GG$). The problem then reduces to the scattering of a particle with a reduced mass $m_0=m_1m_2/(m_1+m_2)$ ($m_0=m/2$ in the case of equal masses of particles), moving with the velocity $\gggg_{21}$ on a \emph{fixed} center.  Trajectories of such  particles look most simple: they are symmetric with respect to the line through the points of closest approach of the molecules and the center of force.

Now we need to specify the two parameters to characterize the initial data for the scattering particle.
For that we use the azimuth angle $\epsilon$ which fixes the orientation of the plane containing the trajectory with respect to an arbitrary fixed plane containing the center of force and parallel to $\gggg_{21}$, and the impact parameter $b$, defined as the distance between the line drawn by $\gggg_{21}$ and the parallel to it line containing the center of force.

Let us consider the incident jet of particles through the area between the two concentric circles of radii $b$ and $b+db$, cut with the angle between $\epsilon$ and $\epsilon+d\epsilon$,
\begin{equation}
	dA_{\rm in}=bdbd\epsilon.
\end{equation}
The particles coming through the area $dA_{\rm in}$ are scattered into the solid angle,
\begin{equation}
	dA_{\rm out}=d^2\ee'=\sin\theta d\theta d\epsilon.
\end{equation}
The differential scattering cross section $\alpha_{12}$ is defined as the ratio between these two areas,
\begin{equation}
	\alpha_{12}=\frac{dA_{\rm in}}{dA_{\rm out}}=\frac{bdbd\epsilon}{\sin\theta d\theta d\epsilon}.
\end{equation}
Thus,
\begin{equation}
	bdbd\epsilon=\alpha_{12}d^2\ee',
\end{equation}
where
\begin{equation}
	\alpha_{21}=\frac{b\left|\frac{\partial b}{\partial \theta}\right|}{\sin\theta}.
\end{equation}
Note that we have written the absolute value of the derivative of the impact parameter with respect to the deflection angle because for repulsive force this function is negative (that is, with the decrease of the impact parameter the particle is "turned back" by the potential, the deflection angle is increased).

Differential scattering cross section thus represents intensity of the incoming flux of particles through the element of the cross-section in terms of the flux of scattered particles into the solid angle. Note that for the central forces, the dependence on the azimuth angle is irrelevant, and the differential scattering cross section depends only on the deflection angle $\theta$ and the magnitude of the relative velocity $g_{21}$. For a specified central potential, the differential cross-section is computed by classical mechanics. We shall discuss a few relevant examples later, but for now we shall keep it unspecified.


\subsubsection{Boltzmann's collision integral}

We shall now proceed with the evaluation of the rate of change of the distribution function, which we represent in a "gain-loss" form,
\begin{equation}
	\mathcal{J}_{\rm B}=\mathcal{J}_{\rm B}^+-\mathcal{J}_{\rm B}^-,
\end{equation}
where
\begin{itemize}
	\item The loss,
	\begin{equation}
		\mathcal{J}_{\rm B}^-(\cc_1,\xx,t)d^3\cc_1 d^3\xx dt,
	\end{equation}
	is the number of collisions of the type $\{\cc_1,\cc_2\}\to\{\cc_1',\cc_2'\}$, resulting in the loss of the particles of the type "1", in the infinitesimal volume of phase space $d^3\cc_1 d\xx$, during the time between $t$ and $t+dt$, and
	\item The gain,
	\begin{equation}
		\mathcal{J}_{\rm B}^+(\cc_1,\xx,t)d^3\cc_1 d^3\xx dt,
	\end{equation}
	is the number of collisions of the type $\{\cc_1',\cc_2'\}\to\{\cc_1,\cc_2\}$, resulting in the gain of the particles of the type "1", in the same infinitesimal phase space volume, for the same time duration.
\end{itemize}
Note that when we say "loss/gain", we do not mean a particle is destroyed or created in the volume element $d\xx$; rather, we mean that a particle, which was in $d^3\cc_1$ before the collision,  has changed its velocity "too much" after the collision, so that is not within  $d^3\cc_1$ any longer (loss), or that after a collision, the velocity of one of the particles changed in such a way that it now belongs to $d^3\cc_1$ (gain). So it is about the loss and gain in the element of the phase space $d^3\cc_1 d^3\xx$.

For the evaluation of the rates $\mathcal{J}_{\rm B}^-$ and $\mathcal{J}_{\rm B}^+$, Boltzmann assumed that any two particles which are about to collide, "do not know anything about each other", neither with respect to their velocities, nor position. In other words, the \emph{molecular chaos assumption} means that velocities of any of the two particles entering an encounter are un-correlated.
With this assumption, we proceed as follows:

\paragraph{The loss}
The number of the type "2" particles which will encounter the $dN_1=f(\cc_1,\xx,t)d^3\cc_1 d^3\xx$
particles of type "1" can be written as,
\begin{equation}
	dN_2=f(\cc_2,\xx,t)d^3\cc_2 \left(|\cc_2-\cc_1|dt \times bdbd\epsilon\right).
\end{equation}
Note the elementary \emph{collision volume} in this expression, $dV_{\rm c}=|\cc_2-\cc_1|dt \times bdbd\epsilon$.
The number of colliding \emph{pairs} is thus estimated as,
\begin{equation}
	dN_{12}=dN_1\times dN_2=\left(f(\cc_1,\xx,t)d^3\cc_1 d^3\xx \right)\left(f(\cc_2,\xx,t)|\cc_2-\cc_1|bdbd\epsilon d^3\cc_2 dt\right).
\end{equation}
Notice that the number of {pairs} of interacting particles is estimated in this expression as a {product} of two one-particle distributions, $dN_{12}\sim f(\cc_1,\xx,t)f(\cc_2,\xx,t)$. This is implied by the molecular chaos assumption: if the colliding particles were assumed to be correlated, then we would need a distribution function of \emph{pairs} of particles.
With the results of the previous section, this expression can be written using the differential scattering cross section,
\begin{equation}
	dN_{12}=\left(f(\cc_1,\xx,t)f(\cc_2,\xx,t) |\cc_2-\cc_1| \alpha_{12}d^2\ee'd^3\cc_2\right) d^3\cc_1 d^3\xx dt.
\end{equation}
Thus, the rate of loss of particles with velocities $\cc_1$ is the sum (integral) of the above expression over all possible velocities $\cc_2$ and  over all the directions of post-collision relative velocity $\ee'$ (the surface of the three-dimensional unit sphere $S^2$):
%
%
\begin{equation}\label{eq:loss}
	\mathcal{J}_{\rm B}^-(\cc_1,\xx,t)=f(\cc_1,\xx,t)\int_{R^3} \int_{S^2}f(\cc_2,\xx,t)|\cc_2-\cc_1|\alpha_{12}d^2\ee' d^3\cc_2.
\end{equation}

\paragraph{The gain}
We now need to construct a similar expression for the rate with which the particles with the velocity $\cc_1$ are produced as the result of collisions of pairs of particles with other velocities.
 The counting process is greatly simplified by the \emph{reversibility} of mechanical motion. Indeed, Newton's equations are reversible in the time. Therefore, if in the direct scattering the initial data for the velocities
$\{\cc_1,\cc_2\}$ was transformed into $\{\cc_1',\cc_2'\}$, then we can take $\{\cc_1',\cc_2'\}$ as the initial data in another (reverse) scattering experiment, and the result of scattering will be known, it is $\{\cc_1,\cc_2\}$.
Precisely this ability to interchange of the past and the future in a mechanical system (time-reversal) is what will be used to count all the pairs of particles which produce the desired velocity $\cc_1$.
We therefore do not need to start a counting from the beginning, as we just need to repeat all of the above by changing non-primed variables into primed variables, and vice versa. Note, however, that we still use the molecular chaos assumption when so doing.
Thus, using the reversibility, we can write the number of pair collisions of the type $\{\cc_1',\cc_2'\}\to\{\cc_1,\cc_2\}$,
\begin{equation}
	dN_{12}'=f(\cc_1',\xx,t)f(\cc_2',\xx,t) |\cc_2'-\cc_1'| \alpha_{12}'d^2\ee d^3\cc_1'd^3\cc_2'd^3\xx dt.
\end{equation}
This expression is transformed in three steps:
As the magnitude of the relative velocity does not change as the result of collision, we have
	\begin{equation}
		dN_{12}'=f(\cc_1',\xx,t)f(\cc_2',\xx,t) |\cc_2-\cc_1| \alpha_{12}'d^2\ee d^3\cc_1'd^3\cc_2' d^3\xx dt.
	\end{equation}
Furthermore, the differential scattering cross section remains same under the inversion of the velocities,
	\begin{equation}
		dN_{12}'=f(\cc_1',\xx,t)f(\cc_2',\xx,t) |\cc_2-\cc_1| \alpha_{12}d^2\ee d^3\cc_1'd^3\cc_2' d^3\xx dt.
	\end{equation}
Finally, the element of the $2+3+3$-dimensional volume, $d^2\ee d^3\cc_1'd^3\cc_2'$ is transformed to the elementary volume $d^2\ee' d^3\cc_1 d^3\cc_2$. First, we compute the Jacobian of the transformation $\{\cc_1,\cc_2\}\to\{\GG,\gggg_{21}\}$,
	\begin{equation}
		J=\frac{\partial(\GG,\gggg_{21})}{\partial(\cc_1,\cc_2)}=1.
	\end{equation}
	Thus, by the definition of the transformation of the elementary volume which includes the determinant of the Jacobian, we have 
		$d^3\GG d^3\gggg_{21}=|J|d^3\cc_1d^3\cc_2=d^3\cc_1d^3\cc_2$,
	and similarly,
		$d^3\GG d^3\gggg_{21}'=d^3\cc_1'd^3\cc_2'$.
	On the other hand,
		$d^3\gggg_{21}=g_{21}^2 dg_{21}d^2\ee$ and $d^3\gggg_{21}'=g_{21}^2 dg_{21}d^2\ee'$.
	Combining these expressions, we have,
	\begin{equation}
		d^2\ee d^3\cc_1'd^3\cc_2'=d^2\ee' d^3\cc_1 d^3\cc_2,
	\end{equation}
	and thus we obtain the final expression for $dN'_{12}$ in the form,
	\begin{equation}
		dN_{12}'=\left(f(\cc_1',\xx,t)f(\cc_2',\xx,t) |\cc_2-\cc_1| \alpha_{12}d^2\ee' d^3\cc_2\right) d^3\cc_1 d^3\xx dt.
	\end{equation}	
As the result, we obtain the rate of gain after integration,
\begin{equation}\label{eq:gain}
	\mathcal{J}_B^+(\cc_1,\xx,t)=\int_{R^3} \int_{S^2}f(\cc_1',\xx,t)f(\cc_2',\xx,t)|\cc_2-\cc_1|\alpha_{12}d^2\ee' d^3\cc_2.
\end{equation}
%
%
Combining the results of the above estimate for the gain (\ref{eq:gain}) and loss (\ref{eq:loss}),
we obtain the rate of change of the one-particle distribution function due to binary collisions,
\begin{equation}\label{eq:collisionintegral}
\mathcal{J}_B	=\int_{R^3} \int_{S^2}\left[f(\cc_1',\xx,t)f(\cc_2',\xx,t)-f(\cc_1,\xx,t)f(\cc_2,\xx,t)\right]K(|\cc_2-\cc_1|,\theta)\ee' d^3\cc_2,
\end{equation}
where the post-collision velocities $\cc_1'$ and $\cc_2'$ are the functions of $\cc_1$, $\cc_2$ and $\ee'$, Eqs.\ (\ref{eq:cprime}) and (\ref{eq:coneprime}), 
and where the function $K$ is the \emph{collision kernel},
\begin{equation}\label{eq:collisionkernel}
	K=|\cc_2-\cc_1|\alpha_{12}(|\cc_2-\cc_1|,\theta).
\end{equation}
This is  the \emph{Boltzmann Stosszahlansatz} or the \emph{Boltzmann collision integral}. 

The collision kernel (\ref{eq:collisionkernel})
 depends on the magnitude of the relative velocity $|\cc_2-\cc_1|$ and on the deflection angle $\theta$. Computation of $K$ is a matter of studying the mechanical scattering problem. Here we mention two cases.
\paragraph{Hard spheres} These are "billiard balls" of a diameter $d$. In this case  the differential cross-section does not depend on the relative velocity of colliding particles. By simple geometrical considerations we obtain
	\begin{equation}\label{eq:HSkernel}
		\alpha_{12}=\frac{d^2}{4},\ K=\frac{d^2}{4}|\cc_2-\cc_1|.
	\end{equation}
\paragraph{Inverse-power law potentials} For any interaction potential $U(r)$, the dependence of the deflection angle on the impact parameter and the relative velocity is given in terms of a quadrature. We do not write it here.
	However, if the potential is inverse proportional to power of distance, $U(r)\sim r^{-n}$ we can infer about the dependence of the differential scattering cross-sections on the velocity even without integrating Newton's equations explicitly.
	It is sufficient to make use of similarity:  if the potential of a mechanical system is a homogeneous function of order $k$,
	that is $U(\alpha r)=\alpha^k U(r)$, then the velocity and the geometric parameters of geometrically similar trajectories are related by
	\begin{equation}
		\left(\frac{v}{V}\right)=\left(\frac{l}{L}\right)^{k/2},
	\end{equation}
	where $v,V$ and $l,L$ are the velocity and characteristic length on two geometrically similar trajectories. This follows immediately from the invariance of Newton's equations under the transformation of space and time $r\to\alpha r$, $t\to \alpha^{(2-k)/2}t$. 
	In our case $k=-n$, thus we have the following relation between the impact parameter and the relative velocity: $b\sim |\cc_2-\cc_1|^{-2/n}$. This implies for the Boltzmann collision kernel
	\begin{equation}\label{MMkernel}
		K=a(\theta)|\cc_2-\cc_1|^{1-4/n},
	\end{equation}
	where function $a$ depends only on the deflection angle.
	We see that in the special case $n=4$ (that is, for the molecular potential inverse proportional to the fourth power of the separation distance in three dimensions), the collision kernel depends only on the deflection angle but not on the relative velocity of particles. This is known as \emph{Maxwell's molecules}, and it was first shown by Maxwell that computations in that case greatly simplify.
	Finally, we mention that function $a(\theta)$ diverges as $\theta\to 0$ because of a large number of particles experiencing \emph{grazing} collisions with almost no deflection. In practice, one often uses a cutoff angle to regularize this divergence. This works for most inverse-power law potentials but not for the Coulomb potential $n=-1$ where grazing collisions have to be treated in a different manner.
	
Summarizing, we can trace the assumptions behind the derivation of the Boltzmann equation by looking once again at the structure of the Boltzmann collision integral (\ref{eq:collisionintegral}):
	\begin{enumerate}
		\item The gas is assumed to be dilute so that only binary collisions of molecules are considered. This is reflected by the quadratic nonlinearity of the collision integral.
		\item Binary collisions are assumed to be local in time and space. That is, the duration of the collision and the effective range of interaction where the trajectories of the particles change appreciably are supposed to be much smaller than any other characteristic time/distance scales of the system. This is reflected by the fact that $\xx$ and $t$ are simply parameters in the collision integral, there is neither space or time derivative in there.
		\item Collisions are supposed to be elastic, which is reflected by the specific post-collision velocities $\cc_1'$ and $\cc_2'$; they are the implication of the momentum and energy conservation in the elastic collision.
		\item Micro-reversibility of the Newton's equation of mechanical interaction is reflected by the structure of the collision kernel; we have explicitly used it to count the pairs of colliding particles by tracing backward in time the scattering trajectories.
		\item The molecular chaos hypothesis is reflected by the specific structure of the gain and loss parts, quadratic in the distribution function.
	\end{enumerate}
\paragraph{Notation convention}
It is customary to use some abbreviation in writing the Boltzmann transport equation. First, we change the notation for the velocity,  $\cc_1\to \cc$ and $\cc_2\to \cc_1$. Second, one usually drops the space and time arguments, and uses the abbreviations
\begin{equation}
	f(\cc,\xx,t)=f,\ f(\cc_1,\xx,t)=f_1,\ f(\cc',\xx,t)=f',\ f(\cc_1',\xx,t)=f_1'.
\end{equation}
With this, the Boltzmann collision integral (\ref{eq:collisionintegral}) is written as,
\begin{equation}
	\label{eq:Bcollisionintegral}
	\mathcal{J}_B=\int_{R^3} \int_{S^2}\left[f'f_1'-ff_1\right]Bd^2\ee' d^3\cc_1,
\end{equation}
while the Boltzmann transport equation in the absence of long-range forces becomes,
\begin{equation}\label{eq:Boltzmann}
	\partial_t f +\cc\cdot\bm{\nabla} f 	= \mathcal{J}_B,
\end{equation}
where $\partial_t=\partial/\partial_t$ and $\bnabla=\partial/\partial \xx$ are shorthand notation for the derivatives in time and space.
\subsection{Transport equations for molecular properties}
\subsubsection{Boltzmann's transport theorem}
A molecular (or microscopic) property can be thought as a generic function of the velocity $\varphi(\cc)$, such as, for example $\varphi=m$ (particle's mass), $\varphi=m\cc$ (particle's momentum), $\varphi=mv^2/2$ (kinetic energy of the particle) and so fort. 
With the help of the distribution function, we define the corresponding \emph{macroscopic densities} (or \emph{fields}) 
by averaging over the entire range of the microscopic velocities according to their distribution function,
\begin{equation}
	\rho^{\varphi}(\xx,t)=\int_{R^3}\varphi(\cc)f(\cc,\xx,t)d^3\cc.
\end{equation}
We can also define the microscopic flux of any microscopic property by multiplying it with $\cc$ so that the \emph{macroscopic flux} $\jj^{\varphi}$ is,
\begin{equation}
	\jj^{\varphi}(\xx,t)=\int_{R^3}\cc\varphi(\cc)f(\cc,\xx,t)d^3\cc.
\end{equation}
When the distribution function evolves in time and space according to the Boltzmann equation, also all the macroscopic densities and fluxes evolve due to $f$.
The transport equation for any macroscopic density is obtained by multiplying the Boltzmann equation with the corresponding microscopic property and integrating over the velocities,
\begin{equation}
	\partial_t \rho^\varphi + \bnabla\cdot\jj^\varphi = 	R^\varphi.
\end{equation}
The left hand side of this equation contains the divergence of the flux $\jj_\varphi$, while the right hand side is the \emph{production rate} due to collisions,
\begin{equation}
	R^\varphi
	=\int_{R^3}\int_{R^3} \int_{S^2}\varphi\left[f'f_1'-ff_1\right]K d^2\ee' d^3\cc_1 d^3\cc.\label{eq:Brate}
\end{equation}
This expression can be written in a more symmetric and telling form by using the symmetries of the with respect to renaming the particles, the time reversal and the combination thereof,
\[ d^2\ee' d^3\cc_1 d^3\cc =  d^2\ee' d^3\cc d^3\cc_1 =   d^2\ee d^3\cc_1' d^3\cc' =   d^2\ee d^3\cc' d^3\cc_1',\]
so that
\begin{align}
R^\varphi  &=	\int_{R^3}\int_{R^3} \int_{S^2}\varphi\left[f'f_1'-ff_1\right]K d^2\ee' d^3\cc_1 d^3\cc\nonumber\\
&= 	\int_{R^3}\int_{R^3} \int_{S^2}\varphi_1\left[f'f_1'-ff_1\right]K d^2\ee' d^3\cc_1 d^3\cc,\nonumber\\
&=  -	\int_{R^3}\int_{R^3} \int_{S^2}\varphi'\left[f'f_1'-ff_1\right]K d^2\ee' d^3\cc_1 d^3\cc,\nonumber\\
&=  -	\int_{R^3}\int_{R^3} \int_{S^2}\varphi_1'\left[f'f_1'-ff_1\right]K d^2\ee' d^3\cc_1 d^3\cc.
\end{align}
Thus, we arrive at Boltzmann's \emph{transport theorem}: The production rate of any molecular property is written in the following symmetrized form,
\begin{equation}
	R^\varphi
	=\frac{1}{4}	\int_{R^3}\int_{R^3} \int_{S^2}(\varphi+\varphi_1-\varphi'-\varphi_1')\left[f'f_1'-ff_1\right]K d^2\ee' d^3\cc_1 d^3\cc.
\end{equation}

\subsubsection{Invariants of collision. Conservation of mass, momentum and energy}

Immediate implication of the Boltzmann's transport theorem is vanishing of the rate whenever the molecular property satisfies the condition,
\begin{equation}\label{eq:collisioninvariant}
	\varphi+\varphi_1-\varphi'-\varphi_1'=0.
\end{equation}
Because each binary collision conserves the number of particles, their momentum (\ref{eq:binarymomentum}) and kinetic energy (\ref{eq:binaryenergy}), 
the most general solution to the \emph{collision invariants} condition is a linear combination of five scalar quantities,
\begin{equation}
	\label{eq:Lininvariant}
	\varphi\in {\rm Lin}\left\{1,\cc,v^2\right\}.
\end{equation}
The conventional basis of the linear subspace (\ref{eq:Lininvariant}) is built by five  molecular properties $m$, $m\cc$ and $mv^2/2$, corresponding to macroscopic densities $\rho$, $\rho\uu$ and $\rho E$, respectively, interpreted as the mass density, the momentum flux and the energy density,
\begin{align}
	\rho(\xx,t) &=m\int_{R^3} f(\cc,\xx,t)d^3\cc, \label{eq:Bdensity}\\
	\rho \uu (\xx,t) &=m\int_{R^3}\cc f(\cc,\xx,t)d^3\cc,\label{eq:Bmomentum}\\
	\rho E(\xx,t) &=(m/2)\int_{R^3} v^2 f(\cc,\xx,t)d^3\cc.\label{eq:Benergy}
\end{align}
%
Therefore, the corresponding terms $R_\varphi$ vanish, and the transport equations for these \emph{locally conserved} fields become conventional statements of the local conservation.
Under suitable boundary conditions (vanishing at infinity, periodic etc), this leads, after integration over the volume of the container, to the \emph{global} mass, momentum and energy conservation.




\subsubsection{Boltzmann's $H$-theorem}
Yet another implication of Boltzmann's transport theorem is manifest when considering the following molecular property,
\begin{equation}
	\varphi=\ln f(\cc,\xx,t)+1.
\end{equation}
The corresponding production rate is,
\begin{align}
	\nonumber \sigma_{B}
	&=\frac{1}{4}\int_{R^3}\int_{R^3} \int_{S^2}(\ln f+\ln f_1-\ln f'-\ln f_1')\left(f'f_1'-ff_1\right)K d^2\ee' d^3\cc_1 d^3\cc\\
	&=\frac{1}{4}\int_{R^3}\int_{R^3} \int_{S^2}\ln \left(\frac{ff_1}{f'f_1'}\right)\left(f'f_1'-ff_1\right)K d^2\ee' d^3\cc_1 d^3\cc.
	\label{eq:Hrate}
\end{align}
Now we denote $X=ff_1$, $Y=f'f_1'$, and notice that the function under integration, $F=\ln\frac{X}{Y}(Y-X)\leq 0$ for any $X>0$, $Y>0$.
Thus, for any $f$, the rate $\sigma$ is non-positive,
\begin{equation}\label{eq:Hratenegative}
	\sigma_B(\xx,t)\leq 0.
\end{equation}
%
Consider the following field, the \emph{Boltzmann $H$-function},
\begin{equation}
	H(\xx,t)=\int_{R^3} f(\cc,\xx,t)\ln f(\cc,\xx,t) d^3\cc.\label{eq:HBoltzmann}
\end{equation}
The \emph{total} $H$-function $\bar{H}$ is obtained after integration of the density (\ref{eq:HBoltzmann}) over the volume $V$. Assuming suitable boundary conditions under which the  flux of the $H$-function vanishes at the boundaries, we find, based on the non-positivity of the $H$-function production rate (\ref{eq:Hrate}),
\begin{equation}
	\frac{d\bar{H}}{dt}=
	\int_{V}\sigma_{B}(\xx,t)d^3\xx\leq 0.
\end{equation}
This is the \emph{$H$-theorem} proven by Boltzmann: $H$-function never increases due to the Boltzmann equation.
This is the spectacular implication of the Boltzmann equation, as it describes irreversible behavior, and the $H$-function  has to be related to the entropy of the system according to second law of thermodynamics (the non-decrease of the entropy with the time).
Identification of the $H$-function with the entropy will be completed below.

\subsubsection{Equilibrium and local equilibrium. The Maxwell-Boltzmann distribution function}
The structure of the Boltzmann collision integral (\ref{eq:collisionintegral}) implies that its zero-point, $\mathcal{J}_B(f)=0$, is achieved on the family of  \emph{local equilibrium} distribution functions $f^{\rm eq}(\cc,\xx,t)$, satisfying the \emph{detail balance condition},
\begin{equation}
	f^{{\rm eq}'}f^{{\rm eq}'}_1=f^{\rm eq}f^{\rm eq}_1,
\end{equation}
or, by taking the logarithm of this expression,
\begin{equation}
	\ln f^{{\rm eq}'}+\ln f^{{\rm eq}'}_1-\ln f^{\rm eq}-\ln f^{\rm eq}_1=0.
\end{equation}
Comparison to Eq.\ (\ref{eq:collisioninvariant}) shows that the logarithm of the local equilibrium distribution function is a linear combination of collision invariants, 
\begin{equation}
	\ln f^{\rm eq}\in\{1,\cc,v^2\}.
\end{equation}
Thus, the generic local equilibrium can be written as a five-parametric subset, 
\begin{equation}\label{eq:MBgen}
	f^{\rm eq}=A\exp\left\{-\frac{(\cc-\blambda)^2}{2\sigma}\right\}.
\end{equation}
We now turn to the definition of the density \eqref{eq:Bdensity}, the momentum \eqref{eq:Bmomentum} and the energy \eqref{eq:Benergy}. For the latter, we  \emph{assume} a relation to the absolute temperature by the caloric equation of state of monatomic ideal gas, 
\begin{align}
	\rho E=\frac{3}{2}\rho RT+\frac{1}{2}\rho u^2,\label{eq:energymonatomic}
\end{align}
where $R$ is gas constant, 
\begin{equation}
	\label{Rconst}
	R=\frac{R_U}{\mu},
\end{equation}
with $R_U$ the universal gas constant and $\mu$ the molar mass.
Using \eqref{eq:MBgen} in \eqref{eq:Bdensity}, \eqref{eq:Bmomentum} and \eqref{eq:Benergy}, the parameters $\{A,\blambda,\sigma\}$ are expressed in terms of the conventional parameters $\{\rho,\uu,T\}$ upon evaluation of the Gauss integrals: $A=({\rho}/{m})\left({2\pi RT}\right)^{-3/2}$, $\blambda=\uu$ and $\sigma=RT$. 
With this re-parameterization, the local equilibrium distribution function reads, 
\begin{equation}\label{eq:BlocalMaxwell}
	f^{\rm eq}=n\left({2\pi RT}\right)^{-3/2}\exp\left\{-\frac{(\cc-\uu)^2}{2RT}\right\},
\end{equation}
where
$n = {\rho}/{m}$
is the number density.
This is the local \emph{Maxwell-Boltzmann distribution function}. Specification "local" means that density, velocity and temperature in this expressions can be arbitrary functions of space and time. The local equilibrium annuls only the right hand side of the Boltzmann equation but not the left hand side, so it is not a solution of the Boltzmann equation. It, however becomes the solution when the density, flow velocity and temperature are constant. Then we speak of the global equilibrium, or just \emph{the equilibrium}.

The Boltzmann $H$-function is easily evaluated \emph{at} the local (or global) equilibrium:
\begin{equation}\label{eq:Heq}
	H^{\rm eq}=\int f^{\rm eq} \ln f^{\rm eq}d^3\cc =-n\left(\frac{3}{2}\ln T-\ln n\right)-\frac{3}{2}n\left(\ln (2\pi R)+1\right).
\end{equation}
Note that this expression does not depend on flow velocity as the consequence of Galilean invariance.
Since the \emph{thermodynamic entropy} concerns the equilibrium states, we consider the case of global equilibrium of $N$ particles, and evaluate the total $H$-function difference between the two states. Integrating \eqref{eq:Heq} over the volume, we find
\begin{equation}\label{eq:DeltaHeq}
		\Delta\bar{H}^{\rm eq}=-N_A\nu \left[\frac{3}{2}\ln\left(\frac{T_2}{T_1}\right)+\ln\left(\frac{V_2}{V_1}\right)\right],
\end{equation}
where $N_A$ and $\nu$ are the Avogadro number and the number of moles of gas, respectively, $N=N_A\nu$.
On the other hand, the \emph{thermodynamic entropy difference} between the two states  of $\nu$ moles of ideal gas with the specific heat at constant volume $c_v=(3/2)R_U$ is,
\begin{equation}\label{eq:DeltaSeq}
	\Delta S=\nu \left[\frac{3}{2}R_U\ln\left(\frac{T_2}{T_1}\right)+R_U\ln\left(\frac{V_2}{V_1}\right)\right].
\end{equation}
Comparing \eqref{eq:DeltaHeq} and \eqref{eq:DeltaSeq}, we see that the thermodynamic entropy difference and the $H$-function difference are related by a dimensional constant and the sign convention, 
\begin{equation}
	\Delta S=-k_B \Delta \bar{H}^{\rm eq},
\end{equation}
where $k_B$ is the \emph{Boltzmann constant}, or "the universal gas constant per particle",
\begin{equation}
	\label{eq:kB}
	k_B=\frac{R_U}{N_A}.
\end{equation}
Note that the matching of the specific heat through the thermodynamic entropy relation proves the assumption already made when choosing the parameterization of the energy \eqref{eq:energymonatomic}. While identification of the thermodynamic entropy can be done only at the equilibrium, the function 
$S(f)=-k_B H(f)$ can be considered as the non-equilibrium entropy density whenever the distribution function $f$ differs from the global equilibrium. The Boltzmann $H$-theorem about the non-positive $H$-function production and non-increase of $H$ can be considered as the specific realization of the thermodynamic concept of the entropy increase due to irreversible processes.

\section{From Boltzmann equation to hydrodynamics\label{sec:frm_bolt_to_hydro}}

\subsection{Fields and fluxes}
In this section, we use Cartesian coordinate notation, $\partial_{\alpha}=\partial/\partial x_{\alpha}$, where $\alpha$ labels the three coordinate directions.
We address in this section the thirteen fields of interest: the density $\rho$, momentum $\rho \bm{u}$, full pressure tensor $\bm{P}$ and energy flux $\bm{Z}$ are defined as follows,
\begin{align}
	\rho&=m\int fd\bm{v},\label{eq:rho}\\
	\rho u_{\alpha}&=m\int v_{\alpha}fd\bm{v},\label{eq:u}\\
	P_{\alpha\beta}&=m\int v_{\alpha}v_{\beta}fd\bm{v},\label{eq:P}\\
	Z_{\alpha}&=\frac{m}{2}\int v_{\alpha}{v^2}fd\bm{v}.\label{eq:S}
\end{align}
Identifying pressure $p$ as
\begin{align}
	p&=\frac{1}{3}\int m(\bm{v}-\bm{u})^2fd\bm{v},\label{eq:pressure_el_kin}
\end{align}
the pressure tensor (\ref{eq:P}) and the energy flux (\ref{eq:S}) are decomposed as follows when measuring particle's velocity relative to the flow velocity $\bm{u}$,

\begin{align}
	P_{\alpha\beta}&=p\delta_{\alpha\beta}+\sigma_{\alpha\beta}+\rho u_{\alpha}u_{\beta},\label{eq:Pexpanded}\\
	Z_{\alpha}&=\left(\frac{5}{2}p+\frac{1}{2}\rho u^2 \right)u_{\alpha}+\sigma_{\alpha\beta}u_{\beta}+q_{\alpha},
\end{align}
where nonequilibrium stress tensor $\bm{\sigma}$ and heat flux $\bm{q}$ are defined as
\begin{align}
	\sigma_{\alpha\beta}&=
	m\int\left[({v}_{\alpha}-{u}_{\alpha})({v}_{\beta}-{u}_{\beta})-\frac{1}{3}\delta_{\alpha\beta}(\bm{v}-\bm{u})^2\right]
	fd\bm{v}, \label{eq:sigma}\\
	{q}_{\alpha}&=\frac{m}{2}\int ({v}_{\alpha}-{u}_{\alpha})(\bm{v}-\bm{u})^2fd\bm{v}.\label{eq:heat}
\end{align}
Rank two symmetric tensor $\bm{\sigma}$ is trace-free. Finally, temperature $T$ is defined by the equation of state of ideal gas,
\begin{align}
	p&=\rho R T.\label{eq:EoS}
\end{align}
%
%
%
Density $\rho$, flow velocity $\bm{u}$, temperature $T$, nonequilibrium stress $\bm{\sigma}$ and heat flux $\bm{q}$ are the \emph{thirteen fields of Grad}.

\subsection{Balance equations for thirteen moments}

Introducing the material derivative along streamline,
\begin{equation}\label{eq:Dtcoord}
	D_t=\partial_t+{u}_{\alpha}\partial_{\alpha},
\end{equation}
we write kinetic equation in the co-moving reference frame,
\begin{equation}\label{eq:BE}
	D_t f=-({v}_{\alpha}-{u}_{\alpha})\partial_{\alpha} f+\mathcal{J}_B.
\end{equation}
Here and below, summation convention is always understood. 
Multiplying (\ref{eq:BE}) with $1$, $v_{\alpha}$, $v_{\alpha}v_{\beta}$ and $v_{\alpha}v^2$ and integrating over velocities, we come to the set of exact balance equations for the thirteen Grad's fields.
Balance equations for the locally conserved fields (mass-momentum-energy) imply
\begin{align}
	D_t \rho&=-\rho(\partial_{\alpha}u_{\alpha}), \label{eq:Dtn}\\
	D_t u_{\alpha}&=-\frac{1}{\rho}\partial_{\alpha}p-\frac{1}{\rho}\partial_{\beta}{\sigma}_{\alpha\beta}, \label{eq:Dtu}\\
	D_t T&=-\frac{2}{3}T(\partial_{\alpha}u_{\alpha})-\frac{2}{3}\left(\frac{T}{p}\right)\sigma_{\alpha\beta}(\partial_{\beta}u_{\alpha})
	-\frac{2}{3}\left(\frac{T}{p}\right)(\partial_{\alpha}q_{\alpha}). \label{eq:DtT}
\end{align}
%
%
The nonequilibrium stress and heat flux are the only nonequilibrium fluxes engaged  in the balance equations (\ref{eq:Dtu}) and (\ref{eq:DtT}), and 
we continue with writing down the exact balance equations for $\bm{\sigma}$ and $\bm{q}$:


\begin{align}
	D_t\sigma_{\alpha\beta}=&-p\left(\partial_{\alpha}u_{\beta}+\partial_{\beta}u_{\alpha}-\frac{2}{3}\delta_{\alpha\beta}\partial_{\gamma}u_{\gamma}\right) -\left(\sigma_{\alpha\gamma}\partial_{\gamma}u_{\beta}+\sigma_{\beta\gamma}\partial_{\gamma}u_{\alpha}
	-\frac{2}{3}\delta_{\alpha\beta}\sigma_{\mu\nu}\partial_{\nu}u_{\mu}\right)\nonumber\\
	&-\sigma_{\alpha\beta}\left(\partial_{\gamma}u_{\gamma}\right)
	-2\partial_{\gamma}\left(Q_{\alpha\beta\gamma}-\frac{1}{3}\delta_{\alpha\beta}q_{\gamma}\right)+R^{\sigma}_{\alpha\beta},
	\label{eq:balances}\\
	\label{eq:balanceq}
	D_t q_{\alpha} &=
	-q_{\alpha}\partial_{\beta}u_{\beta}-q_{\beta}\partial_{\beta}u_{\alpha}
	+\frac{5}{2}RT\partial_{\alpha}p+\frac{5}{2}RT\partial_{\beta}\sigma_{\alpha\beta}
	+\frac{1}{\rho}\sigma_{\alpha\beta}\partial_{\beta}p + \frac{1}{\rho}\sigma_{\alpha\beta}\partial_{\gamma}\sigma_{\gamma\beta}\nonumber\\
	&-\partial_{\beta}T_{\alpha\beta}-2Q_{\alpha\beta\gamma}\partial_{\gamma}u_{\beta}+R^q_{\alpha}.
	%
\end{align}
Here, symmetric rank three tensor $\bm{Q}$ and symmetric rank two tensor $\bm{T}$ are,
\begin{align}
	Q_{\alpha\beta\gamma}&=
	\frac{m}{2}\int ({v}_{\alpha}-{u}_{\alpha})({v}_{\beta}-{u}_{\beta})(v_{\gamma}-u_{\gamma})fd\bm{v},\label{eq:Q}\\
	T_{\alpha\beta}&=\frac{m}{2}\int  (v_{\alpha}-u_{\alpha})(v_{\beta}-u_{\beta})(\vv-\uu)^2fd\bm{v}. \label{eq:T}
\end{align}
For brevity, we refer to them as the $Q$-flux and the $T$-flux, respectively. The $Q$-flux is engaged as a divergence in the stress balance (\ref{eq:balances}) and as a source term in the heat flux balance (\ref{eq:balanceq}) while the $T$-flux contributes as a divergence in the heat flux balance only. The heat flux and the $Q$-flux are connected through,
\begin{align}
	q_{\alpha}&=Q_{\alpha\beta\beta}.
\end{align}
Furthermore, the relaxation terms are defined as rates over collisions,
\begin{align}
	R^{q}_{\alpha}&=\frac{m}{2}\int ({v}_{\alpha}-{u}_{\alpha})(\bm{v}-\bm{u})^2\mathcal{J}_Bd\bm{v},\label{eq:Rq}\\
	R^{\sigma}_{\alpha\beta}&=m\int \left[({v}_{\alpha}-{u}_{\alpha})({v}_{\beta}-{u}_{\beta})-\frac{1}{3}\delta_{\alpha\beta}(\bm{v}-\bm{u})^2\right]\mathcal{J}_Bd\bm{v}.\label{eq:Rs}
\end{align}
Balance equations are identities and cannot be addressed unless a constitutive relation is provided for both the $Q$- and $T$-fluxes, as well as for the collision rates.
%





\subsection{Local equilibrium projection}

The balance equations for the locally conserved fields density  \eqref{eq:Dtn}, flow velocity \eqref{eq:Dtu} and temperature \eqref{eq:DtT} become the equations of hydrodynamics once \emph{constitutive relations} are supplied for the nonequilibrium pressure tensor \eqref{eq:sigma} and heat flux \eqref{eq:heat}. 
Evaluation of these fluxes on the local equilibrium \eqref{eq:BlocalMaxwell} returns,
\begin{equation}
	\label{eq:Efluxes}
	\sigma_{\alpha\beta}[f^{\rm eq}]=0,\ q_{\alpha}[f^{\rm eq}]=0.
\end{equation}
This implies the \emph{Euler's equations} of non-viscous, thermally non-conductive fluid,
\begin{align}
	D_t^E \rho&=-\rho(\partial_{\alpha}u_{\alpha}), \label{eq:EDtn}\\
	D_t^E u_{\alpha}&=-\frac{1}{\rho}\partial_{\alpha}p, \label{eq:EDtu}\\
	D_t^E T&=-\frac{2}{3}T(\partial_{\alpha}u_{\alpha}). \label{eq:EDtT}
\end{align}
The local equilibrium \emph{projection} thus amounts to a neglect of any equilibration process, so that the distribution function never leaves the submanifold of local Maxwellians  while the parameters of the local Maxwellian evolve according to Euler's equations. This can be viewed as an analogy to the ideal quasi-equilibrium reversible processes in thermodynamics. Addressing the equilibration processes towards the local equilibrium states is thus needed to obtain a more realistic picture.

\subsection{Dynamic correction to the local equilibrium projection}

\subsubsection{Estimates from the Maxwell-Boltzmann distribution}
We shall first do a few "back-of-the-envelope" estimates concerning the Boltzmann equation and transport phenomena.
We shall often use $\sim$ instead of $=$ in these estimates, meaning that a more accurate computation shall bring numerical factors of order one but the estimates below will be valid "on the order of magnitude".

The dimension of the distribution function $f$ is
\begin{equation}
	f\sim \frac{1}{{\rm velocity}^3}\times \frac{1}{{\rm volume}} \sim  \frac{1}{(cm/sec)^3}\times \frac{1}{cm^3},
\end{equation}
which is consistent with the fact that
$
dN=f(\cc,\xx, t)d^3\cc d^3\xx
$
is the number of particles in the element of the phase-space volume $d^3\cc d^3\xx\sim (cm/sec)^3\times cm^3$.

We now need some characteristic values for the velocity and the volume to make things non-dimensional.
Let us look at the Maxwell-Boltzmann distribution function \eqref{eq:BlocalMaxwell}.
The factor in front the exponential is $\sim n v_T^{-3}$, where 
\begin{equation}\label{eq:Tspeed}
	v_T=\sqrt{RT},
\end{equation}
has the dimension of the velocity, $v_T\sim (cm/sec)$. We shall call it the \emph{thermal speed}.
Other characteristic velocities may be defined such as 
 the r.m.s. of the velocity fluctuations about the mean velocity $\uu$ is $v_{r.m.s}=\sqrt{3RT}$ or the speed of sound $c_s=\sqrt{5RT/3}$, or the mean absolute value of the velocity fluctuation $v_m=\sqrt{8RT/\pi}$. They all differ by a factor of order one from the thermal speed \eqref{eq:Tspeed}.  
Introducing the reduced \emph{peculiar velocity} (this term is universally used, and means simply the deviation of the molecular velocity from the flow velocity $\uu$),
\begin{equation}\label{eq:Cred}
	\CC=\frac{\cc-\uu}{v_T},
	\end{equation}
the Maxwell-Boltzmann distribution becomes,
\begin{align}
	\mathcal{M}&=\frac{n}{(2\pi)^{3/2}v_T^{3}}e^{-\frac{C^2}{2}}. \label{eq:maxwell}
\end{align}
We shall use $v_T$ to reduce all molecular velocities in the sequel.
Of course, the temperature and the density (and the flow velocity) in the local Maxwell-Boltzmann distribution may vary in space and time, so we are talking about some characteristic number-density and temperature when using them for non-dimensionalization.

\subsubsection{Mean free path}
Mean free path is defined as the average distance traveled by a particel before it comes to a collision with some other particle.
For the hard-sphere model of collision, the number density $n$ and the diameter of the hart-spehere $d$ can be combined to give the quantity with the dimension of the distance,
\begin{equation}
	l_{m.f.p.}\sim \frac{1}{d^2 n}.\label{eq:mfp}
\end{equation}
The estimate of the \emph{mean free path} \eqref{eq:mfp} is intuitively clear, as we expect $l_{m.f.p.}$ to decrease  with the increase of the scattering cross-section $\sim d^2$ and with the increase of the density of scattering centers $n$. More accurate definitions of $l_{m.f.p.}$ are available, in particular, upon the exact estimate of the number of binary collisions from the Bolztmann collision integral \cite{chapman_mathematical_1939} but the above estimate is sufficient for our purpose here.

\subsubsection{Estimates from the Boltzmann collision integral}
We now reduce the Boltzmann collision integral for hard-spheres \eqref{eq:Bcollisionintegral}.
We first make the distribution function dimensionless by reducing it with some typical thermal speed and number density,
	$\bar{f}=v_T^3n^{-1}f$.
Furthermore, we introduce the dimensionless scattering kernel,
	$d^2 v^T\bar{B}=B$,
so that $\bar{B}$ depends only on the reduced relative velocity, $|\bar{\cc_1}-\bar{\cc}|=v_T^{-1}|{\cc_1}-{\cc}|$. Finally, reducing the element $d^3\cc_1=v_T^3d^3\bar{\cc}$, we obtain
\begin{equation}
	\mathcal{J}_B=(d^2 n) v_T (nv_T^{-3}) \bar{\mathcal{J}}_B.
\end{equation}
where $\bar{\mathcal{J}_B}$ is dimension-less,
\begin{equation}
	\label{eq:Bcollisionreduced}
	\bar{\mathcal{J}_B}=\int_{R^3} \int_{S^2}\left[\bar{f}'\bar{f}_1'-\bar{f}\bar{f}_1\right]\bar{K}d^2\ee' d^3\bar{\cc}_1.
\end{equation}
Using the definition of the mean free path \eqref{eq:mfp}, this is also
\begin{equation}\label{eq:Bcollisionscaled}
	{\mathcal{J}_B}=\frac{v_T}{l_{m.f.p.}} (nv_T^{-3}) \bar{\mathcal{J}_B}.
\end{equation}
\subsubsection{Reduced Boltzmann equation}
Now we turn to the left hand side of the Boltzmann equation and introduce the reduced time and space, $t=T\bar{t}$, $\xx=L\bar{\xx}$, with some characteristic macroscopic length $L$ and time $T$. Using the scaled collision integral \eqref{eq:Bcollisionscaled}, and also using the reduced flow velocity, 
${\uu}=(L/T)\bar{\uu}$, the Boltzmann equation is written in the reduced variables,
\begin{equation}
	\left(\frac{L}{Tv_T}\right)\bar{D}_t \bar{f} +\left(\bar{v}_{\alpha}-	\left(\frac{L}{Tv_T}\right)\bar{u}_{\alpha}\right)\bar{\partial}_{\alpha}\bar{f}
	=\left(\frac{L }{ l_{m.f.p.}} \right)\bar{\mathcal{J}}_B.
\end{equation}
The dimensionless quantity, inverse of which multiplies the collision integral is called the \emph{Knudsen number},
\begin{equation}\label{eq:Bknudsen}
	[{\rm Kn}]=\frac{ l_{m.f.p.} }{L}.
\end{equation}
On the other hand, the dimensionless quantity which multiplies the time derivative is called the kinetic \emph{Strouhal number},
\begin{equation}\label{eq:Bstrouhal}
	[{\rm St}]=\frac{L}{Tv_T}.
\end{equation}
With this, the reduced Boltzmann equation is written,
\begin{equation}\label{eq:Bscaled}
		[{\rm St}]\bar{D}_t \bar{f} +\left(\bar{v}_{\alpha}-	[{\rm St}]\bar{u}_{\alpha}\right)\bar{\partial}_{\alpha}\bar{f}
=\frac{1}{[{\rm Kn}]}\bar{\mathcal{J}}_B.
\end{equation}

\subsubsection{Hydrodynamic limit}
The hydrodynamic limit of the Boltzmann equation corresponds to a flows featuring a \emph{small Knudsen number}, when the mean free path is small compared to a typical time-space variation of hydrodynamic fields (local density, flow velocity and temperature).
A typical estimate of Knudsen number for our "daily life" flows is of the order of $[{\rm Kn}]\sim 10^{-5}-10^{-6}$.  In the classical case considered below, the Strouhal number is considered to be of the order one. A related quantity is the Mach number, 
\begin{equation}
	\label{eq:Bmach}
	[{\rm Ma}]=\frac{L}{Tc_s},
\end{equation}
where $c_s=\sqrt{\gamma RT}$, $\gamma=5/3$ is the adiabatic exponent of Boltzmann's gas  (monatomic particles without internal degrees of freedom). Since the speed of sound is of same order as thermal speed, we assume $[{\rm Ma}]\sim [{\rm SM}]$. In summary, the standard hydrodynamic limit assumes the following scaling,
\begin{equation}\label{eq:BKnscaling}
	[{\rm Kn}]\ll 1,\ [{\rm Ma}]\sim[{\rm St}] \sim O(1).
\end{equation}
In that case, the hydrodynamic limit of the Boltzmann equation leads to a compressible flow. 
On the other hand, different other scaling can be also considered. In particular, if the Strouhal number, the Knudsen number but also the Mach number are small and of same order,
\begin{equation}
	[{\rm Kn}]\sim [{\rm St}] \sim [{\rm Ma}] \ll 1,
\end{equation}
then the hydrodynamic limit of the Boltzmann equation corresponds to a nearly-incompressible flow.
Notice, however that this latter scaling requires a really \emph{slow flow}; in a typical situation, $St\sim Ma \sim 10^{-2}-10^{-1}$ are small, but still they are \emph{much} larger than a typical Knudsen number by a few orders of magnitude). 
Before proceeding with the analysis of the hydrodynamic limit in a systematic fashion, we shall consider quantitatively  on the example of viscous transport how the transport phenomena on a conventional macroscopic scale arises from kinetic considerations.

\subsubsection{Elementary derivation of viscosity}
Using the above estimates for the thermal speed and the mean free path, we shall first derive the phenomenon of dissipative transport focusing on the viscosity. Since the mean free path is what is related to the collisions and bearing in mind that all the dissipation comes from the Boltzmann collision integral, we may guess that the viscosity is related to it. Let us therefore
consider a gas flow parallel to the plane $z=0$; the flow velocity is supposed to depend only on the vertical coordinate $z$, thus we have $u_x(z)$, $u_y=u_z=0$. The latter particularly means that there is no net flow in the vertical direction.
The vertical transport of the momentum is therefore effected by the $z$-component of the particle's velocity, and since there is no net velocity in that direction, it can be estimated as being of the order of root-mean-square of the $z$-component of particle's velocity $v_z$, that is, proportional to the thermal speed $v_T$.

In order to estimate the net transport of the $x$-component of momentum through a plane parallel to $(x,y)$-plane (say, the plane $z=0$),
we need to consider two \emph{layers} of gas distanced by the order of mean free path $l_{m.f.p.}$ on both sides of this plane.
Considering the $z=0$ plane, the transferred $x$-momentum at the location $z=-l_{m.f.p.}$ in the positive $z$-direction (towards the $z=0$ plane) per unit of time is thus estimated as
\begin{equation}
	P_{xz}^+\sim  v_T \times\rho u_x(z)|_{z=-l_{m.f.p.}}.
\end{equation}
Note that we could write here a prefactor $1/2$ since it is the half of the of the particles on average which have the positive component of $z$-velocity. We shall omit all such factors in the present rough estimate.

Similarly, the $x$-momentum transported in the \emph{negative} direction from the layer above the $z=0$ plane towards that plane is
\begin{equation}
	P_{xz}^-\sim  - v_T \times\rho u_x(z)|_{z=l_{m.f.p.}}.
\end{equation}
The net transported momentum per unit of time from both the sides of the plane $z=0$ becomes, \emph{assuming the mean free path is small on the scale of a variation of the net velocity},
\begin{equation}
	P_{xz}=P_{xz}^+ + P_{xz}^-\sim  - (\rho v_T l_{m.f.p.}) \frac{du_x(z)}{dz}\bigg|_{z=0}
\end{equation}
This net transported momentum per unit of time is equivalent to the force (per unit area) exerted parallel to $x$.
 This force has the form of the usual viscous stress, where the coefficient of viscosity is found from our "back-of-the-envelope" estimate as
\begin{equation}\label{eq:viscosityelementary}
	\mu\sim  \rho v_T l_{m.f.p.}.
\end{equation}
Note that the usual \emph{kinematic viscosity} $\nu=\mu/\rho$ is $\nu\sim v_T l_{m.f.p.}\sim cm^2/sec$.
Using the above result for the mean free path, we obtain,
\begin{equation}\label{eq:Bviscosityelementary}
	\mu=  \bar{b}\frac{m\sqrt{R T}}{d^2},
\end{equation}
where the non-dimensional coefficient $\bar{b}$ (\emph{pure number}) is of the order one. A rigorous estimate from the Boltzmann equation gives $\bar{b}\approx0.179$, however the dependence of the viscosity coefficient on the temperature, particle's mass and the diameter of the hard-sphere stays as in the estimate \eqref{eq:Bviscosityelementary} obtained by simple argument. From the above elementary consideration we can already derive a classical result that the coefficient of the viscosity is not dependent on the number density $n$, which is a well-verified experimental fact. It is also interesting to note the \emph{inverse} proportionality to the cross-section  of the hard-sphere in the result \eqref{eq:Bviscosityelementary}, as it may seem counter-intuitive at a first glance. Indeed, it seems plausible that "bigger" spheres should result in a "higher" viscosity, contrary to what \eqref{eq:Bviscosityelementary} suggests. However, the right interpretation is restored when one remembers that the mean free path becomes \emph{shorter} if the cross-section is increased.


\subsubsection{Singularly perturbed Boltzmann equation}
In accord with the scaling \eqref{eq:BKnscaling}, we write the Boltzmann equation in the co-moving frame, introducing a formal \emph{large parameter} $1/\epsilon$ in front of the collision integral, 
\begin{equation}
	{D_t f}= -(c_{\beta}-u_{\beta})\partial_{\beta}f+\frac{1}{\epsilon}\mathcal{J}_B.
\end{equation}
Such a form is called a singularly perturbed system because a small parameter $\epsilon$ multiplies the derivative.
For what will follow, it is useful to remember the balance equations for the number (or mass) density, flow momentum, and energy, already mentioned above. 
At this stage, you may already guess that if the distribution function used to close the balance equations is taken as the local Maxwell-Boltzmann distribution, then the non-equilibrium pressure tensor and heat flux vanish,
and the balance equations become the {Euler compressible equations}
where in the right hand side of the flow velocity equation we see the nonlinear advection term and the gradient of the pressure (ideal gas equation of state), while in the right hand side of the temperature equation there are the advection term and the compression work term. We shall obtain this in a more systematic way below when we shall consider a correction to the local Maxwell-Boltzmann distribution due to collisions.

\subsubsection{Normal solutions. The invariance principle}
Following the notion introduced by D. Hilbert (1913), a
\emph{normal solution} is a distribution function which depends on the space and the time only through its (yet unknown) dependence on the locally conserved fields.
\begin{equation}
	f_{nr}(\cc, \xx,t)=f_{nr}(\cc; \rho(\xx,t), \uu(\xx,t), T(\xx,t)).
\end{equation}
Normal solutions satisfy two important conditions.
\paragraph{Consistency conditions} Whatever the dependence on the local fields $\rho(\xx,t)$, $\uu(\xx,t)$, and $T(\xx,t)$ may be found in $f_{nr}$, it has to satisfy the following,
\begin{align}
&	\int  mf_{nr}(\cc; \rho,\uu,T)d^3\cc =\rho, \label{eq:Bconsistn}\\
&	\int \cc mf_{nr}(\cc; \rho,\uu,T)d^3\cc =\rho\uu, \label{eq:Bconsisu}\\
&	\int \frac{mv^2}{2} f_{nr}(\cc; \rho,\uu,T)d^3\cc = \frac{3}{2}\rho RT+\frac{\rho u^2}{2}.\label{eq:BconsistT}
\end{align}
Consistency condition  means that if we evaluate density, momentum and energy on a given normal solution $f_{nr}$, the result must be the same values, as the corresponding arguments of the normal solution.

\paragraph{Invariance condition} So, if the normal solution depends only on the local conservations, at every point in space and at each instant of time, then so does also its time derivative. That means we have to express the {time derivative} of the distribution function through the time derivative of the locally conserved fields (density, flow velocity and temperature).
So, the normal solution should satisfy the invariance condition: \emph{By chain rule of differentiation}

%
\begin{equation}
	\frac{\partial f_{nr}}{\partial \rho}D_t \rho+\frac{\partial f_{nr}}{\partial u_\alpha}D_t u_\alpha+\frac{\partial f_{nr}}{\partial T}D_t T
	=-(v_{\alpha}-u_{\alpha})\partial_{\alpha}f_{nr}+\frac{1}{\epsilon}\mathcal{J}_B(f_{nr}),
\end{equation}
where the time derivative of the locally conserved fields are obtained with the same $f_{nr}$ (that is, the nonequilibrium pressure tensor and heat flux are evaluated on the same normal solution). 
%
%
%
%

\begin{align}\nonumber
	&	\frac{\partial f_{nr}}{\partial \rho}\left(-\rho\partial_{\alpha} u_{\alpha}\right)
		+\frac{\partial f_{nr}}{\partial u_{\alpha}}\cdot\left(-\frac{1}{\rho}\partial_{\alpha} p -\frac{1}{\rho}\partial_{\beta}{\sigma}_{\alpha\beta}[f_{nr}] \right) 
		+\frac{\partial f_{nr}}{\partial T}\left(-\frac{2}{3}T\partial_{\alpha}u_{\alpha}-\frac{2}{3}\left(\frac{T}{p}\right){\sigma}_{\alpha\beta}[f_{nr}]\partial_{\beta}u_{\alpha}
	-\frac{2}{3}\left(\frac{T}{p}\right)\partial_{\alpha}q_{\alpha}[f_{nr}]\right)\nonumber\\
	&	=-(v_{\alpha}-u_{\alpha})\partial_{\alpha} f_{nr}+\frac{1}{\epsilon}\mathcal{J}_B(f_{nr}),
\end{align}
Note that because the dependence of the time derivatives in the left hand side on the distribution function is {linear}, the overall nonlinearity of the invariance condition in $f$ is {quadratic} - because of the quadratic Boltzmann collision on the right but also because of the {quadratic} dependencies through the nonequilibrium pressure tensor  $\bm{\sigma}[f_{nr}]$ and heat flux $\qq[f_{nr}]$ in the balance equations for flow velocity and temperature on the left.
The advantage of this "increase of complexity" is that the "real" time derivative is excluded in favor of space derivatives coming from the {fluxes}. This will immediately pay off below.


\subsubsection{The Chapman--Enskog method}
Owing to the fact that there is a {large parameter} $1/\epsilon$, D. Enskog (1917) suggested to look for the normal solution in terms of a series for the distribution function. This method of solving the invariance equation of the previous section was in fact discovered by Enskog and was made widely known as the Chapman--Enskog method by S. Chapman.
The normal solution is then expanded into a series,
\begin{equation}
	f_{nr}=f^{(0)}+\epsilon f^{(1)}+O(\epsilon^2).
\end{equation}
Accordingly, the non-equilibrium pressure tensor and heat flux are expanded as
\begin{align}
	\sigma_{\alpha\beta} [f_{nr}] &= \sigma_{\alpha\beta} [f^{(0)}]+\epsilon \sigma_{\alpha\beta} [f^{(1)}]+O(\epsilon^2),\\
	q_{\beta} [f_{nr}] &= q_{\beta} [f^{(0)}]+\epsilon q_{\beta} [f^{(1)}]+O(\epsilon^2).
\end{align}
Substituting these two expansions into the left and right hand sides of the invariance equation, and equating terms of same order in $\varepsilon$, we obtain
on the order $\epsilon^{0}$,
\begin{equation}
	\mathcal{J}_B(f^{(0)})=0,
\end{equation}
and on the order $\epsilon$,
\begin{equation}
	\frac{\partial f^{(0)}}{\partial n}D_t^{(0)}\rho+\frac{\partial f^{(0)}}{\partial u_\alpha}D^{(0)}_t u_\alpha+\frac{\partial f^{(0)}}{\partial T}D^{(0)}_t T +(v_{\alpha}-u_{\alpha})\partial_{\alpha}f^{(0)}=Lf^{(1)}.\label{eq:BCE1}
\end{equation}
Here the material derivatives indicated as $"(0)"$ are understood in the sense that the zeroth-order distribution $f^{(0)}$ is used to evaluate the non-equilibrium pressure tensor and the heat flux,
\begin{align}
	D^{(0)}_t \rho&=- \rho\partial_{\alpha}u_\alpha,\\
	D^{(0)}_t u_\alpha &=- \frac{1}{\rho}\partial_\alpha (\rho RT)
	-\frac{1}{\rho}\partial_\beta \sigma_{\alpha\beta}[f^{(0)}],\\
	D^{(0)}_t T&=- \frac{2}{3}T\partial_\alpha u_\alpha-\frac{2}{3R\rho}p_{\alpha\beta}[f^{(0)}]\partial_\beta u_\alpha
	-\frac{2}{3R\rho}\partial_\beta q_\beta[f^{(0)}].
\end{align}
Furthermore, the derivative of the Boltzmann collision operator at the zeroth order distribution, $L$, is the \emph{linearized collision integral},
\begin{equation}\label{eq:L}
	Lf^{(1)}=\int_{R^3} \int_{S^2}\left[f_1^{(0)'}f^{(1)'}+f^{(0)'}f_1^{(1)'}-f_1^{(0)}f^{(1)}-f^{(0)}f_1^{(1)}\right]\bar{B}d^2\ee' d^3\bar{\cc}_1.
\end{equation}
Since we shall deal with only these orders of the \emph{Chapman-Enskog expansion}, we shall not write the next terms of higher-order in $\varepsilon$ (although we shall discuss them briefly at the end).

Finally, we need to expand also the consistency conditions:
\begin{align}
	\int \{m,\ m\cc,\ mv^2/2\}f^{(0)}d^3\cc&=\{\rho,\ \rho\uu,\ (3/2)\rho RT+\rho u^2/2\},\label{eq:CEconsistency0}\\
	\int \{m,\ m\cc,\ mv^2/2\}f^{(k)}d^3\cc&=0,\ k\ge 1. \label{eq:CEconsistency1}
\end{align}

\subsubsection{Euler's compressible flow equations}
On the zeroth order, we must have the velocity distribution which annuls the Boltzmann collision integral; That is it is a local Maxwell-Boltzmann distribution.
Then, the nonequilibrium pressure tensor and the heat flux are vanishing on this approximation,
\begin{equation}
	\sigma_{\alpha\beta}[f^{(0)}]=0,\ q_{\alpha}[f^{(0)}]=0,
\end{equation}
and the closed-form equations for the density, flow velocity and temperature become the Euler compressible flow equations, \eqref{eq:EDtn}, \eqref{eq:EDtu} and \eqref{eq:EDtT},
\begin{align}
	D^{(0)}_t \rho&=D^E_t \rho,\\
	D^{(0)}_t u_\alpha &=D^E_t u_\alpha,\\
	D^{(0)}_t T&=D^E_t T.
\end{align}
But now we have the Euler equations as the lowest-order approximation to the hydrodynamic equations, while these time-derivatives generate the correction which we now discuss.

\subsubsection{Navier-Stokes and Fourier compressible flow equations}
We shall now set up the equation for $f^{(1)}$. You may find it a bit tedious, even though these are just simple algebraic manipulations. So let us do it step-by-step.

\paragraph{Evaluation of the derivatives}

We first compute the derivatives of the local Maxwell-Boltzmann distribution with respect to the five hydrodynamic fields,
\begin{align}
	\frac{\partial f^{\rm eq}}{\partial \rho} &= \frac{1}{\rho}  f^{\rm eq},\label{eq:DnM}\\
	\frac{\partial f^{\rm eq}}{\partial u_\alpha} &= \frac{(c_\alpha-u_\alpha)}{RT}f^{\rm eq},\label{eq:DuM}\\
	\frac{\partial f^{\rm eq}}{\partial T} &= \frac{1}{T}\left(\frac{(\cc-\uu)^2}{2R T}-\frac{3}{2}\right)f^{\rm eq}.\label{eq:DTM}
\end{align}
\paragraph{Evaluation of the Boltzmann equation vector field on the local Maxwell-Boltzmann distribution}
Using the above derivatives, we next compute the term $(v_{\alpha}-u_{\alpha})\partial_{\alpha}f^{\rm eq}$,
\begin{align}
&	(v_{\alpha}-u_{\alpha})\partial_{\alpha}f^{\rm eq}=\nonumber\\
&		\frac{\partial f^{\rm eq}}{\partial \rho}(v_{\alpha}-u_{\alpha})\partial_{\alpha}\rho+
		\frac{\partial f^{\rm eq}}{\partial u_\beta}(v_{\alpha}-u_{\alpha})\partial_{\alpha}u_{\beta}+
		\frac{\partial f^{\rm eq}}{\partial T}(v_{\alpha}-u_{\alpha})\partial_{\alpha}T=\nonumber\\
&	f^{\rm eq}\left\{\frac{1}{\rho}(v_{\beta}-u_{\beta})\partial_{\beta}\rho+
	\frac{1}{RT}(v_{\alpha}-u_{\alpha})(v_{\beta}-u_{\beta})\partial_{\beta}u_{\alpha}
	+ \frac{1}{T}\left(\frac{(\cc-\uu)^2}{2RT}-\frac{3}{2}\right)(v_{\alpha}-u_{\alpha})\partial_{\alpha}T\right\}.
	\label{eq:MDtmicro}
\end{align}
Note that the combinations $\sim v_\alpha v_\beta$ and $\sim v_\alpha v^2$, related to nonequilibrium pressure tensor and heat flux, start popping out in this expression.

\paragraph{Evaluation of the time derivative of the local Maxwell-Boltzmann distribution due to the Euler system}
This is the last place we need to evaluate the derivatives due to the Euler equations;
\begin{align}
&	\frac{\partial f^{\rm eq}}{\partial \rho}D^{(0)}_t \rho+\frac{\partial f^{\rm eq}}{\partial u_\alpha}D^{(0)}_t u_\alpha+\frac{\partial f^{\rm eq}}{\partial T}D^{(0)}_t T =\nonumber \\
&	\frac{\partial f^{\rm eq}}{\partial \rho}D^E_t \rho+\frac{\partial f^{\rm eq}}{\partial u_\alpha}D^E_t u_\alpha+\frac{\partial f^{\rm eq}}{\partial T}D^E_t T =\nonumber \\
&	f^{\rm eq}\frac{1}{\rho}\left(-\rho\partial_\alpha u_\alpha\right) 
	+f^{\rm eq}\frac{(c_\alpha-u_\alpha)}{RT} \left( -\frac{1}{\rho}\partial_\alpha (\rho RT) \right)  
	+f^{\rm eq}\frac{1}{T}\left(\frac{(\cc-\uu)^2}{2RT}-\frac{3}{2}\right)  \left(- \frac{2}{3}T\partial_\alpha u_\alpha\right).
	\label{eq:MDtmacro}
\end{align}
%
%
\paragraph{Invariance defect of the local Maxwell-Boltzmann distribution}
The sum of the two above expressions, \eqref{eq:MDtmicro} and \eqref{eq:MDtmacro}, is so important that it deserves a name of its own. We call it the \emph{defect of invariance}
of the local Maxwell-Boltzmann distribution function with respect to the Boltzmann equation, 

\begin{equation}
	\Delta_{\mathcal{M}}= \frac{\partial f^{\rm eq}}{\partial \rho}D^{(0)}_t\rho+\frac{\partial f^{\rm eq}}{\partial u_\alpha}D^{(0)}_t u_\alpha+\frac{\partial f^{\rm eq}}{\partial T}\partial^{(0)}_t T + (v_{\beta}-u_{\beta})\partial_{\beta}f^{\rm eq}.
\end{equation}
The first Chapman-Enskog equation \eqref{eq:BCE1} now reads, 
\begin{equation}\label{eq:CE_first}
	Lf^{(1)}=\Delta_{\mathcal{M}}.
\end{equation}
The defect of invariance tells us whether or not the local Maxwell-Boltzmann distribution solves the Boltzmann equation. If $\Delta_\mathcal{M}=0$, then the initial condition taken as $f^{\rm eq}$ progresses in time due to the Boltzmann equation without changing its Gaussian shape, only the density, flow velocity and temperature alter  in time and space due to the Euler system. In such case the manifold of the local Maxwell-Boltzmann distributions (all such distributions with different macroscopic parameters) is said to be invariant of the Boltzmann equation, and the first (and higher) order Chapman--Enskog corrections ($f^{(1)}$, $f^{(2)}$ etc) are all vanishing.
So let us compute the invariance  defect.
Adding the above two expressions, we have (\emph{after a lot of cancellations}),
\begin{equation}
	\Delta_{\mathcal{M}} = f^{\rm eq}\left\{ \frac{1}{RT}\left[(c_\alpha-u_\alpha)(c_\beta-u_\beta)-\frac{1}{3}\delta_{\alpha\beta}(\cc-\uu)^2\right]
	(\partial_\alpha u_\beta)        \right\}
	+f^{\rm eq}\left\{ \frac{1}{T}\left(\frac{(\cc-\uu)^2}{2RT}-\frac{5}{2}\right)(c_\beta-u_\beta)\partial_\beta T      \right\}.
	\label{eq:Mdefect}
\end{equation}
Let's briefly discuss this result. The invariance defect consists of two parts. The first part drives the distribution function away from the local Maxwellian due to the inhomogeneity of the flow velocity $\sim \partial_\alpha u_\beta$, which results in the viscosity phenomena.
The second part is due to the  inhomogeneity of the temperature of the fluid $\sim \partial_\beta T $, which results in thermal conduction. Thus, the local Maxwell--Boltzmann distribution can be a solution to the Boltzmann equation only if the velocity of the flow and the temperature field are space-independent. Notice also that the invariance defect does not depend on a space derivative of the  density. This reflects the fact that the non-equilibrium phenomena in the single-component gas do not give rise to any diffusion phenomena (in a mixture of several gases it does). The solution to the Boltzmann equation \emph{may exist} in a form of a local Maxwell--Boltzmann distribution but only if the flow is moving with a uniform velocity and the temperature also stays uniform (the density then obeys $\partial_t \rho=-u_\alpha\partial_\alpha \rho$) but they are of no physical relevance. Further on, note that the invariance defect is Galilean invariant, that is, it depends only on the particles velocity in the reference frame of the flow. Finally, we immediately note that the defect is a truly non-equilibrium driver, as it does not change any value of neither density, nor flow velocity, nor the energy because,
\begin{equation}
	\int\Delta_{\mathcal{M}} d^3\cc = \int\cc \Delta_{\mathcal{M}} d^3\cc =\int v^2 \Delta_{\mathcal{M}} d^3\cc=0.
\end{equation}
What it does change are the \emph{fluxes} of these quantities.
We now need a final step in setting up the first Chapman--Enskog equation, a discussion of the linearized collision integral.


\paragraph{Linearized collision integral}
It is very convenient and customary to write the correction to the local Maxwell-Boltzmann distribution function in the form
\begin{equation}
	f^{(1)}=f^{\rm eq}\varphi.
\end{equation}
Then the linearized Boltzmann collision integral is transformed by noting that the products of the equilibria factor out due to the detail balance, and we have,
\begin{equation}\label{eq:L1}
	Lf^{(1)}=f^{\rm eq}\int_{R^3} \int_{S^2}f_1^{\rm eq}\left[\varphi_1'+\varphi'-\varphi_1-\varphi\right]{K}d^2\ee' d^3{\cc}_1.
\end{equation}
Thus, we can define the linearized Boltzmann {operator} $\Lambda$,
\begin{equation}\label{eq:L2}
	\Lambda \varphi=\int_{R^3} \int_{S^2}f_1^{\rm eq}\left[\varphi_1'+\varphi'-\varphi_1-\varphi\right]{K}d^2\ee' d^3{\cc}_1.
\end{equation}
The \emph{integral operator} $\Lambda$ has the following important properties:
\begin{itemize}
	\item Null-space of $\Lambda$: The null-space  $\ker\Lambda$ is the linear sub-space of summatorial invariants,
	\begin{equation}
		\Lambda \varphi=0\ {\rm if\  and\  only\  if}\ \varphi\in{\rm Lin}\{1,\vv,v^2\}.
	\end{equation}
	\item Let us define a scalar product of functions $\varphi$ and $\psi$ as
	\begin{equation}
		\langle \psi|\varphi\rangle=\int f^{\rm eq} \varphi\psi d^3{\cc}.
	\end{equation}
	Then operator $\Lambda$ is symmetric with respect to this scalar product,
	\begin{equation}\label{eq:Lsym}
		\langle\psi |\Lambda|\varphi\rangle=\langle\varphi |\Lambda|\psi\rangle.
	\end{equation}
	\item Operator  $\Lambda$ is non-positive,
	\begin{equation}
		\langle\varphi |\Lambda|\varphi\rangle \leq 0.
	\end{equation}
	Proof:
	\begin{equation}\label{eq:Lneg}
		\langle\varphi |\Lambda|\varphi\rangle=-\frac{1}{4}\int_{R^3\times R^3} \int_{S^2}f^{\rm eq}f_1^{\rm eq}\left[\varphi_1'+\varphi'-\varphi_1-\varphi\right]^2{K}d^2\ee' d^3{\cc}_1d^3{\cc}\leq 0.
	\end{equation}
\end{itemize}
Note that, the symmetry property of operator $\Lambda$ \eqref{eq:Lsym} and its non-positivity \eqref{eq:Lneg} are the "linearized" versions of the Boltzmann transport theorem and of the $H$-theorem, respectively.

\paragraph{The Chapman--Enskog  equation}
Summarizing, the results of the previous sections, the Chapman-Enskog equation for the first correction can be written, using the reduced peculiar velocity \eqref{eq:Cred}, 
\begin{align}\label{eq:CE1}
	\Lambda\varphi&=(\partial_{\beta}u_{\alpha})\left(C_{\alpha}C_{\beta}-\frac{1}{3}\delta_{\alpha\beta}C^2\right)
	+\sqrt{RT}\left(\frac{\partial_{\alpha}T}{T}\right)C_{\alpha}
	\left(\frac{C^2}{2}-\frac{5}{2}\right).
\end{align}
%
The linear non-homogeneous integral equation \eqref{eq:CE1} is the first Chapman-Enskog equation, the study of which is exhaustively described in \cite{chapman_mathematical_1939}.
For the sake of completeness, we summarize here some major steps.
By Fredholm alternative, the linear non-homogeneous integral equation of the form \eqref{eq:CE1} has a solution if the right hand side is orthogonal to the null-space of the integral operator $\Lambda$.
This solvability condition is satisfied since the defect of the invariance does not alter the density, momentum and energy (see above).
The general solution is then a specific solution to the non-homogeneous integral equation plus a general solution to the homogeneous one, $\Lambda\varphi=0$.
The latter is the linear subspace of summatorial invariants, so we have
\begin{equation}
	\varphi_{gen}=\varphi_{spec}+\varphi_{hom},
\end{equation}
where $\varphi_{hom}\in{\rm Lin}\{1,\vv,v^2\}$ while $\varphi_{spec}$ is orthogonal to conservation laws, $\int \varphi_{spec}f^{\rm eq}\{1,\cc,c^2\}d^3\cc=0$.
However, due to the consistency condition \eqref{eq:CEconsistency1} mentioned at the very beginning of the analysis, we must have $\int f^{(1)}\{1,\cc,c^2\}d^3\cc=0$, therefore we must select $\varphi_{hom}=0$ in the above general solution.
The Chapman-Enskog solution is thus the special solution to the above integral equation \eqref{eq:CE1}.

Now we look at the tensorial dimension of the right hand side of \eqref{eq:CE1} and immediately realize that, since the function $\varphi$ is a scalar, the solution  can depend on the
particle's velocity in only one possible way,
	\begin{align}\label{eq:CE1form}
		\varphi=A(C)C_{\alpha}\sqrt{RT}\left(\frac{\partial_{\alpha}T}{T}\right)+
		B(C)\left(C_{\alpha}C_{\beta}-\frac{1}{3}\delta_{\alpha\beta}C^2\right)(\partial_{\beta}u_{\alpha}),
\end{align}
where two scalar functions, $A$ and $B$ can depend only on the magnitude of reduced velocity, temperature and density while $A$ satisfies the orthogonality condition,
\[\int e^{-\frac{C^2}{2}}A(C)C^2d^3\bm{C}=0.\]
The Chapman-Enskog functions $A$ and $B$ are found from two integral equations,
\begin{align}
	&	\Lambda\left(AC_{\alpha}\right)
	=C_{\alpha}\left(\frac{C^2}{2}-\frac{5}{2}\right),\label{eq:CEA}\\
	&	\Lambda\left(B\left(C_{\alpha}C_{\beta}-\frac{1}{3}\delta_{\alpha\beta}C^2\right)\right)
	=\left(C_{\alpha}C_{\beta}-\frac{1}{3}\delta_{\alpha\beta}C^2\right).\label{eq:CEB}
\end{align}
Apart from Maxwell's molecules, for which $A$ and $B$ do not depend on $\bm{C}$, appropriate technique of solving (\ref{eq:CEA}) and (\ref{eq:CEB}) is based on Sonine polynomial expansion \cite{chapman_mathematical_1939}. Here we shall perform a merely dimensional analysis of \eqref{eq:CEA} and \eqref{eq:CEB} in order to understand  the relation between transport coefficients and specific models of particles interaction.

\paragraph{Dimensional analysis of Chapman--Enskog equations}
Since the Chapman--Enskog solution $\varphi$ \eqref{eq:CE1} is a non-dimensional quantity, the proper dimension of the Chapman--Enskog functions $A$ and $B$ is that of the time, 
 \begin{equation}
 	[A]=[B]\sim{\rm sec}.
 	\end{equation}
 Furthermore, since the right hand side of the Chapman--Enskog equations \eqref{eq:CEA} and \eqref{eq:CEB} is also non-dimensional, we can introduce a \emph{relaxation time} $\tau$  so that 
 \begin{equation}
 	A=\tau \bar{A}, B=\tau\bar{B},
 \end{equation}
 where functions $\bar{A}(C)$ and $\bar{B}(C)$ are dimensionless. 
 
 Evaluation of the relaxation time is based on the dimensional analysis of the linearized collision integral for each specific binary collision mode. 
 For hard spheres, with the collision kernel \eqref{eq:HSkernel}, we write
 \begin{align}\label{eq:HSdim}
 	\Lambda^{HS}  =\left(nd^2\sqrt{RT}\right)\bar{\Lambda}^{HS},
 	\end{align}
 where $\bar{\Lambda}^{HS}$ is a non-dimensional linearized Boltzmann operator for hard spheres,
 \begin{equation}
 	\bar{\Lambda}^{HS} \varphi=\frac{1}{4(2\pi)^{3/2}}\int_{R^3} \int_{S^2}e^{-C_1^2/2}\left[\varphi_1'+\varphi'-\varphi_1-\varphi\right]|\bm{C}_1-\bm{C}|d^2\ee' d^3{\CC}_1.
 \end{equation}
 Thus, for hard spheres, the relaxation time is defined as the inverse of the prefactor in \eqref{eq:HSdim},
\begin{equation}\label{eq:HStau}
	\tau^{HS}=\frac{1}{nd^2\sqrt{RT}}.
\end{equation}
One can readily recognize the earlier estimate of the mean free path in this expression, and the relaxation time is interpreted as an average travel time between binary encounters, with a characteristic thermal speed.  
%
%
%
%
For Maxwell's molecules, with the interaction potential $U=\kappa r^{-4}$, we notice that, since the collision kernel does not depend on the relative velocity (see Eq.\ \eqref{MMkernel} with $n=4$) the only combination of the mass $m$ and the strength of the potential $\kappa$ that gives a proper dimension to the collision kernel is 
\begin{equation}
	K^{MM}\sim \sqrt{\frac{\kappa}{m}}\sim \frac{{\rm cm}^3}{{\rm sec}}.
\end{equation} 
Consequently, the dimensional analysis of the linearized collision operator for Maxwell's molecules gives,
\begin{equation}\label{eq:MMdim}
    \Lambda^{MM}=\left(n\sqrt{\frac{\kappa}{m}}\right)\bar{\Lambda}^{MM},
\end{equation}
where the non-dimensional linearized collision operator is,
\begin{equation}
 	\bar{\Lambda}^{MM} \varphi=\frac{1}{(2\pi)^{3/2}}\int_{R^3} \int_{S^2}e^{-C_1^2/2}\left[\varphi_1'+\varphi'-\varphi_1-\varphi\right]\bar{\alpha}d^2\ee' d^3{\CC}_1,
 \end{equation}
where $\bar{\alpha}(\theta)$ is a non-dimensional function of the deflection angle.
From \eqref{eq:MMdim}, we read off the relaxation time for Maxwell's molecules,
\begin{equation}\label{eq:MMtau}
	\tau^{MM}=\frac{\sqrt{m}}{n\sqrt{\kappa}}.
\end{equation}
It is interesting to remark here that, the mean free path for Maxwell's molecules is proportional to the square root of the temperature, $l_{m.f.p}^{MM}\sim\tau_{MM}v_T\sim \sqrt{k_BT}/(n\sqrt{\kappa})$, while it is temperature-independent for hard spheres, $l_{m.f.p.}^{HS}\sim 1/(nd^2)$. 
In other words, the average distance traveled freely by Maxwell's molecules between collisions increases with the temperature of the gas.
Thus, by the estimate \eqref{eq:viscosityelementary}, the viscosity of the Maxwell's molecules is linear in the temperature,
\begin{equation}
	\label{eq:MMviscosityestimate}
	\mu^{MM}\sim \rho l_{m.f.p.}^{MM}v_T\sim RT\sqrt{\frac{m^3}{\kappa}}.
\end{equation}

With the above dimensional analysis, we write the solution to the first Chapman--Enskog equation by introducing the relaxation time into \eqref{eq:CE1form},	\begin{align}\label{eq:CE1formnodim}
	\varphi=\bar{A}(C)C_{\alpha}\left[\tau\sqrt{RT}\left(\frac{\partial_{\alpha}T}{T}\right)\right]+
	\bar{B}(C)\left(C_{\alpha}C_{\beta}-\frac{1}{3}\delta_{\alpha\beta}C^2\right)\left[\tau\partial_{\beta}u_{\alpha}\right],
\end{align}
where the non-dimensional functions $\bar{A}(C)$ and $\bar{B}(C)$ depend \emph{only} on the magnitude of the reduced peculiar velocity and are special solutions to the reduced (non-dimensional) integral equations,
\begin{align}
	&	2\bar{\Lambda}\left|\bar{A}C_{\alpha}\right\rangle
	=\left|C_{\alpha}\left({C^2}-{5}\right)\right\rangle,\label{eq:CEAnodim}\\
	&	\bar{\Lambda}\left|\bar{B}\left(3C_{\alpha}C_{\beta}-\delta_{\alpha\beta}C^2\right)\right\rangle
	=\left|3C_{\alpha}C_{\beta}-\delta_{\alpha\beta}C^2\right\rangle.\label{eq:CEBnodim}
\end{align}
We shall now proceed with the general form of the first Chapman-Enskog correction \eqref{eq:CE1formnodim} to evaluate the non-equilibrium pressure tensor and the heat flux vector. 
 \paragraph{Viscosity} 
	We are now able to evaluate the nonequilibrium fluxes for a generic collision model. Let's start with the nonequilibrium pressure tensor,
	\begin{align}
		\sigma^{(1)}_{\alpha\beta}&=m\int v_{\alpha}v_{\beta}f^{\rm eq}\varphi d^3\cc.
	\end{align}
	With \eqref{eq:CE1formnodim}, we find
	\begin{align}
		\sigma^{(1)}_{\alpha\beta} = -\mu  \Pi_{\alpha\beta},
	\end{align}
where $\Pi_{\alpha\beta}$ is the \emph{rate-of-shear tensor},
\begin{align}			
	 \Pi_{\alpha\beta}=	{\partial_{\alpha}u_{\beta}+ \partial_{\alpha}u_{\beta}}-\frac{2}{3}\delta_{\alpha\beta}\partial_{\gamma}u_{\gamma},
	\end{align}
%
%
while $\mu$ is the viscosity,
\begin{align}
		\mu = \bar{b}\tau p,
\end{align}
with the coefficient $\bar{b}$ (\emph{pure number}) expressed in terms of the Chapman--Enskog function $\bar{B}$ as,
\begin{align}\label{eq:CEb}
	\bar{b}=-\frac{1}{5}(2\pi)^{-3/2}\int e^{-C^2/2}\bar{B}
	\left(C_{\alpha}C_{\beta}-\frac{1}{3}\delta_{\alpha\beta}C^2\right)\left(C_{\beta}C_{\alpha}-\frac{1}{3}\delta_{\beta\alpha}C^2\right)d^3\CC >0.
	\end{align}
The latter inequality follows immediately from the definition of the function $\bar{B}$ \eqref{eq:CEBnodim} and the entropy production inequality \eqref{eq:Lneg}: 
Indeed, the integral in \eqref{eq:CEb} is proportional to the inner product, \\
$\left\langle \bar{B}\left(C_{\alpha}C_{\beta}-\frac{1}{3}\delta_{\alpha\beta}C^2\right)\bigg|C_{\beta}C_{\alpha}-\frac{1}{3}\delta_{\beta\alpha}C^2\right\rangle.$\\
Hence, from the definition \eqref{eq:CEBnodim}, we have \\
$\left\langle \bar{B}\left(C_{\alpha}C_{\beta}-\frac{1}{3}\delta_{\alpha\beta}C^2\right)\bigg|C_{\beta}C_{\alpha}-\frac{1}{3}\delta_{\beta\alpha}C^2\right\rangle
=\left\langle \bar{B}\left(C_{\alpha}C_{\beta}-\frac{1}{3}\delta_{\alpha\beta}C^2\right)\bigg|\bar{\Lambda}\bigg|\bar{B}\left(C_{\beta}C_{\alpha}-\frac{1}{3}\delta_{\beta\alpha}C^2\right)\right\rangle<0.$

\paragraph{Thermal conduction}
	Similarly, we evaluate the non-equilibrium heat flux,
	\begin{equation}
		q^{(1)}_{\alpha}=\frac{m}{2}\int (\cc-\uu)^2(v_\alpha-u_\alpha)f^{\rm eq}\varphi d^3\cc.
	\end{equation}
	Substituting the solution $\varphi$ \eqref{eq:CE1formnodim}, we get the nonequilibrium heat flux in the form of the \emph{Fourier law},
 \begin{align}\label{eq:CEFourier}
		q^{(1)}_{\alpha} = -\lambda \partial_{\alpha} T,
	\end{align}
 where $\lambda$ is the \emph{coefficient of thermal conduction}, 
 \begin{align}
     \label{eq:CElambda}
     \lambda&=\bar{a}\tau  Rp,
     \end{align}
and the coefficient $\bar{a}$ (\emph{pure number}) is given by a quadrature similar to \eqref{eq:CEb},
\begin{equation}\label{eq:CEa}
	\bar{a}=-\frac{1}{6}(2\pi)^{-3/2}\int e^{-C^2/2}\bar{A}C^4d^3\CC >0.
\end{equation}

\paragraph{Transport coefficients}
Evaluation of the function $B$ for specific collision models is based on a Sonine polynomial expansion, and is exhaustively studied in the classical text by Chapman and Cowling \cite{chapman_mathematical_1939}.
For hard spheres, using the relaxation time \eqref{eq:HStau}, the lowest-order Sonine polynomial expansion results in
\begin{align}\label{eq:HSmu0}
	\mu_0=\bar{b}_0\frac{m\sqrt{RT}}{d^2},
\end{align}
where the $\bar{b}_0$ (pure number) is
\begin{align}\label{eq:HSb0}
	\bar{b}_0=\frac{5}{16\sqrt{\pi}}\approx 0.176.
\end{align}
The dependence of the viscosity on the temperature and on the diameter of hard sphere has been already found by elementary considerations, cf.\ \eqref{eq:Bviscosityelementary}. The lowest-order Sonine polynomial approximation \eqref{eq:HSmu0} is only insignificantly lower than the more accurate result $\bar{b}\approx 0.179$ when more terms of the Sonine polynomial expansion are evaluated \cite{chapman_mathematical_1939}. 
For Maxwell's molecules, using the mean free path \eqref{eq:MMtau}, the viscosity is obtained as
\begin{align}\label{eq:MMmuB}
	\mu=\bar{b}{RT}\sqrt{\frac{{m^3}}{\kappa}},
\end{align}
where the exact value of the constant $\bar{b}$ is \cite{chapman_mathematical_1939},
\begin{align}\label{eq:MMb}
	\bar{b}=\frac{\sqrt{2}}{3\pi A_2(5)}\approx 0.344,
\end{align}
where  $A_2(5)\approx 0.436$ \cite{chapman_mathematical_1939}. Apart from the pure constant $\bar{b}$ \eqref{eq:MMb}, the dependence of the viscosity on the temperature, particle's mass and the strength of the interaction potential strength was already obtained by elementary considerations above, cf.\ Eq.\ \eqref{eq:MMviscosityestimate}.
%
%
The thermal conduction coefficient \eqref{eq:CElambda} can be written in the following suggestive way,
     \begin{align}
      \lambda&=c_p{\rm Pr}^{-1}\mu,
 \end{align}
 where $c_p$ is the specific heat of ideal gas of monatomic molecules, 
 \begin{equation}
     c_p=\frac{5}{2}R,
 \end{equation}
 while ${\rm Pr}$ is the \emph{Prandtl number},
 \begin{equation}
     \label{eq:Pr}
     {\rm Pr}=\frac{c_p\mu}{\lambda}=\frac{\bar{b}}{\bar{a}}.
 \end{equation}
The first-order Sonine polynomial approximation results in the Prandtl number \cite{chapman_mathematical_1939},
\begin{equation}
	\label{eq:Pr0}
	{\rm Pr}_0=\frac{2}{3}.
\end{equation}
For Maxwell molecules, the Prantl number \eqref{eq:Pr0} is the exact result while for hard spheres ${\rm Pr}={2}/{3}$ is only slightly lower than the exact value \cite{chapman_mathematical_1939}.


\paragraph{Correction to the Euler system: The Navier-Stokes equations}

Summarizing the above, the correction to the local equilibrium approximation results in the following balance equations for mass, momentum and energy,  
\begin{align}
	D_t^{NS} \rho&=-\rho(\partial_{\alpha}u_{\alpha}), \label{eq:DtnNS}\\
	D_t^{NS} u_{\alpha}&=-\frac{1}{\rho}\partial_{\alpha}p+\frac{1}{\rho}\partial_{\beta}\left(\varepsilon \mu \Pi_{\alpha\beta}\right), \label{eq:DtuNS}\\
	D_t^{NS} T&=-\frac{2}{3}T(\partial_{\alpha}u_{\alpha})+\frac{2}{3}\left(\frac{T}{p}\right)\left(\varepsilon \mu \Pi_{\alpha\beta}\right)(\partial_{\beta}u_{\alpha})
	+\frac{2}{3}\left(\frac{T}{p}\right)\partial_{\alpha}\left(\varepsilon\lambda\partial_{\alpha}T\right). \label{eq:DtTNS}
\end{align}
Comments are in order:
\begin{itemize}
	\item The correction to the local equilibrium approximation results in the Navier--Stokes--Fourier system for compressible ideal gas, with the caloric equation of state featured by the specific heat at constant volume $c_v=(3/2)R$ and Prantl number close to ${\rm Pr}=2/3$. The result of the derivation has led to the microscopic expressions for the coefficients of viscosity and thermal conductivity which can be computed from any given molecular interaction in the Boltzmann equation.
	\item It is important to keep the smallness parameter $\varepsilon$ in the above expressions for the non-equilibrium fluxes. Higher-order corrections, the Burnett ($\epsilon^2$) and the super-Burnett ($\epsilon^3$) systems remained controversial as they violate the stability of the equilibrium. 
	\item Hilbert's 6th problem: Hilbert proposed to derive "mathematically the limiting processes ... which lead from the atomistic view to the laws of motion of continua" (see, e. g. \cite{saint-raymond_mathematical_2013,gorban_hilberts_2013} and references therein). The issue Hilbert was addressing, namely to use the atomistic theory of his day represented by Boltzmann's kinetic theory of gases and a passage via a limiting process to the continuum theory of compressible Euler system as the Knudsen number approaches zero, or to the  Navier-Stokes-Fourier system, if small corrections are allowed. 
	The problem is however, that without a priori knowledge about solutions to Euler or Navier-Stokes-Fourier equations are, one cannot in general prove that Boltzmann's kinetics converges to these continuum models \cite{saint-raymond_mathematical_2013}. 
	The resolution likely requires addressing other continuum theories, one current candidate is Korteweg-like \cite{gorban_hilberts_2013}. Interested reader is directed to \cite{saint-raymond_mathematical_2013,gorban_hilberts_2013}.
\end{itemize}

\subsection{Lifting the local equilibrium projection: BGK kinetic model}
\label{sec:BGK_B}



The correction to the Euler system considered above explores more of the phase space than it was assumed by the local equilibrium projection.
By measuring the defect of invariance of the local equilibrium we find in which direction the local equilibrium approximation should be corrected in order to take into account the fast motion towards it. There is another way to explore the fast motions: To lift the dynamics to the full phase space by means of a \emph{kinetic model}. 

The lifting of the Euler dynamics which takes place on the local Maxwell
manifold to a kinetics in the whole phase space  is done by the very useful Bhatnagar--Gross--Krook model (BGK),
\begin{equation}
	\label{BGK}
	\partial_t f+\vv\cdot\nabla f=-\frac{1}{\tau}(f-f^{\rm eq}(f)),
\end{equation}
where $\tau>0$ is the relaxation time, and
$f^{\rm eq}(f)$ is a map $f\to f^{\rm eq}$ established by local conservation laws,
\begin{equation}
	\label{consistency}
	\int\{1,\vv,v^2\}(f-f^{\rm eq}(f))d\vv=0.
\end{equation}
The right hand side of Eq.\ (\ref{BGK}),
\begin{equation}
	\label{BGKcollint} \mathcal{J}_{\rm BGK}=-\frac{1}{\tau}(f-f^{\rm eq}(f)),
\end{equation}
is called the BGK collision integral. Proof of the $H$-theorem for the
BGK kinetic equation does not rely anymore on the
microscopic reversibility as in the Boltzmann case, instead, it is
a simple consequence of convexity of the $H$-function, and
of the property of the map (\ref{consistency}):
\begin{align}
	\label{sig}
	\sigma&=-\frac{1}{\tau}
	\int \ln f(f-f^{\rm eq}(f))d^3\vv=
	-\frac{1}{\tau}\int\ln\left(\frac{f}{f^{\rm eq}(f)}\right)(f-f^{\rm eq}(f))d^3\vv\le 0.
\end{align}

When the BGK relaxation model is used instead of the Boltzmann collision operator in finding the correction to the local equilibrium projection, the Chapman--Enskog equation  \eqref{eq:CE_first} becomes,
\begin{equation}
	\label{eq:CE_BGK}
	-\frac{1}{\tau}f^{(1)}=\Delta_{\mathcal{M}}.
\end{equation}
Thus, the Chapman-Enskog solution is given by \eqref{eq:CE1formnodim} with the Chapman--Enskog functions \eqref{eq:CEAnodim} and \eqref{eq:CEBnodim} as
\begin{align}
	\bar{A}_{BGK}&=-\left(\frac{C^2}{2}-\frac{5}{2}\right),\\ \bar{B}_{BGK}&=-1.
\end{align}
The resulting Navier--Stokes--Fourier equations feature the following coefficients of viscosity and thermal conduction,
\begin{align}
	\mu_{BGK}&=\tau p,\\
	\lambda_{BGK}&=\tau c_p p,
	\end{align}
resulting in the BGK Prandtl number,
\begin{equation}\label{eq:BGK_Pr}
	{\rm Pr}_{BGK}=1.
\end{equation} 
The restriction to Prandtl number ${\rm Pr}_{BGK}=1$ arises from the fact that no intermediate states of relaxation towards the local equilibrium are addressed by the BGK kinetic model. This issue shall be considered below. Prior to that, we need to extend the notion of projection onto a wider class of specified states.

\subsection{Grad's thirteen-moments projection}
\subsubsection{Grad's thirteen-moments distribution function}
\label{sec:Grad}
Grad, in his seminal paper of 1949 \cite{grad_kinetic_1949}, derived moment systems by projecting the Boltzmann equation onto an ansatz for the distribution function. Grad considered two sets of moments, and what will be referred to as G13 and G20, with the number indicating how many fields are included. In G13, these are the locally conserved density, momentum and energy plus the nonequilibrium stress tensor and energy flux vector, while G20, a more symmetric system, includes the full third-order flux of the pressure tensor instead of the energy flux vector. 
Grad's method was influential in many ways, far beyond applications to rarefied gas dynamics. It was the touchstone for numerous developments in nonequilibrium thermodynamics (see, e.g. \cite{jou_extended_2010,grmela_hamiltonian_2017,grmela_externally_2017} and references cited therein).
However, the problem with almost any projection on a preselected (often simple) submanifold is that it is not invariant with respect to the the detailed dynamics.
In Grad's context, Grad's distribution function (a polynomial around the local Maxwellian) is not invariant with respect to the Boltzmann equation.
This is certainly not just the feature of Grad's distribution per se. As we have seen above, the local Maxwellian is also not invariant of the Boltzmann equation, and the dynamic correction, well known through the Chapman-Enskog method, delivers the dissipative Navier--Stokes--Fourier terms, missing in the the projection on the local Maxwell manifold (compressible Euler equations). Both correction and lifting of Grad's G13 system shall be considered below.

%
Following Grad \cite{grad_kinetic_1949}, the distribution function providing the closure for the balance equations (\ref{eq:balances}) and (\ref{eq:balanceq}) is written as,
\begin{align}
	\mathcal{G}&=\mathcal{M}+\mathcal{N},\label{eq:G}\\
	\mathcal{N}&=\mathcal{M}\mathcal{G}^{\sigma} +\mathcal{M}\mathcal{G}^q,
\end{align}
where $\mathcal{M}$ is local Maxwellian \eqref{eq:maxwell},
and $\mathcal{N}$ is the nonequilibrium part,
\begin{align}
	\mathcal{G}^{\sigma}&=\frac{\sigma_{\alpha\beta}}{2p}\left(C_{\alpha}C_{\beta}-\frac{1}{3}\delta_{\alpha\beta}C^2\right),\label{eq:G13sigma}\\
	\mathcal{G}^q &=\frac{q_{\alpha}C_{\alpha}}{pv_T}\left(\frac{C^2}{5}-1\right).\label{eq:G13q}
\end{align}
Without engaging a discussion of possible violation of positivity, we consider Grad's function as a submanifold in the space of distribution functions, parameterized with the values of thirteen fields. Grad's system is the natural projection of kinetic equation onto this submanifold.

\subsubsection{Grad's thirteen-moments closure}

Grad's projection amounts to evaluating everything that spoils the closure in the balance equations (\ref{eq:balances}) and (\ref{eq:balanceq}) with Grad's distribution (\ref{eq:G}). 
For the $Q$-fluxes (\ref{eq:Q}), we get, separating the local equilibrium and nonequilibrium contributions,
\begin{align}
	Q_{\alpha\beta\gamma}^{\mathcal{G}}&= Q_{\alpha\beta\gamma}^{\mathcal{M}}+Q_{\alpha\beta\gamma}^{\mathcal{N}},\label{eq:closureQ}\\
	Q_{\alpha\beta\gamma}^{\mathcal{M}}&=0,\label{eq:closureQeq}\\
	Q_{\alpha\beta\gamma}^{\mathcal{N}}&=
	\frac{1}{5}\left(q_{\alpha}\delta_{\beta\gamma}+q_{\beta}\delta_{\alpha\gamma}+q_{\gamma}\delta_{\alpha\beta}\right).\label{eq:closureQnoneq}
\end{align}
In other words, Grad's closure for the $Q$-flux amounts to reducing symmetric rank three tensor to its trace (\ref{eq:closureQnoneq}). 
For the $T$-flux (\ref{eq:T}), one finds
\begin{align}
	T_{\alpha\beta}^{\mathcal{G}}&=T_{\alpha\beta}^{\mathcal{M}}+T_{\alpha\beta}^{\mathcal{N}},\label{eq:closureT}\\
	T_{\alpha\beta}^{\mathcal{M}}&=\frac{5}{2}pRT\delta_{\alpha\beta},\label{eq:closureTeq}	\\
	T_{\alpha\beta}^{\mathcal{N}}&=\frac{7}{2}RT\sigma_{\alpha\beta}\label{eq:closureTnoneq}.
\end{align}
For the collision rates (\ref{eq:Rs}) and (\ref{eq:Rq}), there are several realizations depending on the choice of the collision or relaxation model which we list here in the order of increasing complexity.

\begin{enumerate}
	\item A "poor man's" approach is to use the BGK relaxation time approximation \eqref{BGKcollint};
	then one simply gets
	\begin{align}
		R^{\sigma \mathcal{G}}_{\alpha\beta}&=-\frac{m}{\tau}\int \left[({v}_{\alpha}-{u}_{\alpha})({v}_{\beta}-{u}_{\beta})-\frac{1}{3}\delta_{\alpha\beta}({v}-{u})^2\right] \mathcal{M}\mathcal{G}^{\sigma}d\bm{v}=-\frac{1}{\tau}\sigma_{\alpha\beta},\\
		R^{q\mathcal{G}}_{\alpha} &=-\frac{m}{2\tau}\int ({v}_{\alpha}-{u}_{\alpha})({v}-{u})^2\mathcal{M} \mathcal{G}^{q}d\bm{v}=-\frac{1}{\tau}q_{\alpha}.
	\end{align}
	Similar results are obtained when using most of the relaxation kinetic models available (with more that one relaxation time).
	While certainly far from realistic, relaxation time approximation is useful for analysis of complex situations in order to understand the structure of otherwise involved result.

	\item For the Boltzmann's collision integral \eqref{eq:collisionintegral},
	substituting Grad's distribution function, and taking into account the detail balance, $\mathcal{J}_B(\mathcal{M})=0$, one gets
	\begin{align}\label{eq:linearizationB}
		\mathcal{J}_B(\mathcal{G})=L\mathcal{N}+O(\mathcal{N}^2),
	\end{align}
	where $L$ is the linearized Boltzmann collision integral \eqref{eq:L}.
	%
	Computation of matrix elements of the operator (\ref{eq:L}) greatly simplifies for Maxwell's molecules because in that case
	functions $\mathcal{M}[C_{\alpha}C_{\beta}-\frac{1}{3}\delta_{\alpha\beta}C^2]$ and $\mathcal{M}[C_{\alpha}(C^2-5/2)]$ are eigenfunctions,
	\begin{align}\label{eq:LinBmaxwellS}
		&L\left(\mathcal{M}\left[C_{\alpha}C_{\beta}-\frac{1}{3}\delta_{\alpha\beta}C^2\right]\right)
		=-\frac{p}{\mu}\mathcal{M}\left[C_{\alpha}C_{\beta}-\frac{1}{3}\delta_{\alpha\beta}C^2\right],\\
		&L\left(\mathcal{M}\left[C_{\alpha}\left(\frac{C^2}{2}-\frac{5}{2}\right)\right]\right)=-\frac{2p}{3\mu}\mathcal{M}
		\left[C_{\alpha}\left(\frac{C^2}{2}-\frac{5}{2}\right)\right],\label{eq:LinBmaxwellQ}
	\end{align}
	with $\mu$ the viscosity coefficient of Maxwell's molecules \eqref{eq:MMmuB}.
	With (\ref{eq:LinBmaxwellS}) and (\ref{eq:LinBmaxwellQ}), evaluation of matrix elements reduces to the same integrals as in the relaxation time approximation, and we get 
	\begin{align}\label{eq:elementS}
		R^{\sigma \mathcal{G}}_{\alpha\beta}&={m}\int \left[({v}_{\alpha}-{u}_{\alpha})({v}_{\beta}-{u}_{\beta})-\frac{1}{3}\delta_{\alpha\beta}({v}-{u})^2\right] L(\mathcal{M}\mathcal{G}^\sigma)d\bm{v}=-\frac{p}{\mu}\sigma_{\alpha\beta}\\
		R^{q\mathcal{G}}_{\alpha} &=\frac{m}{2}\int ({v}_{\alpha}-{u}_{\alpha})({v}-{u})^2L(\mathcal{M}\mathcal{G}^q)d\bm{v}=-\frac{2p}{3\mu}q_{\alpha}.
		\label{eq:elementQ}
	\end{align}
	{It should be noted that, for Maxwell's molecules, evaluation of the relaxation rates for various moments can be done in closed form
		without specifying the distribution function, also for the nonlinear collision operator \cite{truesdell_fundamentals_1980,struchtrup_macroscopic_2005}, and relaxation rates (\ref{eq:elementS},\ref{eq:elementQ}) are valid for the full nonlinear case.}
	
	\item For other particle's collision models such as hard spheres or other power law potentials, 
	functions $\mathcal{M}[C_{\alpha}C_{\beta}-\frac{1}{3}\delta_{\alpha\beta}C^2]$ and $\mathcal{M}[C_{\alpha}(C^2-5/2)]$ are not eigenfunctions 
	of linearized collision integral any longer, and evaluation of matrix elements (\ref{eq:elementS}) and (\ref{eq:elementQ}) gives instead 
	\begin{align}
		R^{\sigma \mathcal{G}}_{\alpha\beta}&=-\frac{p}{\mu_0}\sigma_{\alpha\beta},\label{eq:RsGrad}\\
        R^{q \mathcal{G}}_{\alpha}&=-\frac{2p}{3\mu_0}q_{\alpha},\label{eq:RqGrad}
	\end{align}
	where $\mu_0$ is not exact viscosity coefficient but rather the lowest-order Sonine polynomial approximation thereof, cf.\ Eq.\ \eqref{eq:HSmu0}. 
	%
	It is well known that first approximation $\mu_0$ is reasonably close to the exact value, in particular, for hard spheres \cite{chapman_mathematical_1939}.
	This fact, however, does not imply that corresponding eigenfunctions of linearized Boltzmann collision integral are in any sense "close" to those for Maxwell's molecules, and it is therefore  misleading to judge on the quality of Grad's approximation for hard spheres on the basis of viscosity coefficient alone. 
	

	
\end{enumerate}



\subsubsection{Grad's thirteen-moments system}

Substituting Grad's closure relations for the $Q$- and $T$-fluxes (\ref{eq:closureQ}) and  (\ref{eq:closureT}) into balance equations (\ref{eq:balances}) and (\ref{eq:balanceq}), and also using any of the above realizations of the relaxation terms, one arrives at Grad's equations.
For later use, it proves convenient to partition Grad's equations in four parts, three of which regards the non-local in space terms plus the relaxation term. For the stress, we write
\begin{align}
	D^{\mathcal{G}}_t\sigma_{\alpha\beta}&=\dot{\sigma}^{\rm NSF}_{\alpha\beta}+\dot{\sigma}^{\rm lin}_{\alpha\beta}+ \dot{\sigma}^{\rm nlin}_{\alpha\beta}+ R^{\sigma \mathcal{G}}_{\alpha\beta},\label{eq:Gsdotfull}\\
	\dot{\sigma}^{\rm NSF}_{\alpha\beta}&=
	-p\left(\partial_{\alpha}u_{\beta}+\partial_{\beta}u_{\alpha}-\frac{2}{3}\delta_{\alpha\beta}\partial_{\gamma}u_{\gamma}\right)
	\label{eq:sdotNSF}\\
	\dot{\sigma}^{\rm lin}_{\alpha\beta}&=
	-\frac{2}{5}\left(\partial_{\beta}q_{\alpha}+\partial_{\alpha}q_{\beta}-\frac{2}{3}\delta_{\alpha\beta}\partial_{\gamma}q_{\gamma}\right),
	\label{eq:sdotlin}\\
	\dot{\sigma}^{\rm nlin}_{\alpha\beta}&=-\sigma_{\alpha\beta}\left(\partial_{\gamma}u_{\gamma}\right)
	-\left(\sigma_{\alpha\gamma}\partial_{\gamma}u_{\beta}+\sigma_{\beta\gamma}\partial_{\gamma}u_{\alpha}
	-\frac{2}{3}\delta_{\alpha\beta}\sigma_{\mu\nu}\partial_{\nu}u_{\mu}\right).\label{eq:sdotnlin}
\end{align}
A comment on genesis of various terms in Grad's equation (\ref{eq:Gsdotfull}) is in order. 
The first term, $\dot{\sigma}^{\rm NSF}_{\alpha\beta}$ (\ref{eq:sdotNSF}), is designated Navier-Stokes-Fourier (NSF) because it linearly depends on the strain tensor and gives rise to the Navier-Stokes stress in the first-order approximation to G13. This term appears as purely kinematic, that is, it shows up already in the balance equation for the stress before any closure assumption. The second term, $\dot{\sigma}^{\rm lin}_{\alpha\beta}$ (\ref{eq:sdotlin}) is solely produced by the closure relation for the $Q$-flux (\ref{eq:closureQnoneq}). It is indicated as "linear" since it depends linearly on the gradient of the heat flux but not on any gradient of locally conserved fields. Consequently, the term $\dot{\sigma}^{\rm lin}_{\alpha\beta}$ survives  linearization around a global equilibrium state. Finally, the nonlinear term $\dot{\sigma}^{\rm nlin}_{\alpha\beta}$
(\ref{eq:sdotnlin}) is again { purely kinematic} and { independent} of Grad's closure assumption.

Grad's  heat flux equation is decomposed in a similar fashion,
\begin{align}
	&D^{\mathcal{G}}_t q_{\alpha}=\dot{q}^{\rm NSF}_{\alpha}+\dot{q}^{\rm lin}_{\alpha}+ \dot{q}^{\rm nlin}_{\alpha}
	+ R^{q \mathcal{G}}_{\alpha},
	\label{eq:Gqdotfull}\\
	&\dot{q}^{\rm NSF}_{\alpha}=-\frac{5}{2}Rp\partial_{\alpha}T, \label{eq:qdotNSF}\\
	&\dot{q}^{\rm lin}_{\alpha}=-RT\partial_{\beta}\sigma_{\alpha\beta}, \label{eq:qdotlin}\\
	&\dot{q}^{\rm nlin}_{\alpha}=-\frac{7}{5}q_{\alpha}\partial_{\beta}u_{\beta}
	-\frac{7}{5}q_{\beta}\partial_{\beta}u_{\alpha}
	-\frac{2}{5}q_{\beta}\partial_{\alpha}u_{\beta}-\frac{5}{2}R\sigma_{\alpha\beta}\partial_{\beta}T
	+\frac{RT}{\rho}\sigma_{\alpha\beta}\partial_{\beta}\rho
	+\frac{1}{\rho}\sigma_{\alpha\beta}\partial_{\gamma}\sigma_{\gamma\beta}. \label{eq:qdotnlin}
\end{align}
Here, the term $(5/2)RT\partial_{\alpha}p$ in the balance equation (\ref{eq:balanceq}) conspired with the local equilibrium part of the closure (\ref{eq:closureTeq}) to produce the NSF contribution, $\dot{q}^{\rm NSF}_{\alpha}$ (\ref{eq:qdotNSF}). That one gives rise to the Fourier law in the first approximation, through balancing the relaxation term. The term $\dot{q}^{\rm lin}_{\alpha}$ (\ref{eq:qdotlin}) appears with right sign thanks to Grad's closure approximation of the nonequilibrium part of the $T$-flux (\ref{eq:closureTnoneq}). We term it "linear" for the reason explained above, even though it is non-linear through multiplication with the temperature. Similarly to (\ref{eq:sdotlin}), contribution of $\dot{q}^{\rm lin}_{\alpha}$ does not vanish under linearization. Finally, the nonlinear part of Grad's heat flux equation, $\dot{q}^{\rm nlin}_{\alpha}$ (\ref{eq:qdotnlin}) contains a mixture of terms both present already in the balance equation (\ref{eq:balanceq}) and those resulting from the closure assumption. Note that the "most nonlinear" term,  $(1/\rho)\sigma_{\alpha\beta}\partial_{\gamma}\sigma_{\gamma\beta}$ is purely kinematic and is not affected by  Grad's closure.

\paragraph{Hydrodynamics from Grad's equations}
When the smallness parameter is re-introduced into Grad's equations, the first-order correction to the Euler's equations ($\bm{\sigma}^{(0)}=0$, $\qq^{(0)}=0$) is found by balancing the first and the last terms in \eqref{eq:Gsdotfull} and \eqref{eq:Gqdotfull}. Using the estimates for the relaxation rates \eqref{eq:RsGrad} and \eqref{eq:RqGrad}, together with \eqref{eq:sdotNSF} and \eqref{eq:qdotNSF}, we find the nonequilibrium pressure tensor and heat flux as
\begin{align}
		\sigma_{\alpha\beta}^{(1)}&=-\mu_0 \left(\partial_{\alpha}u_{\beta}+\partial_{\beta}u_{\alpha}-\frac{2}{3}\delta_{\alpha\beta}\partial_{\gamma}u_{\gamma}\right),\label{eq:NSGrad}\\
        q_{\alpha}^{(1)}&=-\frac{15R}{4}\mu_0\partial_{\alpha}T.\label{eq:FGrad}
	\end{align}
Thus, we once again derive the Navier--Stokes--Fourier system \eqref{eq:DtnNS}, \eqref{eq:DtuNS} and \eqref{eq:DtTNS}, for the ideal gas with specific heat at constant pressure $c_p=(5/2)R$ and Prandtl number ${\rm Pr}=2/3$. 

Finally, we mention that, even when linearized, Grad's equations (\ref{eq:Gsdotfull}) and (\ref{eq:Gqdotfull}) remain strongly coupled through their respective linear terms, (\ref{eq:sdotlin}) and (\ref{eq:qdotlin}). The linearized Grad's equations can be used to study the hydrodynamic limit beyond the Navier--Stokes--Fourier approximation by \emph{summing up} a rather involved Chapman-Enskog series expansion. Interested reader is directed to \cite{gorban_short-wave_1996,karlin_hydrodynamics_2002}.

\subsubsection{Quasi-equilibrium projection}
\label{sec:QE}

Grad's method and its variants construct closed systems of equations for macroscopic variables when the latter are
represented by moments of the distribution function $f$ (hence their  alternative  name, the moment method). 
A different but closely related construction is the \emph{maximum entropy method} or the \emph{quasi-equilibrium approximation}.  
Let $M=\{M_1,\dots,M_k\}$ be a finite set of moments chosen to describe the macroscopic state,
\begin{equation}
	M_{i}({\xx},t)=\int \mu _{i}({\vv})f({\vv},{\xx},t)d^3{\vv},\ i=1,\dots,k,
	\label{Tri2.1}
\end{equation}
where  $\mu _{1}({\vv}),\ldots ,\mu _{k}({\vv})$ are the corresponding microscopic
densities of  the  moments. We assume that the microscopic densities are linearly independent and that locally conserved fields are included in the set $M$, that is, the linear envelope of $\{\mu _{1}({\vv}),\ldots ,\mu _{k}({\vv})\}$ is included into the linear envelope of $\{\mu _{1}({\vv}),\ldots ,\mu _{k}({\vv})\}$,
\begin{equation}
	\label{eq:EntMax_conservation}
	{\rm Lin}\{1,\vv,v^2\}\subset{\rm Lin}\{\mu _{1}({\vv}),\ldots ,\mu _{k}({\vv})\}.
\end{equation}
The distribution function of the quasi-equilibrium state is defined as the minimizer of the $H$-function \eqref{eq:HBoltzmann} under the constraint of fixed moments $M$,

%
\begin{equation}
f^*(\vv,M)={\rm argmin}\left\{H(f) \left|\int \mu _{1}fd^3{\vv}=M_1,\dots, \int \mu _{k}fd^3{\vv}=M_k\right\}\right.. 
 \label{Tri1.5}
\end{equation} 
where
The family of the quasi-equilibrium distribution functions
$f^{*}({\vv},M)$ \eqref{Tri1.5} 
parametrically depends on the moments $M$ and is a generalization of the local equilibrium distribution function. 
The latter corresponds to the choice of $M$ as a set of the locally conserved moments and is a subset of the distribution functions \eqref{Tri1.5} by the convention \eqref{eq:EntMax_conservation}.
The \emph{quasi-equilibrium projection} is a closed system for the target moments $M$,
\begin{equation}
	\partial_{t} M_{i}+\bnabla\cdot\int{\mu}_{i}(\vv){\vv} f^{*}({\vv},M)d^3\vv=\int{\mu}_{i}(\vv)\mathcal{J}_B(f^{*}({\vv},M))d^3\vv,\ i=1,\dots,k.
	\label{Tri1.6}\end{equation}
%

Important feature of the quasi-equilibrium projection is that it retains the dissipative property of the Boltzmann equation: The 
 quasi-equilibrium $H$-function,
\begin{equation}
	H^{*}(M)=\int f^{*}({\vv},M)\ln f^{*}({\vv},M)d^3\vv,\label{eq:Hqe}
	\end{equation} 
is a non-increasing function due to the moment equations
(\ref{Tri1.6}).
This follows directly  from the solution  of  the  problem
(\ref{Tri1.5}) using  the method of Lagrange multipliers,
\begin{equation}
f^{*} =\exp \sum_{i=1}^k \lambda_{i}  \mu_{i} ({\vv}), \label{eq:QElagrange}
\end{equation} where
$\lambda_{i}$ are Lagrange multipliers. 
Indeed, by noticing that $\lambda_i=\partial H^*/\partial M_i$, the production of the quasi-equilibrium $H$-function \eqref{eq:Hqe} is obtained as,
\begin{align}
	\sigma^*=\sum_{i=1}^k\frac{\partial H^*}{\partial M_i}\int{\mu}_{i}(\vv)\mathcal{J}_B(f^{*}({\vv},M))d^3\vv
	        =\int{\ln}f^{*}({\vv},M)\mathcal{J}_B(f^{*}({\vv},M))d^3\vv\le 0.
	        \label{eq:QE_dissipation}
	\end{align}
A rationale behind the maximum entropy method can be stated as follows: A state of the gas can be described  by  a  finite set of
moments $M$ on a time scale $\theta$ only if all the other (``fast")  moments evolve on a shorter time scale time $\tau\ll\theta$
to their  values  determined  by the chosen set of slow moments $M$, while the slow moments do not change appreciably over the time
scale $\tau$. In the process of the fast relaxation, the $H$-function decreases,  and  in the end  of  this fast relaxation process a
quasi-equilibrium state sets in with  the distribution function being the solution  of the problem (\ref{Tri1.5}).  After the fast processes reach their completion,  the
moments $M$ evolve on the slow time scale by virtue of (\ref{Tri1.6}).

Explicit solution to the minimization problem amounts to expressing the Lagrange multipliers in terms of the moments upon resolving the implicit system of equations,
\begin{align}\label{eq:QEinversion}
	\int \mu_i(\vv)\exp \sum_{j=1}^k \lambda_{j}  \mu_{j} ({\vv})d^3\vv=M_i,\ i=1,\dots,k.
\end{align}
While the dissipation property \eqref{eq:QE_dissipation} of the quasi-equilibrium projection \eqref{Tri1.6} can be demonstrated without solving the equations for the Lagrange multipliers \eqref{eq:QEinversion}, operating equations  \eqref{Tri1.6} does require the knowledge of the quasi-equilibrium distribution function $f^*(\vv,M)$ explicitly. 
To that end, there is exactly \emph{one} case known where the system \eqref{eq:QEinversion} can be inverted in a closed form.
This corresponds to the moments set $M=\{\rho,\rho\uu,\bm{P}\}$ where $\bm{P}$ is the pressure tensor \eqref{eq:P}. Using the representation \eqref{eq:Pexpanded}, we can write \cite{kogan_derivation_1965,lewis_unifying_1967},
\begin{equation}
	f^*(n,\bm{u},\bm{\Pi})=n\frac{\rho^{\frac{3}{2}}}{\left(2\pi\right)^{\frac{3}{2}}\sqrt{\det \bm{\Pi}}}\exp\left(-\frac{\rho}{2}(\bm{v}-\bm{u})^{\dagger}\cdot\bm{\Pi}^{-1}\cdot(\bm{v}-\bm{u})\right),\label{eq:GaussAnisotropic}
\end{equation}
where $\bm{\Pi}=\bm{P}-\rho\uu\uu$ is the pressure tensor in the co-moving reference frame, $\Pi_{\alpha\beta}=p\delta_{\alpha\beta}+\sigma_{\alpha\beta}$, while the non-equilibrium part $\sigma_{\alpha\beta}$ is the trace-free tensor \eqref{eq:sigma}. Vanishing of the latter returns the subset $M^{\rm eq}=\{\rho,\rho\uu,tr[\bm{P}]\}$, while \eqref{eq:GaussAnisotropic} becomes the local Maxwell--Boltzmann distribution function \eqref{eq:BlocalMaxwell}.
The quasi-equilibrium projection defined by the distribution function \eqref{eq:GaussAnisotropic} results in the ten-moment Grad's approximation for a compressible, viscous but heat non-conductive fluid (${\rm Pr}\to\infty$), which is more conveniently written using the pressure rather than the temperature:
\begin{align}
&	D^{*}_t\rho =-\rho\partial_\alpha u_\alpha,\\
&	D^{*}_t u_\alpha =-\frac{1}{\rho}\partial_{\alpha}p-\frac{1}{\rho}\partial_{\beta}{\sigma}_{\alpha\beta},\\
&	D^{*}_t p =-\frac{5}{3}p(\partial_{\alpha}u_\alpha)-\frac{2}{3}\sigma_{\alpha\beta}(\partial_\beta u_\alpha).\\
&	D^{*}_t\sigma_{\alpha\beta}=\dot{\sigma}^{\rm NSF}_{\alpha\beta}+\dot{\sigma}^{\rm nlin}_{\alpha\beta}+ R^{\sigma \mathcal{G}}_{\alpha\beta},\label{eq:G10sdotfull}
	%
\end{align}
where the right hand side of \eqref{eq:G10sdotfull} is given by \eqref{eq:sdotNSF}, \eqref{eq:sdotnlin} and \eqref{eq:elementS}.


\subsubsection{Triangle entropy method} 
A remedy for the explicit construction of quasi-equilibrium approximations for general macroscopic variables has been proposed in \cite{karlin_relaxation_1989,gorban_quasiequilibrium_1991}.
%
Let a set of macroscopic variables be specified as follows: 
First, a subset of linear functionals (moments) $M$ is defined as before, see \eqref{Tri2.1}.
Second, a subset of nonlinear functionals  (in a general case) $N$ is defined 
${N}(f)=\left\{ {N}_{1}(f),\ldots,{N}_{l}(f)\right\}$. 
Examples of nonlinear macroscopic variables are production rates \eqref{eq:Brate}, \eqref{eq:Rq} or \eqref{eq:Rs}.
The totality of the macroscopic variables is the set $\{M,N\}$. 

The triangle entropy method proceeds as follows: First, the quasi-equilibrium approximation is obtained for the subset $M$, as above in \eqref{Tri1.5} to get the quasi-equilibrium distribution function $f^*(\vv,M)$.
%
Second, we seek a correction to  $f^*(\vv,M)$ in the form,
\begin{equation}
	f=f^{*}(1+\varphi ),\label{Tri1.8}
\end{equation}
\noindent where $\varphi $  is  a  deviation  from   the   first quasi-equilibrium approximation due to the macroscopic variables $N$. 
In order to determine $\varphi$, the  second  quasi-equilibrium approximation is constructed. 
Let us denote $\tilde{H}_{f^{*}}(\varphi)$ as  the  quadratic term in the expansion of the $H$-function into powers of $\varphi $  in  the
neighborhood   of   the   first   quasi-equilibrium   state $f^{*}$,
\begin{equation}
	\tilde{H}_{f^{*}}(\varphi)=H^{*}(M)+ \int 	f^{*}({\vv},M)\left(\ln f^{*}({\vv},M)+1\right)\varphi ({\vv}) d^3{\vv}
	+{1\over 2} \int f^{*}({\vv},M)\varphi ^{2}({\vv}) d^3{\vv},
	\label{Tri2.5} \end{equation} 
\noindent where $H^{*}(M)=H(f^{*})$ is the value of the $H$-function  in the quasi-equilibrium state $f^*$ \eqref{eq:Hqe}.
The distribution function   of    the    second
quasi-equilibrium approximation is the solution to the problem,
\begin{align}\label{Tri1.9}
	&\tilde{H}_{f^*}(\varphi) \rightarrow \hbox{min}, \\
	& \int{\mu}_if^*\varphi d^3\vv= 0,\ i=1,\ldots, k, \label{eq:TriHom}\\
	&D_f{N}_{j}\left|_{f^*}\varphi\right. = \Delta N_j, j=1,\ldots,l,\label{eq:TriDiff}
\end{align}
 where $D_f{N}_{j}\left|_{f^*}\right.$ are linear  operators, i.e. the first differential of the  operators $N_j$ at the first quasi-equilibrium $f^*$,
 while $\Delta N_j$ are the deviations of the macroscopic variables from their values at the first quasi-equilibrium state,
 \begin{equation}
 	\Delta N_j=N_j-N_j^*(M).\label{eq:TriDeltaN}
 \end{equation}
Note the importance of the homogeneous constraints \eqref{eq:TriHom} in the minimization problem (\ref{Tri1.9}).
They reflect the condition that the variables $\Delta N$ are ``slow" in the same sense as the variables $M$, at least in a small neighborhood of the first quasi-equilibrium $f^*$.  

Because the optimization functional in the minimization problem (\ref{Tri1.9}) is quadratic in $\varphi$, and thanks to the linearity of the constraints \eqref{eq:TriHom} and \eqref{eq:TriDiff}, the solution is always available in closed form. The resulting  distribution function shall be referred to as the second quasi-equilibrium,
\begin{equation}
	f^*(\vv,M,N)=f^{*}({\vv},M)(1+\varphi^*({\vv},M,N-N^*(M))),
	\label{Tri1.10}
\end{equation}
\noindent in order to make a distinction with the first quasi-equilibrium $f^*(\vv,M)$. By construction, the function $\varphi^*$ depends linearly on the macroscopic variables of the second quasi-equilibrium $\Delta N_j$ \eqref{eq:TriDeltaN}. Assuming that the first quasi-equilibrium can be also obtained in a closed form, the triangle entropy method makes it possible to extend the maximum entropy construction to wider classes of macroscopic variables. Below, we shall consider a few pertinent realizations.


\subsubsection{Grad's projections via triangle entropy method}

Let us consider the example of using the  triangle entropy method, when all the macroscopic variables of the first and of  the
second quasi-equilibrium states are  moments of the distribution function. In other words, the macroscopic variables $M$ of the first quasi-equilibrium remain as in \eqref{Tri2.1}, \eqref{eq:EntMax_conservation}, while the macroscopic variambes $N$ of the second quasi-equilibrium state $N$ are identified with the corresponding microscopic densities $\nu_{1}({\vv}),\ldots ,\nu _{l}({\vv})$,
%
%
\begin{equation}
	N_{j}({\xx},t)=\int \nu _{j}({\vv})f({\vv},{\xx},t) d^3{\vv},\ j=1,\dots,l.
	\label{Tri2.2}
\end{equation} 
Turning to the minimization problem \eqref{Tri1.9}, in order to take the homogeneous constraint \eqref{eq:TriHom} automatically into account, it is convenient to introduce the following structure of inner product: 
\begin{enumerate}
	\item Define a scalar product,
		\begin{equation}
			(\psi _{1},\psi _{2})=\int f^{*}({\vv},M)\psi _{1}({\vv})\psi_{2}({\vv})d^3{\vv}. \label{Tri2.8}\end{equation}
		\item Let
		$E_{\mu}$ be the linear hull of the set  of  moment densities $
		\mu _{1}({\vv}),\ldots,\mu _{k}({\vv})$. 
		Let us construct a 		basis of  $E_{\mu}$ $\{ e_{1}({\vv}),\ldots,e_{r}({\vv} )\}$ that
		is orthonormal in the sense of the scalar product (\ref{Tri2.8}):
		\begin{equation}
			(e_{i},e_{j})=\delta _{ij},\ i,j=1,\ldots ,r, \label{Tri2.9}
		\end{equation} 
	\noindent
		where $\delta _{ij}$ is the Kronecker delta.
	\item Define a projector $\hat{P}^{*}$ on the first
		quasi-equilibrium state,
		\begin{equation}
			\hat{P}^{*}\psi  =\sum_{i=1}^r e_{i}(e_{i},\psi ).
			\label{Tri2.10}\end{equation} The projector $\hat{P}^{*}$ is
		orthogonal: for any pair of functions $\psi _{1}, \psi _{2}$,
		\begin{equation}
			(\hat{P}^{*}\psi _{1},(\hat{1}-\hat{P}^{*})\psi _{2})=0,
			\label{Tri2.11}\end{equation} \noindent where $\hat{1}$ is the unit
		operator. 
\end{enumerate}	
With this,  the homogeneous condition \eqref{eq:TriHom} in the minimization problem  \eqref{Tri1.9} amounts to
		\begin{equation}
			\hat{P}^{*}\varphi =0, \label{Tri2.12}\end{equation} \noindent and
		the expression for the quadratic part of the $H$-function
		(\ref{Tri2.5}) takes  the form,
		\begin{equation}
			\tilde{H}_{f^{*}}(\varphi)=H^{*}(M)+(\ln f^{*}+1,\varphi )+\frac{1}{2}(\varphi,\varphi). \label{Tri2.13}
		\end{equation}
Now, let us note that the function $\ln  f^{*}$  is  invariant with
respect to the action of the projector $\hat{P}^{*}$:
\begin{equation}
	\hat{P}^{*} (\ln f^{*}+1) =\ln f^{*}+1. \label{Tri2.14}
\end{equation}
This follows directly from the solution of the first quasi-equilibrium problem using of the method of Lagrange multipliers \eqref{eq:QElagrange} and also from the assumption that conservation laws are included into the set of moments $M$  \eqref{eq:EntMax_conservation}.
Thus, if the condition
(\ref{Tri2.12}) is satisfied, then from (\ref{Tri2.11}) and
(\ref{Tri2.14}) it follows that $$ (\ln  f^{*}+1,\varphi
)=(\hat{P}^{*}(\ln f^{*}+1),(\hat{1}-\hat{P}^{*})\varphi )=0. $$
Condition (\ref{Tri2.12}) is satisfied automatically, if  the macroscopic variables  $\Delta
N_{i}$ \eqref{eq:TriDeltaN} are defined as follows:
\begin{equation}
	\Delta N_{i}=((\hat{1}-\hat{P}^{*})\nu _{i},\varphi ), i=1,\ldots ,
	l.  \label{Tri2.15}
\end{equation}
Then   the   problem   (\ref{Tri1.9})   of   finding    the second quasiequilibrium state reduces to
\begin{align}
	(\varphi ,\varphi
	)\rightarrow  \min  \hbox{ for }
	((\hat{1}-\hat{P}^{*})\nu _{i},\varphi )=\Delta N_{i},\ 
	i=1,\ldots,l. \label{Tri2.16}
\end{align}

In the remainder of this section we demonstrate how the triangle entropy method is related to Grad's moment method.
To that end, we  take  the  five collision invariants as moment densities of  the first quasi-equilibrium  state,
\begin{equation}
	\mu _{0}=1, \mu _{\alpha}=v_{\alpha}, \ \mu _{4}={mv^{2}
		\over 2}, \label{Tri2.17}\end{equation} 
	\noindent 
Then the first quasi-equilibrium state is characterized with the  local
Maxwell--Boltzmann distribution function $f^{\rm eq}$ \eqref{eq:BlocalMaxwell}.
%
%
%
%
%
Orthogonalization of the set of moment densities  (\ref{Tri2.17}) with the weight \eqref{eq:BlocalMaxwell} delivers one of the possible
orthonormal basis as
\begin{align}
	e_{0}&={5 \over (10n)^{1/2}}-
	{({\vv}-{\uu})^{2} \over (10n)^{1/2}RT},\\
	e_{\alpha}&={(v_{\alpha}-u_{\alpha}) \over (nRT)^{1/2}},
	\label{Tri2.19}  \\
	e_{4}&={({\vv}-{\uu})^{2} \over (15n)^{1/2}RT}.
\end{align}
%
We shall proceed with specific cases of the moments of the second quasi-equilibrium.

\paragraph{Ten-moments Grad's approximation}

For the moment densities of the second quasiequilibrium state, let us
take,
\begin{equation}
	\nu _{\alpha\beta}=mv_{\alpha}v_{\beta}. \label{Tri2.20} 
\end{equation}
\noindent Then
\begin{equation}
	\left(\hat{1}-\hat{P}^{*}\right)\nu _{\alpha\beta}=m(v_{\alpha}-u_{\alpha})(v_{\beta}-u_{\beta})-{1\over
		3}\delta _{\alpha\beta}m({\vv}-{\uu})^{2}, \label{Tri2.21}\end{equation}
\noindent and, since $((\hat{1}-\hat{P}^{(0)})\nu
_{\alpha\beta},(\hat{1}-\hat{P}^{(0)})\nu _{\gamma\kappa})= (\delta _{\alpha\gamma}\delta
_{\beta\kappa}+\delta _{\alpha\kappa}\delta _{\beta\gamma})pRT$,  where $p=nk_{\rm
	B}T$ is  the  pressure,  and $\sigma
_{\alpha\beta}=(f,(\hat{1}-\hat{P}^{*})\nu _{\alpha\beta})$   is the traceless part
of the stress tensor, from  \eqref{Tri2.16} we obtain the distribution
function of the second quasi-equilibrium state in the form,
\begin{equation}
	\mathcal{G}_{10}=f^{\rm eq} \left( 1+{\sigma _{\alpha\beta}\over 2pRT} \left(
	(v_{\alpha}-u_{\alpha})(v_{\beta}-u_{\beta})-{1\over 3} \delta _{\alpha\beta}({\vv}-{\uu})^{2}
	\right)\right). \label{Tri2.22}
\end{equation}
\noindent This is  the  distribution  function  of  the ten-moment Grad's approximation, cf. Eq.\ \eqref{eq:G13sigma}. Note that the Grad's distribution function \eqref{Tri2.22} is the first-order expansion of the quasi-equilibrium distribution \eqref{eq:GaussAnisotropic} in terms of the nonequilibrium pressure tensor $\sigma_{\alpha\beta}$. The Grad's projection established by the ten-moment approximation \eqref{Tri2.22} is identical to the quasi-equilibrium system \eqref{eq:G10sdotfull}.

\paragraph{Thirteen-moments Grad approximation}

In addition  to  (\ref{Tri2.17}), (\ref{Tri2.20}), let us extend the
list of  moment  densities  of the second quasiequilibrium state
with the functions
\begin{equation}
	\xi _{\alpha}={mv_{\alpha}v^{2}\over 2}.
	\label{Tri2.23}\end{equation}

\noindent The corresponding orthogonal complements to the projection
on the first quasi-equilibrium state are

\begin{equation}
	(\hat{1}-\hat{P}^{*})\xi _{\alpha}={m\over 2}(v_{\alpha}-u_{\alpha}) \left(
	({\vv}-{\uu})^{2}-{5RT} \right).
	\label{Tri2.24}\end{equation}

\noindent The moments corresponding to  the  densities
$(\hat{1}-\hat{P}^{(0)})\xi _{\alpha}$  are  the components of the heat
flux vector $q_{\alpha}$,

\begin{equation}
	q_{\alpha}=(\varphi ,(\hat{1}-\hat{P}^{*})\xi _{\alpha}).
	\label{Tri2.25}\end{equation}
\noindent
Since $ ((\hat{1}-\hat{P}^{*})\xi
_{\alpha},(\hat{1}-\hat{P}^{*})\nu _{\beta\gamma})=0$, the constraints $ ((\hat{1}-\hat{P}^{*})\nu _{\alpha\beta},\varphi
)=\sigma _{\alpha\beta}$ and $((\hat{1}-\hat{P}^{*})\xi _{\gamma},\varphi )=q_{\gamma}$
in the problem (\ref{Tri2.16}) are independent,  and Lagrange
multipliers corresponding to $\xi _{\alpha}$ are    ${(1/ 5n)} \left({RT} \right)^{2} q_{\alpha}$.
Thus,  taking into account (\ref{Tri2.17}), (\ref{Tri2.22}),
(\ref{Tri2.24}), we find the distribution function
of the second quasi-equilibrium state,
\begin{equation}
	\mathcal{G}_{13}=f^{\rm eq} \left(1+{\sigma _{\alpha\beta} \over 2pRT}
	\left((v_{\alpha}-u_{\alpha})(v_{\beta}-u_{\beta})-{1\over 3}\delta
	_{\alpha\beta}({\vv}-{\uu})^{2} \right) +
	{q_{\alpha}\over pRT}(v_{\alpha}-u_{\alpha}) \left(
	{({\vv}-{\uu})^{2}\over 5RT} - 1 \right) \right) ,
	\label{Tri2.27} \end{equation}

\noindent which  coincides  with  the  thirteen-moment   Grad
distribution function \eqref{eq:G} \cite{grad_kinetic_1949}.

\paragraph{{Eight- and fifteen-moments Grad's approximations}}
We note two other quasi-equilibrium approximations of interest. 
If only the heat flux \eqref{Tri2.25} is chosen as the slow variable then the second quasi-equilibrium approximation becomes
\begin{equation}
	\mathcal{G}_8=f^{\rm eq} \left(1+
	{q_{\alpha}\over pRT}(v_{\alpha}-u_{\alpha}) \left(
	{({\vv}-{\uu})^{2}\over 5RT} - 1 \right) \right).
	\label{eq:G8} \end{equation}
Distribution function \eqref{eq:G8} is used in the construction of the Shakhov's S-model \cite{shakhov_generalization_1972} to be discussed below.
Grad's projection established by the distribution function \eqref{eq:G8} corresponds to a non-viscous compressible, thermally conductive fluid (${\rm Pr}\to 0$),
\begin{align}
	&	D^{*}_t\rho =-\rho\partial_\alpha u_\alpha,\\
	&	D^{*}_t u_\alpha =-\frac{1}{\rho}\partial_{\alpha}p,\\
	&	D^{*}_t p =-\frac{5}{3}p(\partial_{\alpha}u_\alpha)-\frac{2}{3}(\partial_\alpha q_\alpha).\\
	&	D^{*}_tq_{\alpha}=-\frac{5}{2}Rp\partial_{\alpha}T -\frac{7}{5}q_{\alpha}\partial_{\beta}u_{\beta}
	-\frac{7}{5}q_{\beta}\partial_{\beta}u_{\alpha}
	-\frac{2}{5}q_{\beta}\partial_{\alpha}u_{\beta}+ R^{q \mathcal{G}}_{\alpha}.
	\label{eq:G8sdotfull}
\end{align}
It is interesting to note that, while the ten- and the eight-moments Grad's projections have little practical usefulness by themselves, they nevertheless are helpful in constructing the kinetic models in order to overcome the restriction of the BGK model.
Finally, if the full $Q$-flux \eqref{eq:Q} is used instead of the heat flux, the corresponding fifteen-moments second quasi-equilibrium distribution function is used in the construction of a family of lattice Boltzmann models with variable Prandtl number \cite{frapolli_multispeed_2014}.
Various Grad's projection mentioned above are collected in Table \ref{tab:Grad}.

\begin{table}
	\begin{center}
		\begin{tabular}{l|l|l}
			Projection&Local conservation&Target nonequilibrium moments\\
			$G5$& $\rho$, $\rho\bm{u}$, $tr[\bm{P}]$ & --\\
			$G8$& $\rho$, $\rho\bm{u}$, $tr[\bm{P}]$ & $\bm{q}$\\
			$G10$& $\rho$, $\rho\bm{u}$, $tr[\bm{P}]$ & $\bm{P}-\frac{1}{3}tr[\bm{P}]\bm{I}$ \\
			$G13$& $\rho$, $\rho\bm{u}$, $tr[\bm{P}]$ & $\bm{P}-\frac{1}{3}tr[\bm{P}]\bm{I}$, $\bm{q}$ \\
			$G15$& $\rho$, $\rho\bm{u}$, $tr[\bm{P}]$&$\bm{Q}$\\
			$G20$& $\rho$, $\rho\bm{u}$, $tr[\bm{P}]$ & $\bm{P}-\frac{1}{3}tr[\bm{P}]\bm{I}$, $\bm{Q}$
		\end{tabular}
		\caption{Grad's projections mentioned in this contribution.}
		\label{tab:Grad}
	\end{center}
\end{table}

\subsection{Dynamic correction to Grad's thirteen-moments projection: The R13 system}
 
\subsubsection{Invariance defect  of Grad's thirteen-moments approximation}\label{sec:DeltaG}
Nothing tells us that Grad's closure relations for both the $Q$- and $T$-fluxes, as well as the closure relation for the relaxation term, will stay invariant under kinetic equation (\ref{eq:BE}). One thus needs, first, to quantify the discrepancy between the proposed projection and the real dynamics due to kinetic equation, to understand the physical mechanisms arising from this discrepancy and which were neglected by the projection, and, secondly, to correct the closure on the basis of the kinetic equation. 

Systematic method to derive dynamic correction to Grad's G13 projection from kinetic equations was introduced in \cite{karlin_dynamic_1998} (KGDN thereafter), and has been realized for near-equilibrium conditions. 
The dynamic correction \cite{karlin_dynamic_1998} followed the path similar to that of the correction to the local equilibrium projection discussed above: first, one evaluates the defect of invariance of the projection, and second, one finds the first iteration of the invariance condition in order to compensate the defect in the lowest-order Knudsen number approximation.
Later, taking a different route via a superset moment system with $26$ moments, Struchtrup and Torrilhon \cite{struchtrup_regularization_2003} proposed a nonlinear extension thereof, and coined the name of regularized Grad's system, or R13, which we here stick to.


%
Evaluation of the invariance defect of Grad's approximation is rather straightforward, 
and has been first reported by KGDN \cite{karlin_dynamic_1998} in the linear approximation, and worked out in detail in \cite{karlin_derivation_2018} for the full nonlinear case.
We begin with evaluating the derivatives of Grad's distribution function with respect to all the thirteen fields, while separating the contributions of the Maxwellian already available from (\ref{eq:DnM}), (\ref{eq:DuM}) and (\ref{eq:DTM}) from that of the newly added nonequilibrium part:

\begin{align}
	\frac{\partial \mathcal{G}}{\partial \rho}&=
	\frac{\partial \mathcal{M}}{\partial \rho},\\
	\frac{\partial \mathcal{G}}{\partial u_{\alpha}} &=
	\frac{\partial \mathcal{M}}{\partial u_{\alpha}}+\frac{\partial \mathcal{N}}{\partial u_{\alpha}},\\
	\frac{\partial \mathcal{G}}{\partial T}&=
	\frac{\partial \mathcal{M}}{\partial T}+\frac{\partial \mathcal{N}}{\partial T},
\end{align}
where
\begin{align}
	\frac{\partial \mathcal{N}}{\partial u_{\alpha}}&=\mathcal{M}\left(\mathcal{G}^{\sigma}+\mathcal{G}^{q}\right)\left(\frac{C_{\alpha}}{v_T}\right)
	-\mathcal{M}\left\{\frac{2q_{\alpha}}{5pv_T^2}\left(\frac{C^2}{2}-\frac{5}{2}\right)+\frac{2}{5pv_T^2}q_{\beta}C_{\beta}C_{\alpha}+\frac{\sigma_{\alpha\beta}C_{\beta}}{pv_T}\right\},\\
	\frac{\partial \mathcal{N}}{\partial T}&=
	\mathcal{M}\left\{\mathcal{G}^{\sigma}\left(\frac{C^2-7}{2T}\right)+\mathcal{G}^q\left(\frac{C^2-9}{2T}\right)\right\}
	-\mathcal{M}\left(\frac{q_{\alpha}C_{\alpha}}{pv_TT}\right).
\end{align}
Finally, derivatives with respect to the nonequilibrium fields are
\begin{align}
	\frac{\partial \mathcal{G}}{\partial\sigma_{\alpha\beta}}&=
	\mathcal{M}\left(\frac{1}{2p}\right)\left(C_{\alpha}C_{\beta}-\frac{1}{3}\delta_{\alpha\beta}C^2\right),\label{eq:dGdsigma}\\
	\frac{\partial \mathcal{G}}{\partial q_{\alpha}}&=\mathcal{M} \left( \frac{ C_{\alpha} }{pv_T} \right)\left(\frac{C^2}{5}-1\right).\label{eq:dGdq}
\end{align}
Using these, we write down the invariance defect of Grad's thirteen moment approximation separating contributions of propagation terms from those of the collisions. Following KGDN convention, the former are termed non-local and the latter local:
\begin{align}\label{eq:DGtotal}
	\Delta_{\mathcal{G}}&=\Delta^{\rm loc}_{\mathcal{G}}+\Delta^{\rm nloc}_{\mathcal{G}}.
\end{align}
The local piece reads,
\begin{align}\label{eq:Dloc}
	\Delta^{\rm loc}_{\mathcal{G}}=&\left[\frac{\partial \mathcal{G}}{\partial\sigma_{\alpha\beta}}R_{\alpha\beta}^{\sigma\mathcal{G}}
	+\frac{\partial \mathcal{G}}{\partial q_{\alpha}}R_{\alpha}^{q\mathcal{G}}\right]-\mathcal{J}_B(\mathcal{G})\nonumber\\
	=&\left[\mathcal{M}\left(\frac{1}{2p}\right)\left(C_{\alpha}C_{\beta}-\frac{1}{3}\delta_{\alpha\beta}C^2\right)R_{\alpha\beta}^{\sigma\mathcal{G}}
	-L(\mathcal{M}\mathcal{G}^\sigma)\right]
	+\left[\mathcal{M} \left( \frac{ C_{\alpha} }{pv_T} \right)\left(\frac{C^2}{5}-1\right)R_{\alpha}^{q\mathcal{G}}
	-L(\mathcal{M}\mathcal{G}^{q})\right].
\end{align}
Neglect of quadratic terms here is consistent with their neglect already made in Grad's closure of relaxation terms.

First observation, already exposed in KGDN \cite{karlin_dynamic_1998}, is about vanishing of the local invariance defect in certain cases of collision models:
%
	Let functions $\mathcal{M}[C_{\alpha}C_{\beta}-\frac{1}{3}\delta_{\alpha\beta}C^2]$ and $\mathcal{M}[C_{\alpha}(C^2/2-5/2)]$
	be eigenfunctions of linearized collision integral,
	\begin{align}\label{eq:LinBS}
		&L\left(\mathcal{M}\left[C_{\alpha}C_{\beta}-\frac{1}{3}\delta_{\alpha\beta}C^2\right]\right)
		=-\lambda^{\sigma}\mathcal{M}\left[C_{\alpha}C_{\beta}-\frac{1}{3}\delta_{\alpha\beta}C^2\right],\\
		&L\left(\mathcal{M}\left[C_{\alpha}\left(\frac{C^2}{2}-\frac{5}{2}\right)\right]\right)=-\lambda^{q}\mathcal{M}
		\left[C_{\alpha}\left(\frac{C^2}{2}-\frac{5}{2}\right)\right].
		\label{eq:LinBq}
	\end{align}
	Then the local invariance defect of Grad's approximation vanishes,
	\begin{align}\label{Dlocvanish}
		\Delta_{\mathcal{G}}^{\rm loc}=0.
	\end{align}
Indeed, with the conditions (\ref{eq:LinBS}) and (\ref{eq:LinBq}), the relaxation terms in Grad's approximation become $R_{\alpha\beta}^{\sigma\mathcal{G}}=-\lambda^{\sigma}\sigma_{\alpha\beta}$ and $R_{\alpha\beta}^{q\mathcal{G}}=-\lambda^{q}q_{\alpha}$.
Using these in (\ref{eq:Dloc}), we prove (\ref{Dlocvanish}).

Non-vanishing of local invariance defect for Grad's distribution function is the first apparent difference from the Maxwellian case.
In Grad's case, the relaxation has to be {\it aligned} with the direction dictated by the eigenfunction of the linearized collision integral in order to annihilate the local invariance defect. This is more demanding compared to the local Maxwellian which annuls the local terms in the invariance defect independently of the collision model used. Vanishing of the local invariance defect unifies the relaxation time approximation (and any similar kinetic model) with Maxwell's molecules and is a consequence of the simple fact that all these have the same eigenfunctions of the form (\ref{eq:LinBS}) and (\ref{eq:LinBq}). Vanishing of local invariance defect has no relation to the {\it values} of transport coefficients, and is non-vanishing for any other model, such as hard spheres. 
Now we proceed with the analysis of the nonlocal part of the invariance defect which is {\it independent} of the choice of the collision model.



The nonlocal part of the invariance defect shall be written as a sum of three pieces,
where we distinguish the Navier-Stokes-Fourier, the linear and the nonlinear contributions,
\begin{align}
	\Delta_{\mathcal{G}}^{\rm nloc}&=\Delta_{\mathcal{G}}^{\rm NSF}+\Delta_{\mathcal{G}}^{\rm lin}+\Delta_{\mathcal{G}}^{\rm nlin}.
\end{align}
Each type of contributions is dictated by the partition of the Grad's equations (\ref{eq:Gsdotfull}) and (\ref{eq:Gqdotfull}), and we have,
\begin{align}\label{eq:DNSF}
	\Delta_{\mathcal{G}}^{\rm NSF}=&
	\frac{\partial \mathcal{G}}{\partial \rho}\left[D_t \rho +v_TC_{\beta}\partial_{\beta}\rho\right]
	+\frac{\partial \mathcal{M}}{\partial u_{\alpha}}\left[D_t^{\mathcal{M}} u_{\alpha}+v_TC_{\beta}\partial_{\beta}u_{\alpha}\right]
	+\frac{\partial \mathcal{M}}{\partial T}\left[D_t^{\mathcal{M}} T+v_TC_{\beta}\partial_{\beta}T\right]\nonumber\\
	&+\frac{\partial \mathcal{G}}{\partial\sigma_{\alpha\beta}}\dot{\sigma}^{\rm NSF}_{\alpha\beta}
	+\frac{\partial \mathcal{G}}{\partial q_{\alpha}}\dot{q}^{\rm NSF}_{\alpha},\\
	\label{eq:Dlin}
	\Delta_{\mathcal{G}}^{\rm lin}=&
	\frac{\partial \mathcal{M}}{\partial u_{\alpha}}\left[-\frac{1}{\rho}\partial_{\beta} \sigma_{\alpha\beta}\right]
	+\frac{\partial \mathcal{M}}{\partial T}\left[-\frac{2T}{3p}\partial_{\alpha}q_{\alpha}\right]\nonumber\\
	&+\frac{\partial \mathcal{G}}{\partial\sigma_{\alpha\beta}}\left[\dot{\sigma}^{\rm lin}_{\alpha\beta}+v_TC_{\gamma}\partial_{\gamma}\sigma_{\alpha\beta}\right]
	+\frac{\partial \mathcal{G}}{\partial q_{\alpha}}\left[\dot{q}^{\rm lin}_{\alpha}+v_TC_{\beta}\partial_{\beta}q_{\alpha}\right],\\
	\label{eq:Dnlin}
	\Delta_{\mathcal{G}}^{\rm nlin}=&
	\frac{\partial \mathcal{M}}{\partial T}\left[-\frac{2T}{3p}\sigma_{\alpha\beta}\partial_{\alpha}u_{\beta}\right]
	+\frac{\partial \mathcal{N}}{\partial u_{\alpha}}\left[D_t u_{\alpha}+v_TC_{\beta}\partial_{\beta}u_{\alpha}\right]+\frac{\partial \mathcal{N}}{\partial T}\left[D_t T+v_TC_{\beta}\partial_{\beta}T\right]\nonumber\\
	&+\frac{\partial \mathcal{G}}{\partial\sigma_{\alpha\beta}}\dot{\sigma}^{\rm nlin}_{\alpha\beta}
	+\frac{\partial \mathcal{G}}{\partial q_{\alpha}}\dot{q}^{\rm nlin}_{\alpha}.
\end{align}
%


\subsubsection{The R13 distribution function} The leading-order dynamic correction to Grad's distribution function is written on the basis of the BGK model using a relaxation time $\tau$ as follows:
\begin{align}\label{eq:R13pdf}
	\mathcal{R}=\mathcal{G}+\mathcal{K},
\end{align}
where $\mathcal{G}$ is Grad's distribution and $\mathcal{K}$ is the correction,

\begin{align}\label{eq:Delta_modesR13}
 \mathcal{K}=-\tau	\Delta^{\rm nloc}_\mathcal{G}.
\end{align}

First, we notice that the Navier-Stokes-Fourier part of the invariance defect of Grad's approximation vanishes,
%
	$\Delta_{\mathcal{G}}^{\rm NSF}=0$.
%
Indeed, we notice that the first three terms in (\ref{eq:DNSF}) assemble to the defect of invariance of the local Maxwellian already computed, see Eq.\ \eqref{eq:Mdefect}; evaluating the remaining two terms we get,
\begin{align}
	\Delta_{\mathcal{G}}^{\rm NSF}&=\Delta_{\mathcal{M}}\nonumber\\
	&+\mathcal{M}\left(\frac{1}{2p}\right)\left(C_{\alpha}C_{\beta}-\frac{1}{3}\delta_{\alpha\beta}C^2\right)\left[-p\left(\partial_{\alpha}u_{\beta}+\partial_{\beta}u_{\alpha}-\frac{2}{3}\delta_{\alpha\beta}\partial_{\gamma}u_{\gamma}\right)\right]\nonumber\\
	&+\mathcal{M} \left( \frac{2 C_{\alpha} }{5pv_T} \right)\left(\frac{C^2}{2}-\frac{5}{2}\right)\left[-\frac{5}{2}p\partial_{\alpha}\left(RT\right)\right]\nonumber\\
	&=0.
\end{align}
This observation was first made in KGDN \cite{karlin_dynamic_1998}, and it is not surprising. Indeed, since the Navier--Stokes--Fourier approximation is already fully contained in Grad's equations, there is nothing to correct in Grad's dynamics with respect to the Navier--Stokes--Fourier fluxes.

The remaining therms in the nonlocal defect of invariance of the G13 projection were evaluated in \cite{karlin_dynamic_1998} and  \cite{karlin_derivation_2018}.
The result is written as a combination of eight modes:
\begin{align}\label{eq:modesR13}
	\mathcal{K} & =-\tau\sum_{i=1}^{8}\mathcal{K}^{(i)}.
\end{align}
Each mode has the form,
\begin{align}\label{eq:R13K}
	\mathcal{K}^{(i)}&=\mathcal{M}\bm{P}^{(i)}\bullet \bm{F}^{(i)},
\end{align}
where $\mathcal{M}$ is the local Maxwellian (or {\it mode's amplitude}), $\bm{P}^{(i)}$ is velocity tensor (or {\it mode's direction}),   $\bm{F}^{(i)}$ is the {\it mode's frequency}, while $\bullet$ stands for convolution of tensors. Tensors  $\bm{P}^{(i)}$ are dimensionless and depend only on  $\bm{C}$, particle's velocity relative to the flow $\bm{u}$, reduced by thermal speed, $\bm{C}=(\bm{v}-\bm{u})/v_T$, where 
$v_T=\sqrt{RT}$.
Dimension of the mode's frequencies is inverse of time, $[\bm{F}^{(i)}]\sim 1/{\rm sec}$. 
Modes of R13 are collected in Tab.\ \ref{tab:modes}.
In Tab.\ \ref{tab:modes}, we use shorthand notation for symmetric traceless velocity tensors of rank two, three and four,
\begin{align}
	\notrace{C_{\alpha}C_{\beta}}&=C_{\alpha}C_{\beta}-\frac{1}{3}\delta_{\alpha\beta}C^2,\\
	%
	\notrace{C_{\alpha}C_{\beta}C_{\gamma}}&=C_{\alpha}C_{\beta}C_{\gamma}
	-\frac{1}{5}C^2\left(C_{\alpha}\delta_{\beta\gamma}+C_{\beta}\delta_{\alpha\gamma}+C_{\gamma}\delta_{\alpha\beta}\right),\\
	\notrace{C_{\alpha}C_{\beta}C_{\gamma}C_{\lambda}}&=C_{\alpha}C_{\beta}C_{\gamma}C_{\lambda}
	-\frac{C^4}{15}\left(\delta_{\alpha\beta}\delta_{\gamma\lambda}+\delta_{\alpha\lambda}\delta_{\beta\gamma}+\delta_{\alpha\gamma}\delta_{\beta\lambda}\right).
\end{align}
The three {\it primary modes} $\mathcal{K}^{(1)}$, $\mathcal{K}^{(2)}$ and $\mathcal{K}^{(3)}$ were already identified by KGDN \cite{karlin_dynamic_1998}. 
Corresponding modes frequencies is the only part of the invariance defect that survives linearization around equilibrium \cite{karlin_dynamic_1998}, and thus contribute the conventional \emph{linear nonequilibrium thermodynamics} dissipation to R13. 
The primary modes are accompanied by three nonlinear {\it secondary modes} $\mathcal{K}^{(4)}$,  $\mathcal{K}^{(5)}$ and  $\mathcal{K}^{(6)}$.
Finally, two modes $\mathcal{K}^{(7)}$ and $\mathcal{K}^{(8)}$ are {\it ghost modes}: while they contribute to the R13 distribution function, their projection onto R13 fluxes vanishes, and they are not visible in the R13 balance equations.

\begin{table}[ht!]
	\centering
	\begin{tabular}{l|l|l}
		$\mathcal{K}$ & $\bm{P}$ & $\bm{F}$\\
		1 & $\notrace{\bm{C}\bm{C}\bm{C}}$ & $\frac{\sqrt{RT}}{2p}\notrace{\nabla\bm{\sigma}}
		+\frac{\sqrt{RT}}{2pT}\notrace{\bm{\sigma}\nabla T}
		+\frac{1}{2p\sqrt{RT}}\notrace{\bm{\sigma}D_t \bm{u}}$\\
		2& $(C^2-7)\notrace{\bm{C}\bm{C}}$& $\frac{1}{5p}\notrace{\nabla \bm{q}}
		+\frac{1}{5pRT}\notrace{\bm{q}D_t \bm{u}}
		+\frac{1}{4pT}\bm{\sigma}D_tT$\\
		3& $C^4-10 C^2+ 15 $ & $\frac{1}{15p}\nabla\cdot \bm{q}  + \frac{1}{15p}\bm{\sigma}:\nabla\bm{u}
		+ \frac{1}{15pRT}\bm{q}\cdot D_t\bm{u}$\\
		4 & $(C^2 - 7)\notrace{\bm{C}\bm{C}\bm{C}}$ & $\frac{1}{5p\sqrt{RT}}\notrace{\bm{q}\nabla\bm{u}}$ \\
		5 &  $\notrace{\bm{C}\bm{C}\bm{C}\bm{C}}-{\rm sym}(\notrace{\bm{C}\bm{C}}\bm{U})$ &
		$	\frac{1}{2p}\bm{\sigma}\nabla \bm{u}$ \\
		{6} & ${(C^4-14 C^2 + 35)\bm{C}\bm{C}}$& 
		{$\frac{1}{10pT}
			\bm{q}\nabla T$}\\
		{7} &$(C^2 - 9)\notrace{\bm{C}\bm{C}\bm{C}}$ & $\frac{\sqrt{RT}}{4 p T}\notrace{\bm{\sigma}\nabla T}$\\
		{8} & $(C^4-14 C^2 + 35)\bm{C}$& $\frac{2\bm{q}\cdot\notrace{\nabla \bm{q}}}{25 p\sqrt{RT}}
		-\frac{1}{15 T\sqrt{RT}}\bm{q}\nabla \cdot \bm{q}
		-\frac{1}{15 T\sqrt{RT}}\bm{q}\bm{\sigma}:\nabla \bm{u}
		+\frac{\sqrt{RT}}{10 p T}\bm{\sigma}\cdot\nabla T$
	\end{tabular}
	\caption{Modes of R13. Angular brackets denote symmetrized traceless tensors of rank two, three and four. 
		$D_t=\partial_t +\bm{u}\cdot \nabla$ is the material derivative along streamline; $\bm{u}$ is flow velocity, $p=\rho RT$ is the pressure, $R$ is gas constant, $\rho$ is the density and $T$ is the temperature.}
	\label{tab:modes}
\end{table}

\subsubsection{The R13 equations}
The R13 equations for the nonequilibrium stress tensor $\bm{\sigma}$ and the heat flux $\bm{q}$ are compactly written in  vector notation as follows:
\begin{align}
	D_t^{\mathcal{R}} \bm{\sigma}&=D_t^{\mathcal{G}} \bm{\sigma}-\nabla \cdot \bm{Q}^{\mathcal{K}}, \label{eq:R13dtsigma}\\
	D_t^{\mathcal{R}} \bm{q}&=D_t^{\mathcal{G}}\bm{q}-\nabla \cdot \bm{T}^{\mathcal{K}}-\bm{Q}^{\mathcal{K}}:\nabla\bm{u},\label{eq:R13dtq}
\end{align}
where $D_t$ is the material time derivative along streamline,
\begin{align}\label{eq:Dt}
	D_t=\partial_t + \bm{u}\cdot\nabla,
\end{align}
and $D_t^{\mathcal{G}}$ indicates Grad's contribution. 
The rank three symmetric trace-free  tensor $\bm{Q}^{\mathcal{K}}$ and the  rank two symmetric tensor 
$\bm{T}^{\mathcal{K}}$ are the  R13 {\it fluxes},
\begin{align}
	\bm{Q}^{\mathcal{K}}=&-\nu\left[\frac{3}{2} \notrace{\nabla\bm{\sigma}}
	+\frac{3}{2T}\notrace{\bm{\sigma}\nabla T}
	+\frac{6}{5RT}\notrace{\bm{q}\nabla\bm{u}}\right]
	-\tau\left[\frac{3}{2}\notrace{\bm{\sigma}D_t \bm{u}}\right],
	\label{eq:R13Qflux}\\
	%
	\bm{T}^{\mathcal{K}}=&\notrace{\bm{T}^{\mathcal{K}}}+\frac{1}{3}\bm{U}T^{\mathcal{K}},\label{eq:R13Tfluxtotal}\\
	\notrace{\bm{T}^{\mathcal{K}}}=&-\nu\left[\frac{14}{5}\notrace{\nabla \bm{q}}
	{+\frac{28}{5}\notrace{\bm{q}\nabla\ln T}}+\bm{\sigma}(\nabla\cdot\bm{u})
	+2\notrace{\bm{\sigma}\cdot \bm{S}+\overline{\bm{\sigma}\cdot\bm{S}}}\right]\nonumber\\
	&-\tau\left[\frac{14}{5}\notrace{\bm{q}D_t\bm{u}}+\frac{7}{2}{ R} \bm{\sigma}D_t T\right], 	\label{eq:R13Tflux}\\
	T^{\mathcal{K}}=&-4\nu\left[\nabla\cdot\bm{q} {+\frac{7}{2} \bm{q}\cdot\nabla \ln T} +\bm{\sigma}:\nabla\bm{u}\right]
	-4\tau\bm{q}\cdot D_t\bm{u}. \label{eq:R13Tfluxtrace}
\end{align}
Here angular brackets indicate symmetrized traceless tensors of rank two and three, overline indicates transposition, $\bm{U}$ is unit tensor, $\bm{S}=(\nabla \bm{u}+\overline{\nabla\bm{u}})/2$ is the strain and $\nabla \ln T=T^{-1}\nabla T$. 

Brackets in (\ref{eq:R13Qflux}) and (\ref{eq:R13Tflux})   help to discern  contributions of { two different types}.
The first bracket in (\ref{eq:R13Tflux}) and (\ref{eq:R13Qflux}) is the dissipation flux. First part of dissipation is the {\it linear thermodynamics dissipation flux} (first term in (\ref{eq:R13Qflux}) and (\ref{eq:R13Tflux})). The linear dissipation fluxes are the only contributions in (\ref{eq:R13Qflux}) and (\ref{eq:R13Tfluxtotal}) that survive linearization around equilibrium, as was already shown by KGDN.
The rest of the terms in the first bracket form the {\it nonlinear dissipation flux}, driven by non-uniformity of the macroscopic velocity field and of the temperature. Both types of dissipative fluxes are associated with the kinematic viscosity $\nu=\tau RT$, here in the relaxation time approximation. 

Second bracket in (\ref{eq:R13Qflux}) and (\ref{eq:R13Tflux})  is a remarkably distinct {\it streamline convective flux}, which we term this way because of the material time derivative (\ref{eq:Dt}) participating in their formation. Streamline convective fluxes are
nonlinear and their contribution is  non-dissipative in nature. They are characterized by the {\it relaxation time} $\tau$ rather than by the viscosity. From Tab.\ \ref{tab:modes}, it is visible that streamline convective flux is contributed by primary modes only.
Finally, also the trace $T^{\mathcal{K}}$ \eqref{eq:R13Tfluxtrace} reproduces the said structure.
Together with the balance equations of density, momentum and energy,
\begin{align}
	D_t\rho&=-\rho\nabla\cdot\bm{u},\\
	D_t \bm{u}&=-\frac{1}{\rho}\nabla p - \frac{1}{\rho}\nabla\cdot\bm{\sigma},\\
	D_t T&=-\frac{2}{3}T\nabla\cdot\bm{u}-\frac{2}{3}\left(\frac{T}{p}\right)\bm{\sigma}:\nabla\bm{u}
	-\frac{2}{3}\left(\frac{T}{p}\right)\nabla\cdot\bm{q},
\end{align}
and with  Grad's contribution,
\begin{align}
	D_t^{\mathcal{G}} \bm{\sigma}=&-2p\notrace{\bm{S}}
	-\frac{4}{5}\notrace{\nabla \bm{q}}
	-\bm{\sigma}(\nabla\cdot \bm{u})
	-2\notrace{\bm{\sigma}\cdot\bm{S}}-\frac{1}{\tau}\bm{\sigma},\label{eq:DtsG}\\
	D_t^{\mathcal{G}}\bm{q}=&-\frac{5}{2}R\bm{\Pi}\cdot \nabla T - RT\nabla\cdot\bm{\sigma}-\bm{\sigma}\cdot D_t\bm{u}-\frac{7}{5}\bm{q}(\nabla\cdot\bm{u})
	-\bm{q}\cdot\nabla\bm{u}-\frac{4}{5}\bm{q}\cdot\bm{S}-\frac{1}{\tau}\bm{q},
\end{align}
where $\bm{\Pi}=p\bm{U}+\bm{\sigma}$ is the pressure tensor, \eqref{eq:R13dtsigma} and \eqref{eq:R13dtq} build up the  structure of the R13 system.

	The R13 theory was further refined and extensively studied by Struchtrup and Torrilhon, and their coauthors, in a number of contributions \cite{rana_robust_2013,struchtrup_regularized_2013,struchtrup_derivation_2005,struchtrup_stable_2004,torrilhon_modeling_2016,timokhin_different_2017,struchtrup_h_2007,struchtrup_macroscopic_2005,torrilhon_regularized_2004} that dissect the R13 equations carefully, and show that they can describe all relevant rarefaction phenomena, such as jump and slip at boundaries, Knudsen boundary layers, transpiration flow, thermal stresses, non-temperature-gradient heat flux induced by stresses, damping and dispersion of ultrasound waves and shock structures (for limited Mach numbers). A good summary of this work is referenced and discussed in a recent review \cite{torrilhon_modeling_2016}.


\subsection{Lifting of Grad's and quasi-equilibrium projections: Kinetic models for simple fluid}
\label{sec:kin_models}

\subsubsection{Quasi-equilibrium and related kinetic models}
Lifting of the local equilibrium projection considered in sec.\ \ref{sec:BGK_B} results in the BGK model. 
A variety of kinetic models can be offered by a lifting the 
lifting of the Grad's and related projections. 
We shall discuss now some main classes of kinetic models from this perspective. 
Following \cite{gorban_general_1994}, a kinetic model for a generic quasi-equilibrium approximation $f^*(M)$  \eqref{Tri1.5} can be proposed as follows, 
\begin{equation}
	\partial_t f+\vv\cdot\nabla f=-\frac{1}{\tau}(f-f^*(M))	+\mathcal{J}_B(f^*(M)). \label{gBGK}
\end{equation}
Here $f^*(M)\equiv f^*(M(f))$ is the natural map $f\to f^*(M)$,
\begin{equation}
	\label{gmap}
	\int \mu_i(f-f^*(M)d^3\vv=0,\ i=1,\dots,k,
\end{equation}
and we omit displaying the dependence of the distribution function on the velocity, time and space to simplify notation.
Thus, the first term in the right hand side of equation (\ref{gBGK}) is BGK-like, whereas the second term, the function $\mathcal{J}_B(f^*(M))$, is
the true (Boltzmann) collision integral evaluated on the quasi-equilibrium
	manifold. The latter is crucial: Unlike the true Boltzmann collision integral $\mathcal{J}_B(f)$ which
takes values in the entire phase space of distribution functions, the term 
$\mathcal{J}_B(f^*(M))$ is evaluated only on a relatively "thin" submanifold $f^*$
known a priori, and can be thus {\it pre-computed} to the explicit function
of the moments $M$ and of the velocity $\vv$ (see Ref.\ \cite{gorban_general_1994} for examples).
If the quasi-equilibrium $f^*(M)$ consists only of the local Maxwellians,
then  $\mathcal{J}_B(f^*(M))$ vanishes, and we get back the BGK-model.
In all other cases, the second term in the kinetic model (\ref{gBGK}) is
essential: If it is omitted in equation (\ref{gBGK}) then the
null-space of the resulting collision integral is the entire quasi-equilibrium
manifold $f^*(M)$, and not its local equilibrium submanifold, unlike the case
of the Boltzmann collision integral.

The $H$-theorem for kinetic models (\ref{gBGK}) has the following structure: Let us compute the entropy production $\sigma$:
\begin{align}
	\sigma&=\sigma_1+\sigma_2,\nonumber\\
	\sigma_1&=-\frac{1}{\tau}\int \ln(f) (f-f^*(M)d^3\vv,\nonumber\\
	\sigma_2&=\int \ln(f)J_B(f^*(M))d^3\vv.
\end{align}
Function  $\sigma_1$ is the contribution from the BGK-like relaation term in equation (\ref{gBGK}), and it is  always  non-positive,
again due to the property of the map $f\to f^*(M)$ (\ref{gmap}). 
The second contribution, $\sigma_2$ may be not sign-definite if $f$ is far away from the quasi-equilibrium. 
However, there always exists a non-empty neighborhood
of the quasi-equilibrium manifold, where $\sigma_2\le 0$ \cite{gorban_general_1994} (this is
almost obvious: {\it on} the quasi-equilibrium manifold
$\sigma_2(f^*(M))$ is the  entropy production due to the 
Boltzmann collision integral). Thus, if the relaxation towards
quasi-equilibrium states is fast enough ($\tau$ is sufficiently small), the net entropy production inequality holds,
$\sigma=\sigma_1+\sigma_2\le 0$.

Some further variants of the quasi-equilibrium models (\ref{gBGK}) are possible. 
Let us introduce the \emph{quasi-equilibrium projector} $\mathcal{P}^*$,
\begin{equation}
	\label{eq:QEP}
		\mathcal{P}^*F=\sum_{i=1}^{k}\frac{\partial f^*(M)}{\partial M_i}\int\mu_i Fd^3\vv.
\end{equation}
Instead of the term $\mathcal{J}_B(f^*(M))$ in \eqref{gBGK}, we can use its projection, $\mathcal{P}^*\mathcal{J}_B(f^*(M))$, so that the model \eqref{gBGK} simplifies to 
\begin{equation}\label{eq:gBGK2}
	\partial_t f+\vv\cdot\nabla f=-\frac{1}{\tau}(f-f^*(M)) + \sum_{i=1}^k\frac{\partial f^*(M)}{\partial M_i}R_i^*(M),
\end{equation}
where $R_i^*(M)$ is the quasi-equilibrium production rate of the $i$th moment,
\begin{equation}
	R_i^*(M)=\int \mu_i \mathcal{J}_B(f^*(M)) d^3\vv.
\end{equation}
The variant \eqref{eq:gBGK2} is simpler than the original kinetic model\eqref{gBGK} since the velocity dependence of the corresponding term arises 
in \eqref{eq:gBGK2} only via the quasi-equilibrium distribution function rather than due to the function 
in the function $\mathcal{J}_B(f^*(M))$. Moreover, a further simplification can be suggested in a "multiple relaxation times" form,
\begin{equation}
	\label{eq:gBGK3}
	\partial_t f+\vv\cdot\nabla f=--\frac{1}{\tau}(f-f^*(M)) -\sum_{i=1}^k\frac{\partial f^*(M)}{\partial M_i} \left(\frac{1}{\tau_i}\right)\left(M_i-M_i^{\rm eq}\right),
\end{equation}
where $M_i^{\rm eq}$ denotes the $i$th moment evaluated at the local equilibrium, while $\tau_1,\dots,\tau_k$ are corresponding relaxation times.
Descending the path of simplification from the quasi-equilibrium kinetic model \eqref{gBGK} through \eqref{eq:gBGK2} to \eqref{eq:gBGK3}, the information about the "true" Boltzmann collision integral, which is still manifest in \eqref{gBGK}, is gradually lost and is completely replaced by a relaxation-type form in \eqref{eq:gBGK3}. In the next section, we shall consider a special case of the two-relaxation time kinetic models which are mostly used in applications.


\subsubsection{Two relaxation times quasi-equilibrium models}
The two relaxation times quasi-equilibrium kinetic model is written,


\begin{equation}\label{eq:QEmodel2}
	\partial_t f+\vv\cdot\nabla f=-\frac{1}{\tau_{\rm fast}}(f-f^*(M))-\frac{1}{\tau_{\rm slow}}(f^*(M)-f^{\rm eq}),
\end{equation}
and has the following interpretation: 
The relaxation to the local equilibrium $f^{\rm eq}$ is decomposed into a "fast" relaxation from the current state $f$ to the \emph{intermediate} quasi-equilibrium $f^*$ followed by a "slow" relaxation from the quasi-equilibrium $f^*$ to the local equilibrium $f^{\rm eq}$. 
This is a special reduction of the quasi-equilibrium model \eqref{gBGK}, where the Boltzmann's term $\mathcal{J}_B(f^*(M))$ is replaced with the BGK relaxation \cite{levermore_moment_1996}.
Denoting $\mathcal{J}_{\rm fs}$ the right hand side of \eqref{eq:QEmodel2}, we can write,
\begin{equation}\label{eq:QEmodel2a}
		\mathcal{J}_{\rm fs}=-\left(\frac{1}{\tau_{\rm fast}}-\frac{1}{\tau_{\rm slow}}\right)(f-f^*(M))-\frac{1}{\tau_{\rm slow}}(f-f^{\rm eq}).
\end{equation}
Now, in order to clarify the meaning of the fast-slow decomposition, we compute the entropy production induced by the relaxation term \eqref{eq:QEmodel2a},
\begin{align}
	\sigma&=\sigma_{\rm fast}+\sigma_{\rm slow},\label{eq:HtheoremQEmodel}\\
	\sigma_{\rm fast}&=-\left(\frac{1}{\tau_{\rm fast}}-\frac{1}{\tau_{\rm slow}}\right)\int \ln \left(\frac{f}{f^*(M)}\right)\left(f-f^*(M)\right)d^3\vv,\label{eq:HtheoremQEmodelFAST}\\
		\sigma_{\rm slow}&=-\frac{1}{\tau_{\rm slow}}\int \ln \left(\frac{f}{f^{\rm eq}}\right)\left(f-f^{\rm eq}\right)d^3\vv.\label{eq:HtheoremQEmodelSLOW}
	\end{align}
Clearly, the contribution of the slow relaxation into the entropy production, Eq.\  \eqref{eq:HtheoremQEmodelSLOW}, is always non-positive, $\sigma_{\rm slow}\le0$. 
However, the contribution of the fast relaxation, Eq.\ \eqref{eq:HtheoremQEmodelFAST}, is non-positively definite, $\sigma_{\rm fast}\le0$, only if the fast relaxation time $\tau_{\rm fast}$ is not greater than the slow relaxation time $\tau_{\rm slow}$:
\begin{equation}
	\label{eq:QEtimes}
	\tau_{\rm fast}\le\tau_{\rm slow}.
\end{equation}
The equality $\tau_{\rm fast}=\tau_{\rm slow}$ results in the BGK model. Thus, the compliance to the $H$-theorem for the quasi-equilibrium kinetic model \eqref{eq:QEmodel2} \emph{implies} the fast-slow relaxation times hierarchy \eqref{eq:QEtimes}. 
It is convenient to introduce the ratio of the relaxations times,
\begin{equation}
	{r}_{\rm fs}=\frac{\tau_{\rm fast}}{\tau_{\rm slow}},\ r_{\rm fs}\in[0,1].
\end{equation}
Introducing also a convex linear combination of the quasi-equilibrium and the local equilibrium,
\begin{equation}\label{eq:tildef}
	\tilde{f}_{\rm fs}=\left(1-	r_{\rm fs}\right)f^*(M)+r_{\rm fs}f^{\rm eq}.
\end{equation}
the relaxation term can be written in a BGK-like form,
%
%
\begin{equation}\label{eq:Jqetilde}
		\mathcal{J}_{\rm fs}=-\frac{1}{\tau_{\rm fast}}(f-\tilde{f}_{\rm fs}).
\end{equation}
Realizations of the two relaxation times kinetic model \eqref{eq:QEmodel2} depend on the availability of analytical expressions for the quasi-equilibrium distribution functions. In practice, the "true" quasi-equilibrium can be substituted by Grad's approximation, which is sufficient for many applications. However, the relaxation times hierarchy 	\eqref{eq:QEtimes} must be respected also in such cases. 

\subsubsection{Lifting the eight-moments Grad's projection: Shakhov's S-model}
Using the eight-moments Grad's distribution function \eqref{eq:G8}, we obtain in \eqref{eq:tildef} and \eqref{eq:Jqetilde},
\begin{align}
	\tilde{f}_{\rm fs}&=\left(1-r_{\rm fs}\right)\mathcal{G}_8+r_{\rm fs}f^{\rm eq}\nonumber\\
	&=f^{\rm eq}\left(1+(1-r_{\rm fs}){q_{\alpha}(v_{\alpha}-u_{\alpha})\over pRT} \left(
	{({\vv}-{\bm{u}})^{2}\over 5RT} - 1 \right)\right).\label{eq:Smodel}
\end{align}
The corresponding two relaxation times kinetic model \eqref{eq:QEmodel2} is identified as the \emph{Shakhov's S-model} \cite{shakhov_generalization_1972}. Hydrodynamic limit of the S-model results in the Navier--Stokes--Fourier system with the following viscosity and thermal conductivity,
\begin{align}
	\mu &=\tau_{\rm fast}p,\label{eq:mu8}\\
	\lambda &=\tau_{\rm slow}c_p p.\label{eq:lambda8}
	\end{align}
Thus, the Prandtl number of the S-model satisfies the inequality implied by the fast-slow relaxation hierarchy \eqref{eq:QEtimes},
\begin{equation}
	{\rm Pr}=r_{\rm fs}=\frac{\tau_{\rm fast}}{\tau_{\rm slow}}\le 1.
\end{equation}

\subsubsection{Lifting the ten-moments Grad's projection.}
The ten-moments quasi-equilibrium is available in a closed form, Eq.\ \eqref{eq:GaussAnisotropic}, and hence can be straightforwardly used in the  kinetic model \eqref{eq:QEmodel2}. Here we shall instead consider lifting of the ten-moments Grad's approximation \eqref{Tri2.22}. With \eqref{Tri2.22}, we obtain in \eqref{eq:tildef} and \eqref{eq:Jqetilde},
\begin{align}
	\tilde{f}_{\rm fs}&=\left(1-r_{\rm fs}\right)\mathcal{G}_{10}+r_{\rm fs}f^{\rm eq}\nonumber\\
	&=f^{\rm eq}\left(1+(1-r_{\rm fs}){\sigma _{\alpha\beta}\over 2pRT} \left(
	(v_{\alpha}-u_{\alpha})(v_{\beta}-u_{\beta})-{1\over 3} \delta _{\alpha\beta}({\vv}-{\uu})^{2}\label{eq:QEmodel10}
	\right) \right).
\end{align}
Hydrodynamic limit of the corresponding kinetic model  results in the Navier--Stokes--Fourier system with the following viscosity and thermal conductivity,
\begin{align}
	\mu &=\tau_{\rm slow}p,\label{eq:mu10}\\
	\lambda &=\tau_{\rm fast}c_p p.\label{eq:lambda10}
\end{align}
Note that, the dependence of the transport coefficients on the relaxation times in \eqref{eq:mu10} and \eqref{eq:lambda10} is opposite to that of the S-model, Eqs.\ \eqref{eq:mu8} and \eqref{eq:lambda8}: the fast and slow relaxation times switch their places. Thus, in compliance with the relaxation times hierarchy \eqref{eq:QEtimes}, the Prandl number of the kinetic model based on the ten-moments Grad's approximation satisfies the inequality,
\begin{equation}
	{\rm Pr}=\frac{1}{r_{\rm fs}}=\frac{\tau_{\rm slow}}{\tau_{\rm fast}}\ge 1.
\end{equation}
Results \eqref{eq:mu10}, \eqref{eq:lambda10} remain valid when the full ten-moment quasi-equilibrium distribution function \eqref{eq:GaussAnisotropic} is used in \eqref{eq:QEmodel2}.
We comment that, the S-model \eqref{eq:Smodel} and the model \eqref{eq:QEmodel10} are complementary to each other: The S-model covers the range of the Prandtl number $0<{\rm Pr}\le 1$ while the model \eqref{eq:QEmodel10} corresponds to $1\le{\rm Pr}<\infty$. This is consistent with the corresponding Grad's projections: The ten-moments projection \eqref{Tri2.22} provides a formal limit ${\rm Pr}\to\infty$, cf.\ Eq.\ \eqref{eq:G10sdotfull}, while the eight-moments projection \eqref{eq:G8} provides the limit of a vanishing Prandtl number ${\rm Pr}\to 0$, cf. Eq.\ \eqref{eq:G8sdotfull}. Both models are used in the thermal and compressible lattice Boltzmann models \cite{ansumali_quasi-equilibrium_2007}.

\subsubsection{Holway's ellipsoidal-statistical kinetic model}
\label{sec:ES}
Finally, we mention another option of lifting the quasi-equilibrium projection. Starting with the BGK-like form \eqref{eq:Jqetilde}, 
the attractor $\tilde{f}$ \eqref{eq:tildef} can be replaced by another one, the quasi-equilibrium evaluated on a linear combination between the moments and their equilibrium values. In a general multi-parameter form, and a generic quasi-equilibrium distribution function $f^*(M,\vv)$, this is,  
%
\begin{align}\label{eq:tildefstar}
	\tilde{f}^*&=f^*((1-\alpha_1)M_1+\alpha_1M_1^{\rm eq},\dots,(1-\alpha_k)M_k+\alpha_kM_k^{\rm eq},\vv),
\end{align}
where $\alpha_1,\dots,\alpha_k$ are free parameters. In other words, while the previous lifting operates with linear combinations between the quasi-equilibrium and the local equilibrium distribution functions  \eqref{eq:tildef}, the present one uses the quasi-equilibrium distribution of linear combination of the corresponding moments. Whenever the Grad's approximation is used instead of the true quasi-equilibrium, both approaches result in identical models.

Because only \emph{one} case of a genuine quasi-equilibrium distribution function is known in a closed-form, cf.\ Eq.\ \eqref{eq:GaussAnisotropic}, the unique model of the type \eqref{eq:tildefstar} is the Holway's \emph{ellipsoidal-statistical (ES) model} \cite{holway_kinetic_1965}. 
Introducing in \eqref{eq:GaussAnisotropic} a linear combination between the full pressure tensor  $\bm{\Pi}=p\bONE +\bm{\sigma}$  and its local equilibrium value $p\bONE$, we obtain a one-parametric family of distribution functions,
%
\begin{align}\label{eq:ESqe}
	\tilde{f}^*_{\rm ES}&=n\frac{\rho^{\frac{3}{2}}}{\left(2\pi\right)^{\frac{3}{2}}\sqrt{\det \left[(1-\alpha)\bm{\Pi}+\alpha p\bONE\right]}}\exp\left(-\frac{\rho}{2}(\bm{v}-\bm{u})^{\dagger}\cdot\left[(1-\alpha)\bm{\Pi}+\alpha p\bONE\right]^{-1}\cdot(\bm{v}-\bm{u})\right).
\end{align}
With \eqref{eq:ESqe}, the ES model is written in the BGK-like form,
\begin{equation}\label{eq:ESmodel}
	\partial_t f+\vv\cdot\nabla f=-\frac{1}{\tau}\left(f-\tilde{f}^*_{\rm ES}\right).
\end{equation}
Hydrodynamic limit of the ES kinetic model \eqref{eq:ESmodel} recovers the viscosity and thermal conductivity in the Navier--Stokes--Fourier system,
\begin{align}
	\mu &=\frac{1}{\alpha}\tau p,\label{eq:muES}\\
	\lambda &=\tau c_p p,\label{eq:lambdaES}
\end{align}
which allows to identify the parameter $\alpha$ as the inverse of the Prandtl number,
\begin{equation}\label{eq:alphaES}
	\alpha=\frac{1}{{\rm Pr}}.
	\end{equation}
If the ES distribution function \eqref{eq:ESqe} is expanded to linear order in the nonequilibrium stress tensor $\bm{\sigma}$, the result is identical to \eqref{eq:QEmodel10}, and which makes it possible the matching of the parameters in both cases as $r_{\rm fs}=\alpha$. Thus, the relation between the transport coefficients of the ES model, Eqs.\ \eqref{eq:muES} and \eqref{eq:lambdaES}, is the same as for the case \eqref{eq:QEmodel10}, Eqs.\ \eqref{eq:mu10} and \eqref{eq:lambda10}. We note that, as was shown by \cite{andries_gaussian-bgk_2000}, the $H$-theorem for the ES model can be proven for Prandtl number $2/3\le{\rm Pr}<\infty$.

\subsection{Summary: Projections, corrections and lifting}

The review of some basic, classical aspects of the Boltzmann equation of this section are summarized in Tab.\ \ref{tab:BEsummary}. With this, we emphasize a certain commonality among various approaches. The main building block is an approximation provided by a distribution function, parameterized by a set of macroscopic fields of interest. The projection of the Boltzmann equation provides a starting point of the analysis. The projection can be corrected by improving its invariance property relative to the Boltzmann equation or lifted to a kinetic model. Both approaches are patterned in the lattice Boltzmann realizations to be discussed in the remainder of this contribution.

\begin{table}[ht!]
	\centering
	\begin{tabular}{l|l|l|l}
		Projection & Equation & Correction & Lifting\\
		\hline
		Local equilibrium & Euler & Navier--Stokes--Fourier & BGK\\
		Grad 13           & Grad's 13-moment system         & R13 & \\
		Grad 10           & Grad's 10-moment system         &      & 	\\
		Quasi-equilibrium 10 & Grad's 10-moment system          &                      &      ES model, model \eqref{eq:QEmodel10}	\\
		Grad 8            &                      &      & S-model	
	\end{tabular}
	\caption{Summary of projections, their corrections and liftings.}
	\label{tab:BEsummary}
\end{table}

\pagebreak
\section{Lattice Boltzmann for ideal fluid and related models\label{sec:LBM_intro}}
\label{sec:LBM}
In this section we discuss in detail the lattice Boltzmann method introduced in the early 90's~\cite{mcnamara_use_1988,benzi_lattice_1992,succi_lattice_2002} as an improvement over the lattice gas automata~\cite{frisch_lattice-gas_1986}. In its modern form it is a discrete solver for the Boltzmann equation with the BGK approximation for the collision operator given in Eq.~\eqref{BGK}. After a detailed introduction of discretization strategies in both phase-space and space/time, we will discuss different approximations to external body forces, stability and applicability domain of the classical LBM and possible improvement strategies.
\subsection{Phase-space discretization\label{subsec:phase_space_discrete}}
The first step in deriving a discrete scheme from the Boltzmann-BGK equation, is to discretize the $D$-dimensional space of particles speed. Many different strategies have been adopted in the context of the kinetic theory of gases. Here we review quadrature-based approaches used mainly in the context of the LBM to reduce the continuous space of particles speed into a discrete set of velocities.
\subsubsection{Hermite expansion and Gauss-Hermite quadrature}
One approach to discretise phase-space and derive the corresponding \emph{discrete} EDF consists in expanding it in terms of Hermite polynomials and operating a truncation using Gauss-Hermite quadratures \cite{shan_kinetic_2006,grad_note_1949,grad_kinetic_1949}.\\
Before starting the derivation, let us review the basic concepts of multi-variate Hermite polynomials. More details on the Hermite polynomials can be found in ~\ref{App:Hermite}. They are defined as \cite{grad_note_1949}:
\begin{equation}
    \bm{\mathcal{H}_{n}}\left(\bm{v}\right) = \frac{{\left(-1\right)}^n}{w\left(\bm{v}\right)} \bm{\nabla}_{\bm{v}}^n w\left(\bm{v}\right),
\end{equation}
where $\bm{\mathcal{H}_{n}}$ is a tensor of rank $n$ and $w\left(\bm{v}\right)$ is the normalized weight function defined as:
	\begin{equation}
	    w\left(\bm{v}\right) = {\left(2\pi\right)}^{-D/2} \exp\left({-\frac{ {\bm{v}}^2}{2}}\right),
	\end{equation}
with $D$ the dimension of $\bm{v}$. A function $f$ can then be expanded in terms of Hermite polynomials as:
\begin{equation}
    f = w\left(\bm{v}\right) \sum_{n=0}^\infty \frac{1}{n!}\bm{a_n}:\bm{\mathcal{H}_{n}}\left(\bm{v}\right),
\end{equation}
where ``:'' is the Frobenius inner product and the coefficients tensor $\bm{a_n}$ are computed as:
\begin{equation}
    \bm{a_n} = \int \bm{\mathcal{H}_{n}}\left(\bm{v}\right) f d\bm{\xi}.
\end{equation}
Note that Hermite polynomials are mutually orthogonal with respect to the weighted dot product defined as:
\begin{equation}
    \int \mathcal{H}_{\bm{i}}(\bm{v})w(\bm{v}) \mathcal{H}_{\bm{i}'}(\bm{v}) d\bm{v}  = \begin{cases} \
    0 &\text{if $\bm{i}\neq \bm{i}'$}\\
    n! &\text{if $\bm{i}= \bm{i}'$},
    \end{cases}
\end{equation}
where $\bm{i}$ and $\bm{i}'$ are vectors of size $n$ designating a component of the rank $n$ Hermite polynomial tensor. The infinite set of Hermite polynomials forms a complete orthogonal basis of the weighted space $V_w:=L^2(\mathbb{R}^D; \mathbb{R}, w d\bm{v})$ only under the condition that any function $f\in V_w$ satisfies:
\begin{equation}
    \int f^2(\bm{v}) w(\bm{v}) d\bm{v} < \infty,
\end{equation}
meaning in practice that $f(\bm{v})$ must decay faster than $\sqrt{w(\bm{v})}$ which has implications on the choice of reference temperature in static reference frame-based methods like the LBM. Further discussion on that issue is out of the scope of the present review and will be presented in future publications.\\
The first step in the expansion is the choice of the non-dimensionalization strategy, or reference state. While not necessary in the expansion, this choice is one of the most important steps in the construction of a discrete kinetic scheme as it will play a key role in the final numerical scheme's behavior, especially higher-order moments errors. The recent development of LB models relying on non-symmetrical stencils and adaptive non-dimensionalization is a clear proof of the previous assertion~\cite{sun_lattice-boltzmann_1998,sun_adaptive_2000,sun_three-dimensional_2003,frapolli_lattice_2016,dorschner_particles_2018,saadat_lattice_2019}. For the sake of simplicity, let us re-write the EDF in non-dimensional form as:
	\begin{equation}
	    f^{\rm eq}\left( \bm{v}, \rho, \bm{u}, \theta \right) = \rho {\left(2\pi\theta \right)}^{-D/2} \exp\left[{-\frac{ {\left(\bm{v}-\bm{u}\right)}^2}{2\theta}}\right],
	\end{equation}
where for the remainder of this subsection $\bm{u}$, and $\bm{v}$ are non-dimensionalized with a reference speed of sound $c_s$, $\theta = \frac{k_B T/m}{c_s^2}$ and $c_s=\frac{k_B T_0}{m_0}$, and $T_0$ and $m_0$ are respectively defined as the reference temperature and molecular mass. This results in the following first few Hermite polynomials:
	\begin{subequations}\label{eq:Hermite_polynomials}
		\begin{align}
		\mathcal{H}_{0} &= 1, \\
		\mathcal{H}_{i_1} &= v_{i_1}, \\
		\mathcal{H}_{i_1 i_2}  &= v_{i_1}v_{i_2} - \delta_{i_1 i_2}, \\
		\mathcal{H}_{i_1 i_2 i_3}  &= v_{i_1}v_{i_2}v_{i_3} - \left[v_{i_1}\delta_{i_2 i_3}\right]_{\text{cyc}},\\
		\mathcal{H}_{i_1 i_2 i_3 i_4} &= v_{i_1}v_{i_2}v_{i_3}v_{i_4}+ \left[\delta_{i_1 i_2}\delta_{i_3 i_4}\right]_{\text{cyc}}
		- \left[ v_{i_3}v_{i_4}\delta_{i_1 i_2} \right]_{\text{cyc}},
		\end{align}
	\end{subequations}
where $[]_{\text{cyc}}$ designates cyclic permutations over the involved indexes, and corresponding \emph{isothermal} ($\theta=1$) Hermite coefficients:
	\begin{subequations}
		\begin{align}
		a^{\rm eq}_{0} &= \rho, \\
		a^{\rm eq}_{i_1} &=  \rho u_{i_1}, \\
		a^{\rm eq}_{i_1 i_2} &=  \rho u_{i_1} u_{i_2}, \\
		a^{\rm eq}_{i_1 i_2 i_3} &=  \rho u_{i_1}u_{i_2}u_{i_3}, \\
		a^{\rm eq}_{i_1 i_2 i_3 i_4} &=  \rho u_{i_1}u_{i_2}u_{i_3}u_{i_4}.
		\end{align}
	\end{subequations}
In the context of the classical LBM, the flow is assumed isothermal. The continuous EDF is then expanded as:
\begin{equation}
    f^{\rm eq}\left( \bm{v}, \rho, \bm{u} \right) =  w\left(\bm{v}\right) \sum_{n=0}^\infty \frac{1}{n!}\bm{a_n}^{(eq)}\left(\rho, \bm{u}\right):\bm{\mathcal{H}_{n}}\left(\bm{v}\right).
\end{equation}
As seen here, the expanded EDF still goes over the entire phase-space. Given the form of the EDF and the corresponding moments:
\begin{equation}
    \Pi_{ \underbrace{x\dots x}_{\times p} \underbrace{y\dots y}_{\times q} \underbrace{z\dots z}_{\times r}} = \int {v_x}^p {v_y}^q {v_z}^r f^{\rm eq}\left( \bm{v}, \rho, \bm{u}\right)d\bm{v},
\end{equation}
and using the Hermite expansion, it can be written as:
\begin{equation}\label{eq:moment_integral_hermite}
    \Pi_{ \underbrace{x\dots x}_{\times p} \underbrace{y\dots y}_{\times q} \underbrace{z\dots z}_{\times r}} = \int P^\infty\left(\bm{v},\rho, \bm{u}\right)w\left(\bm{v}\right)d\bm{v},
\end{equation}
where:
\begin{equation}
    P^\infty\left(\bm{v},\rho, \bm{u}\right) = \frac{{v_x}^p {v_y}^q {v_z}^r f^{\rm eq}\left( \bm{v}, \rho, \bm{u} \right)}{w\left(\bm{v}\right)},
\end{equation}
and $P^\infty\left(\bm{v},\rho, \bm{u}\right)$ as defined here is a polynomial function of the variable $\bm{v}$ with order $\infty$ as the Hermite expansion has not yet been truncated. Given that the aim of the LB method is to solve the Boltzmann equation in the hydrodynamic regime one only needs to correctly recover the moments of the EDF involved in the hydrodynamic equations. Furthermore, Hermite polynomials are weighted orthogonal functions and as such higher-order polynomials have no effect on lower-order terms. Given the previously cited arguments, one can limit the Hermite expansion of the EDF:
\begin{equation}
    f^{\rm eq,N}\left( \bm{v}, \rho, \bm{u} \right) =  w\left(\bm{v}\right) \sum_{n=0}^N \frac{1}{n!}\bm{a}_n^{\rm eq}\left(\rho, \bm{u}, \right):\bm{\mathcal{H}}_n\left(\bm{v}\right),
\end{equation}
where $N$ corresponds to the highest-order moment involved in the targeted dynamics. For example, to correctly recover the NS equations for an isothermal flow one needs to correctly recover moments up to order three of the EDF. Now the polynomial $P^\infty$ can be replaced with a finite-order polynomial:
\begin{equation}
    P^M\left(\bm{v}, \rho, \bm{u}\right) = \frac{{v_x}^p {v_y}^q {v_z}^r f^{\rm eq,N}\left( \bm{v}, \rho, \bm{u}\right)}{w\left(\bm{v}\right)},
\end{equation}
where $M=2N$. The integral of Eq.~\eqref{eq:moment_integral_hermite} can be evaluated using a discrete sum through a Gauss-Hermite quadrature as:
\begin{equation}
        \int P^M\left(\bm{v}, \rho, \bm{u}\right) w\left(\bm{v}\right)d\bm{v}\cong \sum_{i=0}^{Q} w_i P^M\left(\bm{c}_i, \rho, \bm{u}\right),
\end{equation}
where $\bm{c}_i$ are discrete non-dimensional abscissae used for the quadrature and $w_i$ are the corresponding weights. According to the fundamental theorem of Gaussian quadratures, choosing the abscissae to be the roots of the orthogonal polynomial of the corresponding degree results in the maximum algebraic degree of precision, namely $2Q-1$. To correctly recover the targeted moments one must have $M\leq2Q-1$. The third-order quadrature (designated by $E^3_{1,5}$ in 1-D) results in the following abscissae: $c_{i}\in \{-\sqrt{3},0,\sqrt{3}\}$ corresponding to the following values $\{-\sqrt{3k_B T_0/m_0},0,\sqrt{3k_B T_0/m_0}\}$ in physical units. While being the most widely applied quadrature order, it is already clear that the third-order quadrature can not correctly recover all the moments appearing at the NS level. More on that issue in the next subsections. The corresponding weights are computed as:
\begin{equation}
    w_i = \frac{n!}{{\left(\mathcal{H}_{n-1}\left(c_{i}\right)\right)}^2}.
\end{equation}
In the multi-variate case, the weights can be computed as the products of the weights in each dimension.
\subsubsection{Product form equilibria and moment matching}
As clearly stated by its name, in this approach one tries to construct a discrete equilibrium by matching the moments appearing in the targeted macroscopic balance equations.\\
To identify the constraints, one first uses the CE analysis. For example, a simple CE analysis at order $\epsilon$ shows that to correctly recover the NS and continuity equations, one needs to exactly match moments up to order three \cite{lin_discrete_2017,hosseini_compressibility_2020}. For example let us consider a 1-D system with only translational degrees of freedom. The following moments need to be correctly recovered:
\begin{subequations}
	\begin{align}
	\Pi_0&=\int_{v_x} \rho\sqrt{\frac{m}{2\pi k_B T}}\exp \left[{-\frac{m{(v_x-u_x)}^2}{2k_B T}}\right] dv_x = \rho,\\
	\Pi_x&=\int_{v_x} v_x \rho\sqrt{\frac{m}{2\pi k_B T}}\exp \left[{-\frac{m{(v_x-u_x)}^2}{2k_B T}}\right] dv_x = \rho u_x,\\
	\Pi_{xx}&=\int_{v_x} v_x^2 \rho\sqrt{\frac{m}{2\pi k_B T}}\exp \left[{-\frac{m{(v_x-u_x)}^2}{2k_B T}}\right] dv_x = \rho \left(u_x^2 + \frac{k_B T}{m}\right),\\
	\Pi_{xxx}&=\int_{v_x} v_x^3 \rho\sqrt{\frac{m}{2\pi k_B T}}\exp \left[{-\frac{m{(v_x-u_x)}^2}{2k_B T}}\right] dv_x = \rho u_x \left(u_x^2 + 3\frac{k_B T}{m}\right).
	\end{align}
\end{subequations}
In the second step of the discrete equilibrium state construction, one chooses a symmetrical stencil (set of discrete velocities) with a number of degrees of freedom equal to the number of constraints \cite{lin_discrete_2017,gan_three-dimensional_2018,gan_lattice_2013}. For example, in the case of the isothermal NS solver, one can either use a four-velocity model or a five-velocity model with an additional constraint to have a unique solution. The discrete equilibrium is then found by solving the following system of equations:
\begin{equation}
\begin{bmatrix} 1 & 1 & 1 & 1 \\
c_0 & c_1 & c_2 & c_3 \\
c_0^2 & c_1^2 & c_2^2 & c_3^2 \\
c_0^3 & c_1^3 & c_2^3 & c_3^3 \end{bmatrix} \begin{bmatrix} f^{\rm eq}_0 \\ f^{\rm eq}_1 \\ f^{\rm eq}_2 \\ f^{\rm eq}_3 \end{bmatrix} = \begin{bmatrix} \rho \\ \rho u_x \\ \rho \left( u_x^2 + \frac{k_B T}{m}\right) \\ \rho u_x \left( u_x^2 + 3\frac{k_B T}{m}\right)  \end{bmatrix},
\end{equation}
where $c_{0-3}$ are the discrete velocities in the stencil and $f^{\rm eq}_{0-3}$ are the unknown discrete equilibria to be found by solving this system. The linear system formed using symmetrical stencils might not always be invertible. As such, for some models one might need to add non-symmetrical components to the system \cite{gan_discrete_2018}.\\
The product form of the EDF is a special realization of the moments matching approach. Considering the standard discrete velocity set $D3Q27$, where $D=3$ stands for three dimensions and $Q=27$ is the number of discrete velocities,
    \begin{equation}\label{eq:d3q27vel}
    	\bm{c}_i=(c_{ix},c_{iy},c_{iz}),\ c_{i\alpha}\in\{-1,0,1\},
    \end{equation}
one first defines a triplet of functions in two variables, $\xi_{\alpha}$ and $\zeta_{\alpha\alpha}$, 
    \begin{align}
    	&	\Psi_{0}(\xi_{\alpha},\zeta_{\alpha\alpha}) = 1 - \zeta_{\alpha\alpha}, 
    	\label{eqn:phi0}
    	\\
    	&	\Psi_{1}(\xi_{\alpha},\zeta_{\alpha\alpha}) = \frac{\xi_{\alpha} + \zeta_{\alpha\alpha}}{2},
    	\label{eqn:phiPlus}
    	\\
    	&	\Psi_{-1}(\xi_{\alpha},\zeta_{\alpha\alpha}) = \frac{-\xi_{\alpha} + \zeta_{\alpha\alpha}}{2},
    	\label{eqn:phis}
    \end{align}
and considers a product-form associated with the discrete velocities $\bm{c}_i$ (\ref{eq:d3q27vel}),
    \begin{equation}\label{eq:prod}
    	\Psi_i= \Psi_{c_{ix}}(\xi_x,\zeta_{xx}) \Psi_{c_{iy}}(\xi_y,\zeta_{yy}) \Psi_{c_{iz}}(\xi_z,\zeta_{zz}).
    \end{equation}
All pertinent populations below are determined by specifying the parameters $\xi_\alpha$ and $\zeta_{\alpha\alpha}$ in the product-form (\ref{eq:prod}). 
The two-dimensional version of the model on the $D2Q9$ lattice is obtained by omitting the $z$-component in all formulas. After matching moments with their continuous counter-parts the parameters are set as,
    \begin{align}
    &\xi_{\alpha}=u_{\alpha},\\
    &\zeta_{\alpha\alpha}=c_s^2+u_{\alpha}^2,
    \end{align} 
and the local equilibrium populations are represented with the product-form \eqref{eq:prod},
    \begin{equation}\label{eq:LBMeq}
        f_i^{\rm eq}=
        \rho\prod_{\alpha=x,y,z}\Psi_{c_{i\alpha}}\left(u_\alpha,c_s^2+u_{\alpha}^2\right).
    \end{equation}
This form of the discrete equilibrium populations, when $c_s^2=k_B T_0/3 m_0$ is equivalent to third-order quadrature-based scheme with a full expansion of the distribution function. As will be seen in the upcoming section this class of equilibrium populations allows to restore Galilean invariance of the Navier-Stokes level shear viscosity but fails to do so for the bulk component.
\subsubsection{Alternative to polynomial equilibria: Entropic equilibria}
In the context of the entropic lattice Boltzmann method as described in \cite{ansumali_minimal_2003}, the discrete equilibrium state is found as the minimizer of a convex discrete entropy functional under mass and momentum conservation constraints. It is this lower number of constraints on moments that allows the scheme to accomodate the entropy constraint. The derivation starts with the roots of the third-order Hermite polynomials as the discrete abscissae and considering the following conservation constraints:
	\begin{equation}
	\sum_\alpha f_i^{\rm eq} = \rho,
	\end{equation}
	\begin{equation}
    \sum_i \bm{c}_i f_i^{\rm eq} = \rho \bm{u},
    \end{equation}
where notations follow those adopted in the previous subsection. The EDF is derived as the function extremizing the discrete entropy function:
    \begin{equation}
    H_{w_i,c_i} = \sum_i f_i \ln\left(\frac{f_i}{w_i}\right),
    \end{equation}
under the previously set constraints. Given the Galilean invariance of the weights the expression for the entropy function is also Galilean invariant \cite{frapolli_lattice_2016}. The EDF can be expressed as:
    \begin{equation}
    f_i^{\rm eq} = w_i \exp\left(\lambda_0\right) \prod_{\alpha=1}^{D}\exp\left(c_{i,\alpha}\lambda_\alpha\right),
    \end{equation}
where $\lambda_0$ and $\lambda_\alpha$ are the Lagrange multipliers associated with constraints on the zeroth and first-order moments. Introducing the following changes of variables, $X = \exp\left(-\lambda_0\right)$ and $Z_\alpha = \exp\left(\lambda_\alpha\right)$ the EDF is re-written as:
    \begin{equation}
    f_i^{\rm eq} = w_i X^{-1} \prod_{\alpha=1}^{D} Z^{c_{i,\alpha}}.
    \end{equation}
Writing down the conservation equations using the new variables for the D2Q9 stencil, the following algebraic system of equations is obtained:
	\begin{subequations}
		\begin{align}
		\rho X &= \sum_i w_i \prod_{\alpha=x,y} Z^{c_{i,\alpha}}, \\
		\rho u_x X &= \sum_i w_i c_{i,x} \prod_{\alpha=x,y} Z^{c_{i,\alpha}}, \\
		\rho u_y X &= \sum_i w_i c_{i,y} \prod_{\alpha=x,y} Z^{c_{i,\alpha}}.
		\end{align}
	\end{subequations}
Solving this system of equation for $Z_x$, $Z_y$ and $X$ and keeping positive roots one gets:
    \begin{equation}
        Z_\alpha = \frac{2u_\alpha + \sqrt{{u_\alpha}^2/c_s^2 + 1}}{1-u_\alpha},
    \end{equation}
    \begin{equation}
        X^{-1} = \rho \prod_{\alpha=x,y} \left(2-\sqrt{{u_\alpha}^2/c_s^2 + 1}\right),
    \end{equation}
and therefore can express the entropic discrete equilibrium as:
    \begin{equation}\label{eq:Entropic_isothermal_EDF}
    f_i^{\rm eq} =  w_i \rho  \prod_{\alpha=x,y} \left(2-\sqrt{{u_\alpha}^2/c_s^2 + 1}\right) {\left(\frac{2u_\alpha + \sqrt{{u_\alpha}^2/c_s^2 + 1}}{1-u_\alpha}\right)}^{c_{i,\alpha}}.
    \end{equation}
It will be shown in the next sections that this approach to constructing the discrete equilibrium populations satisfies a smaller number of constraint on moments, i.e. only of order zero and one, but guarantees unconditional positivity and linear stability of the scheme. 
\subsubsection{Galilean invariance issues on standard lattices}

\begin{figure}[!h]
    \centering
	    \includegraphics{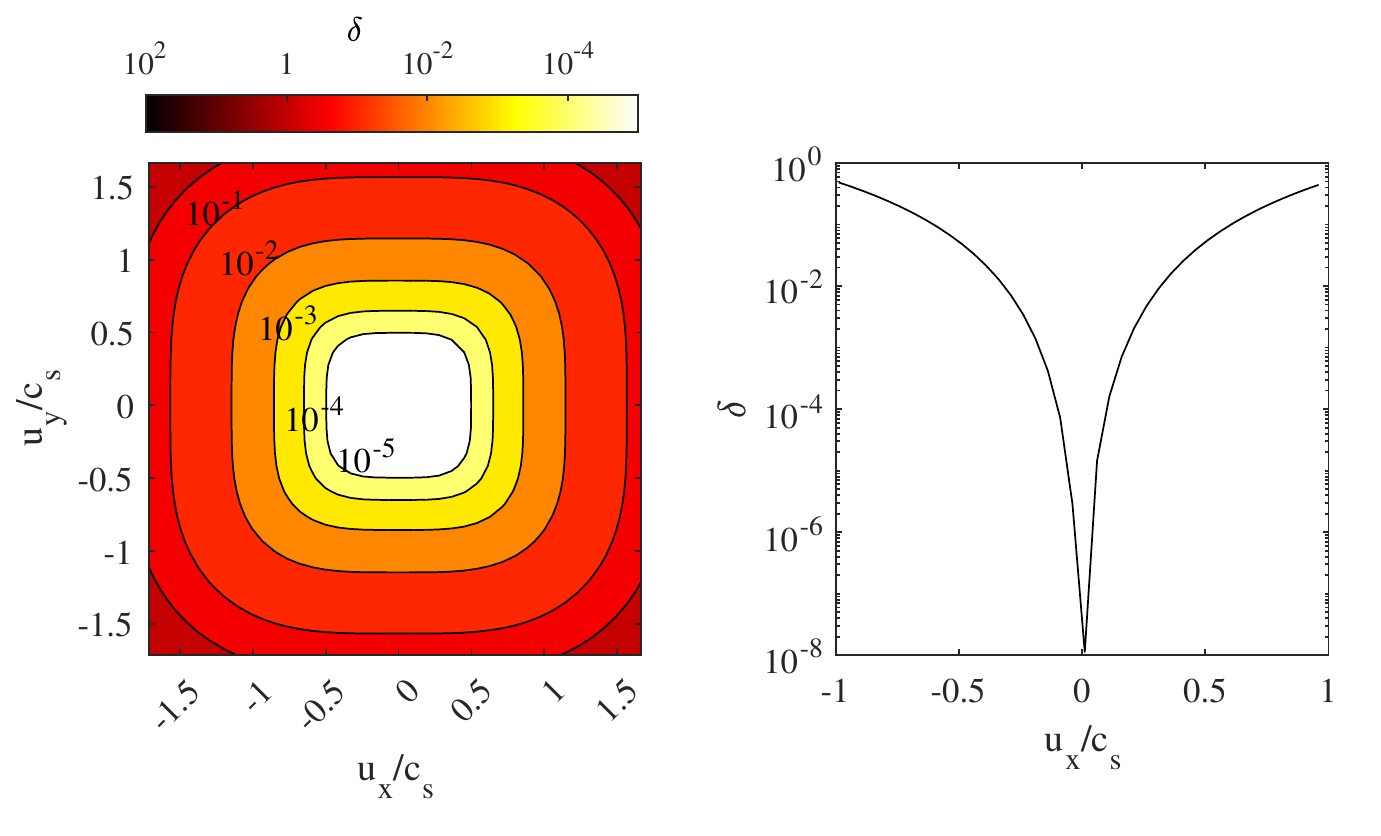}
    \caption{Illustration of deviations in $\Pi^{\rm eq}_{xx}+\Pi^{\rm eq}_{yy}$ moment for the entropic equilibrium. Left: Evolution of normalized error as a function of $u_x/c_s$ and $u_y/c_s$. Right: Normalized error as a function of $u_x/c_s$ with $u_y/c_s=0$.}
    \label{Fig:entropic_eq_Pxx_Pyy}
\end{figure}

\begin{figure}[h!]
	\centering
	\includegraphics[width=12cm,keepaspectratio]{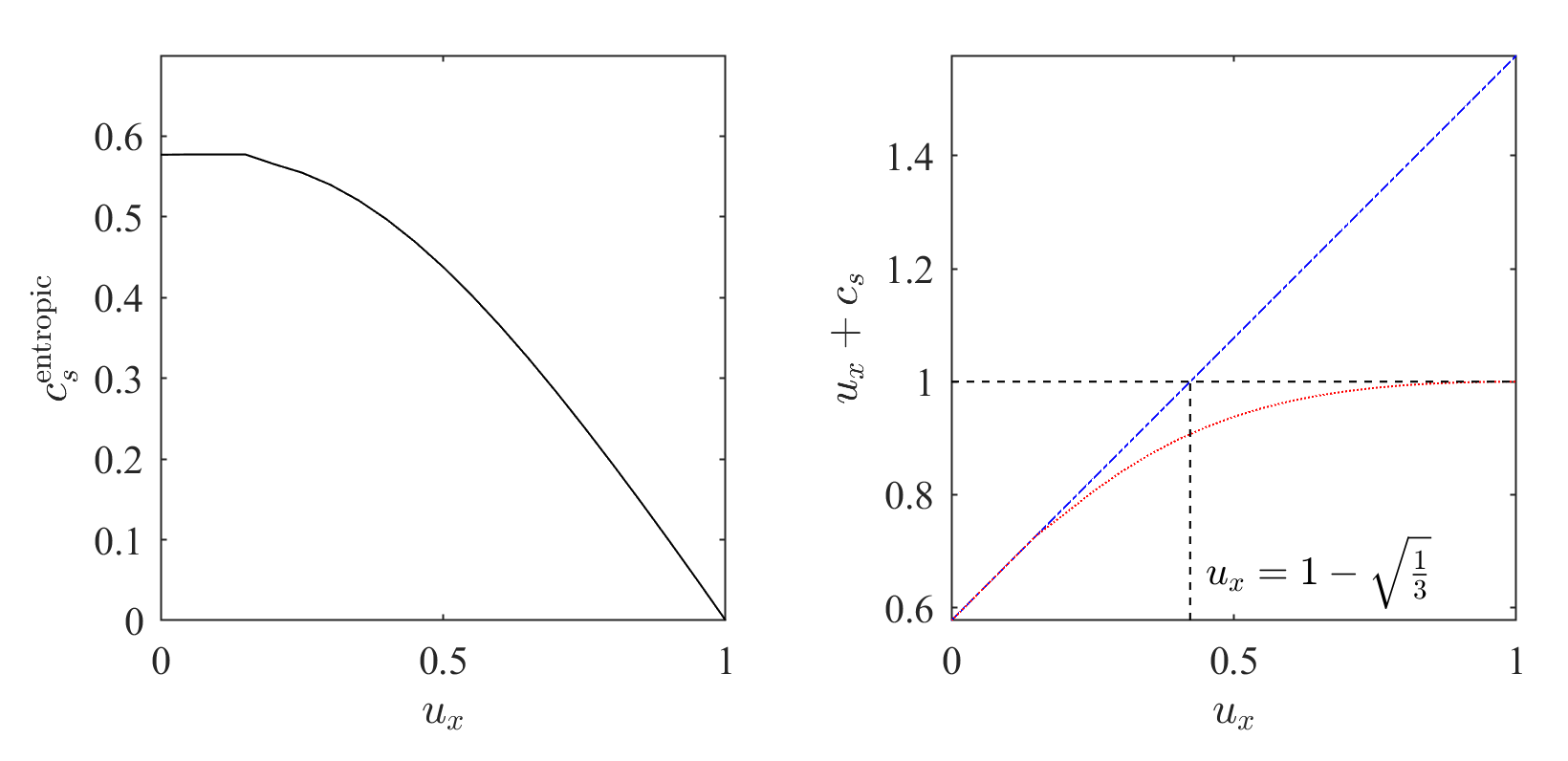}
	\caption{(Left) Non-dimensional sound speed for entropic EDF as a function of velocity $u_x$. (Right) Comparison of the speed of fastest propagating eigen-mode: (blue dotted line) polynomial EDF and (red line) entropic EDF.}
	\label{Fig:sound_speed_entropic_vs_poly}
\end{figure}

\begin{figure}[!h]
    \centering
	    \includegraphics{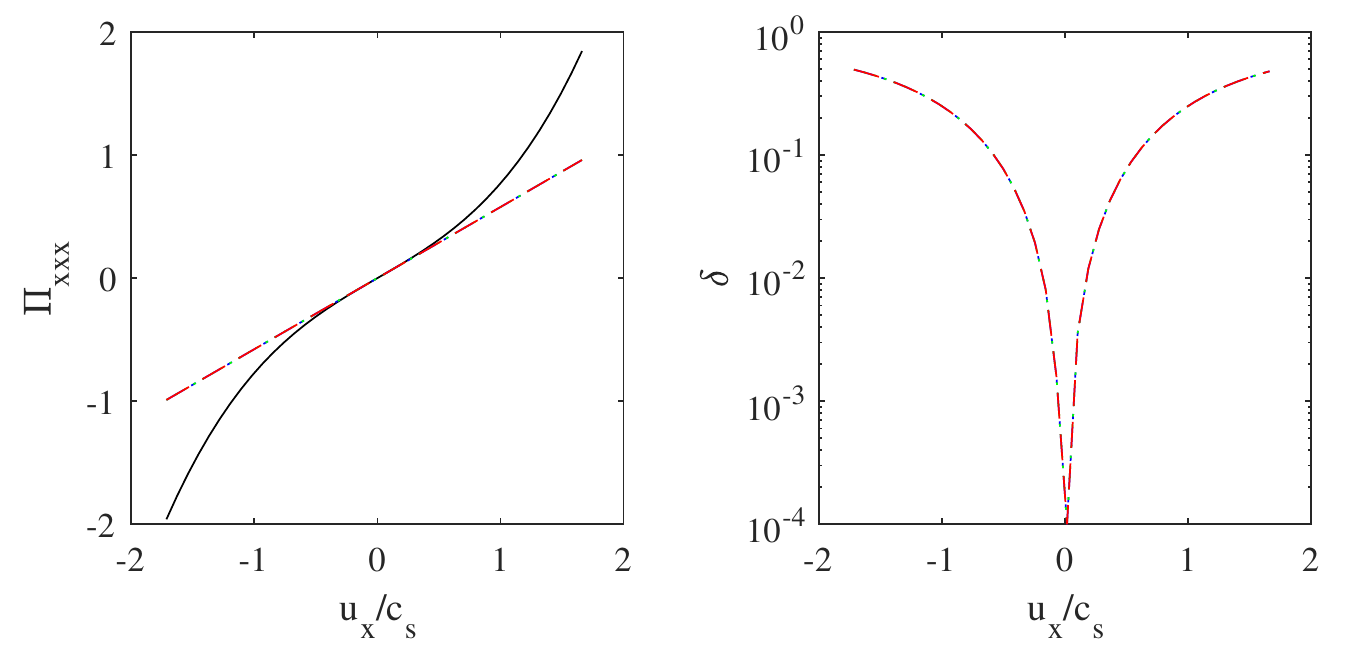}
    \caption{Deviations in the third order moment $\Pi_{xxx}^{\rm eq}$. Left: Moments an Right: Normalized errors. Black solid line: Maxwellian distribution moment, blue dash-dotted line: second-order Hermite expansion, green dotted line: entropic equilibrium and red dashed line: product form equilibrium.}
    \label{Fig:Pixxx_error}
\end{figure}

\begin{figure}[!h]
    \centering
	    \includegraphics{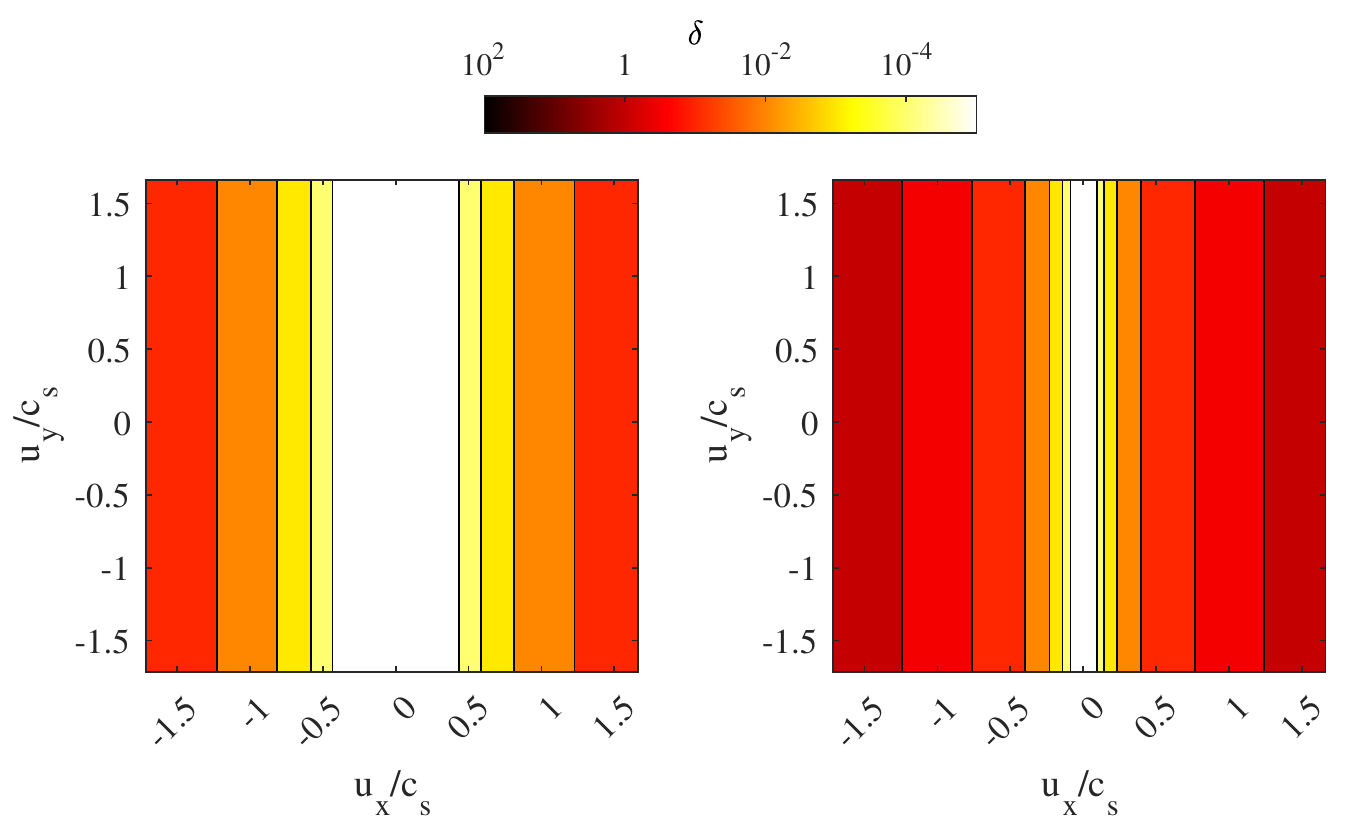}
    \caption{Deviations in the third order moment $\Pi_{xxy}^{\rm eq}$. Left: entropic equilibrium. Right: second-order Hermite expansion.}
    \label{Fig:Pixxy_error}
\end{figure}

\paragraph{Errors in moments of the equilibrium distribution function}
It is well-known that LB formulations based on standard first-neighbor stencils do not exactly recover the NS level dynamics, {i.e.} the stress tensor and, for the entropic equilibrium, the pressure tensor. The former comes from the fact that, due to a lack of symmetry, the third-order moments tensor does not correspond to its phase-space continuous counterpart. While including higher-order components of the Hermite expansion in the EDF or using the product-form can help correct the deviatoric components, consistency of the diagonal components can only be re-established through additional correction terms. Furthermore, use of the entropic approach which enforces smaller number of constraints on moments results in deviations in second-order moments too. To have a better measure of the applicability domain of the LB scheme, we will look at the deviations of these moments from their continuous counterparts for varying Mach numbers. Although readily extendable to other stencils, the D2Q9 stencil will be considered here. Moments of orders two and three of the EDF will be studied through their normalized deviations defined as:
\begin{equation}
\delta = \left\vert 1 - \frac{\sum_{i}c_{i,x}^p c_{i,y}^q f_{i}^{\rm eq}}{\Pi^{\rm MB}_{\underbrace{x\dots x}_{\times p} \underbrace{y\dots y}_{\times q} } }\right\vert,
\end{equation}
where $\Pi^{\rm MB}_{\underbrace{x\dots x}_{\times p} \underbrace{y\dots y}_{\times q} }$ is the continuous moment and $\sum_{i}c_{i,x}^p c_{i,y}^q f_{i}^{\rm eq}$ is the moment of the discrete EDF.\\
First we consider the diagonal components of the second-order moments tensor in the co-moving reference frame. We consider the co-moving reference frame here to focus on the recovered thermodynamic pressure:
\begin{equation}
    \widetilde{\Pi^{\rm MB}_{xx}} + \widetilde{\Pi^{\rm MB}_{yy}} = 2\rho c_s^2.
\end{equation}
The deviations are illustrated in Fig.~\ref{Fig:entropic_eq_Pxx_Pyy}.
While the entropic EDF does not exactly recover the correct trace of the diagonal components of the second-order moments tensor, the deviations are negligible for Mach numbers up to 0.4 which goes above the validity of the weakly compressible flow assumption. Another point worth nothing is the nature of the deviations. The second order central equilibrium moment $\widetilde{\Pi_{\alpha\alpha}^{\rm eq}}$ for the entropic equilibrium is:
\begin{equation}
   \widetilde{\Pi_{\alpha\alpha}^{\rm eq}} = -\frac{\rho(u_\alpha - 1)(6 u_\alpha \sqrt{3 u_\alpha^2 + 1} + 12 u_\alpha^2 - 2 \sqrt{3 u_\alpha^2 + 1} + 4)}{6(2 u_\alpha + \sqrt{3 u_\alpha^2 + 1})},
\end{equation}
indicating that the thermodynamic pressure tensor loses Galilean-invariance and isotropy. In the limit of $u_x,u_y\rightarrow 0$ one recovers the correct pressure, i.e. $\rho c_s^2$. The second-order Hermite-expanded and product form equilibria exactly recover the second-order moments tensor. Another point worth analyzing, is the behavior of sound speed in the entropic EDF. While sound speed is constant for polynomial equilibria it is a function of local speed in the entropic EDF. The behavior of sound speed as a function of local velocity is illustrated in Fig.~\ref{Fig:sound_speed_entropic_vs_poly}.
This behavior shows an interesting property of the entropic model pointing already to a (potentially) unconditional linear stability. Keeping in mind that the fastest eigen-modes in the system can not propagate faster than the lattice links and only considering three \emph{physical} eigen-modes, i.e. $u_x$, $u_x-c_s$ and $u_x+c_s$, one arrives to the following strong condition on linear stability:
\begin{equation}\label{eq:CFL_strict_condition}
    \max(u_x,u_x+c_s,u_x-c_s)\leq\frac{\delta r}{\delta t}.
\end{equation}
For the polynomial equilibria given that sound speed is constant one recovers the following maximum tolerable velocity:
\begin{equation}\label{eq:CFL_lmit_absolute_poly}
    \lvert u_x^{\rm max} \lvert = \frac{\delta r}{\delta t}\left(1-\sqrt{\frac{1}{3}}\right)=0.4226\frac{\delta r}{\delta t},
\end{equation}
which as will be shown in the next section through stability analyses, is indeed the maximum reachable velocity. For the entropic EDF on the other hand, it is observed that the sound speed self-adjusts as a function of local velocity to guarantee Eq.~\eqref{eq:CFL_strict_condition} is always satisfied. At the higher end of the velocity spectrum, i.e. $u_x\rightarrow\frac{\delta r}{\delta t}$ the speed of sound goes to $c_s\rightarrow0$.\\
For the NS level dynamics to be correctly recovered, the components of the third-order moments tensor must also match those of the continuous EDF. However, as observed in Fig.~\ref{Fig:Pixxx_error}, none of the EDFs are able to recover the correct diagonal components for this tensor. This shortcoming is not related to the equilibrium state but, to the limited order of the Gauss-Hermite quadrature used for first-neighbor stencils. In Fig.~\ref{Fig:Pixxx_error}, it is observed that all three EDFs considered there (second-order Hermite expansion, product-form and entropic) have the same moments, i.e. $\Pi_{xxx}^{\rm eq}=\Pi_{x}^{\rm eq}=\rho u_x$.
For the off-diagonal components of the third-order moments tensor however, as shown in Fig.~\ref{Fig:Pixxy_error}, different EDFs result in different behaviors. While Hermite expansions of order higher than three, here the product form, exactly recover the correct moments, the second-order Hermite expansion and entropic EDFs show some deviations. Although not exactly recovering the correct moment the entropic EDF still closely follows its continuous counterpart even for large Mach numbers. This means that the entropic model exhibits less pronounced Galilean invariance in it effective viscosity as compared to the classical LBM with second-order EDF at moderate Mach numbers.
In order to correctly recover the off-diagonal components of the third-order moments tensor in 2-D, the third-order terms of the Hermite expansion must be included. The diagonal components deviations on the other hand can only be accounted for via correction terms discussed in the next paragraph.
\paragraph{Restoring Galilean-invarience of dissipation rate of shear/normal modes}
Given that the EDF of the product-form is equivalent to the fourth-order Hermite-expanded EDF, only the Hermite expansion-based EDFs will be considered here.
A simple CE analysis shows that at the NS level, to match the viscous (non-equilibrium) stress tensor for the continuous Boltzmann equation, moments of orders two and three of the EDF must be exactly recovered. Integrating in phase space, the following continuous second- and third-order moments are in fact recovered~\cite{shan_kinetic_2006}:
\begin{subequations}\label{eq:moments_continuous}
	\begin{align}
	\Pi_{\alpha\beta}^{\rm MB} &= \rho u_{\alpha}u_{\beta} + \rho c_s^2\delta_{\alpha\beta}\theta, \\
	\Pi_{\alpha\beta\gamma}^{\rm MB} &= \rho u_{\alpha}u_{\beta}u_{\gamma} + \rho c_s^2\left[u_{\alpha}\delta_{\beta\gamma}\theta\right]_{\text{cyc}},
	\end{align}
\end{subequations}
while one gets the following moments with the second- and third-order discrete EDFs:
\begin{subequations}\label{eq:moments_discrete_2nd_order}
	\begin{align}
	\Pi^{\rm eq,2}_{\alpha\beta} &= \rho u_{\alpha}u_{\beta} + \rho c_s^2 \delta_{\alpha\beta} \theta, \\
	\Pi^{\rm eq,2}_{\alpha\beta\gamma} &= \rho c_s^2 \left[ u_{\alpha}\delta_{\beta\gamma}\right]_{\text{cyc}}, 
	\end{align}
\end{subequations}
and :
\begin{subequations}\label{eq:moments_discrete_3rd_order}
	\begin{align}
	\Pi^{\rm eq,2}_{\alpha} &= \rho u_{\alpha}u_{\alpha} + \rho c_s^2\delta_{\alpha\beta}\theta, \\
	\Pi^{\rm eq,3}_{\alpha\beta\gamma} &= \rho c_s^2 \delta_{\alpha\beta\gamma}\left[ u_{\alpha}\delta_{\beta\gamma}\right]_{\text{cyc}}  + \rho \left(1-\delta_{\alpha\beta\gamma}\right)\{u_{\alpha}u_{\beta}u_{\gamma} + c_s^2\left[u_{\alpha}\delta_{\beta\gamma}\theta\right]_{\text{cyc}} \}.
	\end{align}
\end{subequations}
The product-form EDF recovers the same second- and third-order moments as the third-order Hermite expansion.
\par To better put forward the shortcoming of the first-order stencil in recovering the NS level terms, let us perform now a brief CE analysis. Introducing the multi-scale expansion into the space and time-continuous system of equations and sorting terms of different orders in $\epsilon$ into separate equations and taking moments of orders zero and one to get mass and momentum balance, at order two in $\varepsilon$ (NS level) the following momentum balance equations are recovered:
\begin{equation}
  \partial_t^{(2)}\rho u_\alpha + \partial_\beta \tau\left(\partial^{(1)}_t \Pi^{\rm eq}_{\alpha\beta}+\partial_\gamma \Pi^{\rm eq}_{\alpha\beta\gamma}\right) + \partial_\beta \tau\left(\sum_i c_{i,\alpha}c_{i,\beta}\Psi^{(1)}_i\right) = 0.
\end{equation}
For the stress tensor to be correctly recovered at this scale one must have:
\begin{equation}\label{eq:correction_1}
  \Psi_i = \frac{w_i}{2c_s^4}\partial_\alpha\mathcal{H}_{i,\beta\gamma}\delta\Pi^{\rm eq}_{\alpha\beta\gamma},
\end{equation}
where $\delta\Pi^{\rm eq}_{\alpha\beta\gamma}$ designates the deviation of the discrete EDF moment from its continuous counterpart. An interesting point to note here is that this form of the correction term is only valid for the time and space continuous discrete Boltzmann system of equations. After discretization in space and time and operation of the change of variable specific to the LBM the correction term will be slightly modified.\\
It is also worth noting that the treatment for third- and fourth-order EDFs differs from that at second-order. Regarding third- and fourth-order EDFs, in 2-D one obtains:
\begin{equation}
  \Psi^{\rm eq,N>2}_i = \frac{w_i}{2c_s^4}\left[\mathcal{H}_{i,xx}\partial_x\delta\Pi^{\rm eq}_{xxx} + \mathcal{H}_{i,yy}\partial_y\delta\Pi^{\rm eq}_{yyy} \right].
\end{equation}
Instead, for the second-order EDF, additional correction terms are required, or one fails to recover correctly deviatoric components of the third-order moments:
\begin{equation}
  \Psi^{\rm eq,2}_i = \Psi^{\rm eq,N>2}_i + \frac{w_i}{c_s^4}\mathcal{H}_{i,xy}\left[\partial_x\left(\delta\Pi^{\rm eq,2}_{xxy}+\delta\Pi^{\rm eq,2}_{xyy}\right) + \partial_y\left(\delta\Pi^{\rm eq,2}_{xxy}+\delta\Pi^{\rm eq,2}_{xyy}\right) \right].
\end{equation}
This type of correction for the third-order equilibrium moments was introduced in~\cite{li_coupling_2012} and later reprised in \cite{feng_three_2015,feng_hybrid_2019,hosseini_compressibility_2020}.
\par Different from the approach taken previously, one can also directly introduce the correction term at order $\epsilon^2$ as proposed in~\cite{prasianakis_lattice_2007,prasianakis_lattice_2008}. In practice this means that the correction would involve a Laplacian and as such be expanded as:
\begin{equation}
    \Psi^{'}_i = \epsilon^2 {\Psi^{'}_i}^{(2)}.
\end{equation}
Re-writing the momentum balance equations at the NS level with this new correction term:
\begin{equation}
  \partial_t^{(2)}\rho u_\alpha + \partial_\beta \tau\left(\partial^{(1)}_t \Pi^{\rm eq}_{\alpha\beta}+\partial_\gamma \Pi^{\rm eq}_{\alpha\beta\gamma}\right) - \sum_i c_{i,\alpha}{\Psi^{'}_i}^{(2)} = 0,
\end{equation}
one gets the following restrictions on the correction term:
\begin{equation}
    \sum_i {\Psi^{'}_i}^{(2)} = 0,
\end{equation}
and:
\begin{equation}
    \sum_i c_{i,\alpha} {\Psi^{'}_i}^{(2)} = \partial_\beta\left(\frac{\mu}{p} \partial_\gamma\delta\Pi^{\rm eq}_{\alpha\beta\gamma}\right).
\end{equation}
The correction term using the second approach can therefore be defined as:
\begin{equation}\label{eq:correction_2}
    {\Psi^{'}_i} = \frac{w_i}{c_s^2}c_{i,\alpha}  \partial_\beta\left(\frac{\mu}{p} \partial_\gamma\delta \Pi^{\rm eq}_{\alpha\beta\gamma}\right).
\end{equation}
Both approaches effectively restore Galilean invariance at the NS level. However, the first approach allows for more flexibility in the treatment of the spatial derivative, and in practice allows for compressible flow simulations in the low supersonic regime. For a detailed comparative study of both approaches and different discretization strategies, interested readers are referred to \cite{hosseini_compressibility_2020}.
\subsection{Space and time discretization: Integration along characteristics\label{subsec:space_time_discrete}}
With the Discrete velocity Boltzmann system of non-homogeneous hyperbolic partial differential equations in hand, the next step is to operate discretization in physical space and time.
\subsubsection{Realization without force and correction term}
Starting from the phase-space discretized form of the Boltzmann equation (a set of $Q$ equations):
\begin{equation}\label{eq:2-3-2-1}
    \partial_t f_i + \bm{c}_i\cdot\bm{\nabla}f_i = \Omega_i,
\end{equation}
where $\Omega_i$ is the collision term, the idea of the Lagrangian approach consists of integrating them along their respective characteristics, which contrary to Lagrangian solvers for the NS equations (given that fluid particle path-lines are space- and time-dependent), results in an exact solution for the advection term. As such integrating the equations from a time $t$ to $t+\delta t$ along the stencil directions one obtains:
\begin{equation}\label{eq:2-3-2-2}
    f_i\left(\bm{r}+\bm{c}_i \delta t, t+\delta t\right) - f_i\left(\bm{r}, t\right) = \int_{t}^{t+\delta t} \Omega_i\left(\bm{r}(t'), t'\right)dt'.
\end{equation}
Obviously within the context of the Lagrangian approach $\delta r/\delta t$ is tied to the abscissae obtained from the Gauss-Hermite quadrature. In the case of the third-order quadrature for the streaming operation to results in space-filling lattices:
\begin{equation}\label{eq:2-3-2-3}
    \lvert\lvert c_{i,\alpha}\lvert\lvert = \sqrt{\frac{3k_B T_0}{m_0}} = \frac{\delta r}{\delta t}.
\end{equation}
Coming back to Eq.~\eqref{eq:2-3-2-2}, it is observed that all the difficulty in this approach lies in the estimation of the collision contribution. A simple first-order explicite approximation would be:
\begin{equation}\label{eq:2-3-2-4_order1}
    \int_{t}^{t+\delta t} \Omega_i\left(\bm{r}(t^{'}), t^{'}\right)dt^{'} = \frac{\delta t}{2}\Omega_i\left(\bm{r}, t\right) + \mathcal{O}\left(\delta t^2\right).
\end{equation}
The classical LBM approach relies on a higher-order alternative; To get a second-order accurate scheme one can use the trapezoidal rule to evaluate the integral:
\begin{equation}\label{eq:2-3-2-4}
    \int_{t}^{t+\delta t} \Omega_i\left(\bm{r}(t^{'}), t^{'}\right)dt^{'} = \frac{\delta t}{2}\Omega_i\left(\bm{r}, t\right) +  \frac{\delta t}{2}\Omega_i\left(\bm{r}+\bm{c}_i\delta t, t+\delta t\right) + \mathcal{O}\left(\delta t^3\right),
\end{equation}
which in turn results in an implicit scheme. To take out the implicitness of the resulting equation, the distribution function is re-defined via the following change of variables:
\begin{equation}\label{eq:2-3-2-5}
    \bar{f}_i = f_i - \frac{\delta t}{2}\Omega_i,
\end{equation}
\begin{equation}\label{eq:2-3-2-6}
\bar{f}^{\rm eq}_i = f^{\rm eq}_i,
\end{equation}
\begin{equation}\label{eq:2-3-2-7}
    \Omega_i = \frac{1}{\tau + \delta t/2}\left(  \bar{f}^{\rm eq}_i - \bar{f}_i\right).
\end{equation}
Using this change of variable and Eqs.~\eqref{eq:2-3-2-2} and \eqref{eq:2-3-2-4} one gets:
\begin{equation}\label{eq:2-3-2-8}
    \bar{f}_i\left(\bm{r}+\bm{c}_i\delta t, t+\delta t\right) - \bar{f}_i\left(\bm{r}, t\right) + \frac{\delta t}{2} \Omega_i\left(\bm{r}+\bm{c}_i\delta t, t+\delta t\right) - \frac{\delta t}{2}\Omega_i\left(\bm{r}, t\right) = \\ \frac{\delta t}{2} \Omega_i\left(\bm{r}+\bm{c}_i\delta t, t+\delta t\right) + \frac{\delta t}{2}\Omega_i\left(\bm{r}, t\right),
\end{equation}
which in turn using Eqs.~\eqref{eq:2-3-2-5}--\eqref{eq:2-3-2-7} results in the classical \emph{collide-stream} algorithm:
\begin{equation}\label{eq:2-3-2-9}
\bar{f}_i\left(\bm{r}+\bm{c}_i\delta t, t+\delta t\right) - \bar{f}_i\left(\bm{r}, t\right) = \frac{\delta t}{\bar{\tau}}\left( f^{\rm eq}_\alpha\left(\bm{r}, t\right) - \bar{f}_i\left(\bm{r}, t\right)\right),
\end{equation}
where $\bar{\tau}$ is defined as:
\begin{equation}\label{eq:2-3-2-10}
\bar{\tau} = \tau + \delta t/2.
\end{equation}
It is also interesting to note that the new distribution functions have the following properties:
\begin{equation}\label{eq:2-3-2-11}
\sum_i \bar{f}_i = \sum_i f_i - \frac{\delta t}{2} \sum_i \Omega_i = \rho,
\end{equation}
\begin{equation}\label{eq:2-3-2-12}
\sum_i \bm{c}_i\bar{f}_i = \sum_i \bm{c}_i f_i - \frac{\delta t}{2} \sum_i \bm{c}_i \Omega_i = \rho \bm{u},
\end{equation}
where we have used the collision-invariance of the zeroth- and first-order moments. More generally for higher-order moments:
\begin{equation}\label{eq:2-3-2-13}
\sum_i \mathcal{H}_{n}\left(\bm{c}_i\right)\bar{f}_i = \sum_i \mathcal{H}_{n}\left(\bm{c}_i\right)f_i - \frac{\delta t}{2} \sum_i \mathcal{H}_{n}\left(\bm{c}_i\right)\Omega_i = \left(1+\frac{\delta t}{2\bar{\tau}}\right)\bm{a}_n - \frac{\delta t}{2\bar{\tau}}\bm{a}^{\rm eq}_n.
\end{equation}
\par While to derive the previous scheme, particle streaming was restricted to be on-grid, it is not a necessary condition for a working LB scheme. For the so-called semi-Lagrangian methods the restriction of Eq.~\eqref{eq:2-3-2-3} is relaxed, resulting in off-lattice propagation. As such, in this formulation the time-evolution operator of Eq.~\eqref{eq:2-3-2-9} is supplemented with an interpolation step to reconstruct the populations at the discrete grid-points:
\begin{equation}\label{eq:2-3-2-14}
\bar{f}_i\left(\bm{r}, t+\delta t\right)= \sum_{\bm{r_j}} A \left(\bm{r},\bm{r_j}\right)\left[\bar{f}_i\left(\bm{r_j}-\bm{c}_i\delta t, t\right) + \frac{\delta t}{\bar{\tau}}\left(  \bar{f}^{\rm eq}_i\left(\bm{r_j}-\bm{c}_i\delta t, t\right) - \bar{f}_i\left(\bm{r_j}-\bm{c}_i\delta t, t\right)\right)\right],
\end{equation}
where $A \left(\bm{r},\bm{r_j}\right)$ are the coefficients involved in the interpolation process and $\bm{r}_j$ are the interpolation stencil points. In practice, this approach has two main advantages: (a) it allows one to use quadratures of order four or five since those result in non-space-filling stencils, they are unusable with the on-lattice solvers~\cite{wilde_semi-lagrangian_2020}, (b) freedom over the choice of the time-step as the streaming does not need to fall on-grid \cite{wilde_multistep_2019,kramer_semi-lagrangian_2017}. On the other hand, the introduction of the interpolation operator strips the LBM from its strictly conservative property. The interpolation-supplemented LBM can only guarantee global mass conservation for uniform grids~\cite{wilde_semi-lagrangian_2020}.
\subsubsection{Correction for the diagonal components of the equilibrium third-order moments tensor}
In LBM schemes taking account of the correction term the system of PDE's changes into:
\begin{equation}
    \partial_t f_i + \bm{c}_i\cdot\bm{\nabla}f_i = \Omega_i + \Psi_i,
\end{equation}
where $\Psi_i$ denotes the correction term derived in Eqs.~\eqref{eq:correction_1} and \eqref{eq:correction_2}. The emergence of this additional term only affects the previously-detailed process of discretization in time and space through the change of variable:
\begin{equation}
    \bar{f}_i = f_i - \frac{\delta t}{2}\Omega_i - \frac{\delta t}{2}\Psi_i,
\end{equation}
\begin{equation}
\bar{f}^{\rm eq}_i = f^{\rm eq}_i,
\end{equation}
\begin{equation}
    \Omega_i = \frac{1}{\tau + \delta t/2}\left(  \bar{f}^{\rm eq}_i - \bar{f}_i\right) - \frac{\delta t/2}{\tau + \delta t/2}\Psi_i,
\end{equation}
which in turn leads to the following final algebraic system:
\begin{equation}
\bar{f}_i\left(\bm{r}+\bm{c}_i\delta t, t+\delta t\right) - \bar{f}_i\left(\bm{r}, t\right) = \frac{\delta t}{\bar{\tau}}\left( f^{\rm eq}_i\left(\bm{r}, t\right) - \bar{f}_i\left(\bm{r}, t\right)\right) + \left(1 - \frac{\delta t}{2\bar{\tau}}\right)\Psi_i.
\end{equation}
This consistent derivation of the \emph{extended} LBM holds for any form of the correction term, whether it is introduced simply as a Hermite-expanded term~\cite{feng_three_2015} or the extended equilibrium approach~\cite{saadat_extended_2021}.
\subsection{Introduction of external body forces\label{subsec:body_forces}}
Introduction of body force contributions in the context of LBM boils down to finding suitable approximations to the body force term $\frac{\bm{F}}{\rho}\cdot\bm{\nabla}_{\bm{v}}f$ in the Boltzmann equation. A large number of approaches and approximations have been devised over the years; We will restrict our review to the most widely used schemes here. For a more in-depth study of different forcing schemes we invite interested readers to look into~\cite{mohamad_critical_2010,huang_forcing_2011,bawazeer_critical_2021}. Given that all approximations can be recast into a generic form made up of a discrete source term $\mathcal{F}_i$ and a redefined \emph{real} velocity $\bm{u}^{\rm real}$ , they will all be presented in that form:
\begin{equation}
    f_i(\bm{r}+\bm{c}_i\delta t,t+\delta t) - f_i(\bm{r},t) = \frac{\delta t}{\bar{\tau}}\left( f_i^{\rm eq}(\rho,\bm{u}) - f_i(\bm{r},t) \right) + \mathcal{F}_i,
\end{equation}
with:
\begin{equation}
    \bm{u} = \frac{1}{\rho}\sum_i \bm{c}_i f_i.
\end{equation}
The aim of the approximation here being recovery of correct hydrodynamic limit, i.e. Euler+NS level dynamic we will have a more detailed look at the moments of $\mathcal{F}_i$ appearing there. It is clear that zeroth- and first-order moments will appear at the Euler level:
\begin{eqnarray}
	\partial_t^{(1)}\rho + \bm{\nabla}\cdot\rho \bm{u} = 0,\label{eq:approach2_continuity1}\\
	\partial_t^{(1)}\rho \bm{u} + \bm{\nabla}\cdot\rho \bm{u}\otimes\bm{u} + \bm{\nabla}\cdot \rho c_s^2\bm{I} + \sum_i \bm{c}_i \mathcal{F}_i = 0,\label{eq:eps1_force_1}
\end{eqnarray}
At the NS level the CE analysis leads to:
\begin{eqnarray}\label{eq:eps2_force_1}
    \partial_t^{(2)}\rho + \frac{1}{2}\bm{\nabla}\cdot\sum_i \bm{c}_i \mathcal{F}_i = 0,\\
	\partial_t^{(2)}\rho \bm{u} + \frac{1}{2} \partial_t^{(1)} \sum_i \bm{c}_i \mathcal{F}_i + \frac{1}{2}\bm{\nabla}\cdot\sum_i \bm{c}_i\otimes\bm{c}_i \mathcal{F}_i - \bm{\nabla}\cdot\left(\frac{1}{2}-\frac{\bar{\tau}}{\delta t}\right) \sum_i \bm{c}_i\otimes\bm{c}_i \mathcal{F}_i
	+ \bm{\nabla}\cdot\left(\frac{1}{2}-\frac{\bar{\tau}}{\delta t}\right)\left(\bm{u}\otimes\bm{F}+\bm{u}\otimes\bm{F}^{\dagger}\right) \nonumber\\
    + \bm{\nabla}\cdot\rho c_s^2\left(\frac{1}{2}-\frac{\bar{\tau}}{\delta t}\right)\left(\bm{\nabla}\bm{u}+\bm{\nabla}\bm{u}^{\dagger}\right) = 0.
\end{eqnarray}
For this system to match the targeted hydrodynamic limit it is obvious that one must have:
\begin{eqnarray}
    \sum_i \bm{c}_i \mathcal{F}_i = \delta t \bm{F},\label{eq:force_condition_1}\\
    \rho\bm{u} + \frac{1}{2}\sum_i \bm{c}_i \mathcal{F}_i = \rho\bm{u}^{\rm real},\label{eq:force_condition_2}\\
    \sum_i \bm{c}_i\otimes\bm{c}_i \mathcal{F}_i = \frac{\delta t^2}{4\rho\bar{\tau}}\bm{F}\otimes\bm{F} + \delta t\left(\bm{u}\otimes\bm{F} + \bm{u}\otimes\bm{F}^{\dagger}\right).\label{eq:force_condition_3}
\end{eqnarray}
\subsubsection{Shan and Chen's forcing scheme}
This approach was initially proposed by Shan and Chen to model multi-phase fluid systems~\cite{shan_lattice_1993}. In Shan and Chen's forcing scheme the velocity used in the computation of the discrete equilibrium is shifted by $\Delta u=\frac{\bar{\tau}\bm{F}}{\rho}$. While in its original form, as presented in~\cite{shan_lattice_1993}, the force contribution only appears in the EDF, the corresponding discrete time-evolution equations can be re-written as:
\begin{equation}\label{eq:shan_chen_body_force}
    f_i(\bm{r}+\bm{c}_i\delta t,t+\delta t) - f_i(\bm{r},t) = \frac{\delta t}{\bar{\tau}}\left( f_i^{\rm eq}(\rho,\bm{u}) - f_i(\bm{r},t) \right) + \underbrace{\frac{\delta t}{\bar{\tau}}\left( f_i^{\rm eq}(\rho,\bm{u}+\Delta\bm{u}) - f_i^{\rm eq}(\rho,\bm{u})\right)}_{\mathcal{F}_i}.
\end{equation}
Additionally the real fluid velocity computation is also affected by the force as:
\begin{equation} \label{eq:shan_chen_body_force_real_vel}
    \bm{u}^{\rm real} = \frac{1}{\rho}\sum_i \bm{c}_i f_i + \frac{\delta t \bm{F}}{2\rho}.
\end{equation}
The source term has the following zeroth- to second-order moments:
\begin{eqnarray}
    \sum_i \mathcal{F}_i &=& 0,\label{eq:SC_force_moments_0}\\
    \sum_i \bm{c}_i \mathcal{F}_i &=& \bm{F}\delta t,\label{eq:SC_force_moments_1}\\
    \sum_i \bm{c}_i\otimes\bm{c}_i \mathcal{F}_i &=& \delta t\left(\bm{F}\otimes\bm{u}+{\bm{F}\otimes\bm{u}}^\dagger\right) + \frac{\delta t \bar{\tau}}{\rho}\bm{F}\otimes\bm{F}.\label{eq:SC_force_moments_2}
\end{eqnarray}
This means that this approach satisfies condition~\eqref{eq:force_condition_1}. However, given the definition of the real velocity,
\begin{equation}
    \rho \bm{u}^{\rm real}\otimes\bm{u}^{\rm real} = \rho \bm{u}\otimes\bm{u} + \frac{\delta t}{2}\left(\bm{u}\otimes\bm{F}+\bm{u}\otimes\bm{F}^{\dagger}\right) + \frac{\delta t^2}{4\rho}\bm{F}\otimes\bm{F},
\end{equation}
it admits an error of the following form in the convective term of the NS level momentum balance equation:
\begin{equation} \label{eq:shan_chen_body_force_error}
    \delta_{\rm convection} = \bm{\nabla}\cdot\frac{\delta t^2}{\rho}\left(\frac{\bar{\tau}^2}{\delta t^2} - \frac{1}{4}\right)\bm{F}\otimes\bm{F}.
\end{equation}
It should be noted that, expanding the body force as a power series of the smallness parameter, this deviation in the second-order moment would scales with $\varepsilon^3$ and therefor not appear formally in the momentum balance equation at the Navier-Stokes level. Nevertheless, for cases involving large force contributions higher-order dynamics can spoil Euler and NS level behavior. Later on, this scheme was improved upon via a Hermite expansion of the force term in the Boltzmann equation~\cite{martys_evaluation_1998}. Using the Hermite expansion of the distribution function, the force term can be expanded as~\cite{martys_evaluation_1998,shan_kinetic_2006}:
\begin{equation}
    \bm{F}\cdot\bm{\nabla}_{\bm{v}}f = \sum_{n=0}\frac{{(-1)}^n}{n!}\bm{F}\otimes\bm{a}_{n-1} :\bm{\nabla}_{\bm{v}}^{n+1} w(\bm{v}) = -w(\bm{v})\sum_{n=0}\frac{1}{(n-1)!} \bm{F}\otimes\bm{a}_{n-1}:\bm{H}_{n}(\bm{v}).
\end{equation}
Absorbing the force in the Boltzmann equation into the equilibrium and using the above-presented Hermite expansion of the body force term a correction to the original Shan and Chen scheme was proposed as:
\begin{equation} \label{eq:shan_chen_body_force_correction}
    \mathcal{F}_i =  \frac{\delta t}{\bar{\tau}}  \left( f_i^{\rm eq}(\rho,\bm{u}+\Delta\bm{u}) - f_i^{\rm eq}(\rho,\bm{u})\right) - \frac{\rho w_i \bar{\tau}}{2}\left(\frac{{(\bm{c}_i\cdot\bm{F})}^2}{c_s^4\rho^2} - \frac{\bm{F}^2}{c_s^2\rho^2}\right),
\end{equation}
which changes the second-order moment of the force term into:
\begin{equation}
    \sum_i \bm{c}_i\otimes\bm{c}_i \mathcal{F}_i = \delta t\left(\bm{u}\otimes\bm{F} + \bm{u}\otimes\bm{F}^{\dagger}\right),
\end{equation}
eventually leading to an error term that is still present but independent from the relaxation time:
\begin{equation}
    \delta_{\rm convection} = -\bm{\nabla}\cdot\frac{\delta t^2}{4\rho}\bm{F}\otimes\bm{F}.
\end{equation}
\subsubsection{Luo's scheme}
As an approximation to the body force term appearing in the Boltzmann expression, and assuming a near equilibrium flow, Luo proposed an expansion similar to that used for the equilibrium~\cite{luo_unified_1998}, i.e.
\begin{equation}
\frac{\bm{F}}{\rho}\cdot\bm{\nabla}_{\bm{v}} f \approx \rho w(\bm{v})\left[ a_0 + \bm{a}_1\cdot\bm{v} + \bm{a}_2:\bm{v}\otimes\bm{v} + \dots\right],
\end{equation}
where expansion coefficients $\bm{a}_i$ are computed via the following constraints on the moments of the body force term:
\begin{eqnarray}
    \int \frac{\bm{F}}{\rho}\cdot\bm{\nabla}_{\bm{v}} f d\bm{v} &=& 0,\\
    \int \bm{v}\frac{\bm{F}}{\rho}\cdot\bm{\nabla}_{\bm{v}} f d\bm{v} &=& -\bm{F},\\
    \int \bm{v}\otimes\bm{v}\frac{\bm{F}}{\rho}\cdot\bm{\nabla}_{\bm{v}} f d\bm{v} &=& -\bm{F}\otimes\bm{u} - {\bm{F}\otimes\bm{u}}^\dagger.
\end{eqnarray}
Applying these constraints and limiting the expansion to order two the following space/time-continuous approximation is obtained:
\begin{equation}
    \frac{\bm{F}}{\rho}\cdot\bm{\nabla}_{\bm{v}} f = -w(\bm{v})\left[ \frac{\bm{F}\cdot(\bm{v}-\bm{u})}{c_s^2} + \frac{(\bm{v}\cdot\bm{u})\bm{v}\cdot\bm{F}}{c_s^4}\right] + \mathcal{O}\left(\bm{v}^3,\bm{u}^2\right),
\end{equation}
which after discretization in phase space results in:
\begin{equation}
    \frac{\bm{F}}{\rho}\cdot\bm{\nabla}_{\bm{v}} f = -w_i\left[ \frac{\bm{F}\cdot(\bm{c}_i-\bm{u})}{c_s^2} + \frac{(\bm{c}_i\cdot\bm{u})\bm{c}_i\cdot\bm{F}}{c_s^4}\right].
\end{equation}
One interesting point worth noting in Luo's forcing scheme is that over the years a point of confusion seems to have been installed in the literature~\cite{mohamad_critical_2010,huang_forcing_2011,bawazeer_critical_2021}; In the original article~\cite{luo_unified_1998}, the author has used a first-order approximation to the collision integral after integration along characteristics, i.e.
\begin{equation}
    \int_t^{t+\delta t} \frac{1}{\tau}\left[f_i^{\rm eq}(\bm{r}+\bm{c}_i t',t') - f_i(\bm{r}+\bm{c}_i t',t')\right] + \mathcal{F}_i(\bm{r}+\bm{c}_i t',t') dt' \approx \frac{\delta t}{\tau}\left[f_i^{\rm eq}(\bm{r},t) - f_i(\bm{r},t)\right] + \delta t \mathcal{F}_i(\bm{r},t),
\end{equation}
naturaly leading to the following time-evolution equations:
\begin{equation}
    f_i(\bm{r}+\bm{c}_i\delta t, t+\delta t) - f_i(\bm{r}, t) = \frac{\delta t}{\tau}\left(f_i^{\rm eq} - f_i\right) + \mathcal{F}_i \delta t,
\end{equation}
where contrary to the second-order approach using the trapezoidal rule, the relaxation time and distribution functions have not been redefined. Most articles following this scheme apply this final expression in the context of the classical lattice Boltzmann model involving a re-definition of the relaxation coefficient and distribution function. Properly transposing Luo's scheme to a second-order LBM would instead lead to:
\begin{equation}\label{eq:luo_body_force}
    \mathcal{F}_i = \frac{\delta t}{\bar{\tau}}\left( f_i^{\rm eq}(\rho,\bm{u}+\frac{\delta t \bm{F}}{2\rho}) - f_i^{\rm eq}(\rho,\bm{u})\right) + \delta t\left(1-\frac{\delta t}{2\bar{\tau}}\right)w_i\left[ \frac{\bm{F}\cdot(\bm{c}_i-\bm{u})}{c_s^2} + \frac{(\bm{c}_i\cdot\bm{u})\bm{c}_i\cdot\bm{F}}{c_s^4}\right]+ \left(1-\frac{\delta t}{2\bar{\tau}}\right)\frac{w_i \delta t^2}{\rho}\left[\frac{{(\bm{c}_i\cdot\bm{F})}^2}{c_s^4} - \frac{\bm{F}^2}{c_s^2}\right],
\end{equation}
with,
\begin{equation} \label{eq:luo_body_force_real_vel}
    \bm{u}^{\rm real} = \frac{1}{\rho}\left(\frac{\bm{F}\delta t}{2} + \sum_i \bm{c}_i f_i \right).
\end{equation}
This leads to an error in the convection term of the form:
\begin{equation} \label{eq:luo_body_force_error}
    \delta_{\rm convection} = \bm{\nabla}\cdot\delta t\left(\bar{\tau}-\frac{\delta t}{2}\right)\frac{\bm{F}\otimes\bm{F}}{\rho}.
\end{equation}
This approach is exactly equivalent to Guo's forcing scheme~\cite{guo_discrete_2002}. A number of articles have reported better stability of this approach as compared to other forcing schemes~\cite{mohamad_critical_2010}, however this is clearly a consequence of the first-order nature of the model as (mis-)used there.
\subsubsection{He et al's scheme}
The next scheme, proposed by He et al. relies on the following fundamental approximation when evaluating the body force contribution~\cite{he_discrete_1998}:
\begin{equation}
    f(\bm{r},t,\bm{v})\approx f^{\rm eq}(\bm{r},t,\bm{v}),
\end{equation}
leading to:
\begin{equation}
    \frac{\bm{F}}{\rho}\cdot\bm{\nabla}_{\bm{v}} f \approx \frac{\bm{F}}{\rho}\cdot\bm{\nabla}_{\bm{v}} f^{\rm eq} = \frac{\bm{F}}{\rho}\cdot \frac{\bm{v}-\bm{u}}{c_s^2}f^{\rm eq},
\end{equation}
which after discretization in phase space, integration along characteristics and using the trapezoidal rule results in:
\begin{equation} \label{eq:he_body_force}
    \mathcal{F}_i = \frac{\delta t}{\bar{\tau}}\left( f_i^{\rm eq}(\rho,\bm{u}+\frac{\delta t \bm{F}}{2\rho}) - f_i^{\rm eq}(\rho,\bm{u})\right)  + \delta t \left(1-\frac{\delta t}{2\bar{\tau}}\right)\frac{\bm{F}}{\rho}\cdot \frac{\bm{c}_i-\bm{u}}{c_s^2}f^{\rm eq}(\rho,\bm{u}+\frac{\delta t \bm{F}}{2\rho}),
\end{equation}
with
\begin{equation} \label{eq:he_body_force_real_vel}
    \bm{u}^{\rm real} = \frac{1}{\rho}\left(\frac{\bm{F}\delta t}{2} + \sum_i \bm{c}_i f_i \right).
\end{equation}
\subsubsection{Guo's scheme}
Guo proposed a modified forcing scheme taking into account so-called \emph{discrete effects} in \cite{guo_discrete_2002}. In essence Guo followed an approach quite similar to Luo to derive the new scheme, i.e. moment matching. Similar to Luo~\cite{luo_unified_1998}, Guo started with a polynomial approximation to the force contribution with coefficients to be fixed by moments constraints. However, at the difference of Luo, moments constraints were extracted from a Chapmann-Enskog analysis of the lattice Boltzmann equations, i.e. after discretization in space and time. Therefore, the remark made in the previous paragraphs about the exact correspondence between Luo's scheme, once a second-order integration along characteristics has been applied, and Guo's discrete forcing scheme is not surprising at all.
\subsubsection{Kupershtokh's scheme}
This approach, also referred to as the exact difference method approximates the force contributions as~\cite{kupershtokh_new_2004,kupershtokh_equations_2009}:
\begin{equation} \label{eq:edm_body_force}
    \mathcal{F}_i = \left( f_i^{\rm eq}(\rho,\bm{u}+\Delta\bm{u}) - f_i^{\rm eq}(\rho,\bm{u})\right),
\end{equation}
with, different from the Shan-Chen scheme, $\Delta u=\frac{\bm{F}\delta t}{\rho}$. Furthermore the velocity $\bm{u}$ is computed as the first-order moment of the distribution function with no additional terms and the \emph{real} fluid velocity, $\bm{u}^{\rm real}$, is computed as $\bm{u}^{\rm real}=\bm{u}+\frac{\bm{F}\delta t}{2\rho}$. At the difference of the Shan-Chen scheme the second-order moment of the source term is:
\begin{equation}
    \sum_i \bm{c}_i\otimes\bm{c}_i \mathcal{F}_i = \delta t\left(\bm{F}\otimes\bm{u}+{\bm{F}\otimes\bm{u}}^\dagger\right) + \frac{\delta t^2}{\rho}\bm{F}\otimes\bm{F},
\end{equation}
which does not show any dependence on the relaxation. As for previously-listed schemes, the exact difference method also admits a convective error of the form:
\begin{equation} \label{eq:edm_body_force_error}
    \delta_{\rm convective} = \bm{\nabla}\delta t\left(\bar{\tau}-\frac{\delta t}{4}\right)\frac{\bm{F}\otimes\bm{F}}{\rho}.
\end{equation}
This deviation can be eliminated by introducing an additional contribution in the source term as:
\begin{equation}\label{eq:edm_body_force_correction}
    \mathcal{F}_i = \left( f_i^{\rm eq}(\rho,\bm{u}+\Delta\bm{u}) - f_i^{\rm eq}(\rho,\bm{u})\right) - \left(1-\frac{1}{4\bar{\tau}}\right) w_i\mathcal{H}_2:\frac{\bm{F}\otimes\bm{F}}{\rho}.
\end{equation}
\par The different strategies to incorporate body forces into the LBM are listed in Table~\ref{tab:forcing_schemes}.
\begin{table}[ht!]
	\centering
	\begin{tabular}{l|l|l|l|l}
		Approach & $\mathcal{F}_i$ & $\bm{u}^{\rm real}$ & Leading-order error & Improvement \\
		\hline
        Shan and Chen~\cite{shan_lattice_1993} & Eq.~\eqref{eq:shan_chen_body_force} & Eq.~\eqref{eq:shan_chen_body_force_real_vel} & Eq.~\eqref{eq:shan_chen_body_force_error} & Eq.~\eqref{eq:shan_chen_body_force_correction} \\
        Luo~\cite{luo_unified_1998} & Eq.~\eqref{eq:luo_body_force} & Eq.~\eqref{eq:luo_body_force_real_vel} & Eq.~\eqref{eq:luo_body_force_error} & None \\
        Guo~\cite{guo_discrete_2002} & Eq.~\eqref{eq:luo_body_force} & Eq.~\eqref{eq:luo_body_force_real_vel} & Eq.~\eqref{eq:luo_body_force_error} & None \\
        EDM ~\cite{kupershtokh_new_2004} & Eq.~\eqref{eq:edm_body_force} & Eq.~\eqref{eq:luo_body_force} & Eq.~\eqref{eq:edm_body_force_error} & Eq.~\eqref{eq:edm_body_force_correction} \\
	\end{tabular}
	\caption{Summary of body force methods in LBM.}
	\label{tab:forcing_schemes}
\end{table}
\subsection{Stability of LB-BGK}
The applicability range of different closures for the discrete equilibrium distribution functions was characterized in previous sections by monitoring deviations of moments of order two and three. Here we discuss the issue of application range with more details via the linear stability and positivity domain of the discrete equilibrium distribution functions. Prior to that we also discuss a numerical artifact brought about by the isothermal assumption of the classical LBM responsible, in part, for the stability of the solver.
\subsubsection{Isothermal closure: spurious bulk viscosity and stabilization of normal modes}
Although widely used for simulation in the incompressible regime, it is well known that the LBM relies instead on an isothermal closure leading to a fixed and finite speed of sound, $c_s$. For a characteristic convective velocity $\mathcal{U}$, in the limit $\mathcal{U}/c_s\rightarrow 0$ it is expected to recover the incompressible flow behavior. This means that contrary to classical incompressible solvers with the Poisson equation as closure for the pressure field, here $\bm{\nabla}\cdot\bm{u}\neq0$ and $c_s\neq\infty$, which in turn points to presence of so-called acoustic eigen-modes that are damped at a rate $\eta$, also referred to as the bulk viscosity. For low-dissipative central-in-space discretization methods like the LBM, one must therefor guarantee positivity of the dissipation rates of both shear and acoustic modes. However, it is well known that for a mono-atomic molecule $\eta$=0. Given that only translational degrees of freedom are accounted for, the LBM is based on a mono-atomic molecule and should lead to zero bulk viscosity. This is readily observed by looking at the non-equilibrium stress tensor of the Boltzmann equation at the NS level:
\begin{equation}
    \Pi_{\alpha\beta}^{(1)} = -\tau\left(\partial_t^{(1)}\int v_\alpha v_\beta f^{\rm eq}d\bm{v} + \partial_\gamma \int v_\gamma v_\alpha v_\beta f^{\rm eq}d\bm{v}\right),
\end{equation}
where the first and second terms can be, after some algebra, re-written as:
\begin{equation}
    \int v_\gamma v_\alpha v_\beta f^{\rm eq}d\bm{v} = \rho u_\alpha u_\beta u_\gamma + \rho \frac{k_B T}{m} \left[ u_{\alpha}\delta_{\beta, \gamma}\right]_{\text{cyc}},
\end{equation}
and
\begin{equation}
    \partial_t^{(1)}\int v_\alpha v_\beta f^{\rm eq}d\bm{v} = \delta_{\alpha\beta}\partial_t^{(1)}\rho \frac{k_B T}{m} + u_\alpha \partial_t^{(1)}\rho u_\beta + u_\beta \partial_t^{(1)}\rho u_\alpha - u_\alpha u_\beta \partial_t^{(1)}\rho.
\end{equation}
Using the mass, momentum and internal energy balance equations,
\begin{equation}
    \partial_t^{(1)}\rho \frac{k_B T}{m} + \partial_\alpha \rho\frac{k_B T}{m} u_\alpha = - \frac{2\rho k_B T}{D m} \partial_\alpha u_\alpha,
\end{equation}
the non-equilibrium stress tensor can be re-written as:
\begin{equation}
    \Pi_{\alpha\beta}^{(1)} = -\tau \rho \frac{k_B T}{m} \left(\partial_\alpha u_\beta + \partial_\beta u_\alpha - \frac{2}{D}\partial_\gamma u_\gamma\right),
\end{equation}
confirming the absence of bulk viscosity. In the classical LBM, an isothermal closure is used for energy/temperature meaning the solvability condition for energy is replaced by $\frac{k_B T}{m}=\frac{k_B T_0}{m_0}$ leading to the following non-equilibrium stress tensor~\cite{dellar_bulk_2001}:
\begin{equation}
    \Pi_{\alpha\beta}^{(1)} = -\tau \rho \frac{k_B T}{m} \left(\partial_\alpha u_\beta + \partial_\beta u_\alpha - \frac{2}{D}\partial_\gamma u_\gamma\right) - \tau \rho \frac{2k_B T}{D m}\partial_\gamma u_\gamma,
\end{equation}
meaning $\eta = \frac{2}{D}\mu$. After discretization in phase space, space and time this property is maintained and one recovers a non-zero bulk viscosity for the classical isothermal LBM:
\begin{equation}
    \eta = \frac{2 c_s^2}{D}\left(\bar{\tau} - \frac{\delta t}{2}\right) + \mathcal{O}({\rm Ma}^2),
\end{equation}
where the Galilean-variant error comes from deviation in the diagonal components of the equilibrium third-order moments which can be eliminated via appropriate correction terms derived in previous sections.
\subsubsection{Positivity of the discrete equilibria}
\begin{figure}[h!]
	\centering
	\includegraphics[width=12cm,keepaspectratio]{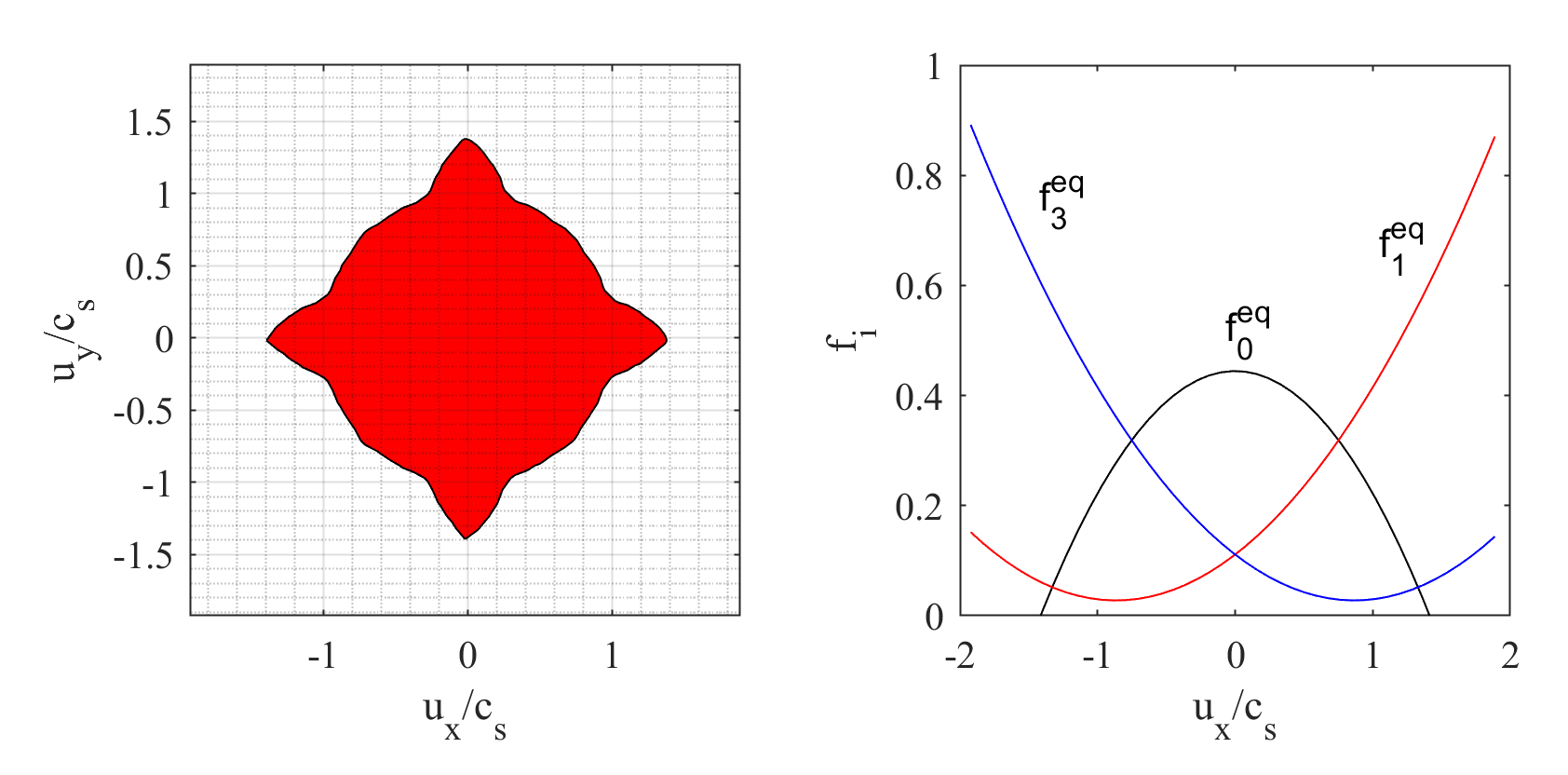}
	\caption{Illustration of positivity of the second-order polynomial EDF. Left: Domain ensuring positivity of discrete equilibrium populations in red. Right: Values of three discrete populations as a function of $u_x/c_s$ for $u_y/c_s=0$.}
	\label{Fig:positivity_HE2}
\end{figure}

\begin{figure}[h!]
	\centering
	\includegraphics[width=12cm,keepaspectratio]{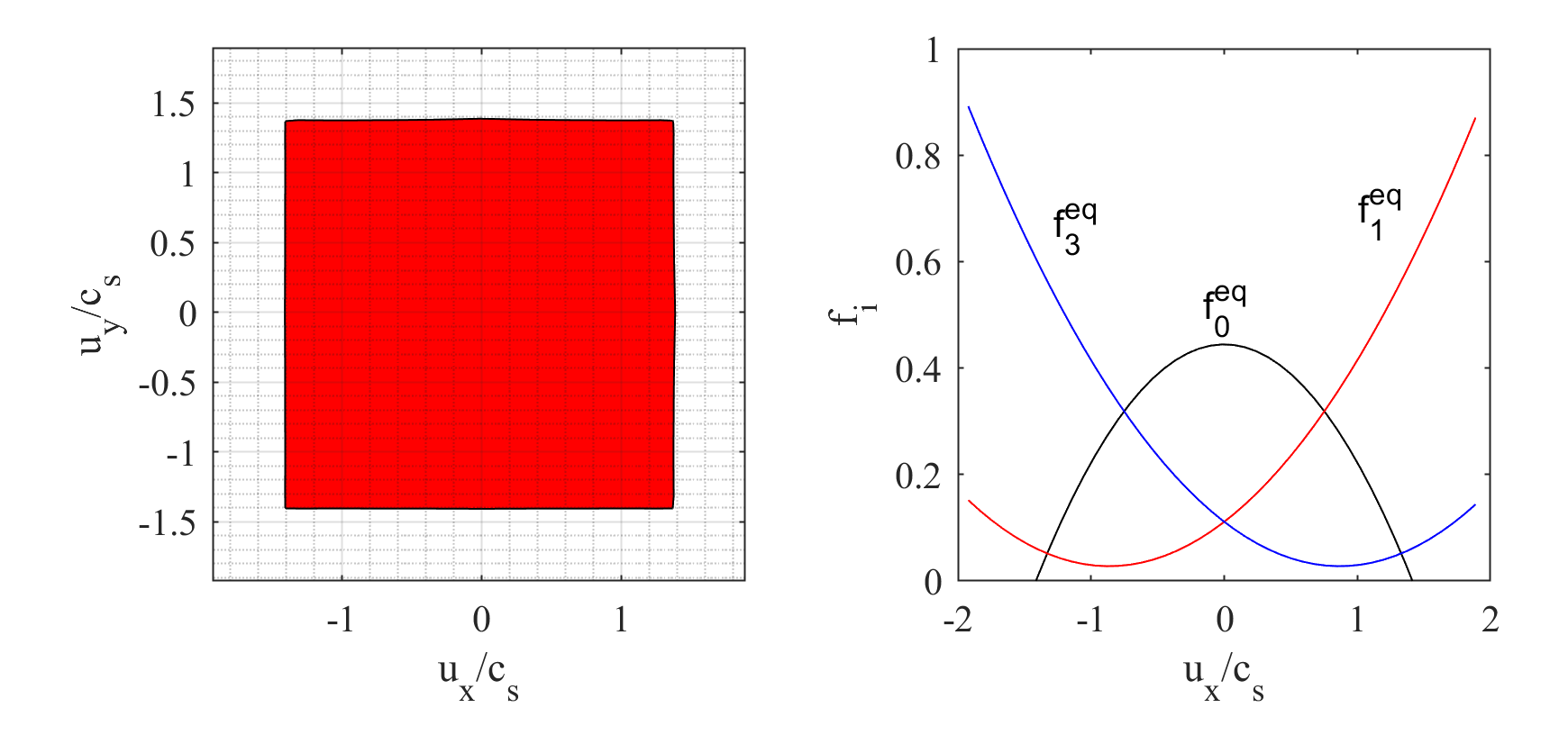}
	\caption{Illustration of positivity of the product-form EDF. Left: Domain ensuring positivity of discrete equilibrium populations in red. Right: Values of three discrete populations as a function of $u_x/c_s$ for $u_y/c_s=0$.}
	\label{Fig:positivity_product_form}
\end{figure}

\begin{figure}[h!]
	\centering
	\includegraphics[width=12cm,keepaspectratio]{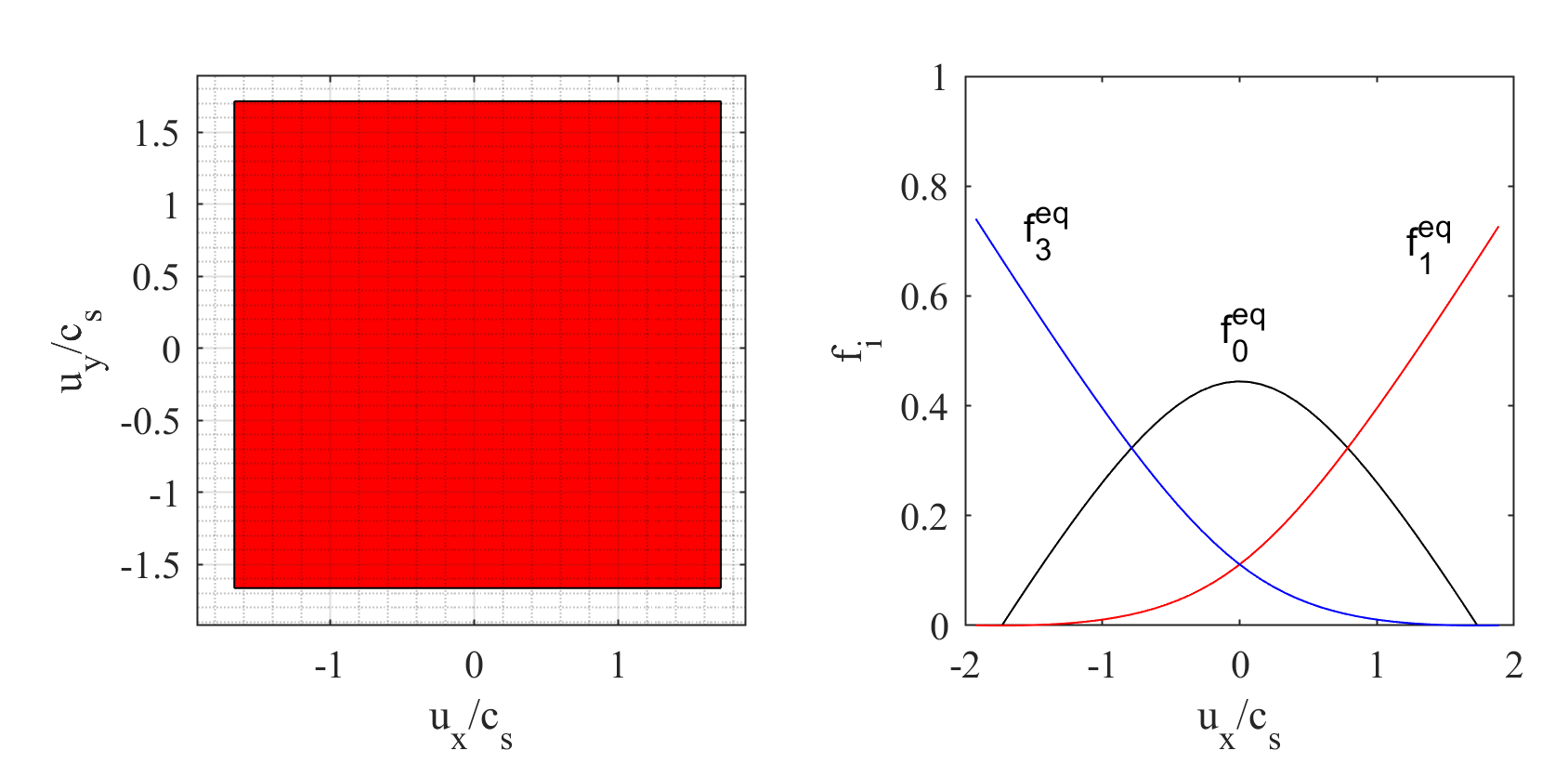}
	\caption{Illustration of positivity of the entropic EDF. Left: Domain ensuring positivity of discrete equilibrium populations in red. Right: Values of three discrete populations as a function of $u_x/c_s$ for $u_y/c_s=0$.}
	\label{Fig:positivity_entorpic}
\end{figure}

Numerical instability in the LBM
has been often tied to the absence of a positivity constraint on the discrete populations~\cite{li_numerical_2004,tosi_numerical_2006,brownlee_stabilization_2006}. In the specific case of LBM for advection-diffusion equations the positivity region of the discrete EDF has been shown to coincide with the linear stability domain in the limit of vanishing diffusion coefficient~\cite{servan-camas_lattice_2008,servan-camas_non-negativity_2009,hosseini_stability_2017}. To that end, before conducting linear stability analyses we look at the positivity domains of different discrete EDFs. The positivity domains of the second-order polynomial, product-form and entropic EDFs are shown in Figs.~\ref{Fig:positivity_HE2}, \ref{Fig:positivity_product_form} and \ref{Fig:positivity_entorpic}.
While the polynomial forms of the EDF, both second-order and product form, do not guarantee positivity of the EDF for all velocities, the entropic EDF ensures that equilibrium populations remain positive for all velocities $-1\leq u_\alpha\delta r/\delta t\leq1$.
However, as will be seen in the next chapter, different from the advection-diffusion LBM the positivity domain is not necessarily a reflection of linear stability.
\subsubsection{Linear stability of discrete solver}
\par The von Neumann (VN) stability analysis aims at studying the time evolution of a perturbation $\bar{f}^{'}_i$ that is injected into the linearized discrete time evolution equations. 
The perturbation is expanded as a combination of standing waves, whose propagation speed and attenuation rate will be obtained as a result of the VN analysis. A positive attenuation rate will result in a growth of the error at the corresponding wave-length and linear instability of the solver for the set of parameters considered ($\bar{\tau}$, $\mathrm{Ma}$, etc). On the contrary, the scheme is linearly stable if it remains negative for all wave-numbers.\\
Furthermore, the spectral behavior and accuracy can be readily analyzed by comparing the spectral dispersions and dissipations to the theoretical modes obtained from the linearized NS equations. The NS theoretical modes for an isothermal flow can be expressed as \cite{chavez-modena_improving_2018}:
	\begin{subequations}
		\begin{align}
		\omega^{\text{shear}} &=\bm{u}\cdot\bm{k} - i \nu\bm{k}^2, \\
		\omega^{\text{acoustic}} &=\left(\bm{u}\pm c_s\right)\cdot\bm{k} - i \left(\frac{D-1}{D}\nu+\frac{\eta}{2\rho}\right)\nu\bm{k}^2,
		\end{align}
	\end{subequations}
where $D$ is the physical dimension of the system and $\bm{k}$ the wave-number vector. As a consequence, the VN stability analysis can be used to evaluate the spectral behavior and linear stability domain of a LBM for a given set of parameters. As such it can be perceived as a tool to objectively evaluate the stabilization properties of different collision models, on the basis of necessary conditions. The latter comes from the fact that the analysis relies on a linearization step and as such gets the sufficient condition for stability only under the linear regime assumption (small amplitude perturbations). It has been widely used in the past to evaluate the stability properties of the LBM. Interested readers are referred to \cite{sterling_stability_1996,worthing_stability_1997,lallemand_theory_2000,hosseini_stability_2017,hosseini_stability_2019,hosseini_extensive_2019,wissocq_extended_2019,chavez-modena_improving_2018}, among other sources.
\par
\begin{figure}[h!]
	\centering
	\includegraphics[width=13cm,keepaspectratio]{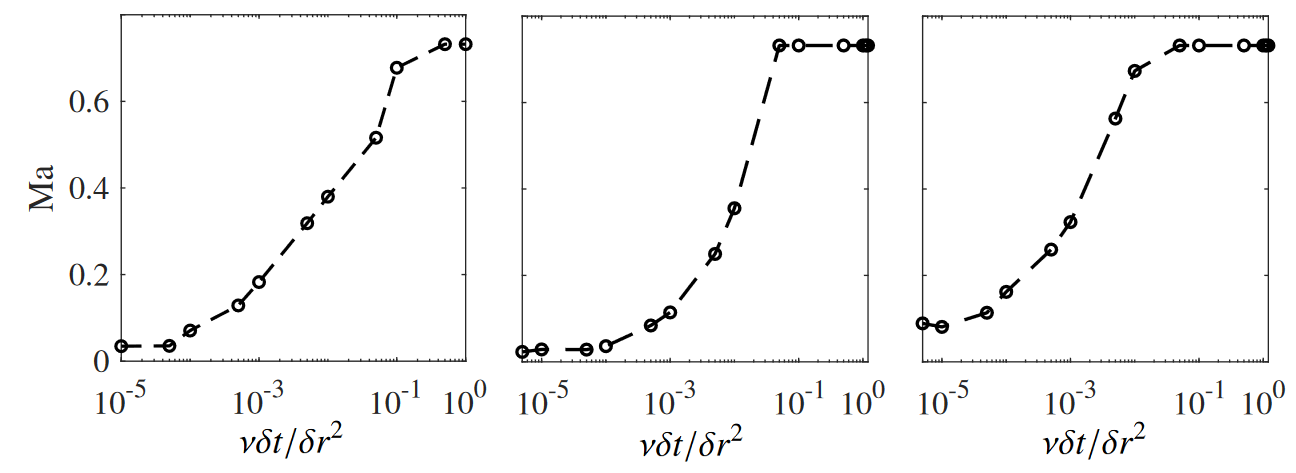}
	\caption{Linear stability domains of SRT collision operator with EDFs of orders (from left to right) two, three and four. The fourth-order expansion is equivalent to the product-form. Reproduced from \cite{hosseini_development_2020}.}
	\label{Fig:SRT_stability_domain}
\end{figure}

\begin{figure}[h!]
	\centering
	\includegraphics[width=13cm,keepaspectratio]{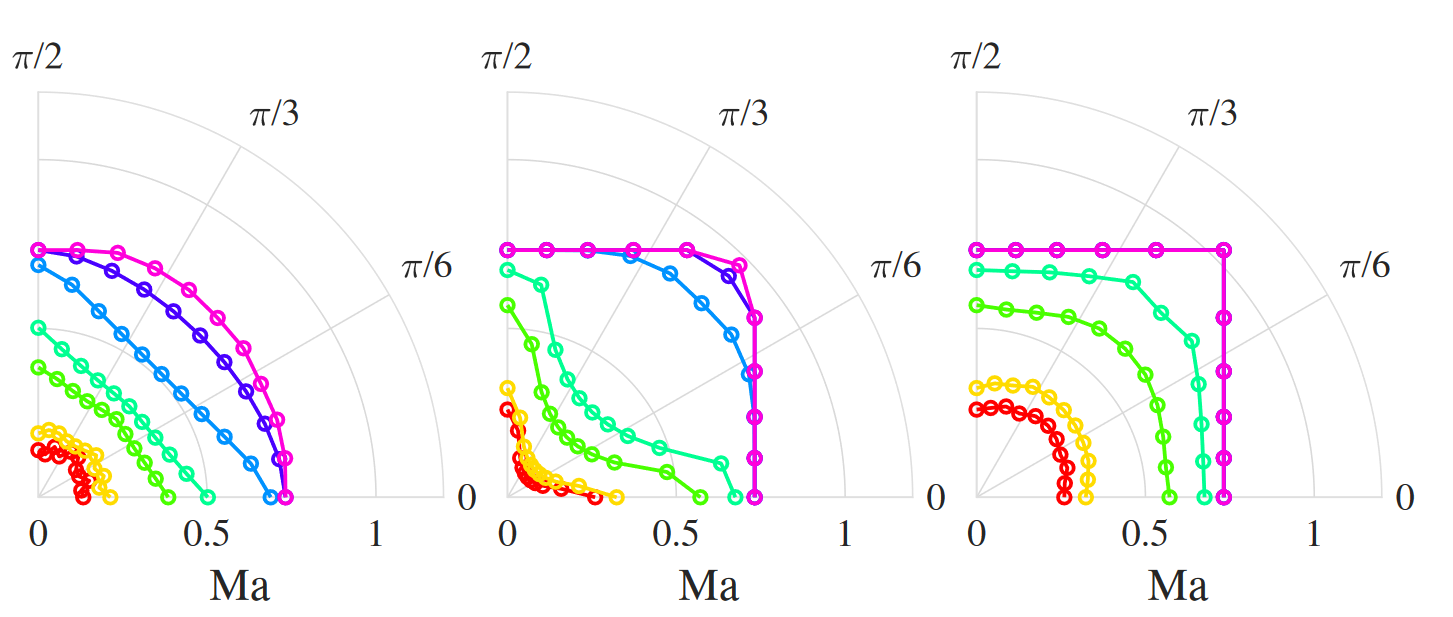}
	\caption{Illustration of anisotropy of linear stability domains for EDFs of orders (from left to right) two, three and four, and for seven different non-dimensional kinematic viscosities, {i.e.} (\protect\lineone)$5\times10^{-4}$, (\protect\linetwo)$1\times10^{-3}$, (\protect\linethree)$5\times10^{-3}$, (\protect\linefour)$0.01$, (\protect\linefive)$0.05$, (\protect\linesix)$0.1$, (\protect\lineseven)$0.5$. Reproduced from \cite{hosseini_development_2020}.}
	\label{Fig:SRT2_stability_area}
\end{figure}

\begin{figure}[h!]
	\centering
	\includegraphics[width=5cm,keepaspectratio]{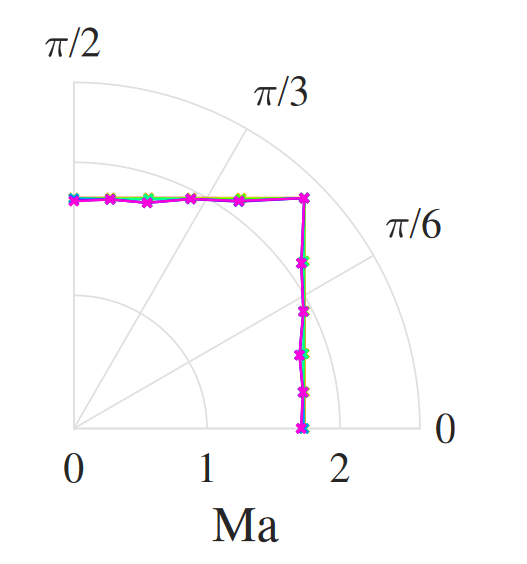}
	\caption{Illustration of linear stability domain for the entropic EDF for seven different non-dimensional kinematic viscosities, {i.e.} (\protect\lineone)$5\times10^{-4}$, (\protect\linetwo)$1\times10^{-3}$, (\protect\linethree)$5\times10^{-3}$, (\protect\linefour)$0.01$, (\protect\linefive)$0.05$, (\protect\linesix)$0.1$, (\protect\lineseven)$0.5$. Reproduced from \cite{hosseini_development_2020} and \cite{hosseini_extensive_2019}.}
	\label{Fig:ESRT_stability_area}
\end{figure}

\begin{figure}[h!]
	\centering
	\includegraphics[width=12cm,keepaspectratio]{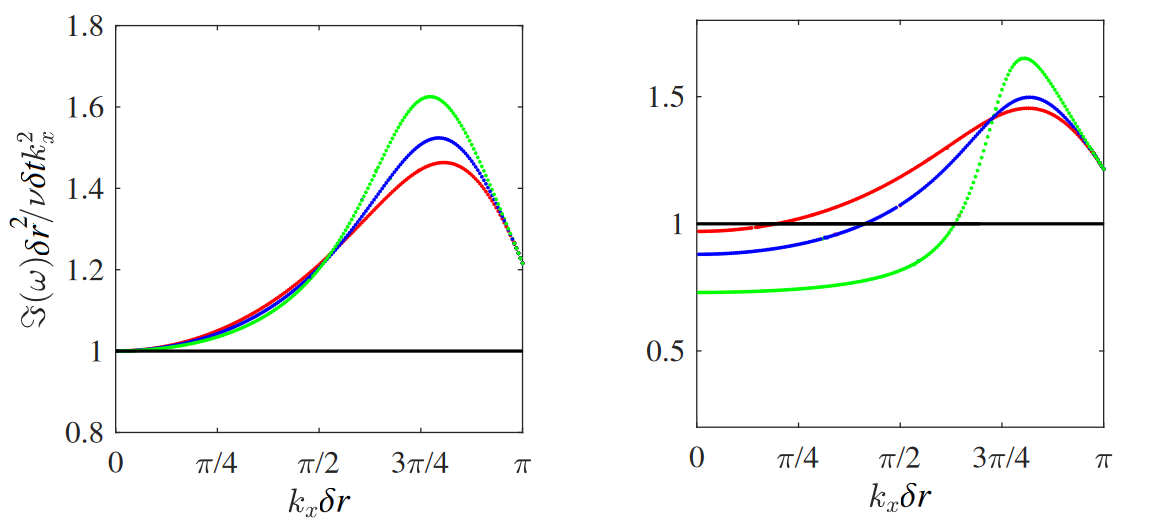}
	\caption{Shear mode dissipation rate (normalized by its physical counterpart) for (left) third- and (right) second-order EDF for three different Mach numbers, {i.e.} (in red) 0.1, (in blue) 0.2 and (in green) 0.3. The continuum reference is shown with a plain black line. Reproduced from \cite{hosseini_development_2020}.}
	\label{Fig:STR2_vs_SRT3_shear_dissipation}
\end{figure}

 As mentioned previously, the equilibrium state is one of the most important components of a kinetic scheme and controls, for the most part, the leading-order dynamics of the system ({i.e.} the macroscopic PDEs of interest), but also the behavior of higher-order (errors, ghost modes) terms. The effects of the EDF on leading-order terms were studied in previous sections. In this subsection, using the VN formalism we look at the effect of the EDF on the linear stability domain. To do so the eigenvalue problem of the VN equations is solved for different values of non-dimensional viscosities, over the entire wave-number space, {i.e.} $k_x$ and $k_y$ with a resolution of 100 points in each direction. The highest Mach number resulting in negative dissipation rates over all wave-numbers is retained as the linear stability limit. These limits are shown in Fig.~\ref{Fig:SRT_stability_domain}.
Looking at those results a number of points are worth noting: For all of these EDFs, regardless of the value of the non-dimensional viscosity (Fourier number), the maximum stable Mach number never goes beyond $\text{Ma}=\sqrt{3}-1\approx0.732$. This confirm the observation in Eq.~\ref{eq:CFL_lmit_absolute_poly}:
\begin{equation}
    {\rm Ma}^{\rm max}_\alpha = \frac{\lvert u_\alpha^{\rm max}\lvert}{c_s} = \sqrt{3} - 1.
\end{equation}
Furthermore while the addition of third-order components appears not to have a large effect on the stability domain, the fourth-order component (which does not affect the terms appearing at the NS level) extends it.\\
Apart from extending the linear stability domain, the addition of the fourth-order component results in more isotropic behavior especially for small values of the non-dimensional viscosity. The directional stability domains obtained with different orders of the EDF are shown in Fig.~\ref{Fig:SRT2_stability_area}.
It is also worth noting that the entropic EDF, is found to come with unconditional linear stability for all values of the Mach number supported by the stencil, {i.e.} $\hbox{Ma}=\sqrt{3}$, even for vanishing viscosities. This unconditional stability can be readily explained by the observations in Fig.~\ref{Fig:sound_speed_entropic_vs_poly}, i.e. self-adjusting sound speed guaranteeing fastest mode propagates at a speed smaller than or equal to $\delta r/\delta t$.
The stability domain is shown in Fig.~\ref{Fig:ESRT_stability_area}. This in turn confirms the effectiveness of the discrete EDF construction approach in guaranteeing linear stability (by enforcing a discrete H-theorem).
Finally, one can readily confirm the assertion made in the previous subsections concerning the effect of third-order Hermite terms on the deviatoric components of the third-order moments tensor by looking at the spectral dissipation of physical modes. The spectral dissipation of the shear modes of the third and second-order EDF for three different Mach numbers are shown in Fig.~\ref{Fig:STR2_vs_SRT3_shear_dissipation}.
It is clearly observed that for the third-order EDF, in the limit of vanishing wave-numbers $k_x/\delta r \rightarrow 0$, the obtained dissipations converge to the correct value regardless of the Mach number. However for the second-order EDF signs of Galilean invariance problems are clearly observed as the continuum limit of shear mode dissipation changes with the Mach number.\\
The results obtained in this subsection also point to the fact that the SRT collision operator becomes practically unusable below non-dimensional viscosities of $10^{-3}-10^{-4}$. Different strategies have been developed to allow simulations at lower non-dimensional viscosities. We will discuss some of these approaches in the next section.
\subsection{Extension of stability domain}

\subsubsection{Relaxation of discrete populations in alternate spaces}
\begin{figure}[h!]
	\centering
	\includegraphics[width=6cm,keepaspectratio]{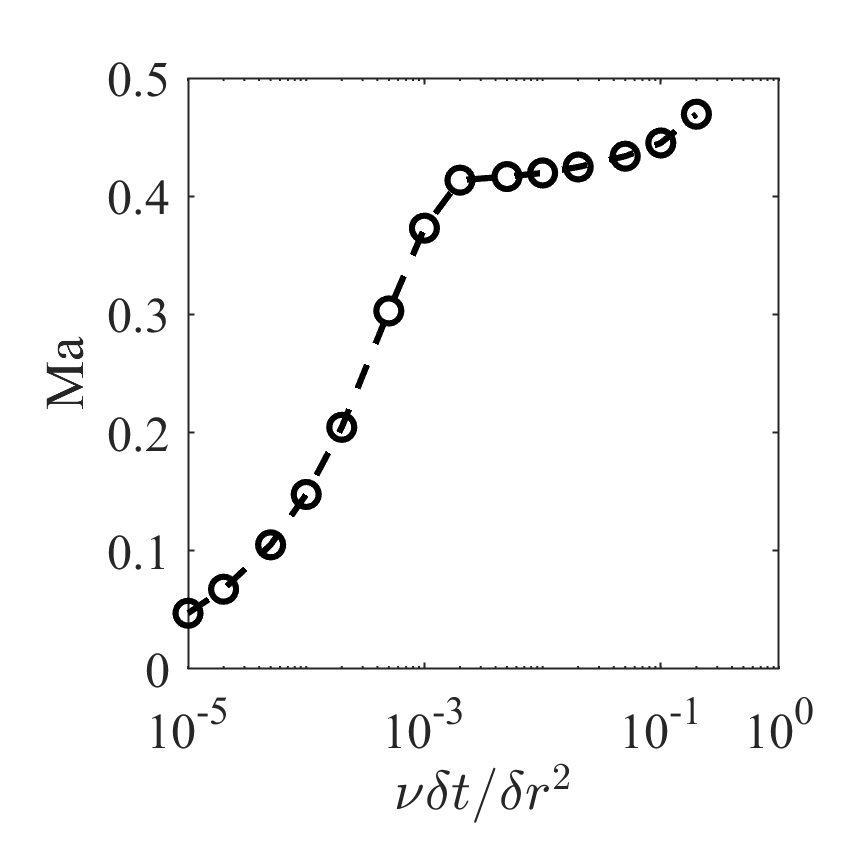}
	\caption{Linear stability domain of the MRT collision operator of~\cite{lallemand_theory_2000} with the corresponding set of optimized relaxation rates. Following the original publication the discrete equilibrium is taken to be a second-order polynomial expansion.}
	\label{Fig:linear_stability_lallemand}
\end{figure}

A first attempt at extending the stability domain of the LB-BGK solver, introduced in the early 2000s' is the so-called Multiple Relaxation Time (MRT) collision model \cite{dhumieres_generalized_1992,coveney_multiplerelaxationtime_2002,lallemand_theory_2003}. The idea behind this approach is to relax the discrete populations in a space other than the discrete populations; In principle this introduces additional parameters independent from the kinematic viscosity, opening the door for a more flexible equilibration path \cite{lallemand_theory_2003,coveney_multiplerelaxationtime_2002}. The added degrees of freedom can be useful both physically and numerically~\cite{dellar_bulk_2001}. In this approach, the BGK collision operator is written as:
\begin{equation}\label{eq:MRT_collision_operator}
    \Omega_i^{\rm MRT} = \bm{M^{-1}}\bm{S}\bm{M}\left( f^{\rm eq}_i - f_i\right),
\end{equation}
where $\bm{M}$ is the transformation matrix from discrete population to the relaxation space such that:
\begin{equation}
    \Pi_i = \sum_{j} M_{ij} f_j,
\end{equation}
where $\Pi_i$ are the moments chosen for the application of the collision operation. As seen in Eq.~\ref{eq:MRT_collision_operator}, using the transformation matrix $\bm{M}$ the discrete populations are taken to momentum space. Then the relaxation matrix $\bm{S}$ is applied and relaxed moments are converted back to discrete populations through $\bm{M^{-1}}$. The relaxation rates tensor $\bm{S}$ is defined as:
\begin{equation}
    \bm{S}={\rm diag}(\frac{1}{\bar{\tau}_0}, \dots, \frac{1}{\bar{\tau}_{Q-1}}),
\end{equation}
where the operator ${\rm diag}$ is defined as:
\begin{equation}
    {\rm diag}(\bm{A}) = (\bm{A}\otimes\bm{1})\circ \bm{I},
\end{equation}
with $\bm{A}$ a given vector, $\bm{1}$ a vector with elements 1, $\bm{I}$ the unitary tensor and $\circ$ the Hadamard product. For a typical DdQq stencil, $q$ moments are needed to span the phase-space. The choice of the relaxation space, is the other important ingredient in this class of models both differentiating between different approaches listed below and controlling the numerical properties of the solver. For instance Lallemand and Luo ~\cite{lallemand_theory_2000} proposed a set of mutually orthogonal moments for the D2Q9 stencil defined as:
\begin{equation}
    \Pi_i\in\{ \Pi_0, 3(\Pi_{xx}+\Pi_{yy})-4\Pi_0, , \Pi_x, 3(\Pi_{xxx}+\Pi_{xxy})-5\Pi_x, \Pi_y, 3(\Pi_{yyy}+\Pi_{xyy})-5\Pi_y, \Pi_{xx}-\Pi_{yy}, \Pi_{xy}\}.
\end{equation}
With this choice of moments for the relaxation it is clear that the relaxation rates of the last two moments corresponds to shear viscosity and that of the second moment to the bulk viscosity while the rest can be freely tuned for stability~\cite{coveney_multiplerelaxationtime_2002}, optimal dispersion \cite{chavez-modena_improving_2018}, fixing the boundary position for the half-way bounce-back boundary condition \cite{pan_evaluation_2006} etc.. It is interesting to note that this specific choice of moments space along with the \emph{optimized} set of \emph{ghost} relaxation rates do not, as opposed to the conclusions of~\cite{lallemand_theory_2000}, actually extend the linear stability domain of the BGK collision operator as shown in Fig.~\ref{Fig:linear_stability_lallemand}.
This might be explained by the fact that in~\cite{lallemand_theory_2000}, the authors assumed the wave-number vector $\bm{k}$ always parallel to the velocity vector $\bm{u}$ which is not a sufficient condition for linear stability. Here the full 2-D space of wave-number vectors has been scanned for all considered velocities resulting in a smaller domain of stability. While the moments chosen as basis for the MRT realization of \cite{lallemand_theory_2000} are orthogonal, other alternatives based on weighted orthogonal moments have also been proposed. For instance the Hermite polynomials form a basis of moments that are mutually orthogonal with respect to the weighted dot-product and are defined as:
\begin{equation}
    \Pi_i \in \{ \Pi_0, \underbrace{\Pi_x, \Pi_y}_{\bm{a_1}}, \overbrace{\Pi_{xy}, \Pi_{xx}-c_s^2, \Pi_{yy}-c_s^2}^{\bm{a_2}}, \underbrace{\Pi_{xxy}-c_s^2\Pi_{y}, \Pi_{xyy}-c_s^2\Pi_{x}}_{\bm{a_3}}, \overbrace{\Pi_{xxyy} - c_s^2\left( \Pi_{xx}+\Pi_{yy}\right) + c_s^4}^{\bm{a_4}} \}.
\end{equation}
Below we will look into two other classes of MRT collision models that present conceptual singularities with respect to other realizations, i.e. the two relaxation time model and central moments-based MRT.
\paragraph{Two relaxation time model} As noted in~\cite{karlin_gibbs_2014}, using the full set of moments leads to a number of free parameters, the ghost moments relaxation coefficients, for which no formal physical closures exist. As previously mentioned, apart from the entropic argument, only \emph{a posteriori} closures based on numerical arguments can be devised for these free parameters. Another way around this issue is to adopt targeted, on specific moments of the distribution function, minimalist MRT models. The TRT (Two Relaxation Time) collision operator developed and proposed by I.~Ginzburg is an example of these minimalist models \cite{ginzburg_equilibrium-type_2005,ginzburg_generic_2005,ginzburg_lattice_2007}. In this collision model the distribution function is decomposed into symmetrical, $f_i^{+}$, and non-symmetrical parts, $f_i^{-}$, defined as~\cite{ginzburg_lattice_2007}:
\begin{subequations}
	\begin{align}
	f_i^{+} &= \frac{f_i + f_{\bar{i}}}{2},\\
	f_i^{-} &= \frac{f_i - f_{\bar{i}}}{2},
	\end{align}
\end{subequations}
resulting in two relaxation times, $\bar{\tau}^+$ and $\bar{\tau}^-$, with the first one tied to the fluid viscosity. The collision operator is then expressed as~\cite{ginzburg_generic_2005}:
\begin{equation}
\Omega_i^{\rm TRT} = \frac{1}{\bar{\tau}^+} \left(f_i^{\rm eq+}-f_i^{+}\right) + \frac{1}{\bar{\tau}^-} \left(f_i^{\rm eq-}-f_i^{-}\right).
\end{equation}
As demonstrated in~\cite{dhumieres_viscosity_2009}, judicious choices of the free parameter, the so-called ``magic values'', can lead to, among other effects, the wall being placed exactly half-way when used with the half-way bounce-back boundary condition. Defining :
\begin{equation}
\Lambda=\left(\frac{\delta t}{\bar{\tau}^+}-\frac{1}{2}\right)\left(\frac{\delta t}{\bar{\tau}^-}-\frac{1}{2}\right),
\end{equation}
it can be shown that setting $\Lambda=3/16$ places the wall half-way \cite{ginzburg_two-relaxation-time_2008}, while $\Lambda=1/6$ and $\Lambda=1/12$ cancel out, respectively, the third- and fourth-order spatial error terms \cite{ginzburg_consistent_2008,ginzburg_magic_2009} and $\Lambda=1/4$ results in optimal stability in that specific relaxation space~\cite{kuzmin_role_2011}.\\
\paragraph{The lattice kinetic scheme}
Another example of a minimalist MRT scheme is that of the so-called LKS (lattice kinetic scheme)~\cite{inamuro_lattice_2002,wang_localized_2015}. This collision model is a TRT scheme in the space of Hermite moments, where second-order moments are relaxed using the fluid viscosity while higher-order moments (three and four) are relaxed using a free parameter~\cite{hosseini_stability_2019}. For the LKS the collision operator is written as \cite{hosseini_theoretical_2019,inamuro_lattice_2002}:
\begin{equation}
\Omega_i^{\rm LKS} = -\frac{1}{\bar{\lambda}}\left(f_i - f^{\rm eq, LKS}_i \right).
\label{eq:LKSColl}
\end{equation}
The second relaxation coefficient $\bar{\lambda}$ is related to the SRT relaxation coefficient through~\cite{wang_localized_2015}:
\begin{equation}
\bar{\lambda}-A = \bar{\tau},
\label{eq:secondRelaxation}
\end{equation}
where  $A$ is a constant fixed by the choice of the free parameter. The EDF is then defined as~\cite{hosseini_stability_2019}:
\begin{equation}\label{eq:EDFLKS_2}
f^{\rm eq, LKS}_i = f^{\rm eq}_i - \frac{A}{\bar{\tau}}\frac{w_{i}}{2 }\bm{a}_{2}^{\rm neq}:\bm{\mathcal{H}}_{2,i}.
\end{equation}
The original regularized lattice Boltzmann method (RLBM), as will be shown in the next sections, is an LKS solver where the free relaxation coefficient is set to 1~\cite{hosseini_stability_2019}. This collision operator has been applied to a variety of configurations ranging from multi-phase \cite{kataoka_numerical_2011} to non-Newtonian flows \cite{yoshino_numerical_2007} and advection-diffusion equations with variable diffusion coefficients \cite{meng_localized_2016,perko_single-relaxation-time_2014}.
\subsubsection{Central moments-based decomposition}
\begin{figure}[h!]
	\centering
	\includegraphics[width=9cm,keepaspectratio]{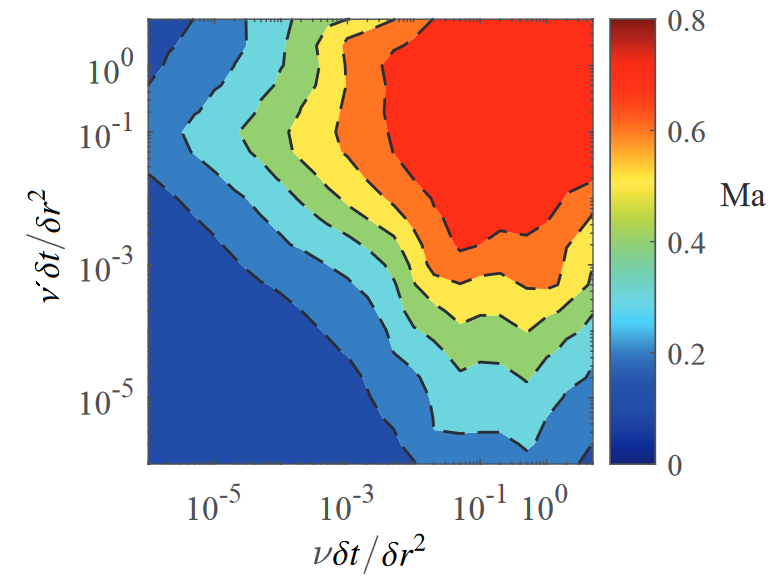}
	\caption{Linear stability domain of the central Hermite MRT collision operator. Here second-order moments relax with the viscosity while higher-order ghost moments relax with another pseudo-viscosity $\nu'$. Reproduced from \cite{hosseini_development_2020}.}
	\label{Fig:linear_stability_CHM_MRT}
\end{figure}

In the Central Moments Multiple Relaxation Time (from here on referred to as CM-MRT) model, while the paradigm is quite similar to the MRT, a different set of moments are used: the central moments, designated by $\widetilde{\Pi}_{\underbrace{x \dots x}_{\times p} \underbrace{y \dots y}_{\times q} \underbrace{z \dots z}_{\times r}}$ and defined as \cite{geier_cascaded_2006,geier_-aliasing_2008}:
\begin{equation}
    \widetilde{\Pi}_{\underbrace{x \dots x}_{\times p} \underbrace{y \dots y}_{\times q} \underbrace{z \dots z}_{\times r}} = \sum_{i} {\left(c_{i,x}-u_x\right)}^p {\left(c_{i,y}-u_y\right)}^q {\left(c_{i,z}-u_z\right)}^r f_i.
\end{equation}
Taking again the example of the D2Q9 stencil with the Hermite coefficients as the projection space and a fourth-order expansion of the EDF results in the following central equilibrium moments \cite{asinari_generalized_2008,hosseini_central_2021,huang_simulation_2022}:
\begin{equation}
    \bm{\widetilde{\Pi}}^{\rm eq} = \{ \rho, 0, 0, 0, 0, 0, 0, 0, 0 \}.
\end{equation}
The stability domain of this collision model is shown in Fig.~\ref{Fig:linear_stability_CHM_MRT}.
From the stability domain one can already see that setting $\nu'=\nu$ which is equivalent to the SRT operator is not the optimal choice in terms of linear stability. An optimal route can be observed at $\frac{\nu'\delta t}{\delta r^2}\approx0.167$ which, as will be seen in the next section, is equivalent to the recursive regularized collision operator. While the Hermite-based moments space does not allow for independent control over bulk viscosity this can readily be achieved by modifying the second-order moments resulting in the following system:
\begin{equation}
    \Pi_i\in\{\widetilde{a}_0, \widetilde{a}_x, \widetilde{a}_y, \widetilde{a}_{xy}, \widetilde{a}_{xx}-\widetilde{a}_{yy}, \widetilde{a}_{xx}+\widetilde{a}_{yy}, \widetilde{a}_{xxy}, \widetilde{a}_{xyy}, \widetilde{a}_{xxyy}\},
\end{equation}
where $\widetilde{a}_i$ are central Hermite coefficients. Here the trace of the second-order moments is relaxed independently allowing for an independent bulk viscosity. It must be noted that when used in combination with the correction for diagonal components of the third-order equilibrium moments the coefficient in front of the correction term must be changed to $1 - \frac{\delta t}{\bar{\tau}+\bar{\tau}_{\eta}}$ where $\bar{\tau}_{\eta}$ is the relaxation time tied to the bulk viscosity~\cite{hosseini_lattice_2022}.
\subsubsection{Closures for the relaxation rates: entropic}
\begin{figure}[htbp!]
	\centering
    \includegraphics{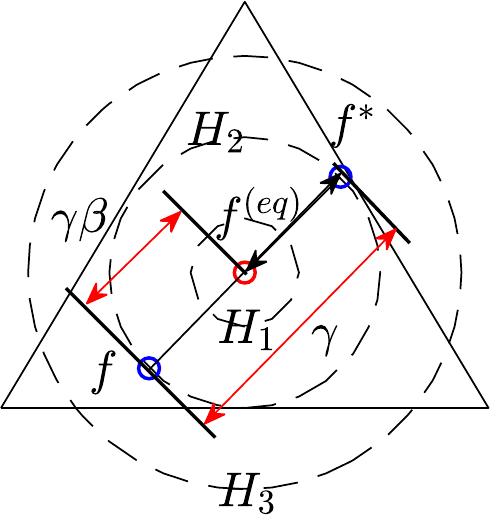}
	\caption{Schematic representation of the relaxation process in the ELBM. Dashed lines represent entropy levels while the triangle illustrates the positivity polytope.}
	\label{Fig:ELBM_relaxation}
\end{figure}

\paragraph{The single relaxation time entropic model}
The original entropic LBM ensures stability of the solver by imposing a monotonous decrease of a discrete entropy function. While a number of different discrete entropy functionals have been proposed in the context of the ELBM \cite{karlin_perfect_1999,karlin_maximum_1998}, the following form has gained the most attention \cite{chikatamarla_entropic_2006,ansumali_entropic_2004,ansumali_single_2002,frapolli_multispeed_2014,boghosian_galilean-invariant_2003}:
\begin{equation}
    H = \sum_i f_i \ln\left(\frac{f_i}{w_i}\right).
\end{equation}
In practice, the monotonicity of the discrete entropy is enforced using a two-step linear reconstruction achieved through the following modified time-evolution equation \cite{chikatamarla_entropic_2006,ansumali_entropic_2004}:
\begin{equation}
    f_i\left(\bm{r}+\bm{c}_i\delta t, t+\delta t\right) - f_i\left(\bm{r}, t\right) = \beta\gamma\left(f^{\rm eq}_i\left(\bm{r}, t\right)-f_i\left(\bm{r}, t\right)\right),
\end{equation}
where $\beta$ is tied to the fluid viscosity as:
\begin{equation}
    \beta = \frac{\delta t}{2\tau+\delta t},
\end{equation}
with $\tau=\nu/c_s^2$, while $\gamma$ is obtained by solving the following system \cite{mazloomi_entropic_2015}:
\begin{equation} \label{Eq:entropic_parameter}
    H\left(f^{*}\right) = H\left(f\right),
\end{equation}
with:
\begin{equation}
    f_i^* = f_i+\gamma\left(f_i^{\rm eq}-f_i\right).
\end{equation}
This two-step reconstruction procedure is illustrated in Fig.~\ref{Fig:ELBM_relaxation}. In the first step, the equal entropy mirror state relative to the equilibrium, $f^{*}$, is found by solving the non-linear equation shown in Eq.~\ref{Eq:entropic_parameter}. As observed there $\gamma$ is the maximum path length not resulting in an increase in entropy \cite{atif_essentially_2017}. It is interesting to note that at thermodynamic equilibrium Eq.~\ref{Eq:entropic_parameter} has the non-trivial root $\gamma=2$ which corresponds to the SRT collision operator \cite{atif_essentially_2017}. In the second step, dissipation is introduced via the parameter $\beta$.
The solution to Eq.~\ref{Eq:entropic_parameter} can be obtained using a Newton-Raphson iterative solver as:
\begin{equation}
    \gamma^{n+1} = \gamma^n - \frac{G^n}{\partial_\gamma G^n},
\end{equation}
with:
\begin{equation}
    G^n = H\left({f^{*}}^n\right) - H\left(f\right),
\end{equation}
and:
\begin{equation}
    \frac{\partial G^n}{\partial \gamma} = \sum_i \left[1+\ln\left(\frac{{f_i^*}^n}{w_i}\right)\right]\left(f_i^{\rm eq}-f_i\right),
\end{equation}
where $\gamma^n$ and $\gamma^{n+1}$ are solutions obtained in the previous and current iterations. The iterative root-finding algorithm being rather expensive (especially when the populations get away from equilibrium) alternative approaches have been developed in recent years~\cite{chikatamarla_entropic_2006,atif_essentially_2017}. These approximate solutions are also useful in the vicinity of equilibrium as the Newton-Raphson solver might diverge there.\\
\paragraph{Extension of the entropic closure to multiple relaxation rate formulations}
The two-relaxation time entropic formulation is realized by writing discrete populations as \cite{karlin_gibbs_2014}:
\begin{equation}
        f_i = k_i + s_i + h_i.
\end{equation}
where the kinematic part $k_i$, represents contributions from conserved moments, $s_i$ contributions from the stress and $h_i$ all higher-order moments contributions. Considering the discrete time evolution equation:
\begin{equation}
    f_i\left(\bm{r}+\bm{c}_i\delta t, t+\delta t\right) = \left(1-\frac{\delta t}{2\bar{\tau}}\right)f_i\left(\bm{r}, t\right) + \frac{\delta t}{2\bar{\tau}} f^{\rm mirr}_i\left(\bm{r}, t\right),
\end{equation}
Considering invariance of conserved moments and physical constraint on the relaxation rate of second-order moments defining $s_i$, the mirror state can be written:
\begin{equation}
        f_i^{\rm mirr} = k_i + \left(2s_i^{\rm eq} - s_i\right) + (1-\gamma)h_i + \gamma h^{\rm eq},
\end{equation}
where the free parameter $\gamma$ here allows independent control over the relaxation rate of higher-order moments. This free parameter is found by minimizing the discrete entropy in the post-collision state, $f'_i$:
\begin{equation}
    \frac{dH(f')}{d\gamma} = 0,
\end{equation}
which upon expansion around equilibrium up to the first non-vanishing order results in~\cite{bosch_entropic_2018}:
\begin{equation}
    \frac{\gamma}{2} = \frac{\bar{\tau}}{\delta t} -\left(1-\frac{\bar{\tau}}{\delta t}\right)\frac{\langle\Delta s\lvert \Delta h\rangle}{\langle\Delta h\lvert \Delta h\rangle},
\end{equation}
where $\Delta s_i= s_i^{\rm eq}-s_i$, $\Delta h_i= h_i^{\rm eq}-h_i$, and the entropic scalar product $\langle\lvert\rangle$ is defined as:
\begin{equation}
    \langle X\lvert Y\rangle=\sum_{i=1}^{Q}\frac{X_i Y_i}{f^*_i}.
\end{equation}
It is interesting to note here that for a moments space where the moments are weighted-orthogonal in the co-moving reference frame, in the absence of body forces, the entropy minimizer free parameter $\gamma\rightarrow 2\bar{\tau}/\delta t$ setting all higher-order moments to the equilibrium state~\cite{hosseini_entropic_2022}.
\subsubsection{The specific case of regularization}
\begin{figure}[h!]
	\centering
	\includegraphics[width=6cm,keepaspectratio]{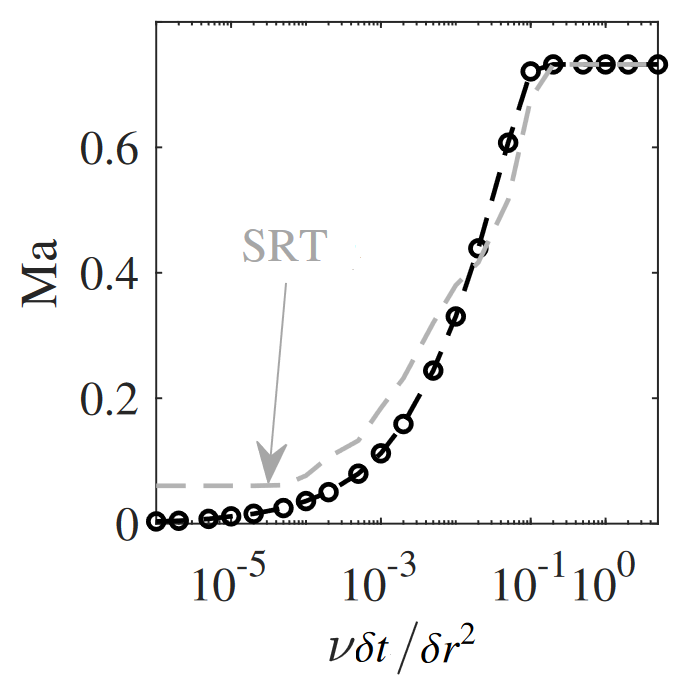}
	\caption{Linear stability domain of the projection-based regularized collision model of \cite{latt_lattice_2006} compared to the the SRT. Reproduced from \cite{hosseini_development_2020}.}
	\label{Fig:linear_stability_PR_Reg}
\end{figure}

The rational behind regularized collision operators is to filter out higher-order components of the distribution functions (in the sense of the Chapman-Enskog expansion). It can be shown that first-order terms appear at the NS level, while higher-order terms intervene at the Burnett, super Burnett etc. scales, not of interest in the context of the LBM. In the context of the regularized collision approach the non-equilibrium part of the distribution function is reconstructed using only first-order contributions, $f_i^{\rm neq}\approx f_i^{(1)}$. The discrete time-evolution equation can be re-expressed as \cite{latt_lattice_2006,latt_hydrodynamic_2007}:
\begin{equation}
f_i\left(\bm{r}+\bm{c}_i\delta t, t+\delta t\right) = f^{\rm eq}_i\left(\bm{r}, t\right) + \left(1-\frac{\delta t}{\bar{\tau}}\right)f^{\rm neq}_i\left(\bm{r}, t\right).
\end{equation}
Following the Hermite expansion used for the EDF, we can express the first-order component of the distribution function as:
\begin{equation}
f_i^{(1)} = w_i \sum_{n=2} \frac{1}{n!}\bm{a}^{(1)}_n:\bm{\mathcal{H}}_n.
\end{equation}
In the original regularized model~\cite{latt_lattice_2006} only the second-order Hermite polynomial was considered for the reconstruction process:
\begin{equation}
f_i^{(1)} = w_i \frac{\bm{a}_2^{(1)}:\bm{\mathcal{H}}_2}{2}.
\end{equation}
The only unknown in this equation is $\bm{a}_2^{(1)}$. In \cite{latt_lattice_2006}, this coefficient is computed as:
\begin{equation}
\bm{a}_2^{(1,PR)}\approx \bm{a}_2^{\rm neq}=\sum_i \bm{\mathcal{H}}_2:\left(f_i-f_i^{\rm eq}\right).
\end{equation}
This approach to reconstructing the non-equilibrium part is commonly referred to as the projection regularization approach. In the context of the classical LB formulation with a second-order polynomial EDF, and given the orthogonality of the independent moments, this collision operator aimed at eliminating non-equilibrium effects of higher-order (kinetic) moments. It is interesting to note that this formulation has a number of shortcomings: (a) Errors in all components of the third-order moments tensor of the EDF (given the absence of higher-order terms in the EDF), and (b) presence of higher-order effects (tied to $f^{(n)}_i$ with $n\geq 2$) coming from the approximation used for $\bm{a}_2^{(1)}$. While initially believed to improve stability, the second-order projection-based regularized collision operator actually reduces the domain of stability as compared to the SRT collision operator as illusrtated in Fig.~\ref{Fig:linear_stability_PR_Reg} via the linear stability domain.
The latter can be, to some extent, cured using a CE-based closure for $\bm{a}_2^{(1)}$. Using this approach it can be shown that \cite{malaspinas_increasing_2015}:
\begin{equation}
\bm{a}_2^{(1,CE)}=-\rho c_s^2 \frac{\bar{\tau}}{\delta t}\left(\bm{\nabla}\bm{u}+{\bm{\nabla}\bm{u}}^T\right).
\end{equation}
This expression can be computed using classical FD approximations \cite{jacob_new_2018}, which has been shown to improve stability at the cost of non-negligible numerical dissipation considerably raising computational costs for direct numerical simulations. Recent publications have proposed to reduce dissipation by considering a weighted combination of the projection- and CE-based approaches~\cite{jacob_new_2018,feng_hybrid_2019}:
\begin{equation}
\bm{a}_2^{(1)} = \sigma \bm{a}_2^{(1,PR)} + \left(1-\sigma\right) \bm{a_2}^{(1,CE)},
\end{equation}
where $\sigma$ is the weight.\\
The first problem with the RLBM of~\cite{latt_lattice_2006}, namely errors in the off-diagonal components of the third-order moments tensor can be accounted for by using third- (or fourth-)order terms in the EDF and using the recursive properties of the off-equilibrium Hermite coefficients~\cite{malaspinas_increasing_2015}, {i.e.}:
\begin{equation}
    {a}^{(1)}_{\alpha_1\alpha_2...\alpha_m} = {a}^{(1)}_{\alpha_1,\alpha_2...\alpha_{m-1}}u_{\alpha_n}+\left[{a}^{(1)}_{\alpha_{n}\alpha_{n-1}}u_{\alpha_1} u_{\alpha_2}...u_{\alpha_{n-2}}\right]_{\text{cyc}}.
\end{equation}
For the D2Q9 stencil, assuming a fourth-order isothermal polynomial EDF the different non-equilibrium Hermite coefficients are computed as:
\begin{subequations}
	\begin{align}
		a^{(1)}_{xyy} &= u_x a^{(1)}_{yy} + 2u_y a^{(1)}_{xy}, \\
		a^{(1)}_{xxy} &= u_y a^{(1)}_{xx} + 2u_x a^{(1)}_{xy}, \\
		a^{(1)}_{xxyy} &= u_y a^{(1)}_{xxy} + {u_x}^2 a^{(1)}_{yy} + u_x u_y a^{(1)}_{xy}.
	\end{align}
\end{subequations}
It can be shown, via simple algebra, that the projection-based recursive regularization approach is equivalent to a central Hermite collision operator with all ghost relaxation rates set to unity.

\pagebreak
\section{Extension to non-ideal fluids\label{sec:non_ideal}}
In this section we discuss introduction of non-ideal contributions to the LBM. Given that all models discussed in the present contribution recover some form of a second-gradient fluid, described first by van der Waals \cite{van_der_waals_thermodynamische_1894}, we start by introducing fundamental aspects of this description of non-ideal fluids. The introduction of the target macroscopic system is followed by a brief overview of kinetic models for the dense fluid regime recovering the second-gradient fluid behavior in the hydrodynamic limit. Most widely used lattice Boltzmann models for isothermal non-ideal fluid flows are then reviewed and a comprehensive analysis of numerical and physical properties is provided.
\subsection{Second-gradient theory and non-ideal fluids thermodynamics\label{sec:vdW}}
\subsubsection{Non-ideal  equation of state: van der Waals}
\begin{figure}[h!]
	\centering
	\includegraphics[width=8cm,keepaspectratio]{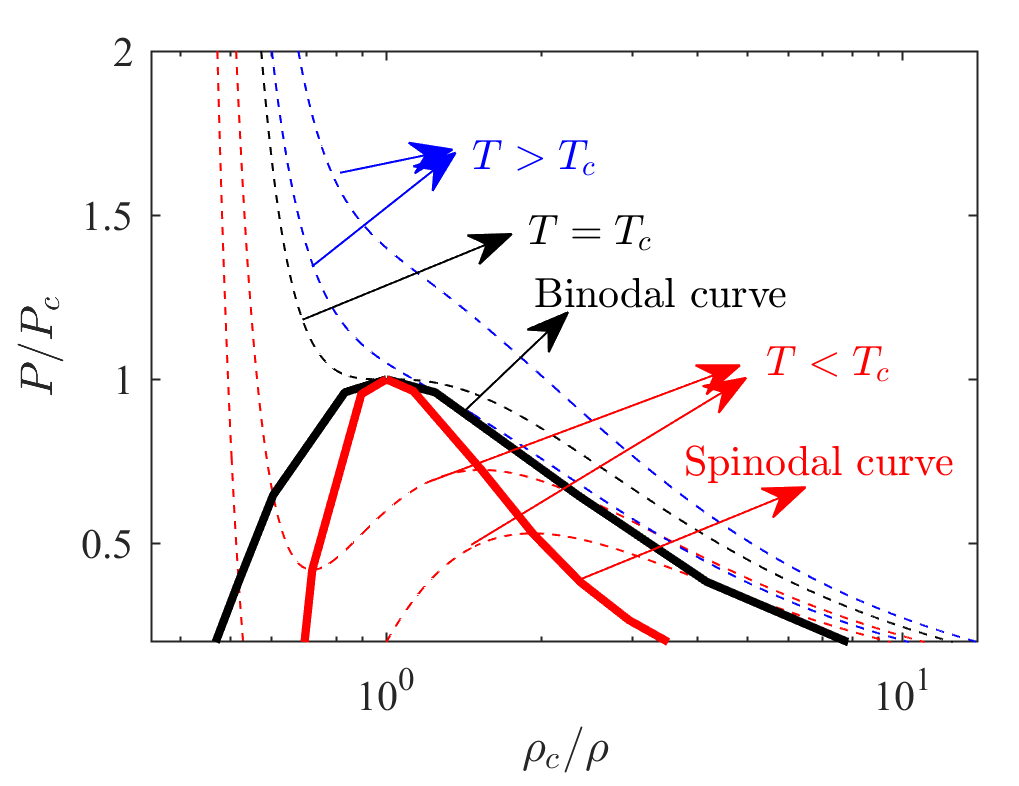}
	\caption{Clapeyron diagram of van der Waals equation of state.}
	\label{Fig:Clapeyron_vdW}
\end{figure}

At rather large pressures or in dense liquid phases the assumption used to derive the ideal gas equation of state, that molecules interact with each-other mostly via local elastic collisions does not hold anymore. The ideal gas law has two shortcomings in these regimes: it neglects the volume occupied by molecules via the local interaction assumption and it does not take into account long-range interactions. In doing so it fails to correctly model the behavior of so-called non-ideal fluids, e.g. co-existence of two phases at a given temperature and existence of a critical temperature $T_c$ above which the fluid become single-phased. The first attempt at a model correctly describing non-ideal fluid thermodynamics was made by van der Waals~\cite{van_der_waals_over_1873} via a cubic equation of state:
\begin{equation}\label{eq:vdW_EoS}
    P = \frac{\rho r T}{1 - b\rho} - a\rho^2,
\end{equation}
where $a$ and $b$ are the long range interaction and volume exclusion constants. The thermodynamic behavior of the van der Waals equation of state is illustrated in Fig.~\ref{Fig:Clapeyron_vdW} via the corresponding Clapeyron diagram.
Looking at the isotherms of this equation of state, three categories can be identified: (a) $T< T_c$ where $\frac{\partial P}{\partial (1/\rho)}=0$ has two roots and $\frac{\partial^2 P}{{\partial (1/\rho)}^2}=0$ one, (b) $T=T_c$ where  $\frac{\partial P}{\partial (1/\rho)}=0$ has a single root which is the same as and $\frac{\partial^2 P}{{\partial (1/\rho)}^2}=0$ and (c) $T>T_c$ where the former equations do not admit any solutions. The transition isotherm, $T_c$, is known as the critical temperature and the combined root of $\frac{\partial P}{\partial (1/\rho)}=0$ and $\frac{\partial^2 P}{{\partial (1/\rho)}^2}=0$ on that isotherm identifies the critical state. The outer envelope shown in Fig.~\ref{Fig:Clapeyron_vdW} is the saturation curve (also called binodal curve) corresponding to the liquid/vapor coexistence densities while the inner envelope is the spinodal curve. In the region within the spinodal envelope, referred to as the spinodal region, the thermodynamic states predicted by the van der Waals equation (more generally all cubic equations of state) are mechanically unstable, $\frac{\partial P}{\partial (1/\rho)}>0$. In the region between the spinodal and binodal envelops, referred to as the binodal region, the fluid is in a pseudo-stable state meaning that it is not mechanically unstable but at the same time it is not a local minimzer of energy. As such when subjected to a small perturbation the fluid leaves the binodal region states towards the binodal curve which corresponds to local energy minima.  In practice, a state in the bi- and spinodal regions corresponds to the coexistence of the liquid and vapor phases, i.e. both branches of the binodal curve, a distinctive feature for an interface. A point of cardinal importance is to identify the liquid and vapor coexistence states on given isotherms below the critical temperature.

\subsubsection{Co-existence densities: Common-tangent construction}
\begin{figure}[h!]
	\centering
	\includegraphics[width=6cm,keepaspectratio]{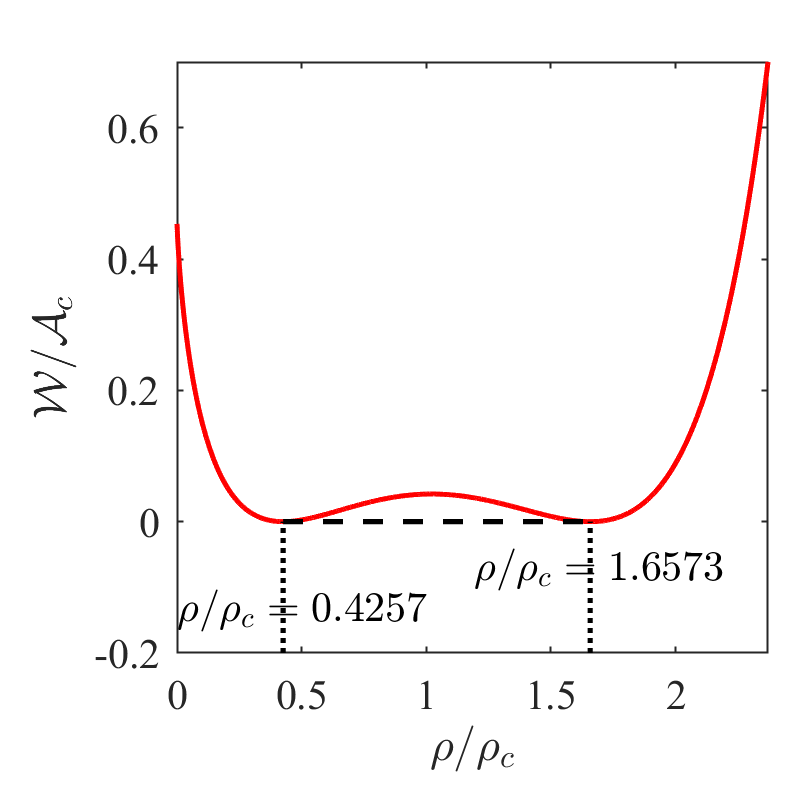}
	\caption{Illustration of the common tangent construction for the van der Waals equation of state at $T/T_c=0.9$. In the $y-$axis $\mathcal{W}(\rho)=\mathcal{A}(\rho) - \frac{\partial \mathcal{A}}{\partial \rho}\bigg\vert_{\rho_v}(\rho - \rho_v) + \mathcal{A}(\rho_v)$. The liquid/vapor coexsitence densities are found using Eqs.\eqref{eq:common_tang_v} and \eqref{eq:common_tang_l}. }
	\label{Fig:common_tangent_vdW}
\end{figure}

For the liquid and vapor states, $\rho_l$ and $\rho_v$, in contact via an interface to coexist one can readily show that both states must have the same pressure, $P$, and chemical potential $\lambda$. A detailed derivation of these conditions will be given in \ref{subsubsec:Gibbs_eq} for a flat interface.
The interface equilibrium conditions can be re-cast in the following form:
\begin{eqnarray}
\frac{\mathcal{A}(\rho_l)-\mathcal{A}(\rho_v)}{\rho_l-\rho_v}&=&\frac{\partial \mathcal{A}}{\partial \rho}\bigg\vert_{\rho_v} \label{eq:common_tang_v},\\
\frac{\mathcal{A}(\rho_l)-\mathcal{A}(\rho_v)}{\rho_l-\rho_v}&=&\frac{\partial \mathcal{A}}{\partial \rho}\bigg\vert_{\rho_l} \label{eq:common_tang_l},
\end{eqnarray}
where we have introduced the free energy $\mathcal{A}$, and $\lambda=\frac{\partial \mathcal{A}}{\partial \rho}$. In simple words, the reformulated equilibrium conditions mean that in a plot of free energy versus density at a given temperature below critical the two minimizers of the bulk free energy, i.e. $\rho_l$ and $\rho_v$, are characterized as two points that can be joined with a line of slope equal to the derivatives of the free energy at those points. This system of equations can be used to systematically find the two stable states of the fluid for a given temperature below critical and is known as the \emph{common tangent} construction method. To further illustrate that consider the bulk free energy of the van der Waals fluid:
\begin{equation}
\mathcal{A} = \rho r T \ln\left(\frac{\rho}{1-b\rho}\right) - a \rho^2.
\end{equation}
Using this expression and conditions of Eqs.\eqref{eq:common_tang_v} and \eqref{eq:common_tang_l} one can find the coexistence densities at tany temperature. The free energy at $\frac{T}{T_c}=0.9$ along with the corresponding local minima and tangents are illustrated in Fig.~\ref{Fig:common_tangent_vdW}.
It is readily observed that the derivatives at the two coexistence densities are the same and equal to the slope of the line joining them. The common tangent approach can therefor be used to systematically construct the liquid/vapor coexistence densities curve.
\subsubsection{Co-existence densities: First-order transition and Maxwell construction}
\begin{figure}[h!]
	\centering
	\includegraphics[width=6cm,keepaspectratio]{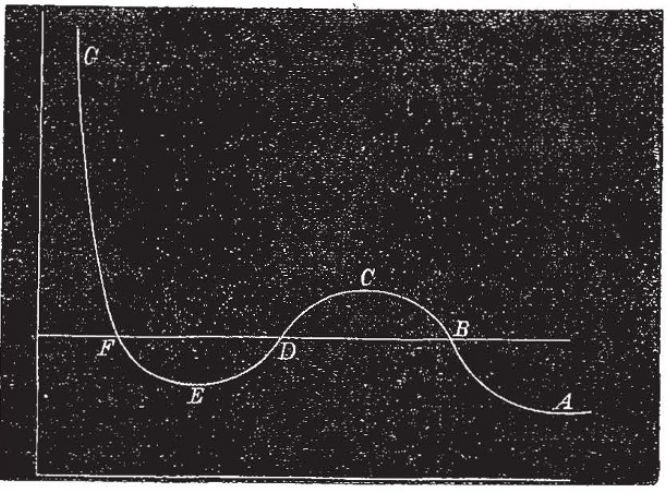}
	\caption{Plot used by Maxwell in \cite{clerk-maxwell_dynamical_1875} to illustrate phase transition. Vertical axis is pressure and horizontal axis is specific volume. The plot is reproduced from \cite{clerk-maxwell_dynamical_1875}.}
	\label{Fig:P_V_Maxwell}
\end{figure}

As seen in the previous paragraph, below the critical point the free energy of non-ideal fluids admits two local minima separated by a local maxima indicating an energy barrier for transition. The concave nature of the free energy between those two stable states indicates that a homogeneous state is disadvantageous in-between them. In \cite{clerk-maxwell_dynamical_1875}, discussing an experiment where pressure variations are studied in a vessel containing a fixed amount of substance while gradually decreasing the volume Maxwell observes that "\emph{We have hitherto supposed the experiment to be conducted in such a way that the density is the same in every part of the medium. This, however, is impossible in practice, as the only condition we can impose on the medium from without is that the whole of the medium shall be contained within a certain vessel. Hence, if it is possible for the medium to arrange itself so that part has one density and part another, we cannot prevent it from doing so. Now the points B and F represent two states of the medium in which the pressure is the same but the density very different. The whole of the medium may pass from the state B to the state F, not through the intermediate states C-D-E, but by small successive portions passing directly from the state B to the state F. In this way the successive states of the medium as a whole will be represented by points on the straight line B-F, the point B representing it when entirely in the rarefied state, and F representing it when entirely condensed. This is what takes place when a gas or vapour is liquefied. Under ordinary circumstances, therefore, the relation between pressure and volume at constant temperature is represented by the broken line A-B-F-G}". The original plot used by Maxwell to illustrate his purpose is shown in Fig.~\ref{Fig:P_V_Maxwell}.
In practice this means that in-between the liquid and vapor states, the fluid is made up of a mixture of those two stable states which overcomes the concavity of free energy leading to instability. This is equivalent to replacing the free energy with its convex hull~\cite{sekerka_thermal_2015}. A simple way of determining the boundaries of this \emph{miscibility} domain is the Maxwell construction. The thermodynamic argument leading to that approach is based on equality of Gibbs enthalpy, $\mathcal{G}=e_{\rm int} + P v + T s$ where $e_{\rm int}$ is internal energy, $v$ the specific volume and $s$ entropy, in the pure vapor and liquid states, i.e.
\begin{equation}
    \mathcal{G}_v - \mathcal{G}_l = \int_{P_v}^{P_l} v dP - \int_{T_v}^{T_l} s dT = 0,
\end{equation}
which for an isothermal process, and using integration by parts leads to:
\begin{equation}
    \int_{v_l}^{v_v} (P - P_{\rm sat})dv = 0.
\end{equation}
This can also be written as:
\begin{equation}\label{eq:MaxConstCond}
    \int_{\rho_v}^{\rho_l} \frac{P - P_{\rm sat}}{\rho^2} d\rho = 0.
\end{equation}
In simple words, the Maxwell construction consists in finding the horizontal line $P=P_{\rm sat}$ in the $P-1/\rho$ diagram intersecting a given isotherm below the critical temperature at three point that guarantees Eq.~\eqref{eq:MaxConstCond}.
\begin{figure}[h!]
	\centering
	\includegraphics[width=12cm,keepaspectratio]{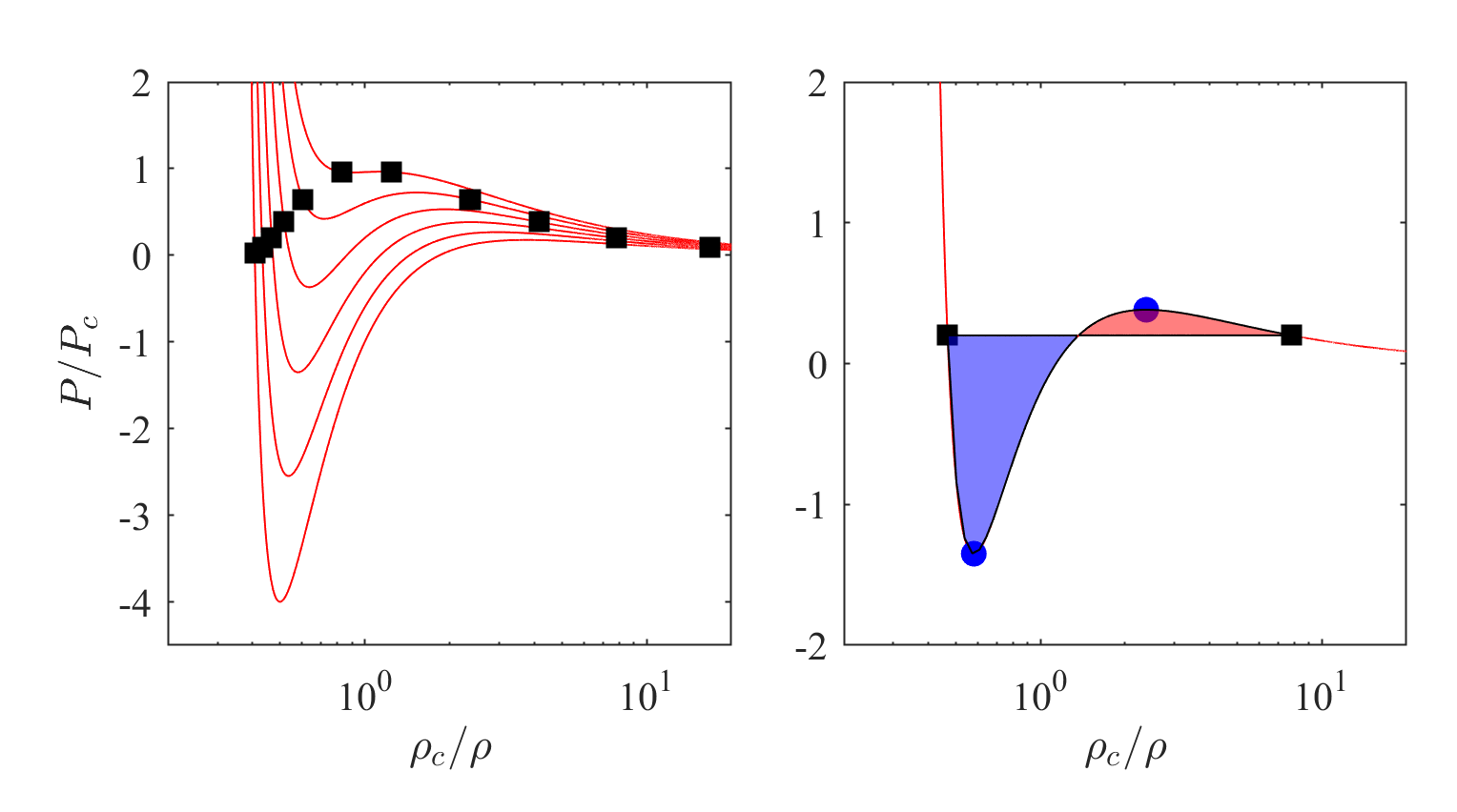}
	\caption{Illustration of the Maxwell construction.  (Left) Isotherms of of the van der Waals equation of state for $T/T_c\in\{0.5,0.6,0.7,0.8,0.9,0.99\}$. Black markers are the liquid/vapor coexistence volumes. (right) Isotherm at $T/T_c=0.7$ with (black square markers) binodal and (blue circular markers) spinodal points. The binodal points guarantee that the red-colored and blue-colored domains have equal area.}
	\label{Fig:Max_Const_vdW}
\end{figure}
The construction process is illustrated in Fig.~\ref{Fig:Max_Const_vdW}. As shown in the left-hand plot of Fig.~\ref{Fig:Max_Const_vdW} for $P_{sp1}<P_{\rm sat}<P_{sp2}$ where $(\rho_{sp1}, P_{sp1})$ and $(\rho_{sp2},P_{sp2})$ are the two spinodal points $P-P_{\rm sat}$ has systematically three non-degenerate roots $1/\rho_1,1/\rho_2,1/\rho_3$ and an inflection point indicating the existence of two zones, $S_1=\int_{1/\rho_1}^{1/\rho_2} (P - P_{\rm sat})d(1/\rho)<0$, shown in red and $S_2=\int_{1/\rho_2}^{1/\rho_3} (P - P_{\rm sat})d(1/\rho)>0$, shown in blue in Fig.~\ref{Fig:Max_Const_vdW}. The choice of pressure $P_{\rm sat}$ that guarantees $S_1-S_2=0$ results in $\rho_1=\rho_l$ and $\rho_2=\rho_v$. We have therefore introduced a second approach to determine the coexistence densities of non-ideal fluids completing our introductory discussion on the thermodynamics of uniform non-ideal fluids. The next step is to introduce the thermodynamic formalism describing non-uniform regions, i.e. interfaces.
\subsubsection{Free energy of non-uniform non-ideal fluid}
In the second-gradient theory as introduced in both \cite{van_der_waals_thermodynamische_1894} and \cite{cahn_free_1958}, the presence of interfaces is conceptualized by endowing the fluid with capillary energy; In practice this means that free energy has both a classical bulk contribution function of local thermodynamic properties and a non-local contribution function of space-derivatives of, for single-component fluids, density:
\begin{equation}
    \mathcal{A}_{\rm vdW} = \mathcal{A}_{\rm vdW}(\rho, \bm{\nabla}\rho, \bm{\nabla}\otimes\bm{\nabla}\rho, \dots).
\end{equation}
Assuming that the non-homogeneities span over distances larger than the characteristic molecular interaction length, the free energy can be developed using a Taylor expansion around the homogeneous state as:
\begin{equation}
    \mathcal{A}_{\rm vdW} = \mathcal{A}_{\rm vdW}\lvert_0 + \bm{\nabla}\rho\cdot\frac{\partial \mathcal{A}_{\rm vdW}}{\partial(\bm{\nabla}\rho)}\bigg\vert_{0} + \bm{\nabla}\bm{\nabla}\rho:\frac{\partial \mathcal{A}_{\rm vdW}}{\partial(\bm{\nabla}\bm{\nabla}\rho)}\bigg\vert_0 + \frac{1}{2} \bm{\nabla}\rho\otimes\bm{\nabla}\rho:\frac{\partial \mathcal{A}_{\rm vdW}}{\partial(\bm{\nabla}\rho)}\bigg\vert_{0}\otimes\frac{\partial \mathcal{A}_{\rm vdW}}{\partial(\bm{\nabla}\rho)}\bigg\vert_{0} + \dots
\end{equation}
where $\lvert_0$ refers to a function evaluated at the homogeneous state. Considering geometrical arguments such as symmetry and rotational-invariance the expression simplifies to the second-gradient free energy as introduced in \cite{van_der_waals_thermodynamische_1894}:
\begin{equation}\label{eq:2nd_grad_energy}
    \mathcal{A}_{\rm vdW} = \mathcal{A} + \frac{\kappa}{2}\lvert\bm{\nabla}\rho\lvert^2,
\end{equation}
with:
\begin{equation}
    \kappa = 2\left[ 2 \frac{\partial^2 \mathcal{A}}{\partial{(\lvert\bm{\nabla}\rho\lvert)}^2}\bigg\vert_0 - \frac{\partial^2 \mathcal{A}}{\partial(\bm{\nabla}^2\rho)\partial\rho}\bigg\vert_0 \right].
\end{equation}
where for the sake of readability we have replaced the bulk free energy $\mathcal{A}_{\rm vdW}\lvert_0$ with $\mathcal{A}$. The second term represents the interface energy while the bulk free energy is solely a function of the local density and temperature \cite{cahn_free_1958,giovangigli_kinetic_2020}.
\subsubsection{Gibbs equilibrium conditions for a flat interface\label{subsubsec:Gibbs_eq}}
The equilibrium state of a 1-D infinite domain, in $x$-direction, made up of the vapor phase in one half and liquid phase in the other with an interface at the center is obtained as the minimizer of free energy under the constraint of constant total mass, i.e.
\begin{equation}\label{eq:minimisation_free_energy}
    \delta \int_{-\infty}^{\infty} \mathcal{L} dx = 0,
\end{equation}
with the Lagrange function $\mathcal{L}$ defined as
\begin{equation}
	\mathcal{L} = \mathcal{A} + \frac{\kappa}{2} {\bigg\lvert \frac{\partial \rho}{\partial x}\bigg\lvert}^2 - \lambda\rho,
\end{equation}
where $\lambda$ is the Lagrange multiplier for the mass constraint. Computing the variation term by term one gets
\begin{equation}
    \int_x \delta \rho \left[\frac{\partial \mathcal{A}}{\partial \rho} - \kappa\frac{\partial^2\rho}{\partial x^2} - \lambda\right] dx,
\end{equation}
which in turn leads to the equilibrium condition:
\begin{equation}\label{eq:EQflat}
    \frac{\partial \mathcal{A}}{\partial \rho} - \kappa\frac{\partial^2\rho}{\partial x^2} - \lambda = 0,
\end{equation}
subject to the boundary conditions: $\rho(\infty)=\rho_l$,  $\rho(-\infty)=\rho_v$. As the gradient of density vanishes at infinity for both phases, we have
\begin{equation}\label{eq:EQchemFLAT}
\frac{\partial\mathcal{A}}{\partial \rho}\bigg\vert_{\rho_l} =\frac{\partial\mathcal{A}}{\partial \rho}\bigg\vert_{\rho_v} =\lambda,
\end{equation}
where we can readily identify $\lambda$ as the bulk chemical potential, leading to the Gibbs equilibrium conditions, i.e. equality of chemical potentials
\begin{equation}
\lambda_l=\frac{\partial \mathcal{A}}{\partial\rho}\bigg\vert_{\rho_l} ,\quad
\lambda_v=\frac{\partial\mathcal{A}}{\partial\rho}\bigg\vert_{\rho_v} ,
\end{equation}
in both phases,
\begin{equation}
\lambda_l=\lambda_v.
\end{equation}
Multiplying \eqref{eq:EQflat} with $\frac{d\rho}{dx}$,
\begin{equation}\label{eq:consFLAT}
\frac{d}{dx}\left[\mathcal{A}-\lambda \rho-\frac{\kappa}{2}\left(\frac{d\rho}{dx}\right)^2\right]=0,
\end{equation}
or 
\begin{equation}\label{eq:PFLAT}
P=-\mathcal{A}+\lambda\rho+\frac{\kappa}{2}\left(\frac{d\rho}{dx}\right)^2={\rm constant}
\end{equation}
Considering this expression at $\pm\infty$, and taking into account that the gradient of the density vanishes, $\left(\frac{d\rho}{dx}\right)^2\to 0$ as $x\to\pm\infty$, we get with the help of chemical potential equality \eqref{eq:EQchemFLAT}:
\begin{equation}\label{eq:EQpressureFLAT}
\left(\rho\frac{\partial\mathcal{A}}{\partial\rho}- \mathcal{A}\right)\bigg\vert_{\rho_l} =\left(\rho\frac{\partial\mathcal{A}}{\partial\rho}-\mathcal{A}\right)\bigg\vert_{\rho_v} 
\end{equation}
This enables to identify the {\it pressure} $P$ (or {\it equation of state, EoS}),
\begin{equation}
P=\rho\frac{\partial \mathcal{A}}{\partial\rho}-\mathcal{A},
\end{equation}
and the result (\ref{eq:EQpressureFLAT}) is nothing but the remaining Gibbs equilibrium condition: The pressure is the same in both the phases,
\begin{equation}
P_l=P_v.
\end{equation}

\subsubsection{Korteweg stress tensor and second-gradient fluid balance equations}
The stress tensor of a non-ideal fluid in 3-D is obtained by extending the constrained minimization problem of \eqref{eq:minimisation_free_energy} to 3-D leading to \cite{anderson_diffuse-interface_1998},
    \begin{equation}
        \bm{T}_{K} =  \bm{\nabla}\otimes\frac{\partial\mathcal{L}}{\partial(\bm{\nabla}\rho)}- \mathcal{L}\bm{I},
    \end{equation}
where $\bm{I}$ is unit tensor and $\mathcal{L}$ is the Lagrange function in 3-D,
    \begin{equation}
        \mathcal{L} = \mathcal{A} + \frac{1}{2}\kappa {\lvert \bm{\nabla}\rho\lvert}^2 - \lambda\rho,
    \end{equation}
and $\lambda$, corresponding to the chemical potential, is,
    \begin{align}
    	\lambda=\frac{\partial\mathcal{A}}{\partial\rho}-\kappa\bm{\nabla}^2\rho.
    \end{align}
This in turn leads to the following Korteweg's stress tensor~\citep{korteweg_sur_1901}:
    \begin{equation}\label{eq:Korteweg_stress}
        \bm{T}_{K} = \left(P-\kappa\rho\bm{\nabla}^2\rho - \frac{1}{2}\kappa {\lvert \bm{\nabla}\rho\lvert}^2 \right)\bm{I} + \kappa \bm{\nabla}\rho\otimes\bm{\nabla}\rho,
    \end{equation}
    where 
    \begin{align}
    	P=\rho \frac{\partial\mathcal{A}}{\partial\rho}-\mathcal{A},
    	\label{eq:pressure}
    \end{align}	
    	 is the thermodynamic pressure, or equation of state. From the local balance equations for mass and momentum one obtains the macroscopic governing laws for an isothermal capillary fluid:
    \begin{align}\label{eq:macro_mass}
      &  \partial_t\rho + \bm{\nabla}\cdot\rho \bm{u} = 0,\\
      &  \partial_t\rho\bm{u} + \bm{\nabla}\cdot\rho \bm{u}\otimes\bm{u} + \bm{\nabla}\cdot\bm{T}= 0,\label{eq:momentum_balance}
    \end{align}
   where $\bm{u}$ is the fluid velocity and the stress tensor $\bm{T}$ is
    \begin{equation}\label{eq:macro_stress}
        \bm{T} = \bm{T}_K + \bm{T}_{\rm NS}.
    \end{equation}
The Navier--Stokes viscous stress tensor reads,
\begin{equation}\label{eq:NS_stress}
	\bm{T}_{\rm NS}=-\mu\bm{S}-\eta (\bm{\nabla}\cdot\bm{u})\bm{I},
\end{equation}
where $\bm{S}$ is the trace-free rate-of-strain tensor,
\begin{equation}\label{eq:viscous_stress}
	\bm{S}=\bm{\nabla}\bm{u} + {\bm{\nabla}\bm{u}}^{\dagger} -\frac{2}{3}(\bm{\nabla}\cdot\bm{u})\bm{I},
\end{equation}
and $\mu$ and $\eta$ are the dynamic and the bulk viscosity, respectively.\\ 
The momentum balance equation \eqref{eq:momentum_balance} can  be recast in the following form,
    \begin{equation}\label{eq:momentum_balance_force}
    		\partial_t\rho\bm{u} + \bm{\nabla}\cdot\rho \bm{u}\otimes\bm{u} + \bm{F}_{\rm K} +\bm{\nabla}\cdot\bm{T}_{\rm NS}=0,
    \end{equation}
where Korteweg's {\it force} $\bm{F}_{\rm K}$ is the divergence of the Korteweg pressure tensor,
    \begin{equation}
        \bm{F}_{\rm K}=\bm{\nabla}\cdot\bm{T}_{\rm K}.
    \end{equation}
Eqs.~\eqref{eq:macro_mass} together with Eqs.~\eqref{eq:vdW_EoS}, \eqref{eq:Korteweg_stress}, \eqref{eq:macro_stress}, \eqref{eq:NS_stress} and \eqref{eq:viscous_stress} describe the dynamics of the \emph{van der Waals fluid} targeted by non-ideal lattice Boltzmann models in the hydrodynamic limit. One of the remarkable features of this description of non-ideal fluids and interfaces is that it can also be recovered from the kinetic theory of gases. In the next section we provide a brief description of simplified kinetic models leading to the second-gradient fluid model.
\subsection{Kinetic models for non-ideal fluid \label{sec:statmech}}
To introduce a kinetic model for non-ideal fluids, one needs to begin with the first Bogolioubov--Born--Green--Kirkwood--Yvon (BBGKY) equation,
\begin{equation}\label{eq:Boltzmann_eq}
	\partial_t f + \bm{v}\cdot\bm{\nabla} f = \mathcal{J}=\int\int  \bm{\nabla} V\left(\lvert \bm{r}-\bm{r}_1\rvert\right)\cdot \frac{\partial}{\partial \bm{v}}f_{2}(\bm{r},\bm{v}, \bm{r}_1,\bm{v}_1,t) d\bm{v}_1d\bm{r}_1,
\end{equation}
where $f(\bm{r},\bm{v},t)$ and $f_{2}(\bm{r},\bm{v},\bm{r}_1,\bm{v}_1,t)$ are the one- and the two-particle distribution functions, respectively, $\bm{r}$, $\bm{r}_1$ and $\bm{v}$, $\bm{v}_1$ are particles position and velocity, while $V$ is a potential of pair interaction. The Boltzmann collision integral of Eq.~\eqref{eq:Bcollisionintegral} is not sufficient for the dense regime as it neglects the volume occupied by the molecules and long range interactions. The BBGKY equation on the other hand make no assumptions on the nature of the interaction as it involves a general non-local interaction potential and dependence on the two-particle distribution function. The main difficulty of the BBGKY is that to solve the one-particle equation one needs the two-particle distribution function which itself is governed by a non-homogeneous hyperbolic partial differential equation with a collision term involving the three-particle distribution function and so on. To operate a truncation at first-order proper approximations for the correlation function and the interaction potential are needed. In the next section we briefly present two of the most widely used models for short- and long-range interaction.
\subsubsection{Hard sphere potential: The Enskog model}
In Enskog's approach, particles are assumed to be hard impenetrable spheres of diameter $d$. This fixes the form of the interaction potential effectively allowing to  re-write the collision integral as~\cite{struchtrup_grads_2019,struchtrup_twenty-six_2022}:
\begin{equation}
    \mathcal{J}_{\rm E} = d^2\int \int \left[f_{2}(\bm{r}+\frac{d}{2}\bm{k},\bm{v}', \bm{r}_1,\bm{v}_1',t) - f_{2}(\bm{r}-\frac{d}{2}\bm{k},\bm{v}, \bm{r}_1,\bm{v}_1,t)\right] \bm{g}\cdot\bm{k}d\bm{k}d\bm{v}_1,
\end{equation}
where $\bm{k} = (\bm{r}_1-\bm{r})/\lvert\bm{r}_1-\bm{r}\lvert$, $\bm{g}=\bm{v}_1-\bm{v}$, $\bm{v}'=\bm{v}+\bm{k}(\bm{g}\cdot\bm{k})$, $\bm{v}_1'=\bm{v}_1-\bm{k}(\bm{g}\cdot\bm{k})$. The second approximation in Enskog's standard theory applies to the two-particle distribution function~\cite{enskog_warmeleitung_1921,chapman_mathematical_1939}:
\begin{eqnarray}\label{eq:SET_two_particles}
    f_{2}(\bm{r}+\frac{d}{2}\bm{k},\bm{v}', \bm{r}_1,\bm{v}_1',t) &=& \chi\left(\bm{r}+\frac{d}{2}\bm{k}\right) f\left(\bm{r},\bm{v}'\right)f\left(\bm{r}+d\bm{k},\bm{v}_1'\right), \\
    f_{2}(\bm{r}-\frac{d}{2}\bm{k},\bm{v}, \bm{r}_1,\bm{v}_1,t) &=& \chi\left(\bm{r}-\frac{d}{2}\bm{k}\right)f\left(\bm{r},\bm{v}\right)f\left(\bm{r}-d\bm{k},\bm{v}_1\right), 
\end{eqnarray}
where $\chi$ is the equilibrium pair correlation function, evaluated at local density taking into account the effect of volume of particles in the collision probability~\cite{chapman_mathematical_1939}. Using both these approximations one recovers Enskog's hard-sphere collision integral~\cite{enskog_warmeleitung_1921,chapman_mathematical_1939},
\begin{equation}\label{eq:enskog_integral}
	\mathcal{J}_{\rm E} = d^2\int \int \left[\chi\left(\bm{r}+\frac{d}{2}\bm{k}\right)f\left(\bm{r},\bm{v}'\right)f\left(\bm{r}+d\bm{k},\bm{v}_1'\right)  - \chi\left(\bm{r}-\frac{d}{2}\bm{k}\right)f\left(\bm{r},\bm{v}\right)f\left(\bm{r}-d\bm{k},\bm{v}_1\right) \right] \bm{g}\cdot\bm{k}d\bm{k}d\bm{v}_1.
\end{equation}
The above integral can be approximated using a Taylor expansion around $\bm{r}$ as,
\begin{eqnarray}
	\chi\left(\bm{r}\pm \frac{d}{2}\bm{k}\right) &=& \chi\left(\bm{r}\right) \pm \frac{d}{2}\bm{k}\cdot\bm{\nabla}\chi\left(\bm{r}\right) + \frac{d^2}{8} \bm{k}\otimes\bm{k}:\bm{\nabla}\bm{\nabla} \chi\left(\bm{r}\right) 
 + \mathcal{O}(d^3),\\
	f\left(\bm{r}\pm d\bm{k},\bm{w}\right) &=& f\left(\bm{r},\bm{w}\right) \pm d\bm{k}\cdot\bm{\nabla} f\left(\bm{r},\bm{w}\right) + \frac{d^2}{2}\bm{k}\otimes\bm{k}:\bm{\nabla}\bm{\nabla} f\left(\bm{r},\bm{w}\right) + {O}(d^3),
\end{eqnarray}
which leads to:
\begin{equation}
	\mathcal{J}_{\rm E} = \chi\mathcal{J}_{\rm B} + \mathcal{J}_{\rm E}^{ (1)} + \mathcal{J}_{\rm E}^{(2)},
\end{equation}
where we have neglected terms of order $3$ and above, $\mathcal{J}_{\rm B}$ is the Boltzmann collision integral for hard-spheres,
\begin{equation}
	\mathcal{J}_{\rm B} = d^2  \int \int \left[f\left(\bm{r},\bm{v}'\right)f\left(\bm{r},\bm{v}_1'\right)  - f\left(\bm{r},\bm{v}\right)f\left(\bm{r},\bm{v}_1\right) \right] \bm{g}\cdot\bm{k}d\bm{k}d\bm{v}_1,
\end{equation}
and $\mathcal{J}_{\rm E}^{ (1)}$ and $\mathcal{J}_{\rm E}^{(2)}$ are the first- and second-order non-local contributions defined as:
\begin{align}
	\mathcal{J}_{\rm E}^{(1)} =& d^3 \chi\left(\bm{r}\right) \int \int \bm{k}\cdot\left[f\left(\bm{r},\bm{v}'\right)\bm{\nabla}f\left(\bm{r},\bm{v}'_1\right)  + f\left(\bm{r},\bm{v}\right)\bm{\nabla}f\left(\bm{r},\bm{v}_1\right) \right] \bm{g}\cdot\bm{k}d\bm{k}d\bm{v}_1 \nonumber\\
	&	+ \frac{d^3}{2} \int \int \bm{k}\cdot\bm{\nabla}\chi\left(\bm{r}\right)\left[f\left(\bm{r},\bm{v}'\right)f\left(\bm{r},\bm{v}'_1\right) + f\left(\bm{r},\bm{v}\right)f\left(\bm{r},\bm{v}_1\right) \right] \bm{g}\cdot\bm{k}d\bm{k}d\bm{v}_1,
\end{align}
and~\cite{kremer_introduction_2010}
\begin{align}
	\mathcal{J}_{\rm E}^{(2)} =& \frac{d^4}{2} \chi\left(\bm{r}\right) \int \int \bm{k}\otimes\bm{k}\left[f\left(\bm{r},\bm{v}'\right) \bm{\nabla}\bm{\nabla} f\left(\bm{r},\bm{v}'_1\right) - f\left(\bm{r},\bm{v}\right) \bm{\nabla}\bm{\nabla}f\left(\bm{r},\bm{v}_1\right) \right] \bm{g}\cdot\bm{k}d\bm{k}d\bm{v}_1 \nonumber\\
    & + \frac{d^4}{2}\int \int \bm{k}\otimes\bm{k}:\bm{\nabla}\chi\left(\bm{r}\right) \otimes \left[f\left(\bm{r},\bm{v}'\right)\bm{\nabla}f\left(\bm{r},\bm{v}'_1\right)  - f\left(\bm{r},\bm{v}\right)\bm{\nabla}f\left(\bm{r},\bm{v}_1\right) \right] \bm{g}\cdot\bm{k}d\bm{k}d\bm{v}_1 \nonumber \\
	&	+ \frac{d^4}{8} \int \int \bm{k}\otimes\bm{k}:\bm{\nabla}\bm{\nabla}\chi\left(\bm{r}\right)\left[f\left(\bm{r},\bm{v}'\right)f\left(\bm{r},\bm{v}'_1\right) - f\left(\bm{r},\bm{v}\right)f\left(\bm{r},\bm{v}_1\right) \right] \bm{g}\cdot\bm{k}d\bm{k}d\bm{v}_1.
\end{align}
Only keeping first-order contributions and evaluating $\mathcal{J}_{\rm E}^{(1)}$ at the local equilibrium $f^{\rm eq}$ \eqref{eq:LM} one gets~\cite{chapman_mathematical_1939},
\begin{align}
	\mathcal{J}_{\rm E}^{(1)} = d^3 \int \int f^{\rm eq}\left(\bm{r},\bm{v}\right)f^{\rm eq}\left(\bm{r},\bm{v}_1\right) \bm{k}\cdot\bm{\nabla}\ln\left[\chi\left(\bm{r}\right)f^{\rm eq}\left(\bm{r},\bm{v}\right)f^{\rm eq}\left(\bm{r},\bm{v}_1\right)\right] \bm{g}\cdot\bm{k}d\bm{k}d\bm{v}_1,
\end{align}
which, after integration in $\bm{v}_1$ and $\bm{k}$, for the isothermal flow results in,
\begin{align}
	\mathcal{J}_{\rm E}^{(1)} =& -b\rho\chi f^{\rm eq}\left[(\bm{v}-\bm{u})\cdot\bm{\nabla}\ln\rho^2\chi T\right] \nonumber\\  
	&-b\rho\chi f^{\rm eq}\left[ \frac{2}{5RT}(\bm{v}-\bm{u})(\bm{v}-\bm{u}):\bm{\nabla}\bm{u} + \left(\frac{1}{5RT}{\lvert\bm{v}-\bm{u}\lvert}^2-1\right)\bm{\nabla}\cdot\bm{u}\right],
	\label{eq:E_approximation}
\end{align}
where $b=2\pi d^3/3m$.
While the phenomenological Enskog's collision integral \cite{enskog_warmeleitung_1921} was used above, the lowest-order approximation Eq.~\eqref{eq:E_approximation} is identical in other versions of hard-sphere kinetic equations such as the revised Enskog theory (RET) \citep{van_beijeren_modified_1973} or kinetic variational theory \citep{karkheck_kinetic_1981}. 
\subsubsection{Long-range interactions: Vlasov model}
Enskog's model only accounts for strong repulsive short-range interactions introducing the excluded volume into the equation of state. However it is not enough to account for long-range attractive interactions leading to spinodal decomposition. A widely used model for long-range attractive contributions of the collision intergal in Eq.~\eqref{eq:Boltzmann_eq} is Vlasov's mean-field approximation. Here we briefly review this model. Assuming absence of correlations, the two-particle distribution function is approximated as
\begin{equation}
	f_{2}(\bm{r},\bm{v}, \bm{r}_1, \bm{v}_1) \approx f(\bm{r},\bm{v}) f(\bm{r}_1, \bm{v}_1),
\end{equation}
while the long-range interaction integral can be simplified,
\begin{equation}\label{eq:LR_o}
	\mathcal{J}_{\rm V} = \frac{\partial f(\bm{r},\bm{v}) }{\partial\bm{v}}\cdot\bm{\nabla}\left[\int_{\lvert\bm{r}_1-\bm{r}\lvert > d} \rho(\bm{r}_1)V(\lvert \bm{r}_1-\bm{r}\lvert) d\bm{r}_1\right].
\end{equation}
With a Taylor expansion around $\bm{r}$, 
\begin{equation}\label{eq:TmcL_density_longrange}
	\rho(\bm{r}_1) = \rho(\bm{r}) + (\bm{r}_1-\bm{r})\cdot\bm{\nabla}\rho(\bm{r}) + \frac{1}{2}(\bm{r}_1-\bm{r})\otimes(\bm{r}_1-\bm{r}):\bm{\nabla}\otimes\bm{\nabla}\rho(\bm{r}) + \mathcal{O}(\bm{\nabla}^3\rho),
\end{equation}
and neglecting higher-order terms, with Eq.~\eqref{eq:LR_o} one recovers the meanfield Valsov long-range molecular interaction,
\begin{equation}\label{eq:Vlasov_collision_approx}
	\mathcal{J}_{\rm V} = -\bm{\nabla}\left[2 a \rho(\bm{r}) +\kappa \bm{\nabla}^2 \rho(\bm{r})\right]\cdot\frac{\partial}{\partial\bm{v}}f(\bm{r},\bm{v}),
\end{equation}
where parameters $a$ and $\kappa$ are, after integration over a unit sphere,
\begin{align}
	a &= -2\pi\int_{d}^{\infty} r^2V(r) dr,\\
	\kappa &= -\frac{2\pi}{3}\int_{d}^{\infty} r^4V(r) dr.
\end{align}
Eq.~\eqref{eq:Boltzmann_eq} together with:
\begin{equation}
    \mathcal{J} = \chi\mathcal{J}_{\rm B} + \mathcal{J}_{\rm E}^{(1)} + \mathcal{J}_{\rm V},
\end{equation}
and the short- and long-range interaction models of Eqs.~\eqref{eq:E_approximation} and \eqref{eq:Vlasov_collision_approx} form the constitutive equations for moderately dense isothermal fluid Enskog-Vlasove model. This kinetic model, under some specific scaling assumptions detailed in \ref{ap:CE_cont}, is shown to recover the second-gradient fluid model at the Euler-Navier-Stokes level with,
\begin{equation}
    P = b\rho^2\chi r T - a\rho^2.
\end{equation}
Motivated by our recently published work~\cite{hosseini_towards_2021}, before entering the discussion of lattice Boltzmann models for non-ideal fluids we introduce next a general kinetic framework for non-ideal fluids that can be easily transposed into the lattice Boltzmann framework.
\subsubsection{General Enkog-Vlasov-BGK kinetic framework for hydrodynamics}
In an attempt to provide a unified an numerically efficient kinetic framework for non-ideal fluids simulation in ~\cite{hosseini_towards_2021} the authors introduced a projector $\mathcal{K}$ onto a local equilibrium at constant temperature \cite{gorban_invariant_2005}, 
    \begin{align}
    	\mathcal{K}\mathcal{J}=
    	\left(\dfrac{\partial f^{\rm eq}}{\partial \rho}-\frac{1}{\rho}\bm{u}\cdot\frac{\partial f^{\rm eq}}{\partial \bm{u}}\right)\int \mathcal{J} d\bm{v}
    	+\frac{1}{\rho}\dfrac{\partial f^{\rm eq}}{\partial \bm{u}}\cdot \int \bm{v}\mathcal{J} d\bm{v}.
    	\label{eq:projector}
    \end{align}
with the property, $\mathcal{K}^2=\mathcal{K}$, which can be verified by a direct computation. The equilibrium attractor is defined as:
    \begin{equation}
	f^{\rm eq}=\frac{\rho}{\left(2\pi P_0/\rho\right)^{D/2}}\exp\left[-\frac{(\bm{v}-\bm{u})^2}{2P_0/\rho}\right],
	\label{eq:eq_ref}
    \end{equation}
where $P_0$ is a freely tunable reference pressure. With the projector of Eq.~\eqref{eq:projector}, the interaction term in Eq.~\eqref{eq:Boltzmann_eq} is split into two parts by writing an identity,
\begin{equation}\label{eq:enskog_projection}
        \mathcal{J} = \left(1-\mathcal{K}\right)\mathcal{J} + \mathcal{K} \mathcal{J}.
\end{equation}
The first term,
\begin{align}
	\mathcal{J}_{\rm loc}=\left(1-\mathcal{K}\right)\mathcal{J},
\end{align}
satisfies the local conservation of both mass and momentum,
\begin{align}
	\mathcal{K}\mathcal{J}_{\rm loc}=0.
\end{align}
It is conventional to model the locally conserving part of the interaction with a single relaxation time Bhatnagar--Gross--Krook (BGK) approximation,
\begin{equation}\label{eq:introduction_of_bgk}
        \mathcal{J}_{\rm loc} \to  \mathcal{J}_{\rm BGK} = -\frac{1}{\tau}\left(1-\mathcal{K}\right)f = -\frac{1}{\tau}\left(f- f^{\rm eq}(\rho,\bm{u},\frac{P_0}{\rho})\right),
\end{equation}
where the relaxation time $\tau$ is a free parameter. The second term in the identity \eqref{eq:enskog_projection},
\begin{align}
	\mathcal{J}_{\rm nloc}=\mathcal{K}\mathcal{J},
\end{align}
satisfies the local mass but not the local momentum conservation. 
After integration by part in the velocity $\bm{v}$ and neglecting boundary integrals, one arrives at
\begin{equation}\label{eq:final_non_kinetic_contributions}
    \mathcal{J}_{\rm nloc} = -\frac{1}{\rho}\frac{\partial f^{\rm eq}}{\partial \bm{u}}\cdot\bm{F}_{\rm nloc},
\end{equation}
where the force $\bm{F}_{\rm nloc}$ reads,
\begin{equation}
    	\bm{F}_{\rm nloc}=\int\int\int \bm{\nabla}V\left(\lvert \bm{r}-\bm{r}_1\rvert\right)f_{2}(\bm{r},\bm{v}, \bm{r}_1,\bm{v}_1,t) d\bm{v}_1d\bm{r}_1d\bm{v}.
    	\label{eq:force_general}
    \end{equation}
Collecting the BGK approximation together with the nonlocal contribution, a generic kinetic  model may be written,
\begin{equation}\label{eq:final_kinetic_model}
	\partial_t f + \bm{v}\cdot\bm{\nabla} f = -\frac{1}{\tau}\left(f - f^{\rm eq}\right) 
	- \frac{1}{\rho}\frac{\partial f^{\rm eq}}{\partial \bm{u}}\cdot\bm{F}_{\rm nloc}.
\end{equation}
Evaluation of the force \eqref{eq:force_general} requires us to specify the particles interaction. One can invoke the Enskog--Vlasov model~\cite{enskog_warmeleitung_1921,vlasov_many-particle_1961} where both hard-sphere collisions and a weak long-range attraction potential contribute to a non-local momentum transfer or any other type of closure. More generally Eq.~\eqref{eq:final_kinetic_model} defines a family of kinetic models with $P_0$ and $\bm{F}_{\rm nloc}$ as tunable parameters to recover the hydrodynamics of interest. In the next section we review different lattice Boltzmann models for non-ideal fluids as closures for $\bm{F}_{\rm nloc}$ and $P_0$ within this general kinetic framework. For the sake of readability, in the remainder of the text we drop ${\rm nloc}$ and simply use $\bm{F}$ for the force. Before moving on to the LBM models we will discuss the hydrodynamic limit of non-ideal kinetic models under conventional scaling and alternative better suited to the recovery of the target macroscopic system.
\subsubsection{Scaling and hydrdynamic limit}
Considering the Enskog--Vlasov--BGK kinetic model introduced in the previous sections, let us introduce the following parameters:
\begin{itemize}
    \item characteristic flow velocity $\mathcal{U}$, 
    \item characteristic flow scale $\mathcal{L}$,
    \item characteristic flow time $\mathcal{T}=\mathcal{L}/\mathcal{U}$,
    \item characteristic density $\bar{\rho}$,
    \item isothermal speed of sound of ideal gas $c_s=\sqrt{rT}$,
    \item kinematic viscosity of the BGK model of ideal gas $\nu=\tau c_s^2$.
\end{itemize}
With the above, the variables are reduced as follows (primes denote non-dimensional variables):
\begin{itemize}
	\item time $t=\mathcal{T}t'$,
	\item space $\bm{r}=\mathcal{L}\bm{r}'$,
	\item flow velocity $\bm{u}=\mathcal{U}\bm{u}'$,
	\item particle velocity $\bm{v}=c_s\bm{v}'$,
	\item density $\rho$=$\bar{\rho}\rho'$,
	\item distribution function $f=\bar{\rho}c_s^{-3}f'$.
\end{itemize}
    Furthermore, the following non-dimensional groups are introduced:
    Viscosity-based Knudsen number
	${\rm Kn}={\tau c_s}/{\mathcal{L}}$, 
    Mach number 
    $	{\rm Ma}={\mathcal{U}}/{c_s}$,
    Enskog number 
	${\rm En}=
	b\bar{\rho}{{\rm Kn}}/{{\rm Ma}}$ and
    Vlasov number 
	${\rm Vs}={a}/{brT}$.\\
With this, the Enskog--Vlasov--BGK kinetic model is written in non-dimensionla form as:
\begin{multline}\label{eq:EV_model_scaled}
		{\rm Ma}\,{\rm Kn}\,\partial_{t}' f' + \bm{v}'\cdot{\rm Kn}\bm{\nabla}' f' = -\left(f' - {f^{\rm eq}}'\right) \\
		- \frac{1}{\rho'}\frac{\partial {f^{\rm eq}}'}{\partial \bm{u}'}\cdot {\rm En}\left[\bm{\nabla}'\left(\chi \left(\rho'\right)^2-{\rm Vs}\,\left(\rho'\right)^2\right){-\left(\frac{\delta}{\mathcal{L}}\right)^2{\rm Vs}\, \left(\rho'\bm{\nabla}'\bm{\nabla}^{'2}\rho'\right)}\right],
\end{multline}
where $\delta$ is the range of the attraction potential. Assuming $d\ll \delta$, we have $a\sim \bar{V}\delta^3$ and $\kappa\sim \bar{V}\delta^5$, where $\bar{V}$ is a characteristic value of the potential, thus
\begin{equation}\label{eq:longestimate}
	\sqrt{\kappa/a}\sim\delta.
\end{equation}
The following scaling assumptions are applied:  Acoustic scaling, ${\rm Ma}\sim 1$; Hydrodynamic scaling,  ${\rm Kn}\sim {\rm En}\sim \delta/\mathcal{L}\sim\epsilon$; Enskog--Vlasov parity,  ${\rm Vs}\sim 1$. In other words, the conventional hydrodynamic limit treats all non-dimensional groups that are inversely proportional to the flow scale $\mathcal{L}$ (${\rm Kn}$, ${\rm En}$ and $\delta/\mathcal{L}$) as a small parameter while the Enskog--Vlasov parity ensures that both the short- and long-range contributions to the pressure are treated on equal footing.
Returning to dimensional variables, we may write, 
\begin{equation}\label{eq:overall_eq}
	\epsilon\partial_{t} f + \bm{v}\cdot\epsilon\bm{\nabla} f = -\left(f - {f^{\rm eq}}\right) - \frac{1}{\rho}\frac{\partial {f^{\rm eq}}}{\partial \bm{u}}\cdot {\left(\epsilon \bm{F}^{(1)}+\epsilon^3\bm{F}^{(3)}\right)},
\end{equation}
where,
\begin{align}
    &	\bm{F}^{(1)}=\bm{\nabla}(P-P_0).\\
    &	\bm{F}^{(3)}=-\kappa\rho\bm{\nabla}\bm{\nabla}^2\rho.\label{eq:F3}
\end{align}
The analysis of the kinetic model under the above-detailed conventional scaling of a small deviation from a uniform state \citep{chapman_mathematical_1939}, 
\begin{equation}\label{eq:scaling_1}
	\bm{\nabla}\to \epsilon\bm{\nabla},\ \partial_t\to\epsilon\partial_t,
\end{equation} 
is detailed in \ref{ap:CE_cont}. To second order in space derivatives, the resulting momentum balance equation reads,
\begin{equation}\label{eq:momentum_balance_force_weak}
	\partial_t\rho\bm{u} + \epsilon\bm{\nabla}\cdot\rho \bm{u}\otimes\bm{u} + \epsilon\bm{\nabla}P +\epsilon\bm{\nabla}\cdot\epsilon\bm{T}_{\rm NS}+O(\epsilon^3)
	=0,
\end{equation}
where the dynamic viscosity $\mu$ and the bulk viscosity $\eta$ in the Navier--Stokes stress tensor (\ref{eq:NS_stress}) are defined by the reference pressure ($D=3$),
\begin{align}
	\mu&=\tau P_0,\label{eq:visc_gen}\\
	\eta&=\left(\frac{5}{3}-\frac{\partial\ln P_0}{\partial\ln\rho}\right)\tau P_0.\label{eq:bulk_visc_gen}
\end{align}
Thus, the momentum balance equation (\ref{eq:momentum_balance_force_weak}) is form-invariant with respect to the choice of reference pressure, provided $P_0$ satisfies a {sub-isentropic condition},
\begin{equation}\label{eq:subisentropic}
    {P_0}\le C\rho^{5/3},
\end{equation}
for some $C>0$. With (\ref{eq:subisentropic}), the bulk viscosity (\ref{eq:bulk_visc_gen}) is positive and vanishes when the reference pressure follows an isentropic process for ideal monatomic gas, $P_0=C\rho^{5/3}$. For example, any polytropic process, $P_0=A\rho^n$, $1\le n\le 5/3$ satisfies the sub-isentropic condition and results in $\eta=(5/3-n)\tau P_0$.
Special case of isothermal process $n=1$ returns $\eta=(2/3)\tau P_0$, and the viscous stress tensor becomes,
\begin{equation}
	\bm{T}_{\rm NS}=-\tau P_0\left(\bm{\nabla}\bm{u}+\bm{\nabla}\bm{u}^\dagger\right).
\end{equation}
On the other hand, when compared to the two-phase momentum equation, the macroscopic limit recovers only the nonideal gas component thereof while missing Korteweg's capillarity contribution. 
Indeed, the third-order term, $\sim\epsilon^3\rho\bm{\nabla}\bm{\nabla}^2\rho$  does {\it not} contribute to the momentum equation (\ref{eq:momentum_balance_force_weak}) under the scaling \eqref{eq:scaling_1}. This is consistent with the well-known results from kinetic theory \citep{chapman_mathematical_1939} and is not surprising:  The scaling \eqref{eq:scaling_1} is essentially based on the Knudsen number, which overrides the relative contribution of the capillarity term by two orders, cf. Appendix \ref{ap:CE_cont}. Thus, under weak non-uniformity assumption \eqref{eq:scaling_1}, the capillarity terms are seen as higher-order, Burnett-level contributions, and cannot appear in the main (first and second) orders in the momentum balance equation \eqref{eq:momentum_balance_force_weak}. In fact, condition \eqref{eq:scaling_1} rules out situations at an interface between phases where gradients of density become large over a relatively short distance. Therefore, in order for the kinetic model to recover in-full the momentum balance (\ref{eq:momentum_balance_force}), a different scaling needs to be applied. \\
To overcome that issue a rescaling of the kinetic model by a {time step} $\delta t$, here merely a characteristic time representing the level of coarse-graining in time, was introduced in \cite{hosseini_towards_2021}. As a preliminary consideration, we evaluate the contribution of the force term over the time step. For a generic force $\bm{F}$, the action of the force on the distribution function can be written as a full derivative in a frame that moves with the local fluid velocity,
\begin{equation}
    	\frac{1}{\rho}\frac{\partial f^{\rm eq}}{\partial \bm{u}}\cdot\bm{F} = \frac{d f^{\rm eq}}{dt}.
\end{equation}
Introducing the \emph{velocity increment}, 
\begin{equation}\label{eq:deltau}
    \delta\bm{u}=  \frac{\bm{F}}{\rho}\delta t,
\end{equation}
and integrating in time, leads to the following contribution to the distribution from the force,
\begin{equation}\label{eq:EDM}
    	\mathcal{F}=\int_{t}^{t+\delta t} \frac{1}{\rho}\frac{\partial f^{\rm eq}}{\partial \bm{u}}\cdot\bm{F} dt.
\end{equation}
With the characteristic values of the flow velocity $\mathcal{U}$, the flow scale $\mathcal{L}$,  the density $\rho$, the force $\mathcal{F}$ and the velocity increment $\delta u$, the following smallness parameter is introduced:
\begin{align}
    &	\frac{\delta u}{\mathcal{U}}\sim \frac{\delta t \mathcal{F}}{\rho \mathcal{U}}\sim\varepsilon,\label{eq:smallu}\\
    & \frac{\delta r}{\mathcal{L}}\sim\varepsilon.\label{eq:resolution}
\end{align}
The first scaling condition (\ref{eq:smallu}) refers to a {\it smallness of velocity increment}, that is, to the smallness of the force action over time $\delta t$. The second scaling condition (\ref{eq:resolution}) is a {\it resolution requirement}. Both conditions are assumed to hold simultaneously. Details of the corresponding multi-scale analysis are provided in \ref{ap:CE}. Unlike the previous results with the classical scaling, the momentum balance includes not only the nonideal gas pressure but also the capillarity term, and is thus consistent with Korteweg's force in the momentum balance. It should be pointed out that the scaling \eqref{eq:smallu} refers to smallness of the increment of the flow velocity rather that to smallness of either the time step or of the force. Thus, rescaling the kinetic model based on the smallness of flow velocity increments results in both the non-ideal gas equation of state and the capillarity revealed at the Euler level $O(\varepsilon)$ of the momentum balance. This is in a contrast to the conventional scaling, which is tied to the non-uniformity and surface tension would appear only at a Burnett level  $O(\epsilon^3)$.
\subsection{Overview of lattice Boltzmann models for non-ideal fluids}
The general kinetic framework of Eq.~\eqref{eq:final_kinetic_model}, regardless of the choice of $P_0$ and $\bm{F}$ can be readily discretized in phase-space, procedure detailed in section \ref{subsec:phase_space_discrete}, and physical space and time, detailed in section \ref{subsec:space_time_discrete}, to yield a discrete time evolution equation of the form:
\begin{equation}
    \bar{f}_i(\bm{r}+\bm{c}_i\delta t, t+\delta t) - \bar{f}_i(\bm{r}, t) = \frac{\delta t}{\bar{\tau}}\left( f_i^{\rm eq}(\rho,\bm{u}) - \bar{f}_i(\bm{r}, t) \right) + \mathcal{F}_i(\bm{F}).
\end{equation}
where $\mathcal{F}_i$ is the discrete form of the force contribution in the kinetic framework that can be treated using any of schemes detailed in section \ref{subsec:body_forces} and $f_i^{\rm eq}$ is the discrete equilibrium function that can be chosen among those discussed in section \ref{subsec:phase_space_discrete}. For a correct recovery of the viscous stress tensor, the redefined relaxation time is tied to the reference pressure $P_0$ as:
\begin{equation}
    \bar{\tau} = \frac{\rho \nu}{P_0} + \frac{\delta t}{2},
\end{equation}
and the bulk viscosity of the model, for the bare single relaxation time collision operator, is:
\begin{equation}
    \eta = P_0 \left(\frac{2+D}{D}-\frac{\partial \ln P_0}{\partial \ln \rho}\right)\left(\bar{\tau}-\frac{\delta t}{2}\right).
\end{equation}
Given that all models of interest in the present manuscript can be fitted within this discrete kinetic model, and that the only point of difference resides in the choice of $\bm{F}$ and $P_0$ we will discuss the different lattice Boltzmann models from that specific perspective.
\subsubsection{Shan and Chen's pseudo-potential model}
In this extension of the classical lattice Boltzmann method for isothermal ideal gases to non-ideal fluids first proposed in \cite{shan_lattice_1993}, the authors introduced a source of non-local interaction through a potential defined as:
\begin{equation}
    V(\bm{r}, \bm{r}') = G(\bm{r}, \bm{r}_1)\psi(\bm{r})\psi(\bm{r}_1),
\end{equation}
where $G(\bm{r}, \bm{r}_1)$ is a Green's function and $\psi$ is an effective number density. For the case of the discrete system, the authors proposed to only consider nearest-neighbour interactions to get the following discrete Green's function:
\begin{equation}
    G(\bm{r}-\bm{r}_1) = \begin{cases} 0, \lvert\bm{r}-\bm{r}_1\lvert\leq 2,\\ G, 0<\lvert\bm{r}-\bm{r}_1\lvert< 2, \end{cases}
\end{equation}
which in turn leads to the discrete non-local momentum source:
\begin{equation}
   \bm{F} = G\psi(\bm{r})\sum_{i=0}^{Q-1} w(\lvert\bm{c}_i\lvert)\psi(\bm{r}+\bm{c}_i\delta t)\bm{c}_i,
\end{equation}
where $G$ is now a constant controlling the interaction strength. Furthermore, this model keeps the isothermal equilibrium of the classical isothermal LBM, therefore effectively setting $P_0=\rho c_s^2$. Computing the Euler-level momentum balance equation it is readily seen that the equation of state changes from the ideal isothermal pressure to:
\begin{equation}
    P = \rho c_s^2 + \frac{G}{2}\psi^2,
\end{equation}
potentially allowing for coexistence of a vapour and liquid phase at a given temperature. The only condition for the stable coexistence of these two phases at temperature $T$ is the existence of two points $\rho_1$ and $\rho_2$, with $\rho_1<\rho_v<\rho_l<\rho_2$, satisfying:
\begin{equation}\label{eq:coexistance_condition}
    \frac{\partial^2 P}{\partial \rho^2}\lvert_{\rho=\rho_1,\rho_2} = \frac{\partial P}{\partial \rho}\lvert_{\rho=\rho_1,\rho_2} = 0.
\end{equation}
For instance setting $\psi\propto\rho$ would lead to a quadratic equation of state and could not satisfy either of the above-listed conditions regardless of the choice of the coefficients. In \cite{shan_lattice_1993}, the authors propose an effective density number of the form:
\begin{equation}
    \psi = \rho_0\left(1-\exp-\frac{\rho}{\rho_0}\right),
\end{equation}
where $\rho_0$ is freely tunable parameter. This form of the effective density number systematically admits two roots to Eq.~\eqref{eq:coexistance_condition} for $G<G_c$, where $G_c$ is the critical interaction strength above which the density degeneracy goes from two to one. Using the two conditions on derivatives of pressure at the critical state:
\begin{eqnarray}
    c_s^2 - 2G_c\rho_0 \exp\left(-\frac{\rho_c}{\rho_0}\right)\left(\exp\left(-\frac{\rho_c}{\rho_0}\right) - 1\right) &=& 0,\\
    -2 G_c \exp\left(-\frac{2\rho_c}{\rho_0}\right)\left(\exp\left(\frac{\rho_c}{\rho_0}\right) - 2\right) &=& 0,
\end{eqnarray}
the critical state is shown to be:
\begin{eqnarray}
    \rho_c &=& \rho_0 \ln 2, \\
    G_c &=& -\frac{2 c_s^2}{\rho_0}.
\end{eqnarray}
While $G$ is routinely assimilated to a pseudo-temperature as $G=-\frac{1}{T}$, see for instance \cite{chen_critical_2014}, a dimensional analysis shows $[G] = \frac{ {\rm m}^5 }{ {\rm kg} {\rm s}^2} $, which indicates $G\propto \frac{rT}{\rho_0}$.\\
From both kinetic theory and the van der Waals fluid theory one would expect two separate non-local contributions to allow for stable phase separation: A short-range repulsive and a long-range attractive interaction. As such the single-term form of the pseudo-potential model should not produce long range interaction meanfield effects such as surface tension. This point can be clarified by writing a Taylor expansion of the discrete non-local momentum source term:
\begin{equation}\label{eq:Taylor_pseudopotential}
    \bm{F} = G\psi(\bm{r})\sum_{i=0}^{Q-1} w(\lvert\bm{c}_i\lvert)\psi(\bm{r}+\bm{c}_i\delta t)\bm{c}_i = G\psi\bm{\nabla}\psi - \frac{G\delta r^2}{3}\psi\bm{\nabla} \Delta\psi + \mathcal{O}(\bm{\nabla}^5).
\end{equation}
It is observed that the discrete non-local interaction is akin to a second-order accurate finite-differences approximation to the the first-order derivative of $\psi$ admitting a leading third-order error that takes on a form similar to surface tension. While naturally generating a non-zero surface tension via this leading-order error term it has the following short-comings: (a) the surface tension coefficient is fixed and (b) different from the Kortwewg stress tensor, van der Waals fluid and Enskog-Vlasov theory the interfacial excess energy is not a function of density but the effective number density, which in principle is not a conserved variable~\cite{he_thermodynamic_2002,hosseini_towards_2021}.
\subsubsection{Free energy model and derivatives}
In its original form the free energy model was first introduced in~\cite{swift_lattice_1995}. Different from the pseudo-potential approach where the model was constructed based on a microscopic interaction argument, here it targets a specific macroscopic system: the van der Waals quasi-local thermodynamics, which results, after minimization of free energy in the Kortweweg pressure tensor:
\begin{equation}
    \bm{T}_{\rm K} = \left(P - \kappa\rho\bm{\nabla}^2\rho - \frac{\kappa}{2}{\lvert \bm{\nabla}\rho\lvert}^2\right)\bm{I} + \kappa \bm{\nabla}\rho\otimes\bm{\nabla}\rho.
\end{equation}
To extend the ideal gas LBM to recover the Korteweg stress tensor the authors derived a discrete equilibrium function by moment matching, considering moments up to order two. For a 1-D system with a three-velocity stencil:
\begin{equation}
\begin{bmatrix} 1 & 1 & 1 \\
c_{-1} & c_{0} & c_{1}\\
c_{-1}^2 & c_0 & c_1 \end{bmatrix} \begin{bmatrix} f^{\rm eq}_{-1} \\ f^{\rm eq}_0 \\ f^{\rm eq}_1 \end{bmatrix} = \begin{bmatrix} \rho \\ \rho u_x \\ \rho u_x^2 + P - \kappa\rho\partial_x^2\rho\end{bmatrix}.
\end{equation}
This is equivalent to setting $P_0 = T_K$ and $\bm{F}=0$. The 2-D version of the model with the standard D2Q9 velocity set can be found in \cite{swift_lattice_1995}. We will refer to this type of approach, introducing non-ideal contribution directly into the second-order moment of the equilibrium as pressure-based model. Later multi-scale analyses showed that this approach is subject to Galilean-variant errors scaling with $\propto{\rm Ma}$~\cite{holdych_improved_1998,inamuro_galilean_2000}. Via modification to the discrete equilibrium function, improved formulation with reduced Galilean-variant errors were proposed~\cite{holdych_improved_1998,inamuro_galilean_2000}. Another way of introducing non-ideal contributions was discussed in \cite{wagner_investigation_2006}: Introducing a Vlasov-like forcing term given by the divergence of the pressure tensor. This approach, referred to as force-based, was first proposed in~\cite{luo_theory_2000} and is equivalent to setting $P_0 = \rho c_s^2$ and $\bm{F} = \bm{\nabla}\cdot(\bm{T}_K - P_0\bm{I})$. While the force-based approach still had Galilean-variant errors, those affecting the viscous shear stress at the NS level could be easily eliminated using a third-order polynomial equilibrium function. The force-based approach was later extended to more complex and realistic configurations through the use of the entropic LBM~\cite{mazloomi_entropic_2015,mazloomi_moqaddam_simulation_2016,mazloomi_moqaddam_drops_2017}.\\
More recently the free energy method was reformulated using the chemical potential~\cite{wen_chemical-potential_2020}; Using the definition of the chemical potential and pressure in the free energy model one can derive the following relationship between the divergence of the stress tensor and gradient of chemical potential:
\begin{equation}
    \bm{\nabla}\cdot \bm{T}_K = \rho\bm{\nabla}\lambda.
\end{equation}
Using the ideal gas LBM, i.e. $P_0=\rho c_s^2$, with a force-based approach to introduce non-ideal contribution the body force is derived as:
\begin{equation}
    \bm{F} = -\rho\bm{\nabla}\lambda + c_s^2\bm{\nabla}\rho,
\end{equation}
where, for instance, for the van de Waals equation of state, see Eq.~\eqref{eq:vdWEoS}:
\begin{equation}
    \lambda = r T \left[ \ln\left(\frac{\rho}{1-b\rho}\right) + \frac{1}{1-b\rho}\right] - 2a\rho -\kappa\bm{\nabla}^2\rho.
\end{equation}
Using this approach, in combination with a strategy to thicken the interface the authors were able to considerably extend the range of accessible density ratios.\\
All lattice Boltzmann models discussed here, at the hydrodynamic scale, target a form of the Korteweg stress tensor with a non-ideal equation of state and surface tension term. However such models come with a limitation: the thickness of the interface which contrary to phase-field models relying on a double-well potential is dictated by physical properties of the considered fluid, i.e. the volume exclusion coefficient, the long-range interaction coefficient and the capillary coefficient. We discuss this limitation along with a solution allowing to go above that scale limitation which has long been overlooked in the literature.
\subsection{Bridging the scale gap: Principle of corresponding states}
As briefly mentioned in the previous sections, the non-ideal fluids models of interest here all come with interface thicknesses solely determined by thermodynamics and properties of considered fluids. This creates a scale gap between the characteristic size of the interface and target characteristic scales of interest for drops and bubble, e.g. $\approx10^{-2}-10^{-3}$~m. We will clarify this size restriction here and discuss efficient solutions allowing to up-scale the solver.
\subsubsection{Dimensional form of equations and restriction by the interface thickness}
\begin{figure}[h!]
	\centering
	\includegraphics[width=8cm,keepaspectratio]{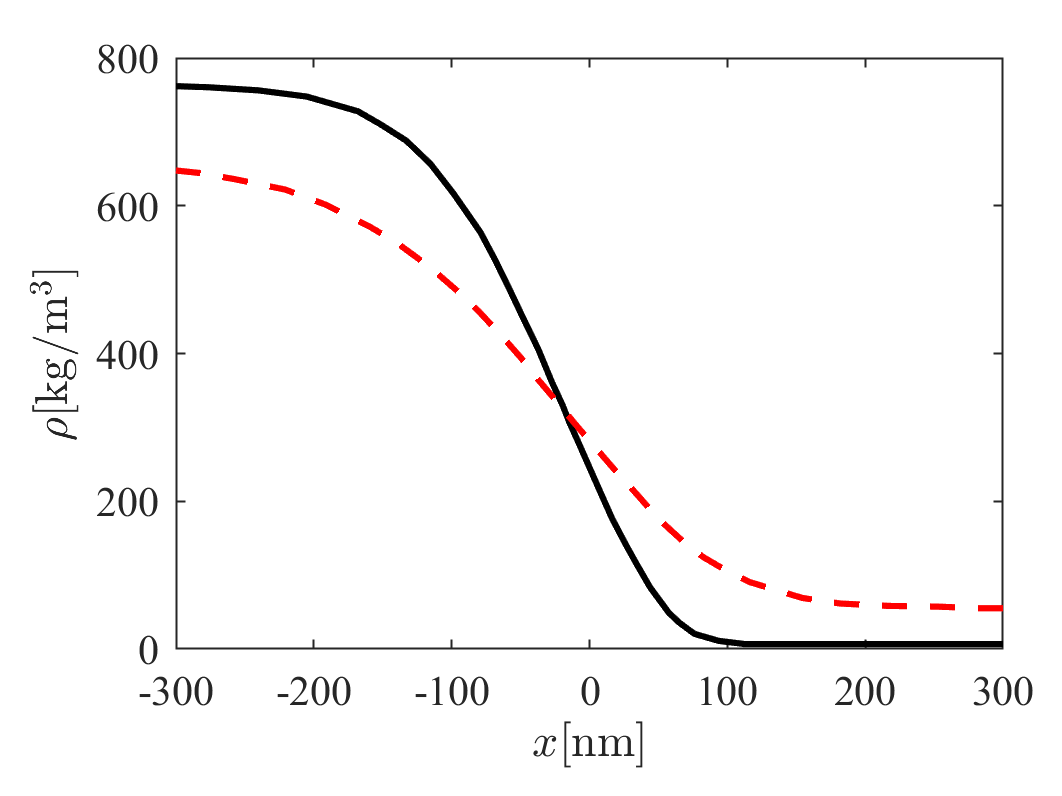}
	\caption{Density profile of ${\rm N}_2$ liquid/vapor interface at (black plain) $T = 63~{\rm K}$ and (red dashed) $T=94.5~{\rm K}$. }
	\label{Fig:density_profile_N2}
\end{figure}

While discussions around second-gradient models for non-ideal fluids are expressed in non-dimensional units, we will show that such models come with characteristic sizes imposed by physics. Consider for that purpose a 1-D isothermal interface; The steady-state Navier-Stokes-Korteweg equation reduces to:
\begin{equation}\label{eq:1dNSK}
    \frac{\partial P}{\partial x} = \kappa \rho \frac{\partial^3 \rho}{\partial x^3},
\end{equation}
with $\lim_{x\rightarrow-\infty} \rho = \rho_l$ and $\lim_{x\rightarrow+\infty} \rho = \rho_v$. While we are considering the classical Korteweg stress tensor here, the same line of reasoning applies to the macroscopic equation recovered by the pseudo-potential method. Here for the sake of simplicity we use the van der Waals equation of state for the thermodynamic pressure. As a result the ordinary differential equation describing the interface has six degrees of freedom, i.e. $a$, $b$, $r$, $\kappa$, $\rho_v$ and $\rho_l$. The former three along with with liquid and vapor phase densities are readily fixed via conditions on the critical state of the considered fluid and the temperature of the system. Consider for instance nitrogen ${\rm N}_2$; To recover proper critical density, pressure and temperature, i.e. $\rho_c=311~{\rm kg}/{\rm m}^3$, $P_c=33.5~{\rm Pa}$ and $T_c=126~K$~\cite{guggenheim_principle_1945}, the excluded volume and attractive force coefficients and specific gas constant, i.e. $b$, $a$ and $r$ should be set to:
\begin{eqnarray}
    a_{{\rm N}_2} &=& 105.28~\frac{{\rm m}^5}{{\rm kg}  {\rm s}^2},\\
    b_{{\rm N}_2} &=& 0.00107~\frac{{\rm m}^3}{{\rm kg}},\\
    r_{{\rm N}_2} &=& 230.99~\frac{{\rm m}^2}{{\rm s}^2 K}.
\end{eqnarray}
To recover the correct critical density, $r$ is set to a value which is different from its ideal counterpart. This is justified by the fact that here to set the constants we have relied on properties of the fluid at the critical state where it does not behave as an ideal gas, see~\cite{winterbone_advanced_1996} for detailed discussion. The remaining degree of freedom, the capillary coefficient $\kappa$, is also a physical property of the considered fluid. A variety of expressions have been proposed for the capillary coefficient both in the context of the kinetic theory of gases, e.g. ~\cite{yang_molecular_1976,lovett_generalized_1973,ornstein_accidental_1914}, or via semi-empirical correlations to match experimental measurements, see \cite{lin_gradient_2007} for detailed review. Here for ${\rm N}_2$ we will set $\kappa=10^{-10}~\frac{{\rm m}^7}{{\rm kg}{\rm s}^2}$~\cite{nayigizente_development_2021}. It is then clear that once the temperature of the system has been fixed, all parameters/variables in Eq.~\eqref{eq:1dNSK} are set by physical closures, also fixing the coexistence densities and density profile at the interface. The liquid/vapor interface at two different temperatures for the ${\rm N}_2$ system is illustrated in Fig.~\ref{Fig:density_profile_N2}.
The interfaces at both temperatures, resulting in density ratio of the order of $10^2$ and $10$, have sizes of the order of $200-400~{\rm nm}$. In practice, considering one needs to resolve the interfaces with at least 3-4 points, the maximum grid-size is limited to $\delta r\approx 100~{\rm nm}$ meaning for configurations of practical interest involving drops, for instance of diameter 1~mm, one would need at least $10^{12}$ grid-points in 3-D. This obviously makes the approach of little to no use for simulations of practical interest. The only workaround would be to rescale the interface thickness, here the limiting charactersitic size, to minimize the required number of grid-points for realistic simulations. In the next section we discuss the principle of corresponding states which provides the theoretical basis for such rescaling operations.
\subsubsection{Extension to realistic-sized systems: Rescaling interface thickness and the principle of corresponding states}
\begin{figure}[h!]
	\centering
	\includegraphics[width=8cm,keepaspectratio]{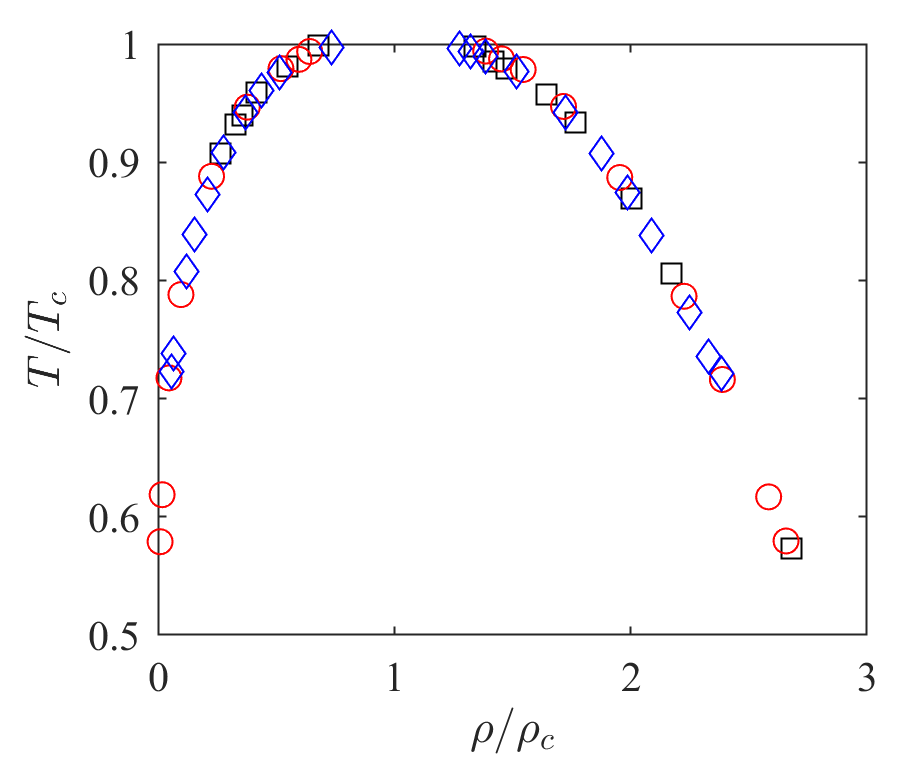}
	\caption{Co-existence densities obtained from experiments as reported in \cite{guggenheim_principle_1945} for three different fluids: (black squares) ${\rm CH}_4$, (red circles) ${\rm N}_2$ and (blue diamonds) ${\rm Xe}$.}
	\label{Fig:coexistence_density_real_guggenheim}
\end{figure}

The principle of corresponding states, introduced for the first time by van der Waals in \cite{van_der_waals_over_1873}, quoting Guggenheim \cite{guggenheim_principle_1945}, \emph{may safely be regarded as the most useful by-product of van der Waals' equation of state.} The principle maintains that all properties that depend on inter-molecular forces are related to the critical properties of the substance in a universal way, regardless of the molecular compound of interest. This observation has had two main practical consequences: (a) \emph{prediction of unknown properties of many fluids from known properties of a few}~\cite{leland_corresponding_1968}, (b) extension of the domain of applicability of second-gradient-based numerical solvers to large systems at acceptable cost. The possibility of the former was put forward first theoretically in \cite{pitzer_corresponding_1939} and confirmed experimentally in Guggenheim's work, illustrated in Fig.~\ref{Fig:coexistence_density_real_guggenheim}, for a family of fluids called \emph{perfect liquids} by Pitzer~\cite{pitzer_corresponding_1939}.
While the principle of corresponding states is an approximation for real fluids, it is an exact property of the van der Waals fluids. This is most notably demonstrated by the fact that the non-dimensional form of the equation of state, non-dimensionalized by critical state properties, does not have any fluid-dependent constant:
\begin{equation}
    P_r = \frac{8 \rho_r T_r}{3 - \rho_r} - 3\rho_r^2,
\end{equation}
where we have introduced reduced variables, e.g. $\rho_r=\frac{\rho}{\rho_c}$. The implication here is that regardless of the fluid considered, the coexistence densities plot in non-dimensional units are universal. To illustrate that four different fluids were considered and parameters $a$, $b$ and $r$ were chosen so as to fit the experimental critical state of each fluid. The corresponding data is summarized in Table~\ref{table:vdW_fluid_properties}.
\begin{table}[!h]
\begin{center}
\begin{tabular}{||c c c c c||}
 \hline
   & ${\rm H}_2{\rm O} $ & Xe & ${\rm N}_2$ & ${\rm CH}_4 $ \\ [0.5ex] 
 \hline\hline
 $T_c~[{\rm K}]$ & 647.14 & 289.8 & 126 & 190.3\\ 
 \hline
 $\rho_c~[{\rm kg}/{\rm m}^3]$ & 322  & 1155 & 311 & 162 \\
 \hline
 $P_c$~[MPa] & & 5.897 & 3.394 & 4.610 \\
 \hline
 \hline
 $a~\left[\frac{{\rm m}^5}{{\rm kg}\cdot {\rm s}^2}\right]$ & 1709  & 13.2614 & 105.272 & 526.978\\
 \hline
 $b~\left[\frac{{\rm m}^3}{{\rm kg}}\right]$ & 0.0017  & $2.886\times10^{-4}$ & 0.0011 & 0.0021\\ 
 \hline
 $r~\left[\frac{\rm m^2}{\rm s^2 K}\right]$ & 196.79 &  46.981 & 230.967 & 398.764 \\
 \hline
\end{tabular}
\caption{Critical state and corresponding van der Waals coefficients for different fluids; Experimental data from \cite{guggenheim_principle_1945} and \cite{sato_sixteen_1991} (for water). }
\label{table:vdW_fluid_properties}
\end{center}
\end{table}
Plugged into the van der Waals equation of state, all fluids, as shown in Fig.~\ref{Fig:coexistence_density_real_guggenheim}, result in exactly the same non-dimensional coexistence densities curve.
\begin{figure}[h!]
	\centering
	\includegraphics[width=8cm,keepaspectratio]{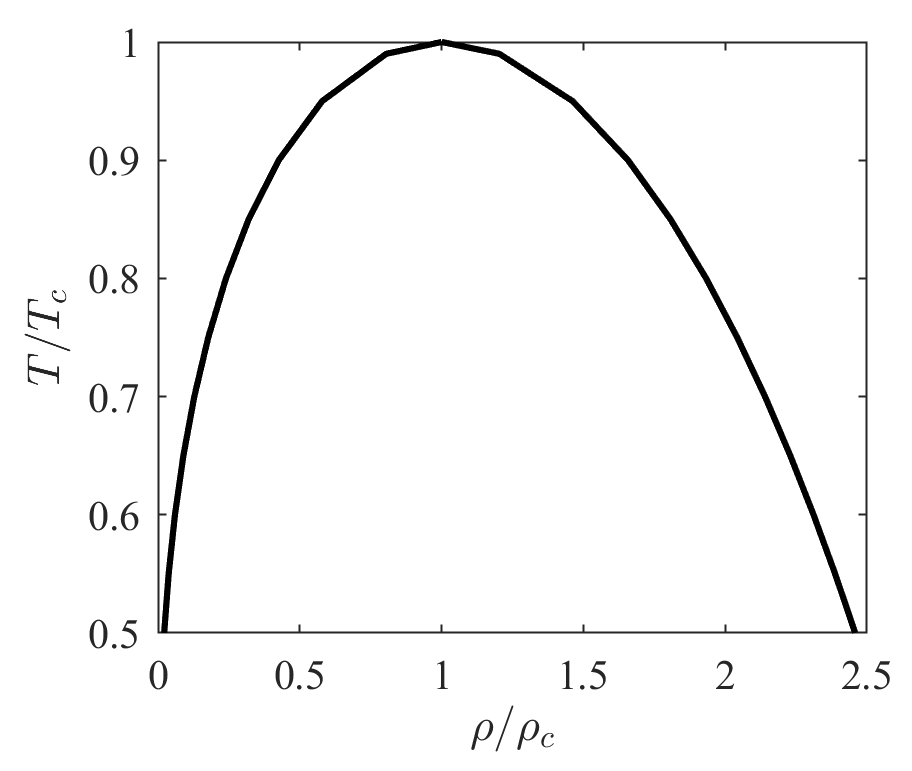}
	\caption{Co-existence densities obtained from the van der Waals equation of state for all fluids of Table~\ref{table:vdW_fluid_properties}. Given that all fluids led to exactly the same curve, only one is shown here.}
	\label{Fig:coexistence_density_vdW}
\end{figure}
Now that we have showed that the density ratio is independent of substance-specific variables let us go back to~\eqref{eq:1dNSK} and discuss simple strategies that would allow one to thicken the interface. To illustrate our purpose we will use the simple demonstration discussed in~\cite{jamet_etude_1998} by considering a 1-D interface of a fluid near the critical state. This consideration allows us to simplify the chemical potential of the van der Waals equation of state as a third-order polynomial~\cite{domb_critical_1996}:
\begin{equation}
    \lambda - \lambda_{\rm sat} = 4A(\rho-\rho_v)(\rho-\rho_l)\left(\rho-\frac{\rho_v+\rho_l}{2}\right),
\end{equation}
with
\begin{equation}
    A = \frac{1}{2\rho_{l,v}{(\rho_l - \rho_v)}^2}\frac{dP}{d\rho},
\end{equation}
leading to the following analytical density profile:
\begin{equation}
    \rho(x) = \frac{\rho_v+\rho_l}{2} + \frac{\rho_l - \rho_v}{2} \tanh\left(  \frac{\rho_l - \rho_v}{2}\sqrt{\frac{2A}{\kappa}}x\right).
\end{equation}
Using this analytical profile the interface thickness, here defined as,
\begin{equation}
    W = \frac{\rho_l - \rho_v}{\max(d\rho/dx)},
\end{equation}
and surface tension, $\sigma$,
\begin{equation}
    \sigma = \kappa \int_{\rho_v}^{\rho_l} \frac{d\rho}{dx} d\rho,
\end{equation}
are readily evaluated to be:
\begin{eqnarray}
    W &=& \frac{4}{\rho_l - \rho_v}\sqrt{\frac{\kappa}{2A}},\label{eq:interface_thickness_control_near_critical}\\
    \sigma &=& \frac{{(\rho_l - \rho_v)}^3}{6}\sqrt{2A\kappa}.
\end{eqnarray}
This means that the interface thickness and surface tension can be modified via different choices of the capillary coefficient and isothermal compressibility, through the coefficient $A$. Combined with the principle of corresponding states, this allows one to modify parameters $\kappa$, $a$ and $b$ at will to reach the desired interface thickness and surface tension at a fixed reduced temperature and maintain the density ratio unchanged. In addition a number of other considerations must be taken into account when using this strategy to thicken the interface:
\begin{itemize}
    \item Interface thickening via modification of $A$ is akin to a rescaling of the isothermal speed of sound -tied to the isothermal compressibility. To remain within the low Mach approximation, assuming a characteristic convective speed $\mathcal{U}$, one must ensure that $\frac{\mathcal{U}}{c_s}<0.3$.
    \item Artificial thickening of the interface will correctly capture the dynamics of the considered interface in the limit of a \emph{sharp} interface. By \emph{sharp} interface we mean $\mathcal{W}\ll\mathcal{L}$, where $\mathcal{L}$ is the characteristic hydrodynamic size and $\mathcal{W}$ the characteristic interface length. The characteristic interface length is $\mathcal{W}=\max(W,\delta_T)$ where $\delta_T$ is the Tolman length characterizing curvature-dependence of the surface tension detailed in section~\ref{subsubsec:Tolman}.
\end{itemize}
It was shown here that the use of the principle of corresponding states in combination with restrictions on the Mach number and the ratio of characteristic interface to hydrodynamic sizes allows for a simple way to thicken the interface and remove the fundamental restriction on grid-size in second-gradient fluid simulations. While the analysis was restricted to the van der Waals equation near critical state it equally applies to other equations of state, the pseudo-potential model and temperatures much lower than the critical temperature. In the authors opinion this is one of the main components of modern lattice Boltzmann models for non-ideal fluids.
\subsection{Numerical artifacts and issues of non-ideal lattice Boltzmann models}

\subsubsection{Deviations in normal stress at interface: The issue of thermodynamic consistency}
\begin{figure}
\centering
	\includegraphics{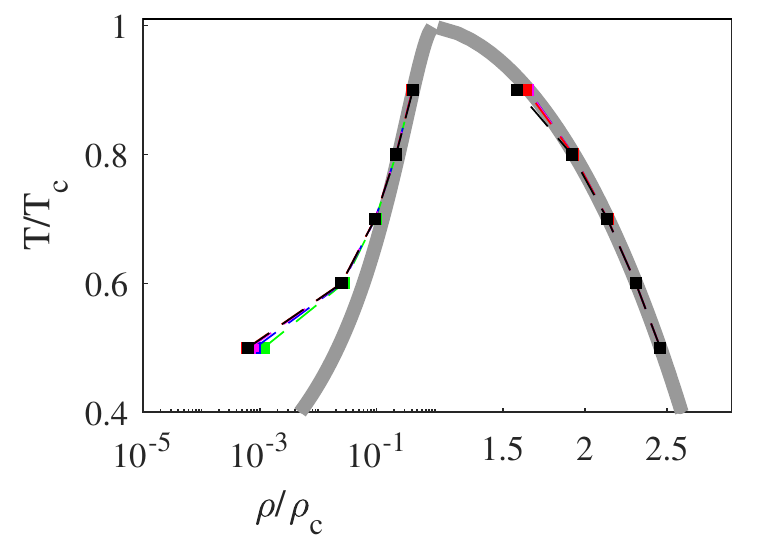}
\caption{ Coexistence densities with van der Waals equation of state as obtained from (grey lines) Maxwell's construction and (markers) simulations with different choices of $a$: (green x) $a=0.0102$, (magenta +) $a=0.0051$, (blue triangles) $a=0.0026$, (red squares) $a=0.0013$ and (black circles) $a=0.00064$. Simulations conducted using the first-neighbour pseudo-potential model with exact difference method to treat the force term. Plot is reproduced from \cite{hosseini_towards_2021}. }
\label{Fig:SC_convergence}
\end{figure}

The issue of \emph{thermodynamic consistency} as referred to in the lattice Boltzmann literature refers to the non-equivalence of the mechanical stability and Maxwell construction conditions when considering the discrete pressure tensor; To illustrate that purpose let us consider a force discretized with the first-neighbor D2Q9 stencil, i.e.
\begin{equation}\label{eq:pseudo_potential_force}
    \bm{F} = 2\sqrt{P-P_0}(\bm{r})\sum_{i=0}^{Q-1} w(\lvert\bm{c}_i\lvert) \sqrt{P-P_0}(\bm{r}+\bm{c}_i\delta t)\bm{c}_i.
\end{equation}
The discrete pressure tensor as introduced in the previous section is:
\begin{equation}
    P(\bm{r}) = P_0(\bm{r}) + \sqrt{P-P_0}(\bm{r})\sum_i w(\lvert\bm{c}_i\lvert) \sqrt{P-P_0}(\bm{r}+\bm{c}_i\delta t) \bm{c}_i\otimes\bm{c}_i,
\end{equation}
which after Taylor expansion results in:
\begin{equation}
    P = \left(P + \frac{\delta r^2}{6}\sqrt{P-P_0}\bm{\nabla}^2\sqrt{P-P_0}\right)\bm{I} + \frac{\delta r^2}{3}\sqrt{P-P_0}\bm{\nabla}\bm{\nabla}\sqrt{P-P_0}.
\end{equation}
Now considering a flat interface normal to the $x$-axis, the normal pressure can be written as:
\begin{equation}
    P_n = P + \frac{\delta r^2}{2}\sqrt{P-P_0}\frac{d^2}{dx^2} \sqrt{P-P_0},
\end{equation}
and further developed into:
\begin{equation}
    P_n = P + \frac{\delta r^2}{4}\sqrt{P-P_0} \frac{d}{d \sqrt{P-P_0}} {\left(\frac{d\sqrt{P-P_0}}{dx}\right)}^2.
\end{equation}
After some algebra one can recover the following equation:
\begin{equation}
    \left(P_n - P\right)\frac{4\frac{d \sqrt{P-P_0}}{d\rho}}{\delta r^2\sqrt{P-P_0}} = \frac{d}{d\rho}\left[ {\left(\frac{d \sqrt{P-P_0}}{d\rho}\right)}^2 {\left(\frac{d\rho}{dx}\right)}^2\right].
\end{equation}
Integrating from the vapor phase to the liquid phase and noting that $d\rho/dx=0$ in both phases one arrives at the following mechanical stability condition:
\begin{equation}\label{eq:discrete_mechanical_stability}
    \int_{\rho_v}^{\rho_l} \left(P_n - P\right) \frac{d \ln (P-P_0)} {d\rho} d\rho = 0.
\end{equation}
This form of the mechanical stability condition only matches the thermodynamic coexistence condition from Maxwell's reconstruction only if:
\begin{equation}
    \frac{ d\ln (P-P_0)}{d\rho} = \frac{1}{\rho^2},
\end{equation}
which is the basis of the proposal formulated in \cite{sbragaglia_consistent_2011} to define a pseudo-potential of the form:
\begin{equation}
    \psi = \sqrt{P-P_0} = \rho_0 \exp-\frac{\rho_0}{\rho}.
\end{equation}
Note that other forms of the non-local contribution will lead to similar mistmatches; For instance consider the free energy form:
\begin{equation}
    \bm{F} = \sum_{i=0}^{Q-1} w(\lvert\bm{c}_i\lvert)(P-P_0)(\bm{r}+\bm{c}_i\delta t)\bm{c}_i.
\end{equation}
The stress normal to the interface will be:
\begin{equation}
    P_n = P + \frac{\delta r^2}{4}(P-P_0) \frac{d}{d (P-P_0)} {\left(\frac{d(P-P_0)}{dx}\right)}^2,
\end{equation}
which leads to exactly the same mechanical equilibrium condition as before. The mismatch between the Maxwell construction condition and discrete mechanical stability condition results in deviations in coexistence density in simulations from those predicted by thermodynamics. This point is illustrated in Fig.~\ref{Fig:SC_convergence}.
Fig.~\ref{Fig:SC_convergence} shows that the lattice Boltzmann model correctly recovers the liquid/vapor coexistence densities closer to the critical temperature; As temperature goes further down, and density ratio increases, coexistence densities in simulations deviate significantly from the Maxwell construction predictions, and that regardless of the choice of the excluded volume and long range interaction coefficients.
\subsubsection{Fixed surface tension}
One point that was noted in Eq.~\eqref{eq:Taylor_pseudopotential} was the emergence of a surface tension-like leading \emph{error} term in the single-neighbour pseudo-potential model. The leading-order error is the reason the original pseudo-potential method of ~\cite{shan_lattice_1993} is able to recover a Korteweg-like surface tension. At the difference of the Korteweg stress tensor, which is a minimizer of the second-gradient free energy under global mass conservation constraint, and where surface tension explicitly appears in the macroscopic equations as an additional term, in the former it is strictly enslaved to the chosen stencil and form of the pseudo-potential and therefore not tunable, see Eq.~\eqref{eq:Taylor_pseudopotential}:
\begin{equation}\label{eq:PP_surface_tension_error}
    \kappa = \frac{G\delta r^2}{3}.
\end{equation}
Introduction of variable capillary coefficient and therefore surface tension has been the topic of a wide number of publications. We will discuss solutions provided in the literature in the next sections.
\subsubsection{The issue of spurious currents at interfaces}
\begin{figure}[h!]
	\centering
	\includegraphics[width=8cm,keepaspectratio]{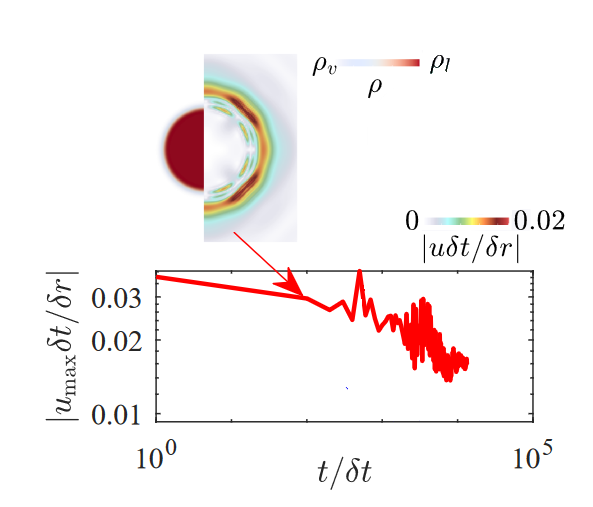}
	\caption{Illustration of spurious currents near liquid/vapor interfaces: Time-evolution of maximum spurious currents along with the density and velocity fields at the converged state as obtained from simulations with SRT collision model and exact difference method, at $\nu\delta t/\delta r^2=0.03$ and $T_r=0.59$. Image reproduced from \cite{hosseini_entropic_2022}. }
	\label{Fig:spurious_currents}
\end{figure}

Common to all multi-phase models, at curved interfaces between the liquid and vapor phases, spurious (often vortical) currents are observed. The magnitude of these spurious currents is often directly tied to the density ratio and interface thickness. The spurious currents are often associated to inbalance of forces at curved interfaces due to discretization errors, most notably the limited degree of isotropy. To illustrate the importance of isotropy consider a 2-D drop surrounded by a vapor phase. The density distribution, and therefore pressure, is only function of $r$ in polar coordinate, i.e. $\partial^n \rho/\partial\theta^n = \partial^n P/\partial \theta^n = 0$ for any n with $n>0$ which results in driving forces of the form, if one considers the Korteweg stress tensor:
\begin{equation}
    \bm{\nabla}P - \kappa\rho\bm{\nabla} \Delta \rho = \frac{\partial P}{\partial r} \bm{e}_r - \kappa\rho \left(\frac{\partial^3 \rho}{\partial r^3} = \frac{2}{r}\frac{\partial^2 \rho}{\partial r^2} - \frac{2}{r^2}\frac{\partial \rho}{\partial r}\right) \bm{e}_r,
\end{equation}
which clearly satisfy rotational invarience. In the absence of rotational forces one expects the velocity field to be rotational-invarient too. Furthermore, at steady state and assuming vanishing velocity far away from the drop the continuity equation
\begin{equation}
    \frac{1}{r}\frac{\partial \rho u_r}{\partial r} = 0,
\end{equation}
leads to:
\begin{equation}
    u_r = 0.
\end{equation}
It can therefore be concluded that isotropy of the driving forces should guarantee the absence of any form of velocity in the case of the static 2-D droplet.\\
Let us now consider the discretized form of the driving force on the D2Q9 stencil, i.e. Eq.~\eqref{eq:pseudo_potential_force}. Operating a Taylor expansion on that stencil one finds:
\begin{multline}\label{eq:Taylor_discrete_force_spurious_current}
    2\sqrt{P-P_0}(\bm{r})\sum_i w(\lvert\bm{c}_i\delta t\lvert) \bm{c}_i \delta t\sqrt{P-P_0}(\bm{r}+\bm{c}_i) = 2\sqrt{P-P_0}\bm{\nabla}\left( 1 + \frac{1}{6}\bm{\nabla}^2 + \frac{1}{72}\bm{\nabla}^2 \bm{\nabla}^2\right) \left(\sqrt{P-P_0}\right) \\ + \frac{\sqrt{P-P_0}}{90}\left( \frac{\partial^5 \sqrt{P-P_0}}{\partial x^5}\bm{e}_x + \frac{\partial^5 \sqrt{P-P_0}}{\partial y^5}\bm{e}_y\right).
\end{multline}
It is clear that the first term is isotropic. The last two terms however are not isotropic meaning this discrete approximation loses isotropy at order five in turn leading to force imbalance which is then countered by non-zero velocity components manifesting around the interface.
A typical result from lattice Boltzmann simulations of a static 2-D drop is shown in Fig.~\ref{Fig:spurious_currents}.
Note that while we have considered the specific form of the body force used in \cite{yuan_equations_2006} to get Eq.~\eqref{eq:Taylor_discrete_force_spurious_current}, the classical form $\bm{\nabla}(P-P_0)$ would also lead to similar behavior, i.e.
\begin{equation}\label{eq:Taylor_discrete_force_free_energy}
    \sum_i w(\lvert\bm{c}_i \delta t\lvert) \bm{c}_i \delta t(P-P_0)(\bm{r}+\bm{c}_i) = \bm{\nabla}\left( 1 + \frac{1}{6}\bm{\nabla}^2 + \frac{1}{72}\bm{\nabla}^2 \bm{\nabla}^2\right) \left(P-P_0\right) + \frac{1}{180}\left( \frac{\partial^5 (P-P_0)}{\partial x^5}\bm{e}_x + \frac{\partial^5 (P-P_0)}{\partial y^5}\bm{e}_y\right).
\end{equation}
also generating non-isotropic terms forcing spurious currents at curved interfaces. The only difference is in the form of non-isotropic errors that appear at order 5. A more detailed discussion of the errors of different forms of the forcing term will be provided in later sections.
\subsubsection{Galilean invariance of viscous dissipation rates and stability}
\begin{figure}
	    \centering
		    \includegraphics{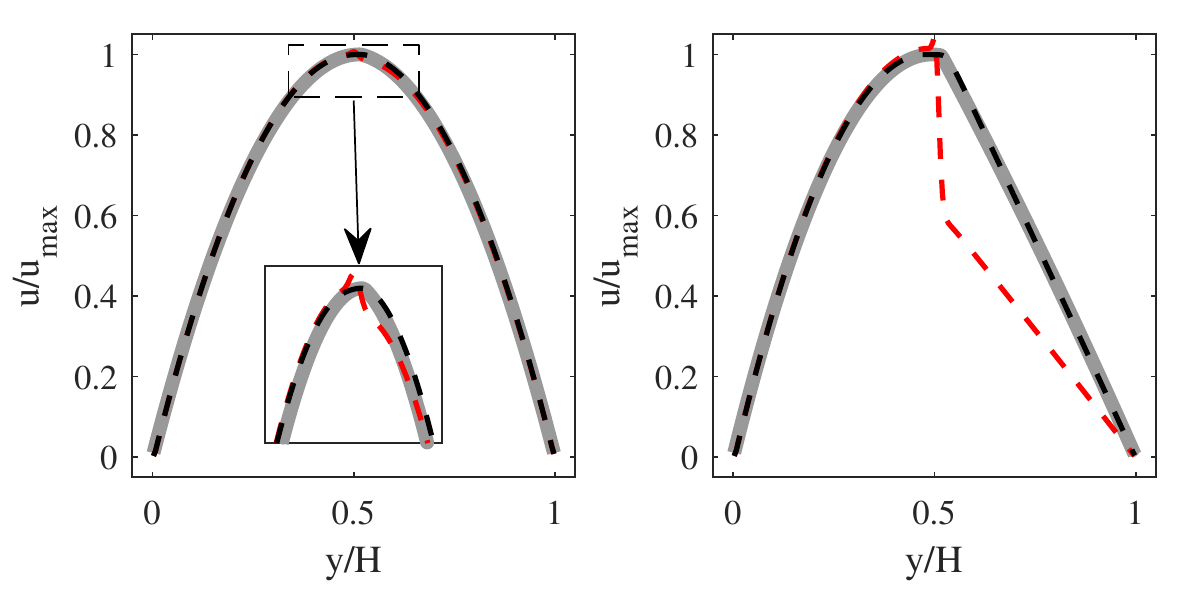}
	    \caption{Steady-state velocity profiles for the layered Poiseuille flow. Left: Configuration (a), $T_r=0.77$, $\rho_l/\rho_v=10.1$, $\mu_l/\mu_v=10.1$ ; Right: Configuration (b), and (right) $T_r=0.36$,  $\rho_l/\rho_v=1030$ and $\mu_l/\mu_v=11.3$. Grey plain line: analytical solution; Black dashed line: Free energy LBM with product-form equilibrium; Red dashed line: Free energy LBM with conventional second-order equilibrium. Results are taken from \cite{hosseini_towards_2021}.}
	    \label{Fig:Poiseuille_R10_R1000}
\end{figure}

For iso-thermal multi-phase flow simulations, even assuming the temperature in the discrete equilibrium is set to the lattice temperature, and that the Mach number is kept very low Galilean-variant errors in the viscous stress tensor, both deviatoric and diagonal components can be quite pronounced. For instance, consider the case of errors in deviatoric components of the third-order equilibrium moments tensor for a second-order equilibrium:
\begin{equation}
    \delta \Pi_{\alpha\beta\gamma}^{\rm eq} = \rho u_\alpha u_\beta u_\gamma + (\rho c_s^2 - P_0) [u_\alpha \delta_{\beta\gamma}]_{\rm cyc},
\end{equation}
Assuming $P_0=\rho c_s^2$ is guaranteed everywhere the second term disappears leaving a term third-order in Mach number. In the case of ideal fluid simulations given that density variation are small the third-order scaling of this term with Mach number allows one to neglect it for small Mach number simulations. However for non-ideal fluids, at liquid/vapor interface the density gradient can get quite large making this error term non-negligible even for small Mach numbers. In practice, the erroneous viscous stress tensor can lead to pronounced velocity jumps at the interface as it does not guarantee continuity of viscous momentum flux. For errors in the deviatoric components, i.e. viscous shear stress, this effect can be readily observed with a case as simple as the two-layer Poiseuille flow. This case consists of a rectangular domain filled with the liquid phase at the bottom and the vapor phase on top. The flow is driven by a body force. Top and bottom are subject to no-slip boundary conditions while the inlet and outlet are fixed by periodicity. Running this case with models relying on second-order polynomial equilibria one recovers results such as those shown in Fig.~\ref{Fig:Poiseuille_R10_R1000}. Note that here the maximum non-dimensional velocities are quite small.
As for ideal gas solvers, it is observed that including third-order terms in the discrete equilibrium eliminates this numerical artifact. This effect is common to \emph{all} multi-phase LBM, regardless of the formulation and the collision model~\cite{leclaire_enhanced_2013,fakhari_mass-conserving_2016}. For instance, it has been documented and studied for the color-gradient~\cite{leclaire_enhanced_2013} and phase-field formulations~\cite{fakhari_mass-conserving_2016,zu_phase-field-based_2013,fakhari_improved_2017,mitchell_development_2018}.\\
While the effect of third-order moments error on Galilean-invariance of the viscous stress tensor is quite well-documented its effect on the continuity equation have never been discussed. The LBM is known to guarantee \emph{global} conservation of mass to be exact. However locally one does not exactly recover the continuity equation. A third-order perturbation analysis shows that at order $\varepsilon^3$ the zeroth-order moment satisfies~\cite{hosseini_theoretical_2019}:
\begin{equation}
    \partial_t^{(3)}\rho - \frac{1}{12}\bm{\nabla}\bm{\nabla}:\left[\partial_t^{(1)}\Pi^{\rm eq}_2 + \bm{\nabla}\cdot\Pi^{\rm eq}_3\right] = 0.
\end{equation}
The second term in that equation shows that there is a third-order deviation in the continuity equation recovered by the LBM. However looking at the form of that deviation term one observes that it is nothing but the $\varepsilon$-level balance equation of the second-order moments restored to its correct form by the correction discussed in this section. We can therefore conclude that apart from restoring Galilean-invariance to the viscous stress the correction discussed here also removes third-order errors in the mass balance equation.
\par All lattice Boltzmann models targeting non-ideal fluids, as all other classical discrete solvers, are subject to discretization errors coming from both the lattice Boltzmann solver and the body force term. The specific form and properties of these higher-order error terms manifests in different forms, i.e.
\begin{itemize}
    \item Magnitude of discretization error normal to interface: As shown in Eq.~\eqref{eq:discrete_mechanical_stability} this affects the mechanical stability condition at the discrete level. The mismatch between the discrete-level mechanical stability condition with the Maxwell construction leads to errors in the coexistence densities even for flat interfaces as illustrated in Fig.~\ref{Fig:SC_convergence}.
    \item Spurious surface tension-like terms: As shown in Eqs.~\eqref{eq:Taylor_discrete_force_spurious_current} and \eqref{eq:Taylor_discrete_force_free_energy} a classical first-neighbour discretization is equivalent to a second-order central finite differences an leads to third-order errors with a surface tension-like structure. In the case of the pseudo-potential method this \emph{error} is the reason one observes surface tension. However on the downside the capillary coefficient is fixed as shown in Eq.~\eqref{eq:PP_surface_tension_error}.
    \item Non-isotropic terms: In 2- and 3-D a look at the form of the error terms recovered at orders three and above shows that non-isotropic terms appear; For instance for the first-neighbour discretization non-isotropic terms appear first at order five. This leads to the formation of spurious currents in the vicinity of curved interfaces. The magnitude of these spurious currents is directly proportional to density ratio and inversely proportional to interface thickness.
    \item Errors affecting viscous stress tensor: The use of a second-order polynomial equilibrium leads to errors in the deviatoric components of the equilibrium third-order moments tensor in turn leading to errors in the effective shear viscosity. Even at very low velocities this error is quite pronounced around liquid/vapor interfaces.
\end{itemize}
All these issues lead to practical limitation such as a maximum reachable density and/or viscosity ratio in simulations. A number of of improvements have been proposed over the year to relax constraints on density and viscosity ratio. We will review major improvements in the next section.
\subsection{Improvements and enhanced models for non-ideal fluids}

\subsubsection{Equations of state}
\begin{figure}[h!]
	\centering
	\includegraphics[width=8cm,keepaspectratio]{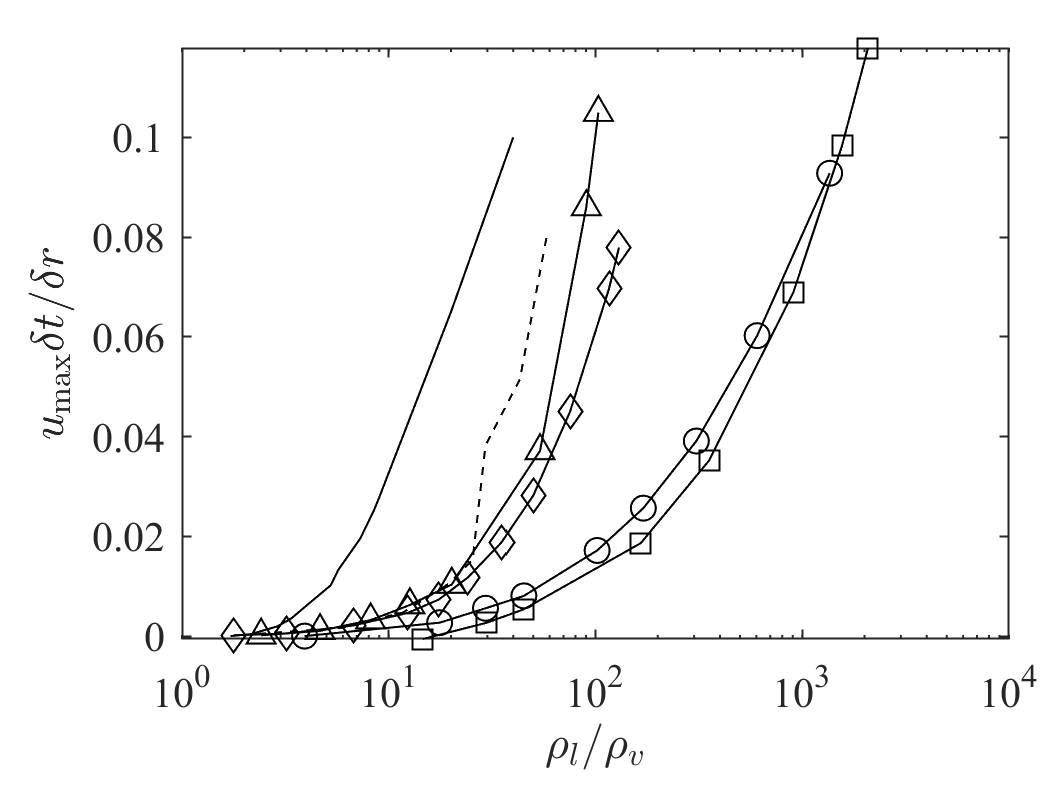}
	\caption{Maximum spurious currents as a function of density ratio for different equations of state: (plain line) Shan-Chen, (dashed line) van der Waals, (plain line with circular markers) Carnahan-Sterling, (plain line with square markers) Peng-Robinson, (plain line with diamond markers) Riedlich-Kwong and (plain line with triangle markers) Riedlich-Kwong-Soave. Simulation have been conducted with the realization of \cite{yuan_equations_2006}. Simulations have been conducted with the same choices of $a$ and $b$ as those reported in \cite{yuan_equations_2006}. }
	\label{Fig:spurious_currents_EoS}
\end{figure}

\begin{figure}
	\centering
	\includegraphics{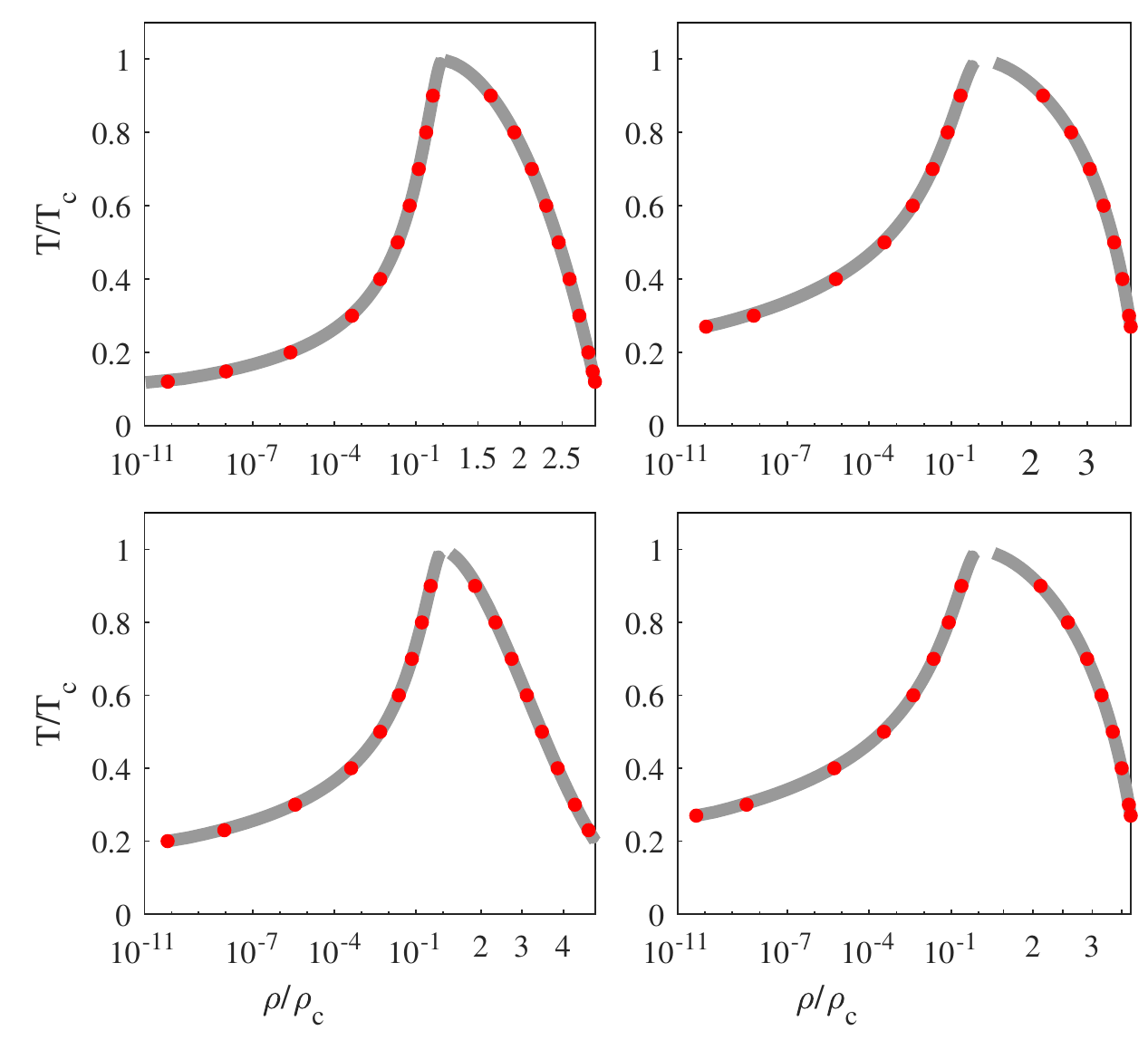}
	\caption{Liquid-vapor coexistence for various equations of state. Gray lines: Maxwell's equal-area construction; Red symbol:  Simulation.  Top left: van der Waals \eqref{eq:vdWEoS} ($a=0.000159$, $b=0.0952$);  Top right: Peng--Robinson \eqref{eq:PREoS} ($a=0.000159$, $b=0.0952$); Bottom left: Carnahan--Starling \eqref{eq:CSEos} ($a=0.000868$, $b=4$); Bottom right: Riedlich--Kwong--Soave \eqref{eq:RKSEoS} ($a=0.000159$, $b=0.0952$). For all simulation $\tilde{\kappa}=0.02$. Results reproduced from \cite{hosseini_towards_2021}. }
	\label{Fig:Coexistence}
\end{figure}

\begin{figure}[h!]
	\centering
	\includegraphics[width=7cm,keepaspectratio]{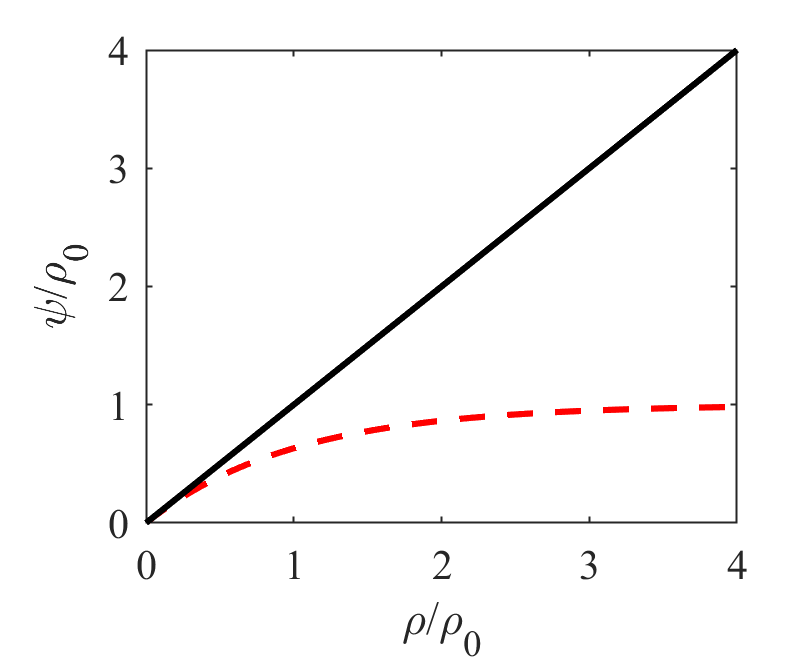}
	\caption{Behavior of potential of Eq.~\ref{eq:SC1} as a function of local density; Black line: $\psi/\rho_0=\rho/\rho_0$, red dashed line: Eq.~\ref{eq:SC1}.}
	\label{Fig:SC_potential_vs_density}
\end{figure}

\begin{figure}[h!]
	\centering
	\includegraphics[width=7cm,keepaspectratio]{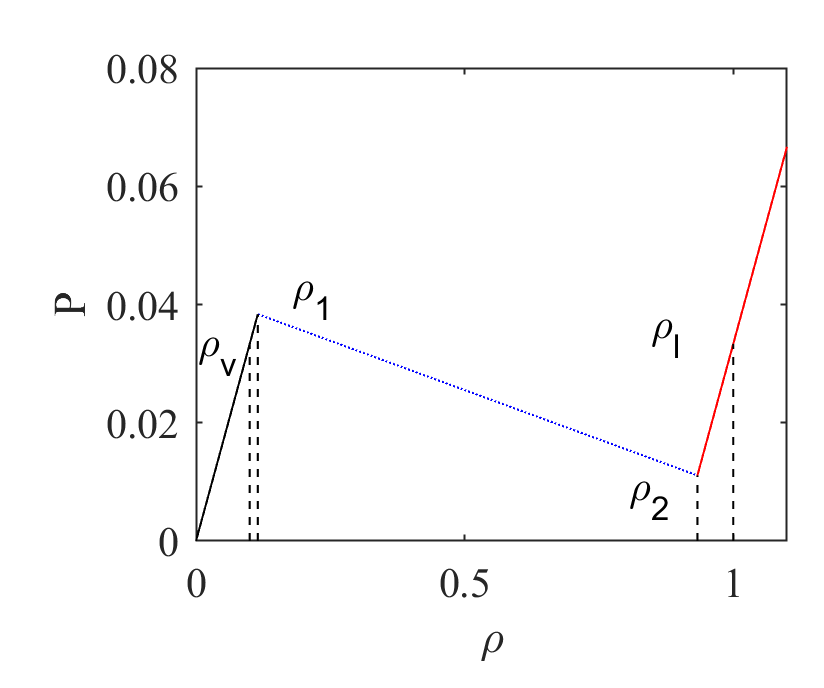}
	\caption{$P-\rho$ diagram for piece-wise linear equation of state, Eq.~\ref{eq:piece_wise_linear}. The density ratio here is set to 10.}
	\label{Fig:piece_wise_linear_eos}
\end{figure}

One approach to reduce spurious currents and access larger coexistence densities that was proposed early on was the use of different equations of state starting with the work of~\cite{yuan_equations_2006}. The authors compared the maximum spurious currents as a function of density ratio for different equations of states. The effect of the choice of equation of state on maximum spurious currents as a function of density ratio is shown in Fig.~\ref{Fig:spurious_currents_EoS}
While initially such proposals were limited to classical cubic equations of states, later many authors proposed \emph{tailored} equations of state designed solely to allow phase separation at the target density ratio and minimize numerical artifacts. Here we provide a brief overview of these equations of state.
\paragraph{Cubic equations of state}
This family of equations of state are widely used in both thermodynamics and non-ideal fluid simulations. The most well-known equation of state in that family was proposed by van der Waals~\cite{van_der_waals_over_1873},
\begin{equation}\label{eq:vdWEoS}
       P = \frac{\rho R T}{1-b\rho} - a \rho^2,
\end{equation}
where parameters $a$ and $b$ are related to critical temperature $T_c$ and pressure $P_c$ as,
\begin{equation}
        a=\frac{27}{64}\frac{R^2T_c^2}{P_c},\ b=\frac{1}{8}\frac{RT_c}{P_c}.
\end{equation}
The Peng--Robinson EoS \cite{peng_new_1976},
\begin{equation}\label{eq:PREoS}
        P = \frac{\rho R T}{1-b\rho} - \frac{a \alpha(T) \rho^2}{1+2\rho b - b^2 \rho^2},
\end{equation}
with
\begin{equation}
        \alpha(T) = \left[1 + (0.37464 + 1.54226\omega' - 0.26992 \omega'^2) \left(1-\sqrt{T/T_c}\right)\right]^2,
\end{equation}
where  $\omega'$ the acentric factor ($\omega'=0.344$ for water), and
\begin{equation}
        a=0.45724\frac{R^2T_c^2}{P_c},\ b=0.0778\frac{R T_c}{P_c}.
\end{equation}
The Riedlich--Kwong--Soave EoS~\citep{redlich_thermodynamics_1949,soave_equilibrium_1972},
\begin{eqnarray}\label{eq:RKSEoS}
        P = \frac{\rho R T}{1-b\rho} - \frac{a \alpha(T) \rho^2}{1+\rho b},
\end{eqnarray}
with
\begin{equation}
        \alpha(T) = \left[1 + (0.480 + 1.574\omega' - 0.176 \omega'^2) \left(1-\sqrt{T/T_c}\right)\right]^2,
\end{equation}
and
\begin{equation}
        a=0.42748\frac{R^2T_c^2}{P_c},\ b=0.08664\frac{R T_c}{P_c},
\end{equation}
and the Carnahan--Starling EoS~\citep{carnahan_equation_1969},
\begin{equation}\label{eq:CSEos}
        P = \rho R T \frac{1+b\rho/4 + {(b\rho/4)}^2 - {(b\rho/4)}^3}{{(1-b\rho/4)}^3} - a\rho^2,
\end{equation}
with
\begin{equation}
        a=0.4963\frac{R^2T_c^2}{P_c},\ b=0.18727\frac{R T_c}{P_c}.
\end{equation}
The co-existence densities of these four EoS as obtained using the Maxwell reconstruction method and simulations reportes in \cite{hosseini_towards_2021} are shown in Fig.~\ref{Fig:Coexistence}.
\paragraph{Shan-Chen equations of state}
In Shan-Chen-type equations of state the thermodynamic pressure is defined as:
\begin{equation}
    P_{\rm SC} = \rho c_s^2 + \frac{G}{2}\psi^2,
\end{equation}
where $\psi$ is the non-local interaction potential. To derive an optimal form for the potential function a number of constraints have to be taken into account; First, using the mechanical equilibrium condition one should arrives at the following so-called \emph{thermodynamic consistency} condition on the potential~\cite{benzi_mesoscopic_2006}:
\begin{equation}\label{eq:SC_thermo_consistency}
    \int_{\rho_l}^{\rho_v}\left[P - \rho c_s^2 - \frac{G}{2}\psi^2\right]\frac{1}{\rho^2}d\rho = 0,
\end{equation}
which is only \emph{strictly} satisfied with $\psi\propto \rho$~\cite{he_thermodynamic_2002}. However, a potential interaction of that form would not allow for co-existence of two phases of different densities. Second, given that the original pseudo-potential has only an attractive non-local contribution term, as noted in \cite{benzi_mesoscopic_2006,sbragaglia_generalized_2007}, to mimic the hard-core repulsive interaction of real molecules and prevent collapse of the liquid phase the potential must saturate at large densities, $\psi\rightarrow{\rm cst}$. Under these constraints a family of potentials were proposed, to satisfy the previously-listed conditions as closely as possible. For instance in~\cite{shan_lattice_1993} the authors used a potential defined as:
\begin{equation}
    \psi = \rho_0 \left(1 - \exp\left(-\frac{\rho}{\rho_0}\right)\right),
\end{equation}\label{eq:SC1}
where $\rho_0$ is a tunable constant. Another form was proposed in~\cite{shan_simulation_1994}:
\begin{equation}\label{eq:SC2}
    \psi = \rho_0 \exp \left(-\frac{\rho_0}{\rho}\right).
\end{equation}
The behavior of the pseudo-potential of Eq.~\ref{eq:SC1} is illustrated in Fig.~\ref{Fig:SC_potential_vs_density}.
A expected in the limit of $\rho/\rho_0\rightarrow 0$ one recovers $\psi\propto\rho$ while on the other end of the spectrum, $\rho/\rho_0\rightarrow\infty$ the potential saturates. This clearly shows that while the saturation allows for stable phases at large density ratios it deviates considerably from $\propto\rho$ making \emph{thermodynamic inconsistencies} in the sense of Eq.~\eqref{eq:SC_thermo_consistency} pronounced.\\
\paragraph{Taylored equations of state for improved numerical properties}
In~\cite{colosqui_mesoscopic_2012} the authors proposed a new family of equations of state to improve numerical properties of the pseudo-potential method, namely better compliance with thermo-mechanical consistency. The piece-wise linear equation of state proposed in~\cite{colosqui_mesoscopic_2012} consists of defining the pressure as:
\begin{equation}\label{eq:piece_wise_linear}
	    P =
    \begin{cases}
    \rho c_{s,v}^2 &\text{if} \rho\leq\rho_1\\
    \rho_1 c_{s,v}^2 + \left( \rho - \rho_1\right) c_{s,m}^2 &\text{if} \rho_1\leq\rho\leq\rho_2\\
    \rho_1 c_{s,v}^2 + \left( \rho_2 - \rho_1\right) c_{s,m}^2 + \left( \rho - \rho_2\right) c_{s,l}^2 &\text{if} \rho_2\leq\rho
    \end{cases}
\end{equation}
where in practice $c_{s,v}$, $c_{s,l}$ and $c_{s,m}$ are chosen \emph{a priori}. The choice of these three variables allows one to fix speed of sound in both the liquid and vapor phases and the interface thickness independently. Once they are fixed, the remaining free parameters, i.e. $\rho_1$ and $\rho_2$ are closed via two conditions, namely mechanical:
\begin{equation}
	    \int_{\rho_v}^{\rho_l}\frac{dP}{d\rho} d\rho = 0,
\end{equation}
   and chemical potential balance:
   \begin{equation}
        \int_{\rho_v}^{\rho_l}  \frac{1}{\rho}\frac{dP}{d\rho} d\rho = 0.
\end{equation}
This equation of state is illustrated in Fig.~\ref{Fig:piece_wise_linear_eos}.
With this approach, the speed of sound in the liquid and vapor phase are tunable parameters and as such can be set to be as close to each-other as possible to minimize compressibility effects in the vapor phase and allow for larger CFL numbers. The piece-wise linear equation of state is now routinely used in combination with the pseudo-potential formulation to model multi-phase flows with density ratios of the order of $10^{3}$, see for instance~\cite{fei_modeling_2019,li_lattice_2013}.\\
Another alternative proposed by~\cite{peng_attainment_2020}, argues that the main issue of cubic equations of state is the \emph{van der Waal loop} section of the $P-V$ diagram. In this region cubic equations of state predict negative compressibility. To better deal with that region the authors proposed an apprioach that consists of using the classical cubic equations of state, e.g. van der Waal Peng-Robinson etc, in the liquid and vapor branch and replacing it with a tailored third-order polynomial in the van der Waals loop region $\rho_v<\rho<\rho_l$, i.e. both the binodal and spinodal regions:
\begin{equation}
	    P =
    \begin{cases}
    P_{\rm EoS}(\rho) &\text{if} \rho\leq\rho_v\\
    P_0(\rho_v) + \theta_m(\rho-\rho_v)(\rho-\rho_l)(\rho-\rho_m) &\text{if} \rho_v\leq\rho\leq\rho_l\\
    P_{\rm EoS}(\rho) &\text{if} \rho_l\leq\rho
    \end{cases}
\end{equation}
The use of the cubic polynomial in the van der Waals loop comes with four free coefficients. Conditions on continuity of pressure and isothermal sound speed (or compressibility) at the vapor and liquid densities would in principle close all free parameters and lead to the original cubic equation of state selected in the vapor and liquid branches. As such, and as detailed in the article, this approach can only have potential to improve interface properties if one of the previously listed physical conditions are neglected. The removal of one of the conditions, sound speed in one of the bulk phases would in principle allow control over the shape in the van der Waals loop region.\\
While the choice of the equation of state can be an efficient tool in reducing spurious currents and errors in coexistence densities, the way these non-ideal contributions are introduced into the kinetic model is not unique and can han dramatic effects on errors and stability.
\subsubsection{Partitioning of pressure contributions}
\begin{figure}[h!]
	\centering
	\includegraphics[width=7cm,keepaspectratio]{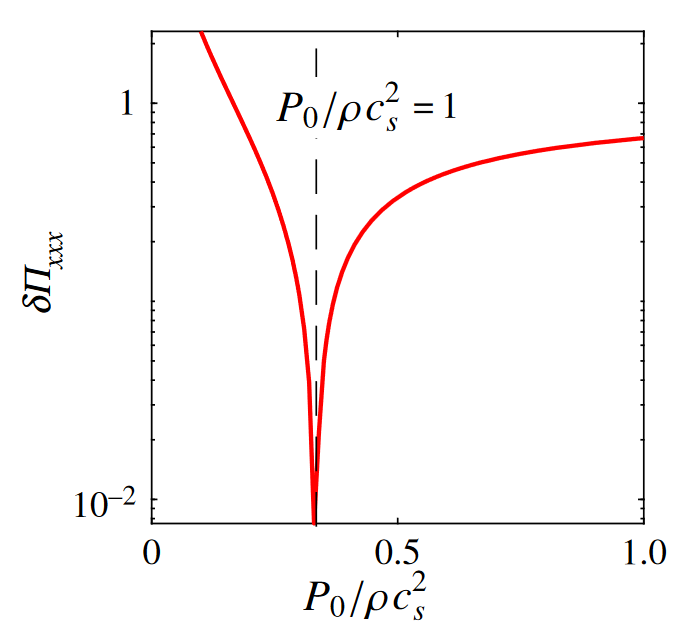}
	\caption{Error in the diagonal components of the third-order moments tensor as a function of the reference pressure.}
	\label{Fig:third_order_error_temp}
\end{figure}

\begin{figure}[h!]
	\centering
	\includegraphics[width=8cm,keepaspectratio]{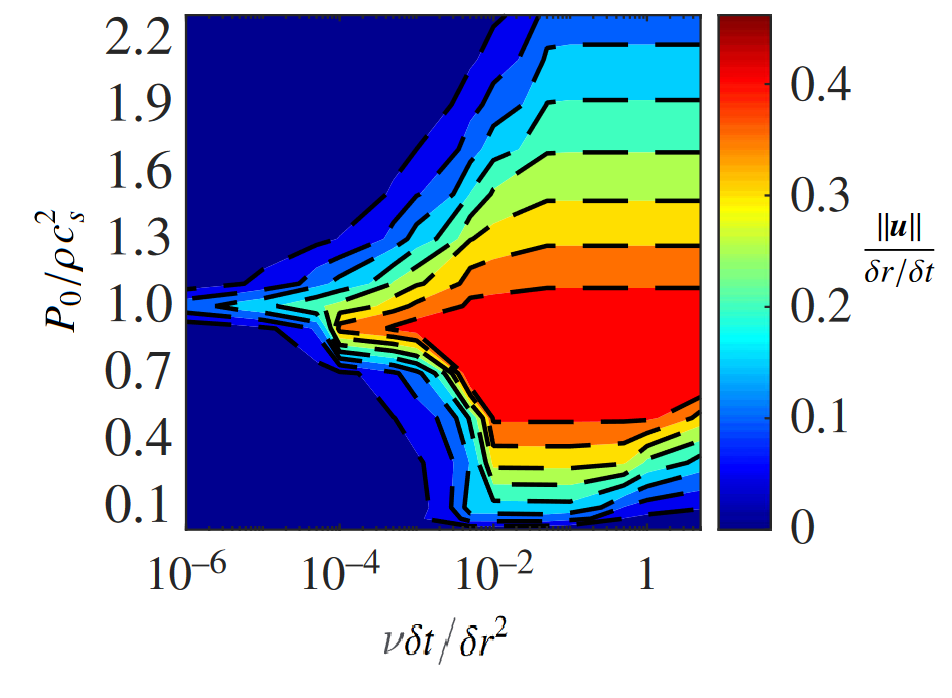}
	\caption{Linear stability domain of the SRT collision operator with a product-form equilibrium as a function of the reference pressure $P_0$.}
	\label{Fig:linear_stability_SRT_temp}
\end{figure}

In previous sections we introduced a general kinetic framework for non-ideal fluids and discussed the invarience of the recovered macroscopic equations with respect to the choice of reference pressure $P_0$ enforce by the BGK collision operator attractor. Here we discuss different possible choices of $P_0$ and its repercussions on numerics. 
\paragraph{The classical route: equilibrium at stencil reference temperature}
The most widely used partition approach in the litterateur consists of fixing $P_0$ appearing in the equilibrium at a value optimizing numerical properties of the lattice Boltzmann solver. For instance for a model based on a third-order Gauss-Hermite quadrature the optimal value of $P_0$ is:
\begin{equation}
    P_0 = \rho \frac{\delta r^2}{3\delta t^2}.
\end{equation}
Setting the reference pressure in that way has two main advantages; As discussed earlier the classical third-order quadrature admits a deviation in the diagonal components of the third-order equilibrium moment that has a third-order dependence on velocity and first-order dependence on $P_0/\rho$:
\begin{equation}\label{eq:third_order_error_temp}
	\delta \Pi_{\alpha\alpha\alpha}^{\rm eq} = \rho u_\alpha \left( u_\alpha^2 + 3(P_0/\rho - c_s^2)\right).
\end{equation}
Therefore small deviations of $P_0/\rho$ from the optimal temperature $c_s^2$ can considerably increase the Galilean-variant errors in the dissipation rate of normal modes, as illustrated in Fig.~\ref{Fig:third_order_error_temp}.
The second advantage of this choice of partition is stability domain. As illustrated in Fig.~\ref{Fig:linear_stability_SRT_temp} via the linear stability domain best results in terms of maximum non-dimensional speed and minimum non-dimensional viscosity are achieved at $P_0/\rho c_s^2=1$.
Both above and below $P_0/\rho c_s^2=1$ the stability domain is considerably reduced making realistic simulations with acceptable CFL numbers practically impossible.\\
The vast majority of publications treating of isothermal non-ideal fluids rely on this type of partition. For instance in the pseudo-potential model with the Shan-Chen equation of state:
\begin{equation}
    P_0 = \rho c_s^2,
\end{equation}
and
\begin{equation}
    P-P_0 = \frac{G}{2}\psi^2.
\end{equation}
In the free energy formulation, this is equivalent to the force-based approach with~\cite{mazloomi_entropic_2015}:
\begin{equation}
    \bm{T}_K- P_0\bm{I} = \left(P - \rho c_s^2 - \kappa\rho\bm{\nabla}^2\rho - \frac{\kappa}{2}{\lvert \bm{\nabla}\rho\lvert}^2\right)\bm{I} + \kappa \bm{\nabla}\rho\otimes\bm{\nabla}\rho.
\end{equation}
This specific partition of pressure is the most widely used approach in the literature, mainly because of the numerical properties enumerated above.
\paragraph{Full pressure used in equilibrium attractor}
As mentioned earlier, the idea of introducing the full thermodynamic pressure into the equilibrium function of a lattice Boltzmann solver started with the first free energy LBM~\cite{swift_lattice_1995,swift_lattice_1996} where the non-ideal equation of state and surface tension were introduced into the discrete equilibrium via moments matching. A similar construction was also discussed in~\cite{martys_energy_1999,martys_aclassical_2001}. The authors argued that this approach has the advantage of guaranteeing mass conservation locally. Furthermore, this approach has the additional advantage of reducing derivatives in the stress tensor by one order and therefore making the overall scheme more local. The first attempt presented in~\cite{swift_lattice_1995} was based on a second-order polynomial discrete equilibrium and subject to Galilean-variant errors in both the diagonal and deviatoric components of the viscous stress tensor. Consider a two-phase fluid in the incompressible regime with a density ratio of only 10: $10 P_0/\rho_l= P_0/\rho_v$. This means that inevitably at least one of the phases will have $P_0/\rho$ considerably larger or smaller than $c_s^2$ resulting in diminished stability domain, see Fig.~\ref{Fig:linear_stability_SRT_temp}. Furthermore, without corrections the viscous stress tensor is subject to errors scaling with $\propto (P_0/\rho - c_s^2)$, see Eq.~\eqref{eq:third_order_error_temp}. In effect this means that the application of such a model would be limited to very low density ratios. As a way to overcome this issue, and minimize errors related to deviation from the reference frame a realization based on the Particles-on-Demand was proposed and used in~\cite{reyhanian_thermokinetic_2020}.\\

\subsubsection{Non-local thermodynamic pressure force contribution: Using mathematical identities to reduce discretization errors}
\begin{figure}[h!]
	\centering
	\includegraphics[width=13cm,keepaspectratio]{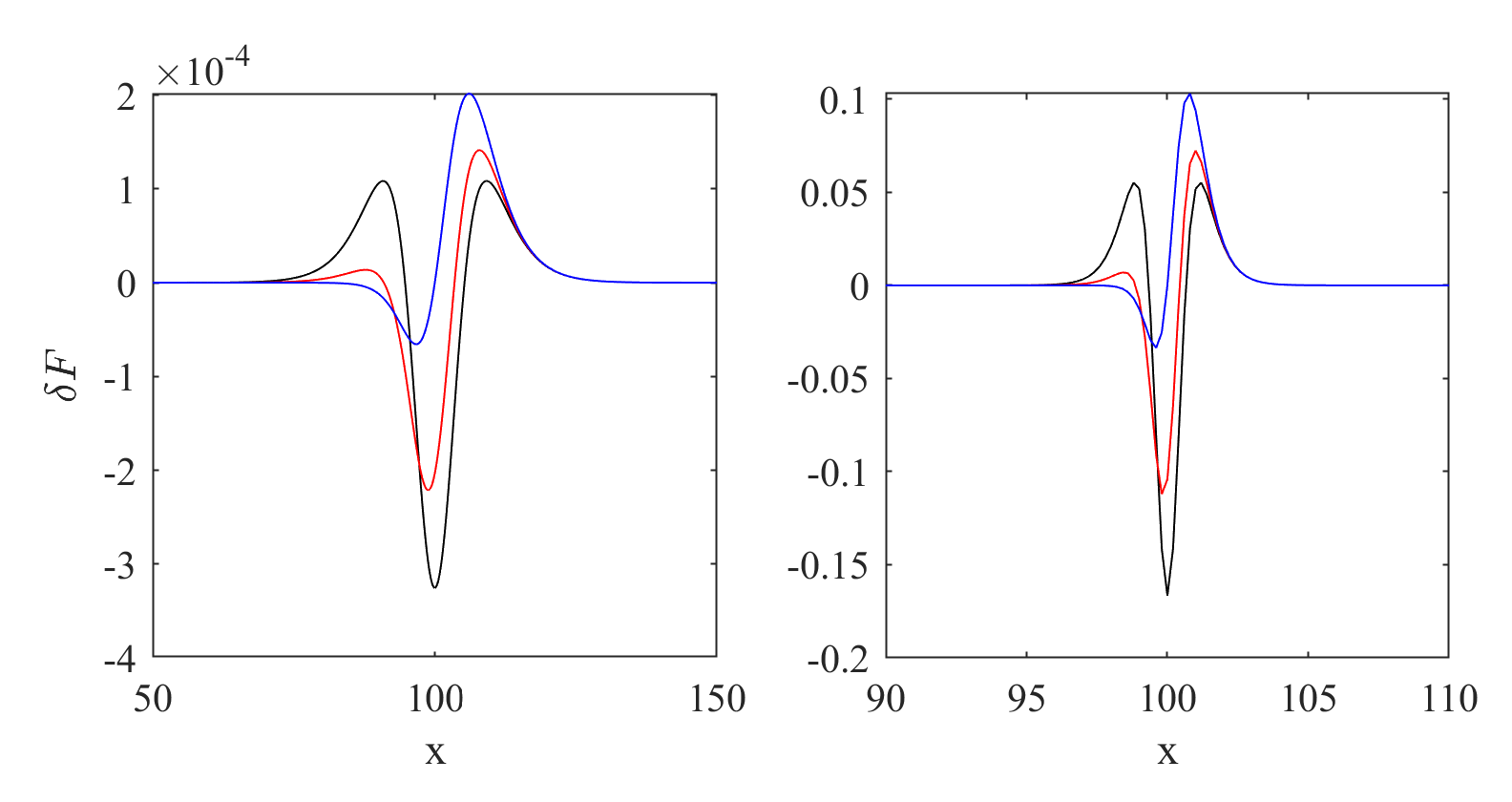}
	\caption{Profiles of leading-order error terms for different realizations of the non-local pressure contributions for (left) $\sigma=8$ and (right) $\sigma=1$. Black line: $\bm{F}_1$, red line: $\bm{F}_2$ and blue line: $\bm{F}_3$.}
	\label{Fig:force_realization_errors}
\end{figure}

The first step in the realization of the force is the way it is treated. It can be introduced as is ~\cite{zhang_lattice_2003}, i.e.
\begin{equation}\label{eq:general_non_ideal_force}
    \bm{F} = \bm{\nabla}(P - P_0),
\end{equation}
or using an identity:
\begin{equation}\label{eq:force_form_2}
    \bm{\nabla}(P-P_0) = 2\sqrt{P-P_0}\bm{\nabla}\sqrt{P-P_0}, { \rm for } P\geq P_0
\end{equation}
This approach, for general non-ideal equations of state was employed in~\cite{yuan_equations_2006} following the original form of the pseudo-potential model~\cite{shan_lattice_1993}. It should be noted that in the case of the original model this equality held unconditionally because the Shan-Chen equations of state were strictly positive. Positivity can become an issue especially near the liquid branch spinodal point where pressure is usually minimal.\\
As a way to extend the domain of accessible density ratios by decreasing the magnitude of leading-order error terms and reduce spurious currents a weighed combination of these two approaches was also proposed ~\cite{kupershtokh_equations_2009,gong_numerical_2012}:
\begin{equation}
    \bm{F} = \gamma \bm{\nabla}(P-P_0) + 2(1-\gamma) \sqrt{P-P_0}\bm{\nabla}\sqrt{P-P_0},
\end{equation}
where the weight $\gamma$ becomes a tuning parameter to better match co-existence densities or reduce spurious currents. Another possibility, not discussed in the literature, would be to rewrite the non-ideal pressure contribution as:
\begin{equation}\label{eq:force_form_3}
    \bm{\nabla}(P-P_0) = (P-P_0)\bm{\nabla}\ln(P-P_0), { \rm for } P\geq P_0.
\end{equation}
To illustrate the effect of the way this term is introduced and provide a simple comparison of their performances let us consider a simple 1-D interface:
\begin{equation}
    P-P_0 \propto \frac{1}{2}\left(1+\tanh\frac{x-x_0}{\sigma}\right),
\end{equation}
located at $x_0$ of thickness $\sigma$. Assuming a second-order accurate finite difference discretization of the contribution, errors are governed by the third- and higher order derivatives. Denoting approaches of Eqs.~\eqref{eq:general_non_ideal_force}, \eqref{eq:force_form_2} and \eqref{eq:force_form_3} respectively with $\bm{F}_1$, $\bm{F}_2$ and $\bm{F}_3$ and assuming zero surface tension, the leading order error of pressure normal to the interface would scale as:
\begin{eqnarray}
    \delta \bm{F}_1 &=& \frac{1}{3!}\partial_x^3 (P-P_0)+ \mathcal{O} (\partial_x^5),\\
    \delta \bm{F}_2 &=& \frac{2\sqrt{P-P_0}}{3!}\partial_x^3 \sqrt{P-P_0} + \mathcal{O} (\partial_x^5),\\
    \delta \bm{F}_3 &=& \frac{P-P_0}{3!}\partial_x^3 \ln(P-P_0) + \mathcal{O} (\partial_x^5).
\end{eqnarray}
The leading-order errors for the considered interface, for two different interface thicknesses illustrating the diffuse and near-sharp cases, are shown in Fig.~\ref{Fig:force_realization_errors}.
Comparing the different errors one can already observe that $\bm{F}_1$ leads to the largest interfacial errors while $\bm{F}_3$ introduces the lowest amount of deviations both in the diffuse and sharp interface configurations.\\
\subsubsection{Higher order discretization: Leading order error and isotropy}
\begin{figure}[h!]
	\centering
	\includegraphics[width=8cm,keepaspectratio]{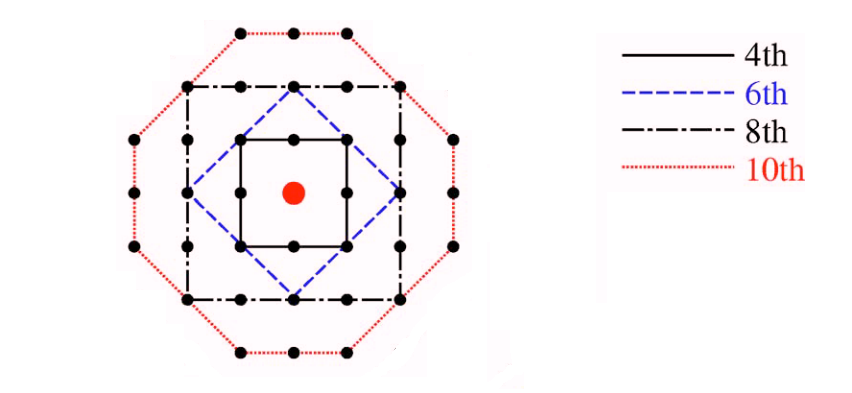}
	\caption{Illustration of different discrete stencils for different orders of isotropy. Weights are listed in Table~\ref{table:stencils_and_weights}. Figure partially reproduced from \cite{sbragaglia_generalized_2007}. }
	\label{Fig:stencils_discretization}
\end{figure}

Once the form of the non-local pressure term has been determined it needs to be discretized; For the sake of readability we will only consider the form $\bm{F}_2$ here. The finite difference discretized form can be written in general as:
\begin{equation}
    \bm{F}_2 = 2\sqrt{P-P_0}(\bm{r})\sum_{i=0}^{Q} w(\lvert\bm{c}_i\lvert) \bm{c}_i \sqrt{P-P_0}(\bm{r}+\bm{c}_i \delta t) , 
\end{equation}
where $\bm{c}_i$ defines the stencil used for the discretization and $w(\lvert\bm{c}_i\lvert)$ are the associated weights. For instance, considering a simple second-order central difference approximation $\bm{c}_i\in\{(1,0),(0,1),(-1,0),(0,-1)\}$ and $w=1/2$. From classical theory of finite differences method and Taylor expansion the order of accuracy of the approximation can be arbitrarily increased by relying on larger stencils. The weights $w(\lvert\bm{c}_i\lvert)$ on such classical stencils for different order of accuracy are listed in Table~\ref{table:classical_stencils_weights}.
\begin{table}
\begin{center}
\begin{tabular}{||c | c c c c||} 
 \hline
  Order of accuracy & $w(1)$ & $w(2)$ & $w(3)$ & $w(4)$\\ [0.5ex] 
 \hline\hline
 $2$ & $1/2$ &  &  & \\ 
 \hline
 $4$ & $2/3$ & $-1/12$ &  & \\ 
 \hline
 $6$ & $3/4$ & $-3/20$ & $1/60$ & \\ 
 \hline
 $8$ & $4/5$ & $-1/5$ & $4/105$ & $-1/280$ \\[1ex] 
 \hline
\end{tabular}
\end{center}
\label{table:classical_stencils_weights}
\caption{List of weights for classical stencils with different orders of accuracy. The notation $w(k)$ refers to the weight of all stencil components of size $k$.}
\end{table}
While the use of such higher order schemes reduces the errors in coexistence density stemming from the thermodynamic inconsistency issue it does necessarily improve the isotropy of the discrete approximation.\\
Applying the Taylor expansion to the general form of the discretized force~\cite{sbragaglia_generalized_2007}:
\begin{equation}
    \bm{F}_{2,\alpha} = 2\sqrt{P-P_0}(\bm{r}) \mathcal{E}_{\alpha_1}^{(2)}\partial_{\alpha_2} \sqrt{P-P_0}(\bm{r}) + \frac{1}{3!} \mathcal{E}_{\alpha_1\alpha_2\alpha_3\alpha_4}^{(4)} \partial_{\alpha_1\alpha_2\alpha_3} \sqrt{P-P_0}(\bm{r}) + \dots
\end{equation}
where
\begin{equation}
    \mathcal{E}_{\alpha_1,\dots,\alpha_m}^{(m)} = \sum_i w(\lvert\bm{c}_i\lvert) c_{i,\alpha_1}\dots c_{i,\alpha_m},
\end{equation}
with $\mathcal{E}^{2n+1} = 0$. The even-order tensors can be re-written as:
\begin{equation}
    \mathcal{E}_{\alpha_1 \dots \alpha_{2n}}^{(2n)} = \mathcal{C}^{(2n)} \Delta_{\alpha_1 \dots \alpha_{2n}}^{(2n)},
\end{equation}
where $\Delta_{\alpha_1 \dots \alpha_{2n}}^{(2n)}$ can be computed using the following recursion relation:
\begin{eqnarray}
    \Delta_{\alpha_1 \alpha_2}^{(2)} &=& \delta_{\alpha_1 \alpha_2},\\
    \Delta_{\alpha_1 \alpha_2 \alpha_3 \alpha_4}^{(4)} &=& \delta_{\alpha_1 \alpha_2}\delta_{\alpha_3 \alpha_4} + \delta_{\alpha_4 \alpha_1}\delta_{\alpha_2 \alpha_3} + \delta_{\alpha_2 \alpha_3}\delta_{\alpha_4 \alpha_1},\\
    \Delta_{\alpha_1 \dots \alpha_{2n}}^{(2n)} &=& \sum_{j=2}^{2n} \delta_{\alpha_1 \alpha_j} \Delta_{\alpha_2\dots \alpha_{j-1} \alpha_{j+1}\dots \alpha_{2n}}.
\end{eqnarray}
This general expansion of the discrete forcing term allows to analyze both the leading-order errors and the degree of isotropy of the discrete approximation. The latter  is an important point to consider as the continuous form of the force is isotropic. Any non-isotropic errors in the discrete approximation would lead to force imbalance on curved interfaces and lead to spurious currents. The use of higher-order stencils can therefore help reduce the spurious currents stemming from these force imbalances. In \cite{sbragaglia_generalized_2007} the authors have derived stencils of different order of isotropy with corresponding weights. These are listed in Table~\ref{table:stencils_and_weights} for a 2-D system.
\begin{table}
\begin{center}
\begin{tabular}{||c | c c c c c c c||} 
 \hline
   & $w(1)$ & $w(\sqrt{2})$ & $w(2)$ & $w(\sqrt{5})$ & $w(2\sqrt{2})$ & $w(3)$ & $w(\sqrt{10})$\\ [0.5ex] 
 \hline\hline
 $E^{(4)}$ & $1/3$ & $1/12$ & & & & & \\ 
 \hline
 $E^{(6)}$ & $4/15$ & $1/15$ & $1/120$ & & & & \\ 
 \hline
 $E^{(8)}$ & $4/21$ & $4/45$ & $1/60$ & $2/315$ & $1/5040$ & & \\ 
 \hline
 $E^{(10)}$ & $262/1785$ & $93/1190$ & $7/340$ & $6/595$ & $9/9520$ & $2/5355$ & $1/7140$\\[1ex] 
 \hline
\end{tabular}
\end{center}
\label{table:stencils_and_weights}
\caption{List of weights for stencils with different orders of isotropy. The notation $E^{(2n)}$ refers to a stencil with order of isotropy $2n$. For weights the notation $w(k)$ refers to the weight of all stencil components of size $k$. The corresponding stencils are illustrated in Fig.~\ref{Fig:stencils_discretization}. }
\end{table}
The stencils corresponding to these weights are shown in Fig.~\ref{Fig:stencils_discretization}.
It was shown in this section that one can reduce discretization errors at interfaces, both magnitude and isotropy, via higher order approximations relying on larger discretization stencils. Larger discretization stencils also mean larger number of discrete operations per grid-point and larger degrees of non-locality of such operations which in turn translate into both computational and communication (for parallel simulations on clusters with distributed memory) overhead. 
\subsubsection{Thickening interface}
\begin{figure}[h!]
	\centering
	\includegraphics[width=5cm,keepaspectratio]{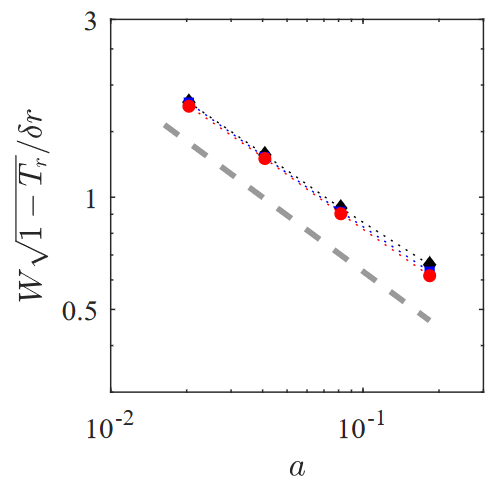}
	\caption{Effect of the choice of $a$ on the interface width. Diamond, square and circle: Simulation for $T_r= 0.98, 0.99, 0.995$, respectively. Simulation conducted using van der Waals equation of state. Plot reproduced from \cite{hosseini_towards_2021}. }
	\label{Fig:interface_thickness_vs_a}
\end{figure}

\begin{figure}[h!]
	\centering
	\includegraphics[width=6cm,keepaspectratio]{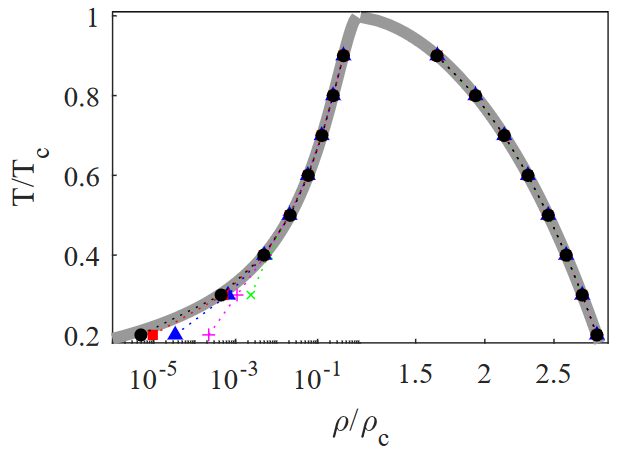}
	\caption{Coexistence densities as obtained from (grey lines) Maxwell's construction and (markers) simulations with different choices of $a$: (green x) $a=0.0102$, (magenta +) $a=0.0051$, (blue triangles) $a=0.0026$, (red squares) $a=0.0013$ and (black circles) $a=0.00064$ Simulation conducted using van der Waals equation of state. Plot reproduced from \cite{hosseini_towards_2021}. }
	\label{Fig:coexistence_vs_a}
\end{figure}

\begin{figure}[h!]
	\centering
	\includegraphics[width=5cm,keepaspectratio]{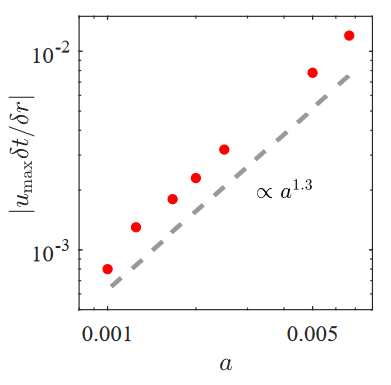}
	\caption{Maximum spurious currents for different values of $a$ at $T_r=0.59$. Simulation conducted using Peng-Robinson equation of state. Plot reproduced from \cite{hosseini_entropic_2022}. }
	\label{Fig:spurious_currents_vs_a}
\end{figure}

Artificial thickening of the interface is another strategy to both reduce errors in coexistence densities and spurious currents. This strategy is motivated by the invariance of the non-dimensional coexistence densities with respect substance-specific coefficients as demonstrated through the principle of corresponding states and and the fact that errors at interfaces are function of higher-order derivatives of the pressure field. In the context of cubic equations of state, the simple analysis leading to \eqref{eq:interface_thickness_control_near_critical} showed that the interface thickness can in principle be controlled through the capillary coefficient and isothermal compressibility tied itself to the long range interaction coefficient $a$. Plugging the van der Waals equation of state into Eq.~\eqref{eq:interface_thickness_control_near_critical} one would expect the interface thickness to scale with $\propto1/\sqrt{a}$. This scaling is readily demonstrated via simulations close to the critical point, shown in Fig.~\ref{Fig:interface_thickness_vs_a}.
Carrying out simulations with different values of $a$ it can easily be shown that this increase in interface thickness allows for better resolution of interfaces and therefore reduced deviations in discrete stress tensor. The coexistence densities as obtained from simulations with different choices of $a$ are shown in Fig.~\ref{Fig:coexistence_vs_a}.
It is observed that increasing the interface thickness one converges to the coexistence density prediction of the Maxwell construction even at very large density ratios.\\
As discussed in previous section, another aspect of errors in the stress tensor that need attention are non-isotropic effects leading to spurious currents. As for errors for flat interfaces these deviations functions of higher-order derivatives of the pressure field and such should reduce with thicker interfaces. The effect of the choice of $a$ controlling interface thickness on spurious currents for the discrete model of \cite{hosseini_towards_2021} is illustrated in Fig.~\ref{Fig:spurious_currents_vs_a}.
It is clearly observed that smaller values of $a$, which as seen before scale with interface thickness as $W\propto 1/\sqrt{a}$, lead to smaller spurious currents scaling as $\propto a^{1.3}$  for the discrete model of \cite{hosseini_towards_2021} meaning they scale with interface thickness as $\propto W^{-2.6}$. Such scaling relationships can also be obtained for other discrete stencils.\\
While changing the isothermal compressibility in combination with capillary coefficient allows to thicken the interface independently from surface tension it comes with a limitation: It changes the isothermal speed of sound in both liquid and vapor phases. Lower values of $a$ lead to smaller speeds of sound in both phases. This in turn means that for the isothermal simulation to be close to the low Mach regime one would need to run simulation at lower convective CFL conditions leading to much smaller time-steps and additional computational costs. 
\subsubsection{Independent control over surface tension}
A solution to the fixed nature of the surface tension recovered by the original pseudo-potential model was provided in \cite{sbragaglia_generalized_2007,shan_pressure_2008} where the authors introduced the concept of dual/multi-range pseudo-potential. In its simplest form a second layer of neighbours is added to the stencil considered for the discretization of the pressure term:
\begin{equation}
    \bm{F}_2 = 2\sqrt{P-P_0}(\bm{r}) \sum_i w(\lvert\bm{c}_i\lvert) \bm{c}_i \left[ \mathcal{G}_1\sqrt{P-P_0}(\bm{r}+\bm{c}_i \delta t) + \mathcal{G}_2\sqrt{P-P_0}(\bm{r}+2\bm{c}_i \delta t)\right],
\end{equation}
where $\mathcal{G}_1$ and $\mathcal{G}_2$ are coefficients to be determined below. Taylor-expanding this term:
\begin{equation}
    \bm{F}_2 = 2(\mathcal{G}_1 + 2\mathcal{G}_2)\sqrt{P-P_0}\bm{\nabla}\sqrt{P-P_0} + 2\sqrt{P-P_0}\frac{\mathcal{G}_1 + 8\mathcal{G}_2}{6}\bm{\nabla}\bm{\nabla}^2\sqrt{P-P_0}.
\end{equation}
To correctly recover the thermodynamic pressure term one must satisfy $\mathcal{G}_1+2\mathcal{G}_2=1$. Using this condition the force can be re-written as:
\begin{equation}
    \bm{F}_2 = \bm{\nabla}\lvert P-P_0\lvert + \sqrt{P-P_0}\frac{1 + 6\mathcal{G}_2}{3}\bm{\nabla}\bm{\nabla}^2\sqrt{P-P_0},
\end{equation}
meaning the surface tension coefficient can now be tuned with $\mathcal{G}_2$, $\kappa=\frac{1+6\mathcal{G}_2}{3}$. Alternatives to the dual-range model have also been proposed in the past years to limit the effect of the choice of $\mathcal{G}_2$ on the mechanical stability conditions. For instance in \cite{li_achieving_2013} the authors introduce an additional source term in the discrete equation:
\begin{equation}
    \mathcal{Q}_i = \frac{w_i}{2 c_s^4\bar{\tau}} \mathcal{H}_2:\kappa' \sqrt{P-P_0}(\bm{r}+\bm{c}_i) \delta t \sum_{i'=0}^{Q} w(\lvert \bm{c}_{i'}\lvert) \bm{c}_{i'}\otimes \bm{c}_{i'} \left[\sqrt{P-P_0}(\bm{r}+\bm{c}_{i'}\delta t) - \sqrt{P-P_0}(\bm{r})\right],
\end{equation}
where $\kappa'$ is the parameter that controls the surface tension. While allowing a variable coefficient in front of the surface tension-like term, the final stress tensor is still different from the Koerteweg tensor.\\
An extension of the dual range approach to recover a consistent Korteweg stress tensor fourth-order accurate in space, guaranteeing the surface tension is not polluted by leading order errors from the discretization of the thermodynamic pressure term was proposed in \cite{hosseini_towards_2021}:
\begin{multline}
    \bm{F}_2 = 2\sqrt{P-P_0}(\bm{r}) \sum_i w(\lvert\bm{c}_i\lvert) \bm{c}_i \left[ \mathcal{G}_1\sqrt{P-P_0}(\bm{r}+\bm{c}_i\delta t) + \mathcal{G}_2\sqrt{P-P_0}(\bm{r}+2\bm{c}_i\delta t)\right] \\
    + \rho(\bm{r}) \sum_i w(\lvert\bm{c}_i\lvert) \bm{c}_i \left[ \mathcal{G}_3\rho(\bm{r}+\bm{c}_i\delta t) + \mathcal{G}_4\rho(\bm{r}+2\bm{c}_i\delta t)\right],
\end{multline}
where $\mathcal{G}_1+2\mathcal{G}_2=1$ and $\mathcal{G}_1+8\mathcal{G}_2=0$ guarantee correct recovery of the thermodynamic pressure term while $\mathcal{G}_3+2\mathcal{G}_4=0$ and $\mathcal{G}_3+8\mathcal{G}_4=6\kappa$ recover the correct Korteweg surface tension.
\subsubsection{The discrete pressure tensor}
\begin{figure}
	    \centering
		    \includegraphics{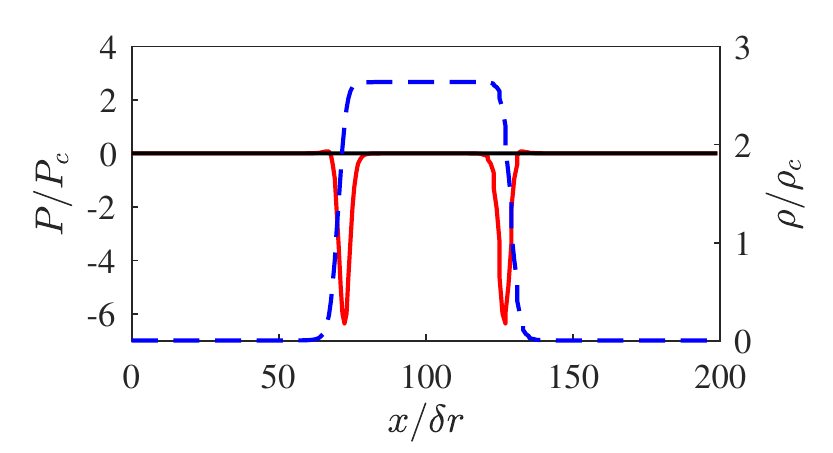}
	    \caption{Pressure distribution from a simulation at $T_r=0.36$, corresponding to $P_r=0.0022$ and $\rho_l/\rho_v=10^3$. Black line: Evaluation using the discrete pressure tensor; Red line: Evaluation using continuous pressure tensor. Dashed blue line: Density profile. 
	    }
	    \label{Fig:discrete_pressure}
    \end{figure}

As shown through Taylor expansion in the previous section, the continuous and discrete pressure tensors are not exactly the same. In~\cite{shan_pressure_2008} proposed an approach to evaluate the discrete pressure tensor as momentum flux through a surface; Given an infinitesimal area $d\bm{S}$ and the force acting through that surface $d\bm{F}$ the pressure tensor is defined as:
\begin{equation}
    d\bm{F} = P\cdot d\bm{S}.
\end{equation}
Integrating over a closed control volume one arrives at the fact that the surface integral of the pressure has to match total force acting on the control volume:
\begin{equation}
    \int \bm{P}\cdot d\bm{S} = \int \bm{F} dV,
\end{equation}
which in discrete form reduces to:
\begin{equation}
    \sum \bm{P}\cdot\bm{S} = \sum \bm{F}.
\end{equation}
Considering for the sake of simplicity a specific discrete velocity direction, i.e. $\bm{c}_i \delta t$, it can be seen that the number of force vectors across a unit vertical surface element $dS = \bm{e}_x$ is $c_{ix} \delta t$ and that across a unit horizontal surface element is $c_{iy} \delta t$. In simplest case where all pressure contributions from the force have a strength $F$ the pressure contribution is then simply obtained as $\bm{c}_i \otimes \bm{c}_i \delta t^2 F$. For a force field where each contribution has a different magnitude one uses the averaged force, where averaging is operated on all force contributions crossing the surface area. This results, for the simplest first neighbor stencil and the classical pseudo-potential interaction in:
\begin{equation}
        \bm{P} = -\frac{G}{2} \psi(\bm{r})\sum_{i=0}^{8} w(\lvert\bm{c}_i\lvert) \bm{c}_i\otimes \bm{c}_i \psi(\bm{r}+ \bm{c}_i\delta t).
\end{equation}
Following that same construction logic write the contribution for a more complex force, like he one used in \cite{hosseini_towards_2021} as
    \begin{align}
        \bm{F} &= \bm{F}^{A} + \bm{F}^{B} + \bm{F}^{C} + \bm{F}^{D},\\
            \bm{F}^{A} &= \pm\frac{8}{3} \sqrt{P-P_0}(\bm{r}) \sum_{i=0}^{Q-1} w(\lvert\bm{c}_i\lvert) \bm{c}_{i} \sqrt{P-P_0}(\bm{r}+\bm{c}_i\delta t),\\
        \bm{F}^{B} &= \mp \frac{1}{3} \sqrt{P-P_0}(\bm{r}) \sum_{i=0}^{Q-1}  w(\lvert\bm{c}_i\lvert)\bm{c}_{i}\sqrt{P-P_0}(\bm{r}+2\bm{c}_i\delta t),\\
        \bm{F}^{C} &= 2\tilde{\kappa} \rho(\bm{r}) \sum_{i=0}^{Q-1} w(\lvert\bm{c}_i\lvert)\bm{c}_{i}\rho(\bm{r}+\bm{c}_i\delta t),\\
        \bm{F}^{D} &= -\tilde{\kappa} \rho(\bm{r}) \sum_{i=0}^{Q-1} w(\lvert\bm{c}_i\lvert)\bm{c}_{i}\rho(\bm{r}+2\bm{c}_i\delta t).
    \end{align}
    The pressure tensor contributions from forces $\bm{F}^{A}$ and $\bm{F}^{C}$ can be readily written as:
    \begin{eqnarray}\label{eq:discrete_pressure_psi1}
        \bm{P}^{A} &=& \mp\frac{8}{6} \sqrt{P-P_0}(\bm{r}) \sum_{i=0}^{Q-1} w(\lvert\bm{c}_i\lvert) \bm{c}_{i}\otimes\bm{c}_{i} \sqrt{P-P_0}(\bm{r}+\bm{c}_i\delta t),\\
        \bm{P}^{C} &=& -\tilde{\kappa} \rho(\bm{r}) \sum_{i=0}^{Q-1} w(\lvert\bm{c}_i\lvert) \bm{c}_{i}\otimes\bm{c}_{i} \rho(\bm{r}+\bm{c}_i\delta t),
    \end{eqnarray}
    while  $\bm{F}^{B}$ and $\bm{F}^{D}$ contribute to the pressure tensor as follows:
    \begin{align}\label{eq:discrete_pressure_psi2}
        \bm{P}^{B} =& \pm \frac{1}{6} \left[\sqrt{P-P_0}(\bm{r}) \sum_{i=0}^{Q-1} w(\lvert\bm{c}_i\lvert) \bm{c}_{i}\otimes\bm{c}_{i} \sqrt{P-P_0}(\bm{r}+2\bm{c}_i\delta t) \right.
         + \left. \sum_{i=0}^{Q-1} w(\lvert\bm{c}_i\lvert)\bm{c}_{i}\otimes\bm{c}_{i} \psi(\bm{r}-\bm{c}_i\delta t)\sqrt{P-P_0}(\bm{r}+\bm{c}_i\delta t)\right],\\
        \bm{P}^{D} =& \frac{\tilde{\kappa}}{2} \left[\rho(\bm{r}) \sum_{i=0}^{Q-1} w(\lvert\bm{c}_i\lvert)\bm{c}_{i}\otimes\bm{c}_{i} \rho(\bm{r}+2\bm{c}_i\delta t) \right. 
         + \left. \sum_{i=0}^{Q-1} w(\lvert\bm{c}_i\lvert)\bm{c}_{i}\otimes\bm{c}_{i} \rho(\bm{r}-\bm{c}_i\delta t)\rho(\bm{r}+\bm{c}_i\delta t)\right].
    \end{align}
    These expressions allow to compute the discrete pressure tensor with high accuracy. Fig.\ \ref{Fig:discrete_pressure} shows the distribution of the normal pressure, $P_{xx}$, in a flat interface simulation as computed from both the discrete and continuous pressure tensors, 
    \begin{equation}\label{eq:pressure_cont}
        P_{xx}^{\rm cont} = P + \kappa\left(\partial_x^2 \rho - \frac{1}{2}{\lvert\partial_x \rho\lvert}^2\right),
    \end{equation}
     While the discrete evaluation method correctly results in a uniform pressure distribution throughout the domain, also across the interface, the continuous approximation evaluated using a finite differences approximation fails to do so, indicating errors due to higher-order terms. This points to the necessity of using the discrete pressure tensor for evaluation of sensitive quantities especially for sharper interfaces.
\subsubsection{Evaluating the effective surface tension\label{subsubsec:ev_surf_ten}}
\begin{figure}
	\centering
	\includegraphics{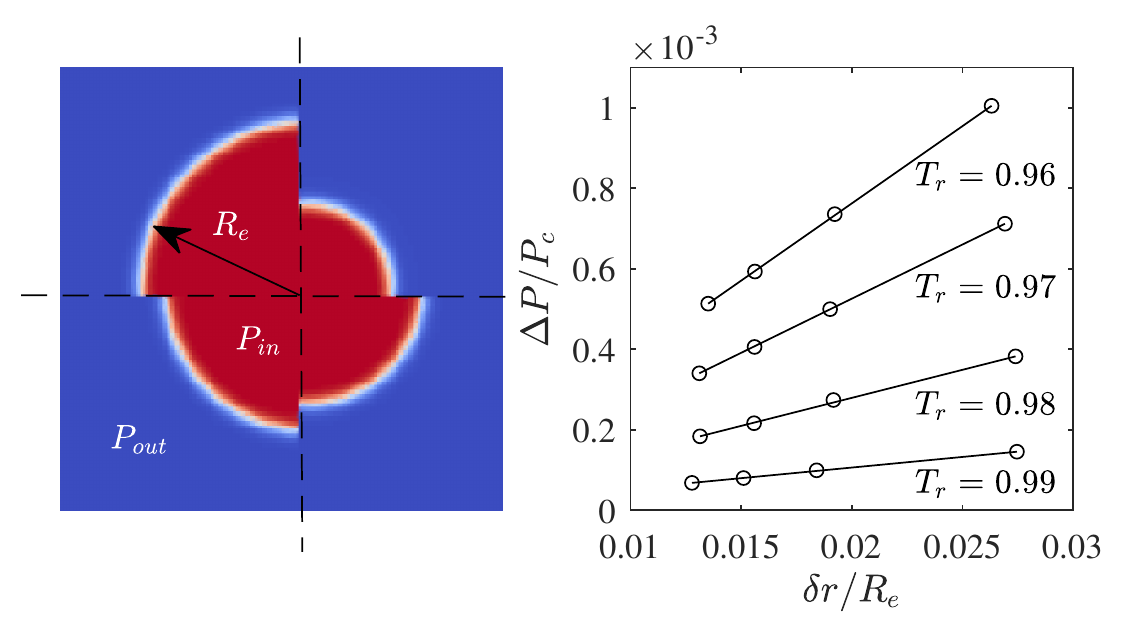}
	\caption{Left: Circular $D=2$ drop configurations; Right: 
		Pressure difference scaling with drop radius for $T_r=0.99,0.98,0.97$ and $0.96$.
		The pressure difference is defined as $\Delta P=P_{\rm in}-P_{\rm out}$. The slope of the straight line is the surface tension coefficient. Results are reproduced from \cite{hosseini_towards_2021}.}
	\label{Fig:SurfTenTempSlopes}
\end{figure}

\paragraph{Laplace's law}
In this approach, simulations of circular/spherical liquid drops of different radii surrounded with vapour are carried out. The corresponding surface tension coefficient is then evaluated using the Laplace law in a form, 
\begin{equation}\label{eq:Laplace_gen_1}
	\Delta P = \frac{(D-1)\sigma}{R},
\end{equation} 
where $R$ is the drop radius. $\Delta P$ can readily be computed by extracting the pressure at the center of the drop $P_{\rm in}$ and a point in the vapor phase far away from the drop $P_{\rm out}$ as $\Delta P=P_{\rm in} - R_{\rm out}$. For simulations carried out using diffuse interface formulations the notion of drop radius becomes ambiguous. To that end, for consistent analysis of results one has to introduce the notion of dividing surfaces, more specifically the \emph{equimolar} surface here, proposed by Gibbs~\cite{gibbs_equilibrium_1874}. A brief reminder of Gibbs' theory of dividing surfaces is in order. The total mass in both the diffuse and sharp interface pictures can be written as:
\begin{equation}\label{eq:mass_sharp}
	\int_{V} \rho dV = \rho_l V_l + \rho_v V_v + \Gamma,
\end{equation}
where $\rho_l V_l$ and $\rho_v V_v$ are the masses in the bulk liquid and vapor phases in the sharp interface picture, while $\Gamma$ is the  excess mass on a dividing surface $\Sigma$, or mass \emph{adsorbance} \citep{gibbs_equilibrium_1874}. By requiring that no mass be stored on the dividing surface we get the definition of the \emph{equimolar} surface:
\begin{equation}\label{eq:equimolar_adsorbance} 
	\Gamma = 0.
\end{equation}
The family of dividing surfaces in the case of drop or bubble are concentric spheres ($D=3$) or concentric circles ($D=2$) parameterized by their radius $R$.
In particular, for a two-dimensional drop, the mass adsorbance can be written as a function of the radius of the dividing circle,
\begin{equation}
	\Gamma(R)= \int_0^{2\pi}\int_0^{\infty}(\rho(r) - \rho_v)r drd\varphi - \int_0^{2\pi}\int_0^{R}(\rho_l - \rho_v)r drd\varphi,
\end{equation}
while the zero-adsorbtion condition \eqref{eq:equimolar_adsorbance}, $\Gamma(R_e)=0$, implies the {equimolar} radius $R_e$,
\begin{equation}\label{eq:equmol2Ddrop}
	R_e = \sqrt{\frac{\int_0^{\infty}(\rho(r) - \rho_v)r dr}{\left(\rho_l - \rho_v\right)}}.
\end{equation}
This definition can then be used to replace the radius in Eq.~\ref{eq:Laplace_gen}. Once both drop radii and pressure differences are known the surface tension can be extracted as the slope of $\Delta P = \sigma \frac{D-1}{R_e}$.  This is illustrated in Fig.~\ref{Fig:SurfTenTempSlopes}.
\paragraph{Kirkwood approach for flat interface}
The surface tension coefficient of a flat interface can be evaluated using its mechanical definition~\cite{kirkwood_statistical_1949} as,
\begin{equation}\label{eq:surface_tension_int}
        \sigma = \int_{-\infty}^{+\infty} \left(P_{xx} - P_{yy}\right)dx,
\end{equation}
where here the interface has been considered normal to the $x$-axis in a two-dimensional simulation setup. The normal $P_{xx}$ and the tangential $P_{yy}$ components of the discrete pressure tensor can be computed from the continuous approximation or using the discrete pressure tensor introduced in the previous section.
\subsection{Fluid-solid interaction: wetting properties}

A first indication on the treatment of a solid boundary for second-gradient fluids, under the assumption of short range-only interaction, was given by Cahn (critical point wetting). Considering a control volume with a solid surface at the boundary the total free energy is defined as:
\begin{equation}
    \mathcal{A} = \int_V \left[\mathcal{A}_0 + \frac{1}{2}\kappa{\lvert \bm{\nabla}\rho\lvert}^2 \right] dV + \int \mathcal{A}_{\rm w}(\rho_{\rm w}) dS,
\end{equation}
where $V$ is the considered control volume, $S$ the solid surface and $\mathcal{A}_{\rm w}$ a surface free energy function which depends only on the fluid density at the solid surface $\rho_{\rm w}$. The derivative of this functional is obtained as:
\begin{equation}
    \delta \mathcal{A} = \int_V \delta \rho \left[ \frac{\partial \mathcal{A}_0}{\partial \rho} - \kappa \bm{\nabla}\cdot\bm{\nabla}\rho \right] dV + \int \delta \rho_{\rm w} \left[\kappa \bm{n}\cdot\bm{\nabla}\rho + \frac{\partial\mathcal{A}_{\rm w}}{\partial \rho_{\rm w}} \right] dS,
\end{equation}
where $\bm{n}$ is the unit vector normal to the solid surface. Minimzation of the total free energy with respect to $\rho_{\rm w}$ leads to:
\begin{equation}
    \kappa \bm{n}\cdot\bm{\nabla}\rho = \frac{d \mathcal{A}_{\rm w}}{d\rho_{\rm w}}.
\end{equation}

\subsubsection{Static contact angle}
\begin{figure}[h!]
	\centering
	\includegraphics[width=8cm,keepaspectratio]{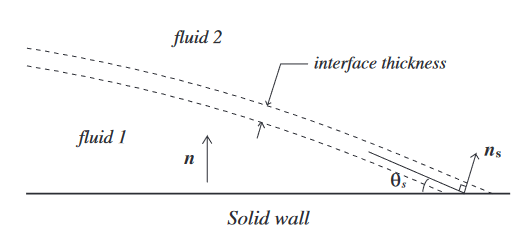}
	\caption{Illustration of contact line in geometrical approach to contact angle.}
	\label{Fig:geometrical_wetting}
\end{figure}

One of the first approaches to impose wetting boundary conditions for non-ideal fluids in the context of the pseudo-potential method was proposed in \cite{martys_simulation_1996}. Following the  structure of the non-local pseudo-potential interaction term they proposed an additional contribution modeling wall interaction of the form:
\begin{equation}
    \bm{F}_{\rm wall} = -\rho(\bm{r})\mathcal{G}_{\rm wall} \sum_i w_i \bm{c}_i\delta ts(\rm{r}+\bm{c}_i \delta t),
\end{equation}
where $s$ is an indicator function equal to one in a solid cell and zero in a fluid cell and $\mathcal{G}){\rm wall}$ is the interaction strength parameter controlling the contact angle. Sukop and Thorn \cite{sukop_lattice_2006} proposed a slightly modified form of the wall interaction term as \cite{raiskinmaki_spreading_2000,raiskinmaki_lattice-boltzmann_2002}:
\begin{equation}
    \bm{F}_{\rm wall} = -\psi(\bm{r})\mathcal{G}_{\rm wall} \sum_i w_i \bm{c}_i\delta ts(\rm{r}+\bm{c}_i \delta t).
\end{equation}
Later on Kang et al. proposed a modified version of Martys and Chen's approach to setting wetting boundary conditions as \cite{kang_displacement_2002}:
\begin{equation}
    \bm{F}_{\rm wall} = -\rho(\bm{r})\mathcal{G}_{\rm wall} \sum_i w_i \bm{c}_i\delta t\rho_{\rm wall}(\rm{r}+\bm{c}_i \delta t).
\end{equation}
The wall is modeled as a phase with a constant density $\rho_{\rm wall}$. As with previous schemes $\mathcal{G}_{\rm wall}$ is used to set the contact angle. Alternatively, Benzi et al proposed a slightly modified form, fully consistent with the bulk non-local interaction \cite{benzi_mesoscopic_2006,sbragaglia_surface_2006}:
\begin{equation}
    \bm{F}_{\rm wall} = -\psi(\bm{r})\mathcal{G} \sum_i w_i \bm{c}_i\delta t\psi_{\rm wall}(\rm{r}+\bm{c}_i \delta t),
\end{equation}
where the only free parameter is $\psi_{\rm wall}$ which controls adhesion of the liquid/vapour phase to the solid boundary, i.e. for $\psi_{\rm wall}\rightarrow\psi_l$ the contact angle goes to zeros while $\psi_{\rm wall}\rightarrow\psi_v$ the contact angles goes to 180.
\begin{figure}
	\centering
	\includegraphics{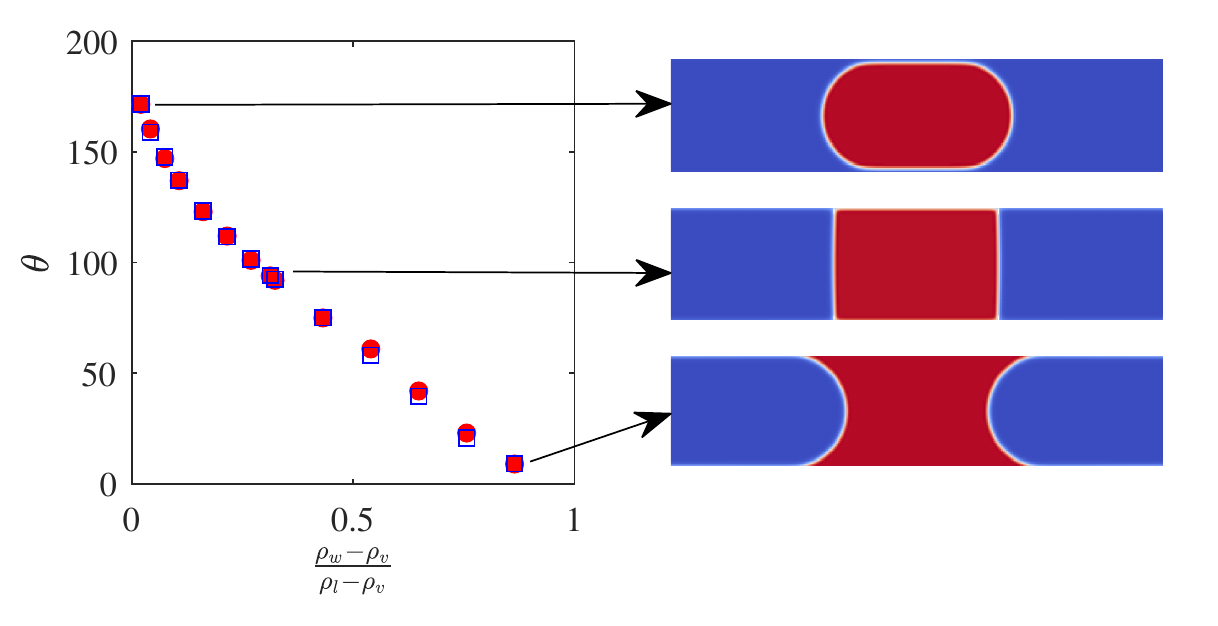}
	\caption{Static contact angles as (blue squares) obtained from the Young--Laplace equation and (red circles) measured directly from the simulations using the approach of \cite{benzi_mesoscopic_2006,sbragaglia_surface_2006}. Image is taken from \cite{hosseini_towards_2021}.}
	\label{Fig:CAResults}
\end{figure}
\paragraph{Geometrical approach}
In parallel with the previously-listed approaches, alternative geometrical approaches widely in use for phase-field based approaches have also been developed. Ding and Spelt proposed a geometrical approach to implement contact angles in phase-field methods as~\cite{ding_wetting_2007}:
\begin{equation}\label{eq:geometrical_approach_CA}
    \tan\left(\frac{\pi}{2}-\theta_s\right) = \frac{-\bm{n}\cdot\bm{\nabla}\psi}{\lvert \bm{\nabla}\psi - (\bm{n}\cdot\bm{\nabla})\bm{n} \lvert},
\end{equation}
which once discretized leads to~\cite{ding_wetting_2007}:
\begin{equation}
    \psi(x,-\delta y) = \psi(x,\delta y) + \tan\left(\frac{\pi}{2}-\theta_s\right)\lvert \psi(x+\delta x,0) - \psi(x-\delta x,0)\lvert,
\end{equation}
where assuming a flat solid interface perpendicular to the $y$-axis at $y=0$, $(x,-\delta y)$ designates a ghost layer within the solid. $\theta_s$ is the contact angle to be imposed. They showed that the geometrical approach can fix the slop of the liquid-gas interface to a value consistent with the imposed contact angle given that there are enough points within the interface, i.e. typically 4-8 grid-points~\cite{ding_wetting_2007}. The approach is illustrated in Fig.~\ref{Fig:geometrical_wetting}.
The geometrical approach of Ding and Spelt was transposed into the pseudo-potential formulation in \cite{hu_contact_2016} for flat interfaces. Later on, in \cite{li_implementation_2019} it was extended to curved solid boundaries. While here we have used the pseudo-potential $\psi$ as the order parameter in defining the contact angle, this approach can readily be extended to free energy methods by replacing $\psi$ with the fluid density.
\subsubsection{Contact angle hysteresis}
\begin{figure}[h!]
	\centering
	\includegraphics[width=8cm,keepaspectratio]{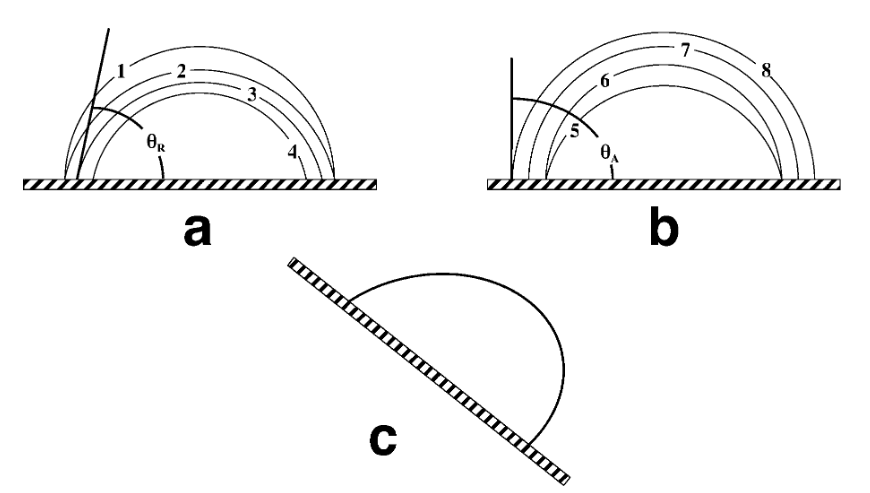}
	\caption{Illustration of (a) advancing and (b) receding drop on solid surface. In (c) a drop sliding on a surface exhibiting both receding and advancing contact angles in shown. Image is taken from \cite{{gao_contact_2006}}.}
	\label{Fig:contact_angle_hysteresis}
\end{figure}

To illustrate the meaning of contact angle angle hysteresis let us take the example detailed in~\cite{gao_contact_2006}. Consider a small droplet resting on a solid surface. If the droplet is allowed to evaporate or if liquid is slowly withdrawn from the droplet with a syringe, over time both the volume and contact angle will decrease maintaining the same contact area, up until the onset of recession. The drop will then recede with a constant contact angle, $\theta_R$, characteristic of both surface chemistry and topography. Now the opposite scenario: if the surface is cooled down below dew point leading to liquid condensing on the drop or if liquid is slowly added to the droplet the droplet both volume and contact angle will initially increase until the drop starts to advance. The drop will advance at a constant angle $\theta_A$ which is also determined by the characteristics of the solid surface. Both cases are illustrated in Fig.~\ref{Fig:contact_angle_hysteresis}. A metastable droplet can form with any angle between these two angles. The interval between the two is referred to as the contact angle hysteresis.
A simple realization of the contact angle hysteresis was proposed and used in \cite{ding_onset_2008} in the context of the geometrical approach to setting the contact angle. There the authors dynamically adjusted the contact angle as a function of the speed of the contact line, $u_{\rm CL}$:
\begin{equation}
    \begin{cases}
        \theta_s = \theta_A, & \forall u_{\rm CL}>0,\\
        \theta_s = \theta_R, & \forall u_{\rm CL}<0,\\
        \theta_s = \theta_s, &  u_{\rm CL}=0.
    \end{cases}
\end{equation}
In practice, the contact angle is evaluated at every time-steps. If the measure angle is within the hysteresis windows it is left unaltered. If it is outside the hysteresis windows it is set to either the advancing or receding contact angle, depending on the speed of the contact line. Similar approaches have been adopted in the contact of the LBM~\cite{ba_color-gradient_2013,zhang_evaporation_2021,yang_effects_2020}. For instance in \cite{wang_scheme_2013} the authors used a similar feedback-based strategy without taking into account the contact line speed. The authors set $\theta=\theta_A$ if the measured angle was larger than $\theta_A$ and $\theta=\theta_R$ if its was smaller than $\theta_R$. While this approach performed well for contact line motion on flat substrates and flow in tubes~\cite{liu_lattice_2015}, as noted in \cite{qin_lattice_2021}, it resulted in un-physical behavior for isothermal drying of droplets. The correct behavior was restored by taking into account the contact line velocity~\cite{qin_lattice_2021}. One important point to note about contact angle hysteresis implementation is that to the authors knowledge all attempts at enforcing a hysteresis window have relied on the geometrical approach of Eq.~\ref{eq:geometrical_approach_CA}. This is due to the fact that to dynamically set the contact angle to $\theta_R$ and $\theta_A$ there must be a way to estimate the corresponding boundary condition \emph{a priori}. Other approaches to setting wetting conditions in the context of LBM do not establish a clear relationship between fluid/solid interaction force and contact angle. 
\subsection{Assessment of thermo-physical properties of models}
\subsubsection{Speed of sound and compressibility}
\begin{figure}[h!]
	\centering
	\includegraphics[width=13cm,keepaspectratio]{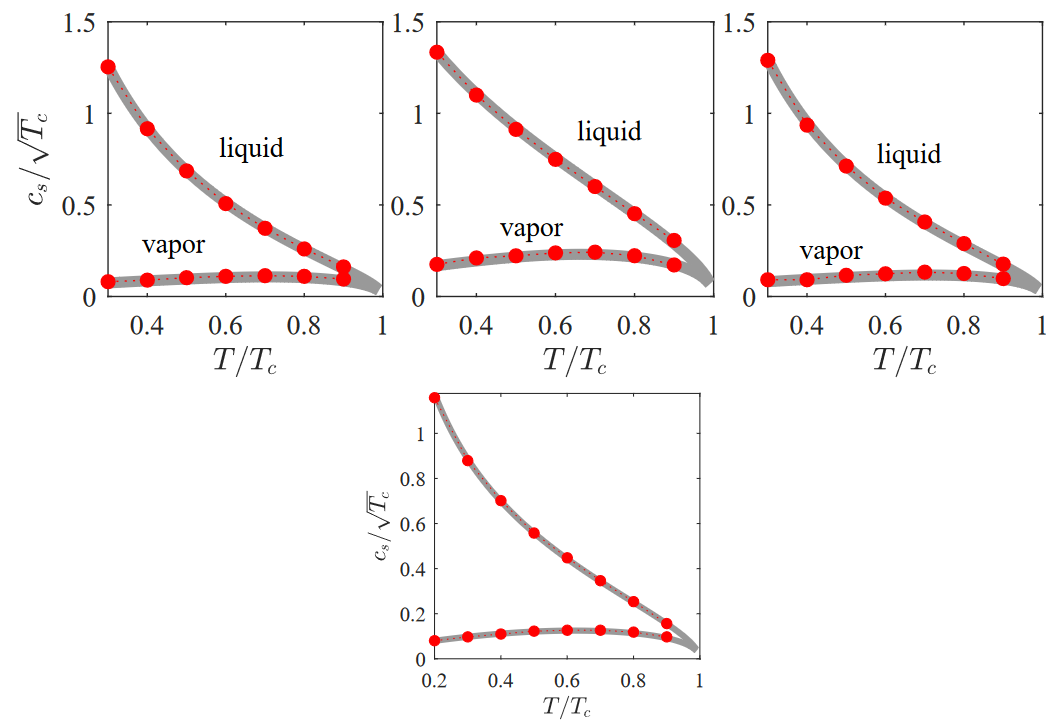}
	\caption{Isothermal sound speed for various equations of state. Top row, from left to right:
	    Peng--Robinson, 
	    Carnahan--Starling and Riedlich--Kwong--Soave;
	    Bottom row: van der Waals.
	    Grey plain lines: Theory; Symbol: Simulations. Plot reproduced from \cite{hosseini_towards_2021}.}
	\label{Fig:speed_of_sound_classical_EoS}
\end{figure}

\begin{figure}[h!]
	\centering
	\includegraphics[width=10cm,keepaspectratio]{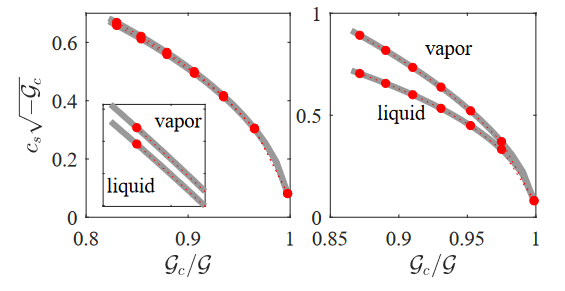}
	\caption{Isothermal sound speed for two Shan-Chen equations of state. Grey plain lines: Theory; Symbol: Simulations. Plot reproduced from \cite{hosseini_towards_2021}.}
	\label{Fig:speed_of_sound_SC_EoS}
\end{figure}

Different from hydrodynamic pressure-based formulation such as those employed in combination with Allen-Cahn interface tracking methods, all LBMs for weakly compressible non-ideal fluids recover the isothermal speed of sound of the implemented equation of state. The analytical isothermal sound speed can be obtained via the derivative of the pressure with respect to density at constant temperature, i.e. $c_s=\sqrt{\partial_\rho P\lvert_{T}}$. For instance for the van der Waals equation of state the isothermal sound speed is,
\begin{equation}\label{eq:sound_vdW}
	c_s = \sqrt{\frac{\partial P}{\partial \rho}\bigg|_{T}} = \sqrt{\frac{RT}{{(b\rho-1)}^2} - 2a\rho}.
\end{equation}
In simulations the sound speed can be measured by modeling the evolution of a pressure step-function and monitoring the position of the pressure front over time in a quasi-one-dimensional simulation at different temperatures. The sound speed for different cubic equations of state at different temperature in both liquid and vapor phases as obtained from simulations are compared to analytical data in Fig.~\ref{Fig:speed_of_sound_classical_EoS}.
Similar to cubic equations of state, Shan-Chen-type equations also admit isothermal sound speeds that can be obtained both analytically and from simulations. The results for two different equations of state are shown in Fig.~\ref{Fig:speed_of_sound_SC_EoS}.
There are two interesting points to note about the latter equations of state: (a) The difference in sound speed between the liquid and vapor branches is less pronounced than cubic equations of state making them interesting for simulations targeting the incompressible regime and (b) the sound speed in the vapor branch is always higher than that in the liquid branch which is not quite physical.\\
In general it must be noted that simulations targeting non-ideal equations of state in the limit of the incompressible regime are more challenging than for ideal gases. In ideal gases to correctly recover the incompressible limit separation of scales between shear and normal mode speeds must be ensured, i.e. $u_{\rm max}\ll c_s$. Furthermore, given the explicit nature of the LBM solvers the time step size is limited by the fastest traveling eigen-mode, i.e. speed of sound: $c_s<\delta r/\delta t$. In non-ideal systems, to ensure weak compressibility in both phases one must have $u_{\rm max}\ll\min(c_{s,l},c_{s,v})$ and $\max(c_{s,l},c_{s,v})<\delta r/\delta t$ to ensure stability of the explicit solver. Given difference of scale of sound speed in the liquid and vapor phases, especially at larger density ratios, this can become extremely prohibitive. In most practical applications where only the behavior of the liquid phase is of interest the first condition is made weaker, i.e. $u_{\rm max}\ll c_{s,l}$, allowing for larger time-steps and introducing more pronounced compressibility effects into the vapor phase.

\subsubsection{Meanfield scaling laws: Interface thickness}
\begin{figure}
	\centering
	\includegraphics{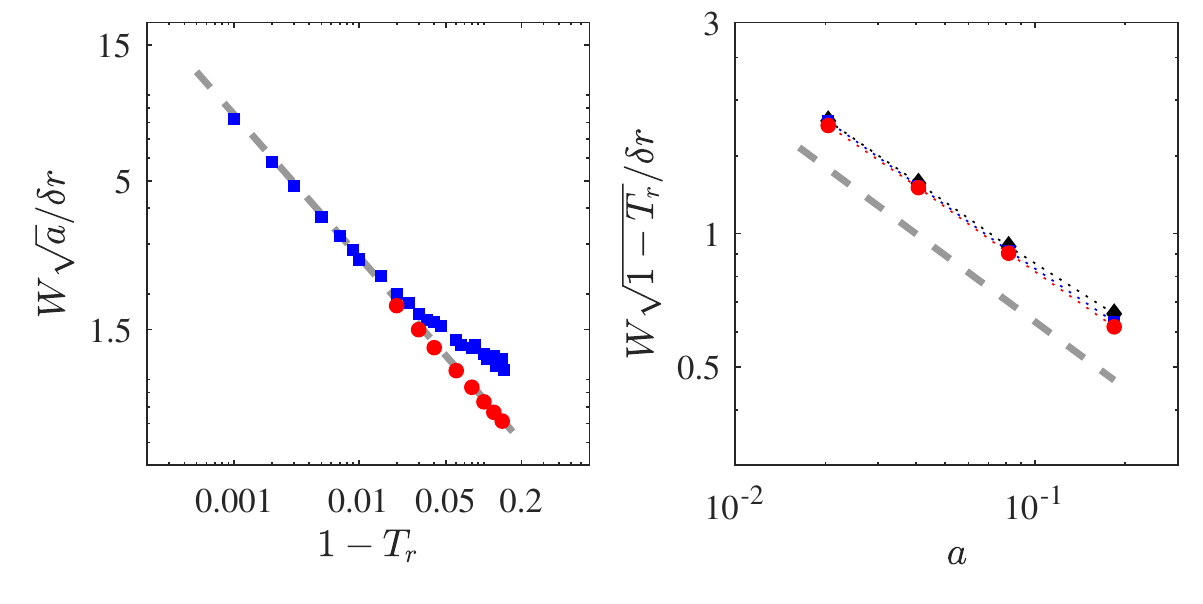}
	\caption{(Left) Interface width as a function of temperature. Blue square:  Simulation with $a=0.184$;  Red circle: Simulation with $a=0.02$;  Grey dashed line: Theoretical scaling \eqref{eq:Wscaling}. (Right) Effect of the choice of $a$ on the interface width. Diamond, square and circle: Simulation for $T_r= 0.98, 0.99, 0.995$, respectively. Plot reproduced from \cite{hosseini_towards_2021}.}
	\label{Fig:WidthTemp}
\end{figure}

Different from sharp interface methods, in diffuse interface approaches the liquid-vapor interface has a non-zero thickness. The thickness of the interface can be defined in many different ways; Here we use a definition for bearing numerical information as to how well the stiff gradients are resolved on a given mesh, making it directly related to the velocity increment per time-step:
\begin{equation}\label{eq:interface_W}
	W = \frac{\rho_l-\rho_v}{\max\lvert\bm{\nabla}\rho\lvert},
\end{equation}
where $\rho_l$ and $\rho_v$ are densities of saturated liquid and vapor, respectively. It is observed that in the limit of a sharp interface, i.e. resolved with $\delta r$, $\delta r/W\to 1$. On the other end of the spectrum for $\delta r/W\to 0$, akin to $\varepsilon\to 0$, one expects to recover the hydrodynamic limit, i.e. meanfield behavior. In that limit, surface tension is known to vanish as the temperature approaches the critical, cf. Eq.~\eqref{eq:vdw_surface_tension}, while the interface diverges as $T\to T_c$. As noted by \cite{widom_surface_1965}, the van der Waals theory predicts the temperature scaling of the interface width as,
\begin{align}\label{eq:Wscaling}
	W(T_r)\propto (1-T_r)^{-1/2}.
\end{align} 
For a well-posed numerical solver for any of the non-ideal fluid models discusses here, e.g. Free energy, pseudo-potential etc, moving from $\delta r/W \approx 1$ where numerical artefacts are known to dominate towards $\delta r/W\to 0$ one expects to recover the behavior predicted by Eq.~\ref{eq:Wscaling}. Articles treating of that issues in the context of non-ideal fluid LBM are rather scarce. In \cite{hosseini_towards_2021} the authors studied the evolution of interface thickness for different reduced temperatures using the van der Waals equation of state to probe the consistency of the solver against the second-gradient fluid theory. To that end simulations of flat interfaces were carried out in a range of reduced temperatures $T_r$ near the critical point, and corresponding interface widths $W(T_r)$ \eqref{eq:interface_W} were measured. Results are shown in Fig.~\ref{Fig:WidthTemp}.
Furthermore, to probe the $W/\delta r\to 1$ and $W/\delta r\to 0$ limits simulations were carried out for different values of $a$ allowing to re-scale the interface thickness. This is equivalent to reducing the grid-size for a fixed substance or changing the substance for a fixed grid size. As noted by many authors in the literature \cite{jamet_second_2001}, the parameter $a$ can be used to control the interface thickness, at a given density ratio, leaving the ratios and Maxwell construction unaffected. In agreement with the equivalent states theory, the right hand side plot in Fig.~\ref{Fig:WidthTemp} points to the universality of the scaling of the interface width near the critical point regardless of the choice of $a$. In addition it is interesting to note that for a fixed grid-size $\delta r$, as $(1-T_r)\to1$, $\delta r/W\to 1$ (equivalent to the scaling parameter $\varepsilon$ introduced in the multi-scale analysis) indicating deviation from the thermodymically converged state. This is illustrated by the deviation of the numerical interface thickness, starting at $T_r\approx0.98$ from the theoretical predictions. Lowering the value of $a$, i.e. rescaling the interface by a factor $1/\sqrt{a}$ and therefor lowering $\varepsilon$, it is observed that interface is again well-resolved and the scaling \eqref{eq:Wscaling} restored.

\subsubsection{Meanfield scaling laws: Surface tension}
\begin{figure}
	\centering
	\includegraphics{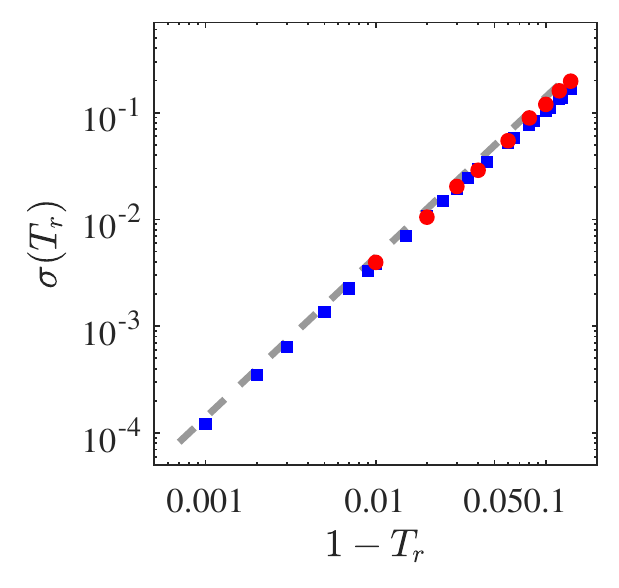}
	\caption{Temperature dependence of the surface tension coefficient near the critical point. Dashed grey line: Theory, Eq.~ \ref{eq:vdw_surface_tension}; Red circles: Simulation results using Laplace's law; Blue squares: surface tension coefficient computed for a flat interface. Results are taken from \cite{hosseini_towards_2021}.}
	\label{Fig:SurfTenTemp}
\end{figure}

Surface tension at liquid-vapour interface decreases with increasing temperature and vanishes at the critical point \citep{guggenheim_principle_1945}. For the van der Waals equation of state, the surface tension coefficient $\sigma$ follows a scaling law as $T_r\to 1$  \citep{van_der_waals_thermodynamische_1894,blokhuis_thermodynamic_2006},
\begin{equation}\label{eq:vdw_surface_tension}
	\sigma = \frac{16a}{27b^2}\sqrt{\frac{\kappa}{a}} {\left(1-T_r\right)}^{3/2}.
\end{equation}
A few publications have discussed the temperature scaling of the surface tension in the limit of a flat interface using the pseudo-potential method with a van der Waals equation of state~\cite{kupershtokh_equations_2009} and the Shan-Chen equations of state~\cite{lulli_mesoscale_2022} and the free energy approach with the van der Waals equation of state~\cite{hosseini_towards_2021}. All have shown that these models are able to recover the correct scaling with temperature close to the critical temperature. As an example the results reported in \cite{hosseini_towards_2021} are shown in Fig.~\ref{Fig:SurfTenTemp}.
It is clearly observed that the surface tensions agree very well with Eq.~\ref{eq:vdw_surface_tension}, provided that $W \ll R_{e}$. For larger interface thickness curvature-dependence come into play which will be discussed in the next section.

\subsubsection{Meanfield scaling laws: Tolman length\label{subsubsec:Tolman}}
In subsection~\ref{subsubsec:ev_surf_ten}, to measure surface tension using Laplace's law we made use of the equimolar dividing surface of radius $R_e$, Eq.~\ref{eq:equmol2Ddrop}. Further discussion on the non-uniqueness of the choice of dividing surface and curvature-dependence of surface tension is in order. Following \cite{gibbs_equilibrium_1874}, the free energy of a drop or bubble separated from the surrounding vapour or liquid by a dividing circle ($D=2$) or sphere ($D=3$) of length or area $\Sigma$ is, $A=U-TS+\sigma \Sigma$, where $U$ and $S$ are the internal energy and entropy of bulk phases while the last term is the adsorbance of free energy.
The equilibrium condition requires vanishing of the variation $\delta A$; for the isothermal case we have,
\begin{equation}
	\delta A=-P_{l,v}\delta V_{l,v}-P_{v,l}\delta V_{v,l} +\Sigma\delta \sigma+\sigma\delta \Sigma=0,
\end{equation}
where $P_{l,v}$ and $P_{v,l}$ are the pressures inside and outside the liquid drop or vapour bubble.
Using {$\delta V_{l,v}=-\delta V_{v,l}=2(D-1)\pi R^{D-1} \delta R$} and {$\delta \Sigma=2(D-1)^2\pi R^{D-2} \delta R$} leads to a generalized Laplace law,
\begin{equation}\label{eq:Laplace_gen}
	\Delta P = \frac{(D-1)\sigma(R)}{R} + \frac{d \sigma(R)}{dR}.
\end{equation}
The derivative of surface tension $d\sigma/dR$ is termed a \emph{notional} derivative by some authors \cite{blokhuis_pressure_1992} in order to stress that it refers to arbitrariness of the dividing surface. Apart from the equimolar surface \eqref{eq:equmol2Ddrop}, the \emph{surface of tension} is another possible choice to lift the ambiguity of the dividing surface. 
The notional derivative vanishes at the surface of tension,
\begin{equation}\label{eq:condition_Rs}
	\frac{d\sigma}{dR}\bigg|_{R=R_s}=0,
\end{equation}
thereby reducing the generalized Laplace law \eqref{eq:Laplace_gen} to a standard form,
\begin{equation}\label{eq:Laplace_surface_of_tension}
	\Delta P = \frac{(D-1)\sigma(R_s)}{R_s}.
\end{equation}
Integrating Eq.~\ref{eq:Laplace_gen} from $R_s$ to $R$, and eliminating $\Delta P$ using  \eqref{eq:Laplace_surface_of_tension}, one obtains  analytic expression for the notional surface tension $\sigma(R)$ relative to its minimum $\sigma_s$ at the surface of tension $R_s$,
\begin{equation}\label{eq:surface_of_tension_drop}
	\frac{\sigma(R)}{\sigma_s} = \frac{1}{D}{\left(\frac{R_s}{R}\right)}^{D-1} + \frac{D-1}{D}\left(\frac{R}{R_s}\right).
\end{equation}

While the notional derivative $d\sigma(R)/dR$ vanishes at the surface of tension, the same does not hold for the surface tension \emph{at} the surface of tension, $d\sigma_s/dR_s\neq0$. In other words,  surface tension $\sigma_s$ depends on the curvature of the surface of tension. In~\cite{tolman_effect_1949} the authors characterized the curvature-dependence of surface tension by the Tolman length: For sufficiently large $R_s$, the leading-order curvature-dependence of the surface tension may be written \citep{tolman_effect_1949},
\begin{equation}\label{eq:tolman_length}
	\sigma(R_s) \approx \sigma_0\left(1 \mp \frac{(D-1)\delta_T}{R_s}\right),
\end{equation}
where $\delta_T$ is the Tolman length and $\sigma_0$ is the flat interface surface tension coefficient. Here, the negative (positive) sign corresponds to drops (bubbles), respectively.

\begin{figure}
	\centering
	\includegraphics{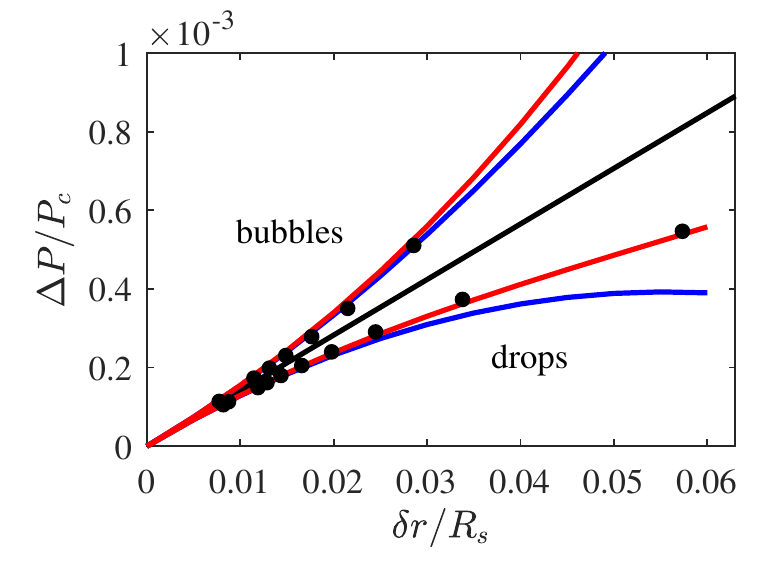}
	\caption{Pressure difference scaling with surface of tension radius $R_s$ for liquid drops and vapour bubbles. Symbol: Simulation data for van der Waals equation of state with $a=0.18$ and $b=0.095$. 
		Black line: Laplace law with $\sigma=\sigma_0$; Blue lines: Best fit with Eq.~\ref{eq:tolman_length} used to compute Tolman length $\delta_T$; Red lines: Best fit with the second-order Helfrich expansion Eq~\eqref{eq:Helfrich}. Plot reproduced from \cite{hosseini_towards_2021}.}
	\label{Fig:TolmanLaplace}
\end{figure}

\begin{figure}
	\centering
	\includegraphics{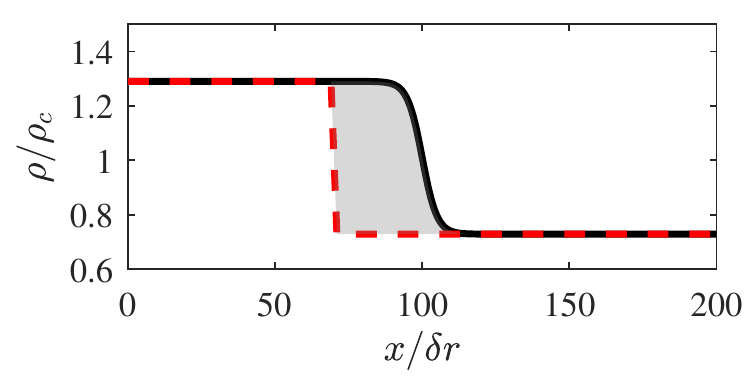}
	\caption{Example of mass adsorbance for a flat interface. Black continuous line: density profile at $T_r=0.98$;
		Red dashed line: Sharp interface profile with the dividing surface at $X/\delta r$=70. Grey area represents the mass   adsorbance $\Gamma(X)$ \eqref{eq:Gamma_flat}. Plot reproduced from \cite{hosseini_towards_2021}.}
	\label{Fig:DivSurfIllustration}
\end{figure}

\begin{figure}
	\centering
	\includegraphics{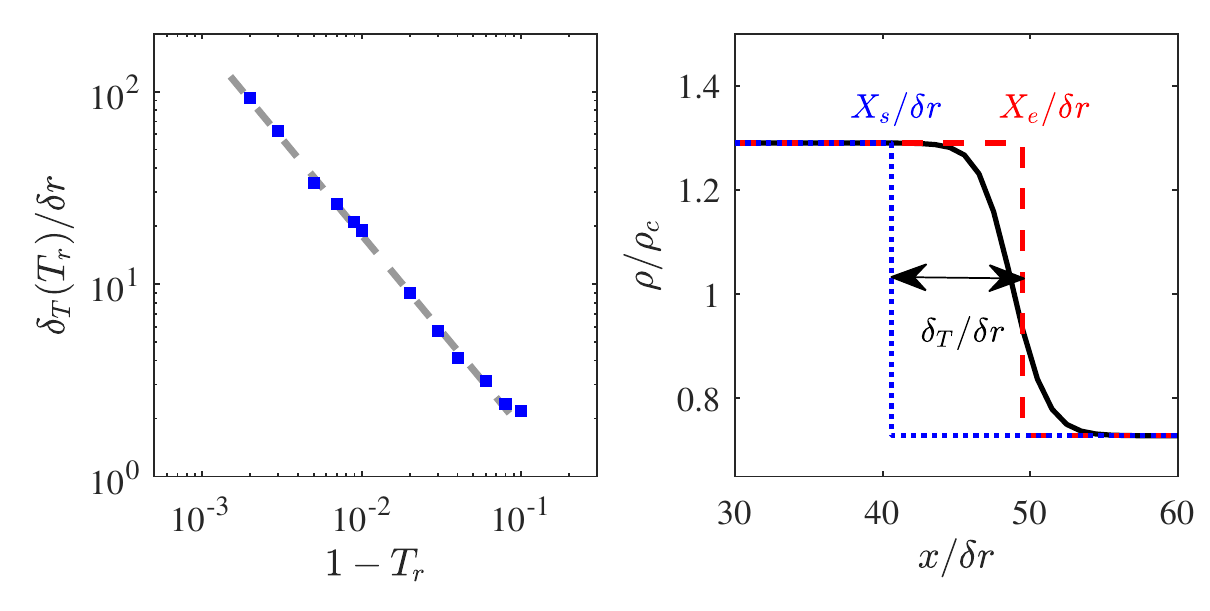}
	\caption{(Left) Temperature dependence of Tolman length $\delta_T$.
		Results from the flat interface simulation for van der Waals fluid are shown with blue squares while the grey dashed line represents  theoretical scaling \eqref{eq:TolmanT}. (Right) Surface of tension, equimolar surface and Tolman length for flat interface. Continuous black line: Density profile at $T_r=0.98$; Red dashed line:  Sharp approximation with the equimolar surface as the dividing surface; Blue dotted line:
		Sharp approximation with the surface of tension as the dividing surface. Distance between the surface of tension and the equimolar surface: Tolman length. In all simulations $a=0.18$ and $b=0.095$. Plot reproduced from \cite{hosseini_towards_2021}.}
	\label{Fig:TolmanTempScale}
\end{figure}

While the leading-order Tolman correction Eq.~\ref{eq:tolman_length} improves agreement with data at moderate $R_s$, it is not sufficient for smaller drops and bubbles. Higher-order terms in the inverse powers of $R_s$, important for droplets or bubbles of small radius, are neglected by Eq.~\ref{eq:tolman_length} and were addressed by \cite{helfrich_steric_1978}:
\begin{equation}\label{eq:Helfrich}
	\sigma = \sigma_0 \mp \sigma_0\frac{(D-1)\delta_T}{R_s} + \frac{k{(D-1)}^2}{2R_s^2} + \frac{\bar{k}(D-2)}{R_s^2} + \dots,
\end{equation}
where $k$ and $\bar{k}$ are the bending and Gaussian rigidities; note that the latter vanishes for $D=2$.\\
The curvature-dependence of surface tension and Tolman length have also been discussed in a limited number of articles using the lattice Boltzmann method~\cite{lulli_mesoscale_2022,hosseini_towards_2021}. The results of a systematic study of the Laplace law for droplets of different sizes as reported in \cite{hosseini_towards_2021} are shown in Fig.~\ref{Fig:TolmanLaplace}.
Comparing the leading-order Tolman model Eq.~\ref{eq:tolman_length} with the simulation. The values of $R_s$ are plotted against the pressure difference for different drop and bubble sizes in Fig.\ \ref{Fig:TolmanLaplace}. It is clear that, for smaller drops and bubbles, the pressure difference deviates from the Laplace law with constant $\sigma_s=\sigma_0$, indicating a curvature-dependent surface tension.
Fitting the data points with Eq.~\ref{eq:tolman_length}, the Tolman length can be extracted from the simulation. In Fig.~\ref{Fig:TolmanLaplace} $\delta_T=9\delta r$ for both drops and bubbles, at the reduced temperature $T_r=0.98$. Taking the second-order term of Eq.~\ref{eq:Helfrich} into account, the best fit in Fig.~\ref{Fig:TolmanLaplace} results in bending rigidity $k=1.049\times10^{5}\sigma_0\delta r^2$.\\
Another consequence of the second-gradient thermodynamics, as shown by \cite{blokhuis_pressure_1992}, is that the Tolman length scales with the reduced temperature as,
\begin{equation}\label{eq:TolmanT}
	\delta_T\propto{(1-T_r)}^{-1}.
\end{equation}
The scaling was also extracted in for the pseudo-potential and free energy models close to critical point~\cite{lulli_mesoscale_2022,hosseini_towards_2021}. There the Tolman length was extracted in the limit of a flat interface for different temperatures. The Tolman length in the limit of flat interface is the distance between the surface of tension and the equimolar surface~\cite{blokhuis_pressure_1992,blokhuis_thermodynamic_2006},
\begin{equation}\label{eq:Tolman_flat}
	\delta_T=X_e - X_s.
\end{equation}
An example from the simulation is presented in Fig.~\ref{Fig:TolmanTempScale}. In this case, location of the surface of tension $X_s$ can be found as the normalized first-order moment of the {normal stress difference}
\cite{rao_location_1979},
\begin{equation}\label{eq:surface_of_tension_flat}
	X_s = \frac{\int_{-\infty}^{\infty} x(P_{xx} - P_{yy}) dx}{\int_{-\infty}^{\infty} (P_{xx} - P_{yy}) dx}.
\end{equation}
With the dividing surface as a vertical straight line at $X$ in two dimensions, the mass adsorbance $\Gamma(X)$ is defined as (see example in Fig.\ \ref{Fig:DivSurfIllustration}),
\begin{equation}\label{eq:Gamma_flat}
	\Gamma(X)= \int_{-\infty}^{\infty} \left[\rho(x) - \rho_v - (\rho_l-\rho_v) H(x-X)\right] dx.
\end{equation}
Similar to the case of cylindrical symmetry considered in subsection~\ref{subsubsec:ev_surf_ten} the equimolar surface is found by annihilating the mass adsorbance, $\Gamma(X_e)=0$. The results as reported in \cite{hosseini_towards_2021} are shown in Fig.~\ref{Fig:TolmanTempScale}.
\par After a detailed overview of different classes of models, shortcomings and numerical artifacts, solutions proposed in the literature and a comprehensive assessment of interface properties the next section will discuss some of the applications of non-ideal lattice Boltzmann models.
\pagebreak
\section{Illustration of applications\label{sec:applications}}
Different from alternatives such as phase-field solvers or the color-gradient approach, non-ideal fluid LBMs have been mostly used for cases involving low or moderate Weber numbers. Below a few area where such solvers are widely in use are listed.
\subsection{Drop interaction with solid substrates}
One area of application where non-ideal fluid LBM has been widely and successfully applied is studies involving drop interaction with flat and complex solid substrates at low and moderate Weber numbers. We provide an overview of recent research in that area.
\subsubsection{Impact on non-wetting surfaces}
\begin{figure}[h!]
	\centering
	\includegraphics[scale=0.9]{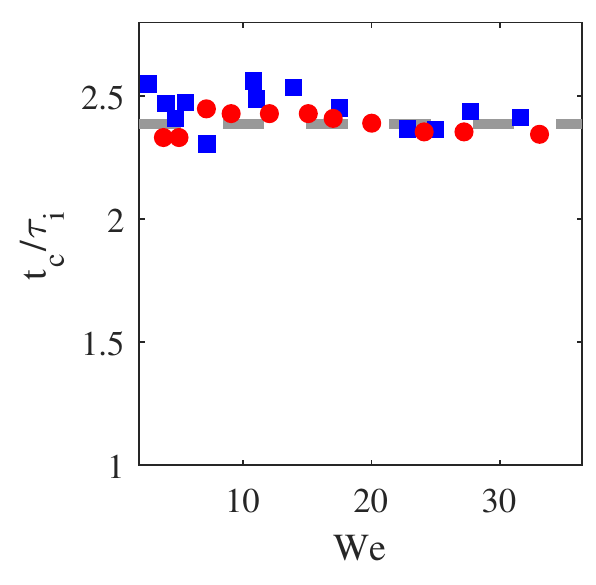}
	\caption{Drop contact times on flat non-wetting surface (contact angle $\theta$=165 for different Weber numbers as obtained from simulations and experiments. Simulations results are shown with red circular markers while experimental data reported by~\cite{gauthier_water_2015} are illustrated with blue square markers. The dashed grey line represents the average contact time as obtained from simulations, $\bar{t_c}/\tau_i=2.4$. This plot is reproduced from \cite{hosseini_towards_2021}.}
	\label{Fig:contact_times}
\end{figure}

Dynamics of drops impacting flat non-wetting surfaces is a topic that has attracted a lot of attention in recent years. Extensive studies have shown that the contact time on such surfaces is independent of the Weber number, ${\rm We}=\rho_l D_0 U_0^2/\sigma$, and only scales with the inertio-capillary time, $\tau_i=\sqrt{\rho_l D_0^3/8\sigma}$, meaning for a given initial diameter $D_0$ the contact time is unaffected by impact velocity~\cite{gauthier_water_2015,richard_bouncing_2000,quere_non-sticking_2005,richard_contact_2002} prompting researchers to proposed a simplistic description of its dynamic via an analogy with a single harmonic oscillator of characteristic mass $\rho_l D_0^3/8$ and spring constant $\sigma$~\cite{okumura_water_2003}. Although different from the case of an oscillating drop studied by Rayleigh, in the sense of asymmetry of the drop dynamic due to the presence of the wall, the scaling of the contact time agrees with his predictions. The coefficient in front of the scaling law is however observed to be different; Rayleigh's analysis led to a coefficient of $\pi/\sqrt{2}\approx 2.2$ while experimental studies of drops impacting non-wetting surfaces led to $2.5\pm0.2$~\cite{richard_contact_2002}. A number of LBM-based studies have looked into the Weber-independence of contact time on non-wetting surfaces, see for instance~\cite{mazloomi_moqaddam_drops_2017,hosseini_towards_2021}, and showed that non-ideal fluid LBMs correctly capture both the scaling and coefficient. An example in shown in Fig.~\ref{Fig:contact_times}.
\par
\begin{figure}[h!]
	\centering
	\includegraphics[width=11cm,keepaspectratio]{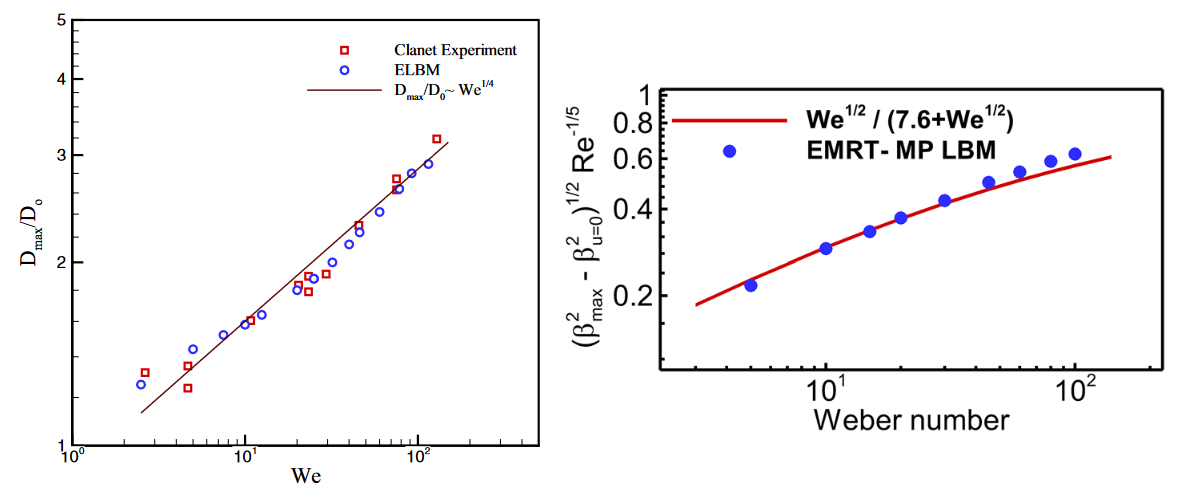}
	\caption{(Left) Reduced maximum diameter of drop as a function of the Weber number. Circles show LBM simulations and squares experimental data from \cite{clanet_maximal_2004}. The solid line corresponds to the scaling $\propto {\rm We}^{1/4}$. This plot is reproduced from \cite{mazloomi_entropic_2015-1}. (Right) Rescaled maximum spreading ratio for viscous drops as a function of Weber number. Circles: LBM simulations. solid line: scaling of Eq.~\ref{eq:max_spread_law_viscous}. This plot is reproduced from \cite{qin_entropic_2018}. }
	\label{Fig:max_spreading_diameter}
\end{figure}

 Another aspect of drops impact non-wetting surfaces, widely discussed in the literature is the diameter of the drop at the maximum spreading state and its dependence on the Weber and Reynolds numbers. A correlation widely used and accepted in the literature in the limit of vanishing Ohnesorge numbers, i.e. ${\rm Oh}=\sqrt{\rm We}/{\rm Re}$, is the one proposed in \cite{clanet_maximal_2004}: $\frac{D_{\rm max}}{D_0}\propto {\rm We}^{1/4}$. Other correlations taking into account viscous dissipation for non-vanishing Oh numbers have also been proposed in the literature, for instance~\cite{lee_universal_2016}:
\begin{equation}\label{eq:max_spread_law_viscous}
    \sqrt{{\left(\frac{D_{\rm max}}{D_0}\right)} - {{\left(\frac{D_{\rm max,0}}{D_0}\right)}}} {\rm Re}^{-1/5} = \frac{{\rm We}^{1/2}}{A + {\rm We}^{1/2}},
\end{equation}
where $D_{\rm max,0}$ is the maximum spreading diameter in the limit of zero impact velocity and $A$ is constant. Both these correlations have been matched with LBM simulations. Results are shown in Fig.~\ref{Fig:max_spreading_diameter}.
\subsubsection{Pancake bouncing}
\begin{figure}
	\centering
	\includegraphics[scale=0.55]{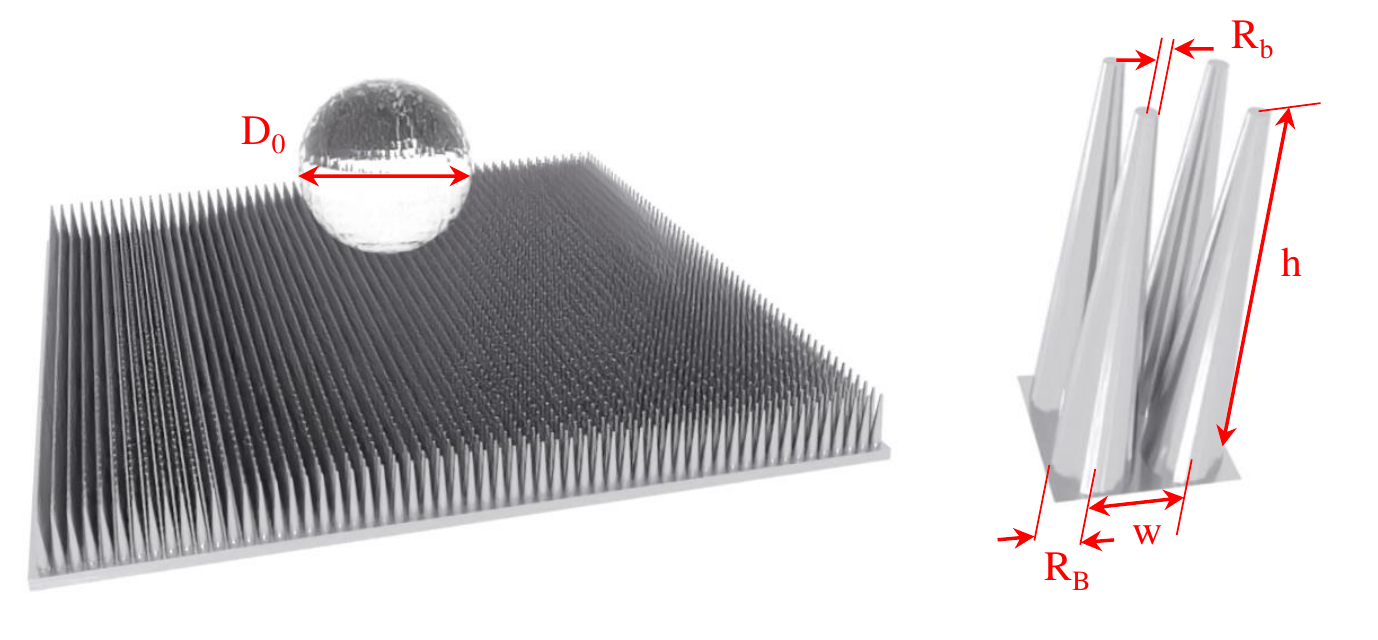}
	\caption{Illustration of the geometry of tapered posts. Image is reproduced from \cite{hosseini_towards_2021}.}
	\label{Fig:posts_geometry}
\end{figure}

\begin{figure}
	\centering
	\includegraphics[scale=0.6]{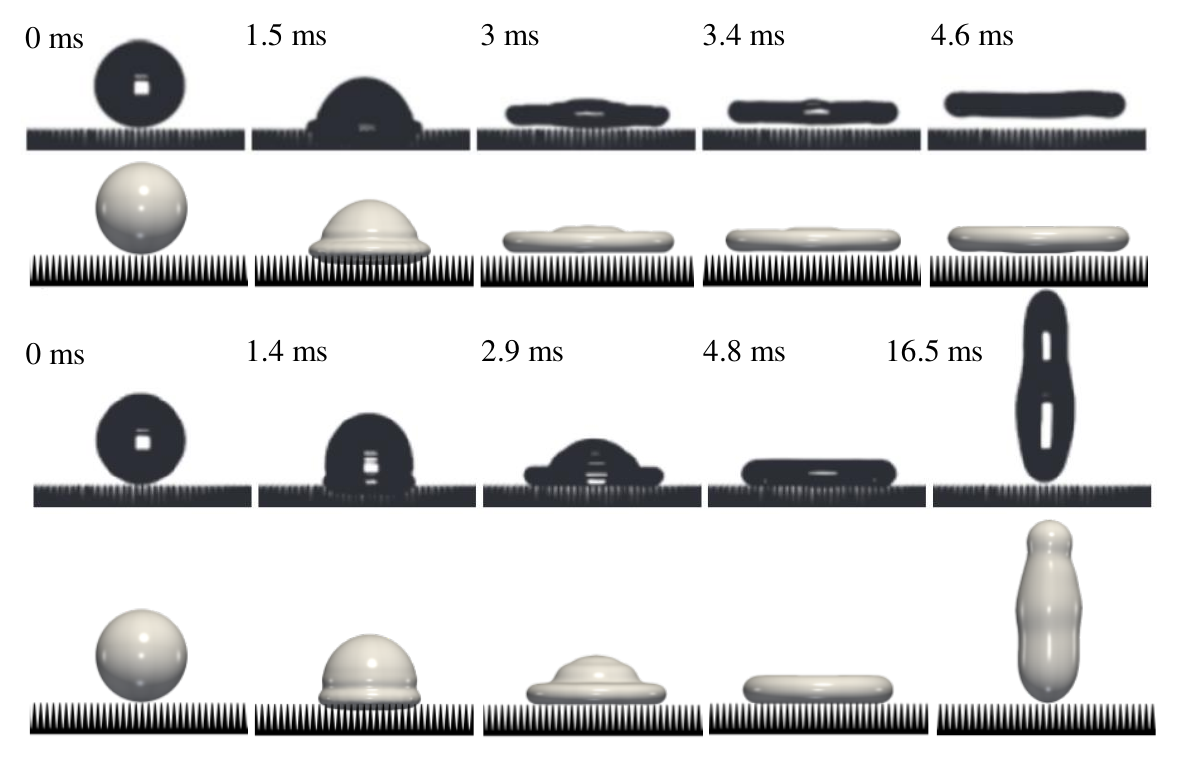}
	\caption{Drop impacting tapered posts at different Weber numbers (first and second rows) We=28.2 with pancake bouncing and (third and fourth rows) We=14.2. The first and third rows are experiments from~\cite{liu_pancake_2014} while the second and fourth rows are from simulations. This image is reproduced from \cite{hosseini_towards_2021}.}
	\label{Fig:PostImpactShots}
\end{figure}

\begin{figure}
	\centering
	\includegraphics[scale=0.75]{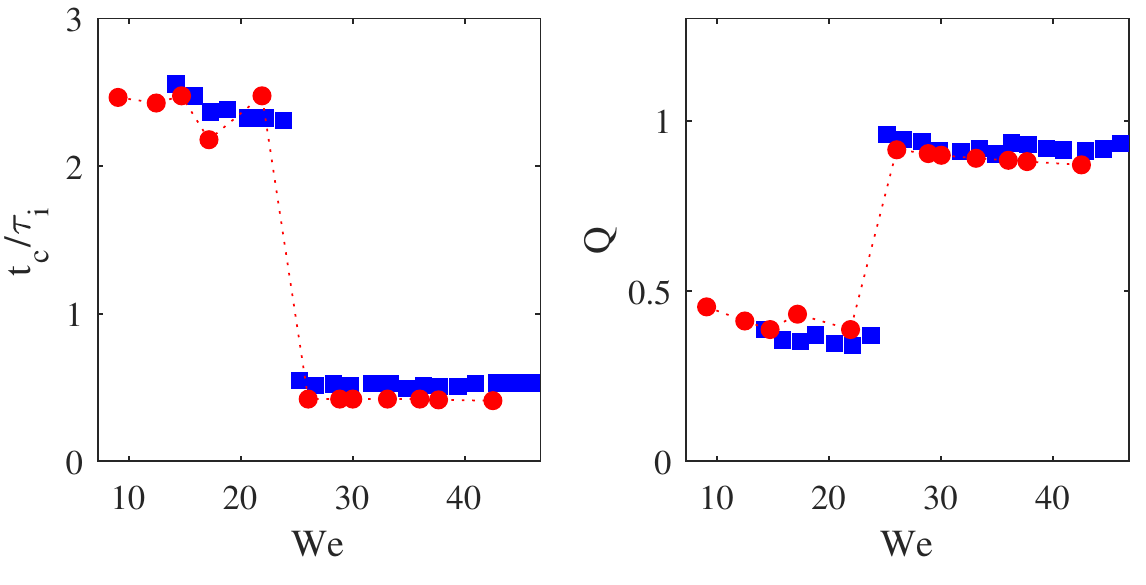}
	\caption{(left) Drop contact times and (right) pancake quality at rebound on tapered posts for different Weber numbers as obtained from simulations and experiments. Simulations results are shown with red circular markers while experimental data reported by~\cite{liu_pancake_2014} are illustrated with blue square markers. This plot is reproduced from \cite{hosseini_towards_2021}.}
	\label{Fig:contact_times_posts}
\end{figure}

To further reduce the drop contact times, a number of different strategies have been devised during the past decades. 
Recently, \cite{liu_pancake_2014} proposed to use macroscopic structures, in the form of tapered posts shown in Fig.~\ref{Fig:posts_geometry}, to reduce the contact time.
It has been shown that above a certain threshold Weber number these structures can decrease the contact time by approximately $75$ percent. This mechanism is also known as pancake bouncing, due to the pancake-like shape of the drop at take off. The dynamics of drops impacting these tapered posts both below and above the said-threshold is illustrated in Fig.~\ref{Fig:PostImpactShots}.
The first numerical study of this phenomenon was conducted using the LBM free energy method in \cite{moevius_pancake_2014} in 2-D. Later a detailed numerical study of pancake bouncing using LBM was presented in~\cite{mazloomi_moqaddam_drops_2017}, for density ratio of the order of $10^2$. Similar studies were also conducted in \cite{hosseini_towards_2021} for larger density ratios, i.e. $10^3$. All simulations were shown recover the drop contact time and capture the threshold Weber number were impact transitions in pancake mode.
The results are illustrated in Fig.~\ref{Fig:contact_times_posts}.
\subsubsection{Other approaches to reduce contact time via macro-structures}
\begin{figure}[h!]
	\centering
	\includegraphics[width=14cm,keepaspectratio]{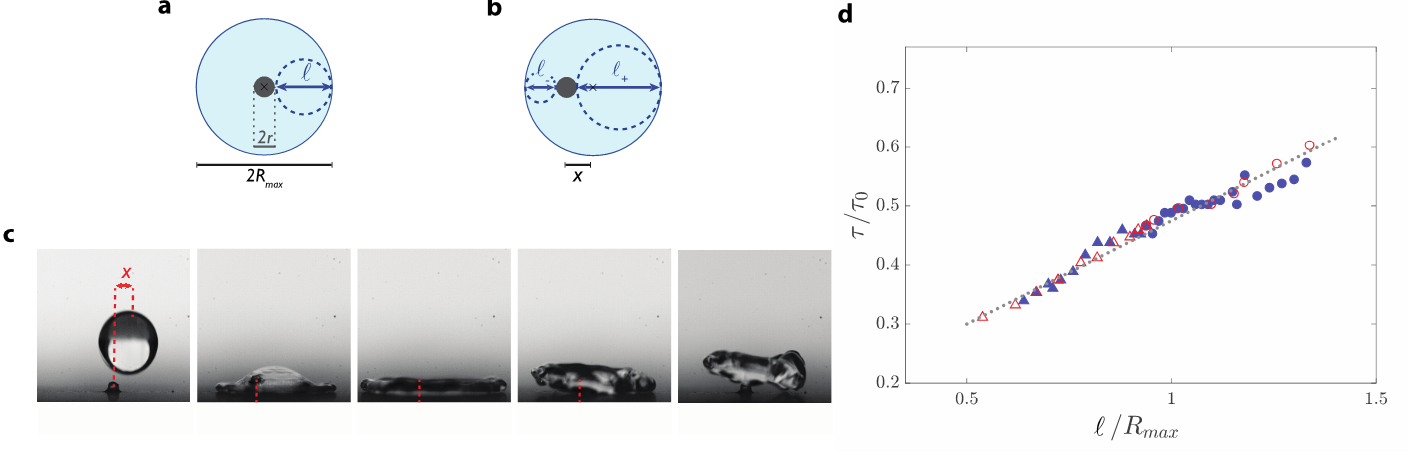}
	\caption{(a) Sketch of water droplet impacting a point-like defect at maximum spreading. (c) Water droplet (R = 1.3 mm) impacting at V = 1.3 m/s a bead with diameter r = 200 µm, with off-centering x = 0.7 mm. (d) Contact time normalized by inertio-capillary time as a function of the normalized blob size. Dots are experimental data and red circles simulations. These plots are reproduced from \cite{chantelot_water_2018}.}
	\label{Fig:drop_ring_bouncing}
\end{figure}

As another simple approach to further reduce the contact time of drops on non-wetting surfaces via the introduction of a singular defect on the substrate in the form of small glass bead of a radius of 200~$\mu$m. For a drop impact the bead at its center and subsequently breaking at it center right after maximum spreading this approach was shown to reduce the contact time by a factor of two~\cite{chantelot_water_2018}. The breaking of the lamella at its center after maximum spreading causes retraction both from the edges and the center eventually leading to a ring-like shape at the time of take off. The reduced take-off time is therefor related to a reduction in retraction time which is a consequence of the reduction of the corresponding characteristic size. The formation of two so-called \emph{blobs} and corresponding characteristic size are illustrated in Fig.~\ref{Fig:drop_ring_bouncing}. The effect of this characteristic size on contact time was extracted via systematic experimental runs and numerical simulations with a free energy LBM. As shown in Fig.~\ref{Fig:drop_ring_bouncing} simulation results were in very good agreement with experimental observations.
Non-ideal LBM solvers have been used for a wide variety of studies involving drop interaction with solids at low and moderate Weber numbers, such as impact on curved surfaces~\cite{liu_symmetry_2015}, mesh arrays~\cite{kooij_sprays_2019,wang_droplet_2020} or perforated flat substrates~\cite{wang_lattice_2020}.
\subsection{Flow in porous media}
Another area where non-ideal LBM solvers have been widely and successfully applied is flow in porous media. A wide variety of industrial processes and physics involve non-ideal fluids (or multi-phase flows) in porous media. The first attempt at modeling the flow of multiple phases in a realistic porous geometry was documented in \cite{martys_simulation_1996} where the authors modeled both wetting and non-wetting liquid invasion of a porous geometry extracted from high resolution microtomography images of a Fontainebleau sandstone. Since then non-ideal LBM's have been extended many more complexe configurations and physics. Here we briefly discuss two main areas, namely water transport in proton exchange membrane fuel cells and isothermal drying
\subsubsection{Water transport in proton exchange membrane fuel cells}
Proton-exchange membrane fuel cells are a promising class of fuel cells mainly developed for transport applications and first introduced in the 60's by General Electrics \cite{uttamchandani_wireless_2017}. While most major manufacturers are close to the commercialization stage, water management remains one of the outstanding issues limiting efficiency and durability. Issues related to water are encountered in the gas diffusion layer and reactant channel \cite{kandlikar_two-phase_2014}. A number of studies with LBM on liquid water transport mechanisms in both the gas diffusion layer and reactant channel have been conducted. In \cite{yang_improved_2021} the authors studied the effect of capillary pressure and contact angle on the water invasion dynamics into Toray-090 gas diffusion layers. The studied gas diffusion layers were made up of carbon fibers coated with Polytetrafluoroethylene (PTFE) for enhanced hydrophobicity with different weight percentages of PTFE.
\begin{figure}[h!]
	\centering
	\includegraphics[width=14cm,keepaspectratio]{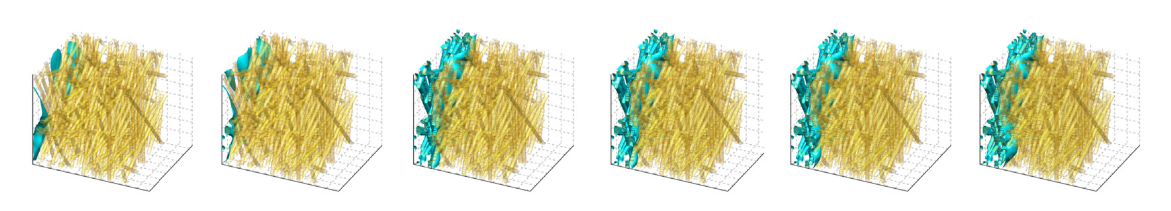}
	\caption{Snapshots of time evolution of liquid water transport inside GDL with 10 percent PTFE content. Image is reproduced from \cite{yang_improved_2021}. }
	\label{Fig:gdl_flow_water}
\end{figure}
In their studies the authors observed that saturation levels within the gas diffusion layer slowly increased with capillary pressure until a threshold were it grows and reaches rapidly full saturation. The threshold level was shown to change as function of the apparent contact angle controlled by the weight percentage of PTFE. The evolution of liquid water front within the gas diffusion layer over time is illustrated in Fig.~\ref{Fig:gdl_flow_water}.\\
Another aspect of water management studied in \cite{yang_dynamic_2021} is the transport dynamics of water droplets in the gas channel bound by the gas diffusion layer. Usually due to the roughness of the gas diffusion layer droplets moving in the gas channel can be subject to pining which can block the channel. The authors conducted systematic studies on the effect of inertia, considering both magnitude and direction.
\begin{figure}[h!]
	\centering
	\includegraphics[width=10cm,keepaspectratio]{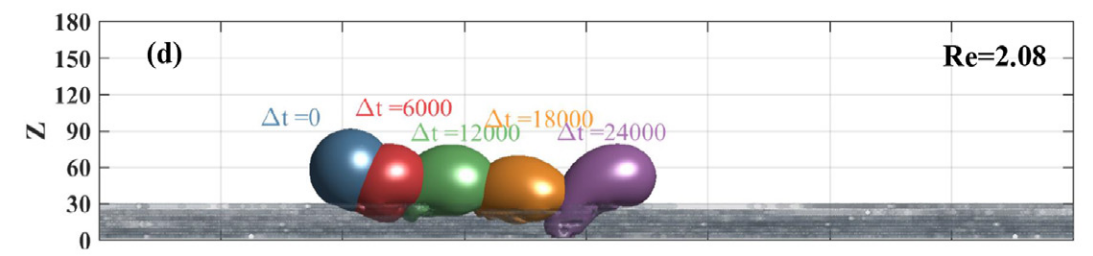}
	\caption{Image reproduced from \cite{yang_dynamic_2021}.}
	\label{Fig:gc_flow_water}
\end{figure}
As expected it was observed that smaller contact angles reduced pining effects and penetration into the gas diffusion layer. Inertia effect with a non-zero component normal to the gas diffusion layer surface was also observe to increase chances of penetration into the gas diffusion layer and even break-up of the drop as illustrated in Fig.~\ref{Fig:gc_flow_water}.\\
\subsubsection{Isothermal drying}
\begin{figure}[h!]
	\centering
	\includegraphics[width=10cm,keepaspectratio]{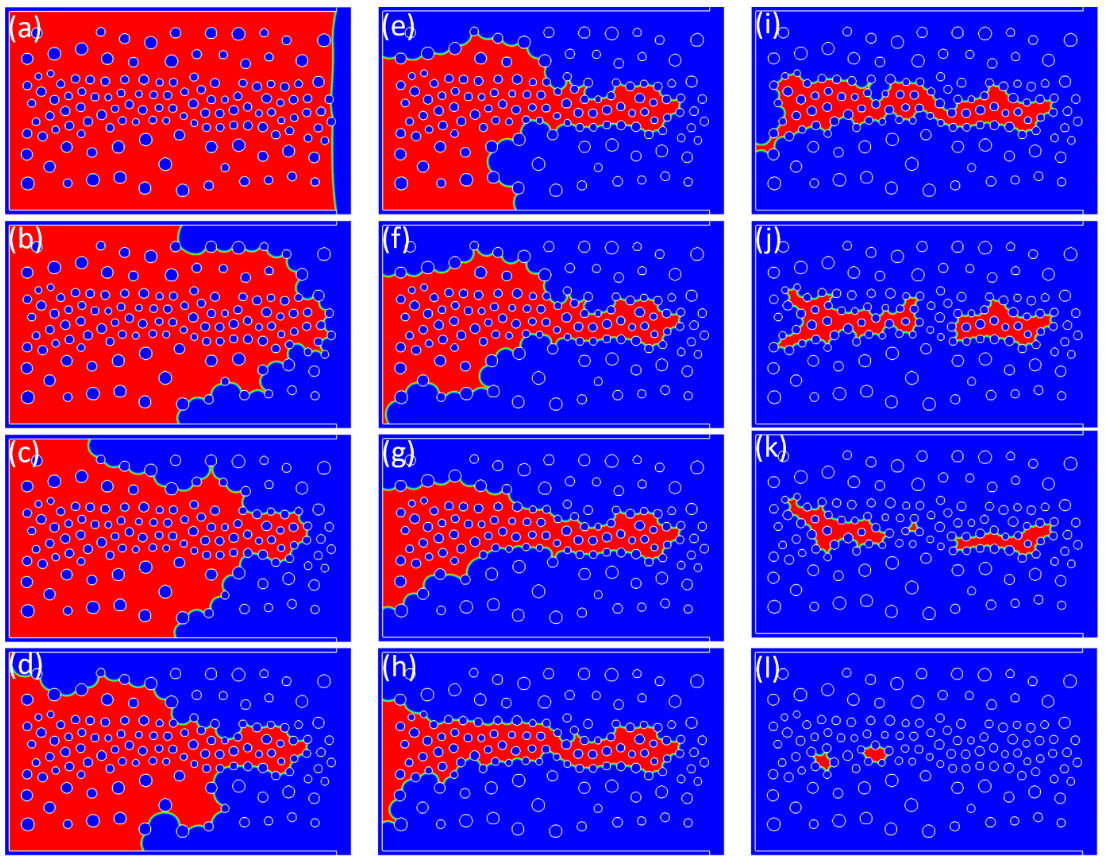}
	\caption{Sequential liquid configurations during drying of a bi-modal porous system obtained with LBM. Image reproduced from \cite{qin_lattice_2021}.}
	\label{Fig:drying_bimodal_porous_media}
\end{figure}

\begin{figure}[h!]
	\centering
	\includegraphics[width=10cm,keepaspectratio]{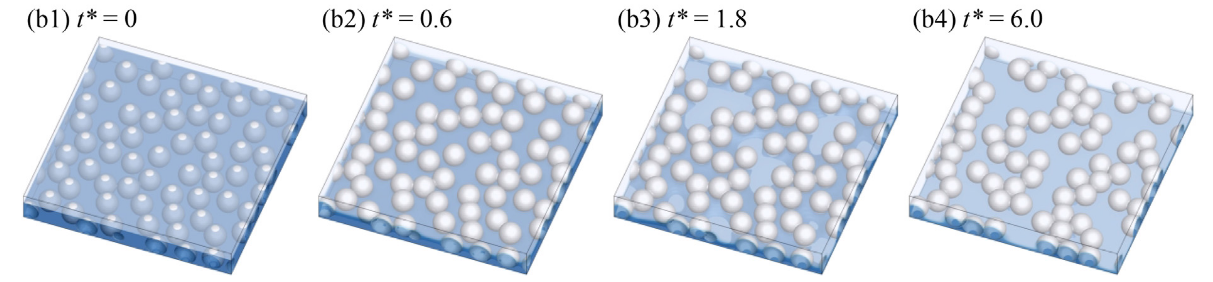}
	\caption{Snapshots of particle arrangement during the drying of a colloidal suspension containing 60 particles with contact angle 90. Image reproduced from \cite{mino_numerical_2022}.}
	\label{Fig:particle_drying}
\end{figure}

Drying in complex porous media is encountered in many set-ups of interest in science and engineering, such as food preservation, coating or volatile hydrocarbons recovery from reservoirs. Traditional description of the drying process at
the continuum scale rely on phenomenological closures that come with many limitations. More detailed descriptions like the Pore Network models have, to some extent, helped to refine the continuum models however they have shortcomings in using the true geometry at the pore scale, and in considering capillary instabilities and film-effects. That is why in the past 30 years, pore-scale studies of drying in porous media have received more attention. On par with global trend, this topic has also been widely considered and studied with the LBM. For instance, in \cite{zhao_drying_2021} and \cite{qin_lattice_2021,zhao_pore-scale_2022} the authors investigated the effect of contact angle hysteresis on isothermal drying dynamics in porous media and proposed a hybrid solver coupling pore-scale lattice Boltzmann simulations to the pore network model. An illustration of the simulation conducted in \cite{qin_lattice_2021} is given in Fig.~\ref{Fig:drying_bimodal_porous_media}.
In \cite{mino_numerical_2022}  the authors coupled the free energy model to a smoothed-profile LBM to model isothermal drying of colloidal suspensions and studied the aggregation of colloidal particles with different wetabilities under isothermal drying. The simulations are illustrated in Fig.~\ref{Fig:particle_drying}.
The authors observed that particles with higher wettabilities aggregated more slowly, but more significantly. They justified the observations by capillary interactions such as capillary flotation and immersion forces between the particles. 
\par As shown in this section, the non-ideal LBMs have been used for a wide variety of applications covering drop interaction with different substrates, liquid invasion of porous media with a special interest for fuel cells and drying in porous geometries. The specific numerical properties of this class of model have made them quite efficient and attractive for flows involving interaction with complex solids, liquid-liquid interactions such as coalescence and binary collisions, all in the low and mid-Weber and Reynolds regimes. Considerable effort is still being put on extending such approaches to larger Weber number configurations such as droplet splashing or primary and secondary break-up in liquid jets. At this point in time, to the authors knowledge, no notable success has been reported in that area.\\
\par Non-ideal fluid models in LBM are not limited to single component isothermal flow configurations. Considerable effort has been put in recent years to extend these models to thermal and multi-component flows. Some of these extensions will be briefly discussed in the next section

\pagebreak
\section{Extension to more complex physics\label{sec:comp_and_multcomp}}
The non-ideal LBM, while initially developed for isothermal single-component fluids, have been extended to thermal and multi-component flows. Below we present a brief overview of some of the developements in that area.
\subsection{Thermal flows with evaporation}
Extension of non-ideal LBMs to non-isothermal flows can be categorized as pertaining to one of two main classes: (a) Hybrid and passive scalar-based approaches and (b) kinetic methods.
\subsubsection{The hybrid and passive scalar approaches}
\paragraph{The form of the energy balance equation}
Two main routes have been taken to arrive at the final form of the energy/temperature balance equation, namely from the entropy or the energy balance equations.
In the hybrid approach the LBM solver for the density and momentum fields is coupled to classical, either finite differences or finite volumes, solvers for the following evolution equation for the temperature field~\cite{li_lattice_2015}:
\begin{equation}
    \partial_t T + \bm{u}\cdot\bm{\nabla}T + \frac{T}{\rho c_v} \frac{\partial P}{\partial T} \bm{\nabla}\cdot\bm{u} - \frac{1}{\rho c_v}\bm{\nabla}\cdot\lambda\bm{\nabla}T= 0.
\end{equation}
\paragraph{Solver for the energy balance equation}
Once the target form of the energy balance equation has been determined an additional solver has to be added/coupled to the flow solver. There are typically two possible routes: (a) LBM advection-diffusion type solvers or (b) classical finite differences or finite volumes solvers.\\
In the first approach, one relies on an additional distribution function $g_i$ to solve an advection-diffusion type partial differential equation. In its simplest form the advection-diffusion LBM with~\cite{biferale_convection_2012}:
\begin{equation}
    \sum_i g_i = T,
\end{equation}
as the conserved moment leads to a PDE of the following form:
\begin{equation}
    \partial_t T + \bm{\nabla}\cdot \bm{u} T - \bm{\nabla}\cdot D \bm{\nabla} T = 0.
\end{equation}
This form of the balance equation poses a number of issues; The advection term $\bm{u}\cdot\bm{\nabla}T$ is only valid in the limit of a divergence-free velocity field, i.e. $T\bm{\nabla}\cdot\bm{u}$, which given that the compressible fluid equations are targeted does not hold in general. Furthermore, the diffusion term $\bm{\nabla}\cdot D\bm{\nabla}T$ with $D=\frac{\lambda}{\rho c_v}$ only holds for $\rho={\rm const}$ and $c_p={\rm const}$. Finally the latent heat release term in Eq.~() is missing here.\\
To circumvent these shortcomings different authors have proposed corrections to be included as source terms in the LBM equation~\cite{hazi_modeling_2008,hazi_bubble_2009,gong_lattice_2012,wang_thermal_2022}. For instance, noting:
\begin{equation}
    \bm{u}\cdot\bm{\nabla}T - \bm{\nabla}\cdot\bm{u}T = -T\bm{\nabla}\cdot\bm{u},
\end{equation}
and,
\begin{equation}
    \frac{1}{\rho c_v}\bm{\nabla}\cdot\lambda\bm{\nabla}T - \bm{\nabla}\cdot\frac{\lambda}{\rho c_v}\bm{\nabla}T = -\lambda \bm{\nabla}T\cdot\bm{\nabla}\frac{1}{\rho c_v},
\end{equation}
a source term of the following form must be included:
\begin{equation}
    S = T\left(1 - \frac{1}{\rho c_v}\frac{\partial P}{\partial T}\right)\bm{\nabla}\cdot\bm{u} + \lambda \bm{\nabla}T\cdot\bm{\nabla}\frac{1}{\rho c_v},
\end{equation}
leading to the following LBM equation:
\begin{equation}
    \bar{g}_i(\bm{r}+\bm{c}_i\delta t, t+\delta t) - \bar{g}_i(\bm{r}, t) = \frac{1}{\bar{\tau}}\left(g_i^{\rm eq}(\bm{r}, t) - \bar{g}_i(\bm{r}, t)\right) + \left(1-\frac{1}{2\bar{\tau}}\right) w_i S,
\end{equation}
with:
\begin{equation}
    T = \sum_i \bar{g}_i + \frac{\delta t}{2}S.
\end{equation}
An alternative to the above-described approach, discussed in \cite{li_effect_2014} is to set:
\begin{equation}
    \sum_i g_i = \rho c_v T,
\end{equation}
with
\begin{equation}
    g_i^{\rm eq} = c_v T f_i^{\rm eq},
\end{equation}
which in combination with a correction term of the form:
\begin{equation}
    S = \frac{c_v T}{c_s^2}\bm{c}_i\cdot\bm{F}, 
\end{equation}
where $\bm{F}$ is the non-local force in the evolution equation of the distribution function $f_i$ results in the following conservative form of the energy balance equation:
\begin{equation}
    \partial_t \rho c_v T + \bm{\nabla}\cdot\rho c_v T\bm{u} - \bm{\nabla}\cdot\lambda\bm{\nabla}T = 0.
\end{equation}
Comparing this equation to the target balance law for non-ideal fluids one observes that the latent heat release due to phase change is missing here.\\
Different variants of the passive scalar approach have been used in a wide number of publications targeting thermal non-ideal fluid flows \cite{hazi_modeling_2008,hazi_bubble_2009,wang_thermal_2022}.\\
An alternative to this approach is to use classical discretization  methods such as finite differences or finite volumes to solve the energy balance equation. One of the first documented attempts at modeling thermal non-ideal fluid flows using such a hybrid model was reported in~\cite{zhang_lattice_2003}. In recent years most authors have opted for a second-order central isotropic finite differences discretization in space, see \cite{kumar_isotropic_2004,thampi_isotropic_2013,ramadugu_lattice_2013,banerjee_isotropic_2017} for more detail, and second-order Runge-Kutta scheme for time-stepping. The hybrid scheme is widely used in the literature for a variety of configurations and physics, see for instance \cite{qin_study_2019,wang_droplet_2022,fei_droplet_2022,fei_mesoscopic_2020,saito_lattice_2021}.
\subsubsection{The kinetic route}
\paragraph{Introduction of generic framework}
We will first introduce elements of a generic kinetic framework for non-ideal fluids. This model is introduced for the first time here and will be detailed in upcoming publications. The kinetic framework of Eq.~\ref{eq:final_kinetic_model} introduced for iso-thermal flows can be readily extended to compressible flows with energy balance by adding energy to the list of conserved variable in the definition of the projector onto the local equilibrium manifold:
     \begin{equation}
         \mathcal{K}\mathcal{J} = \frac{\partial f^{\rm eq}}{\partial \Pi_{\rho}}\int\mathcal{J}d\bm{v}d\mathcal{I} + \frac{\partial f^{\rm eq}}{\partial \Pi_{\bm{u}}}\int\bm{v}\mathcal{J}d\bm{v}d\mathcal{I} +\frac{\partial f^{\rm eq}}{\partial \Pi_{\bm{\mathcal{E}}}}\int\left(\bm{v}^2+\mathcal{I}^2\right)\mathcal{J}d\bm{v}d\mathcal{I},
     \end{equation}
with
    \begin{equation}
        \Pi_{\rho}=\rho,~\Pi_{\bm{u}}=\rho\bm{u},~\Pi_{\mathcal{E}}=\rho\left[\bm{u}^2+(D+\delta)RT\right],
    \end{equation}
where to allow for variable specific heat ratios we have introduced the space of additional non-translational degrees of freedom $\mathcal{I}$. The equilibrium distribution function is modified accordingly as:
    \begin{equation}\label{eq:LM}
    	f^{\rm eq}=\frac{\rho}{\left(2\pi RT\right)^{(D+\delta)/2}}\exp\left[-\frac{(\bm{v}-\bm{u})^2+\mathcal{I}^2}{2RT}\right].
    \end{equation}
Here $\delta$ is the the number of non-translational degrees of freedom. Application of this projector to the collision term results in:
    \begin{equation}\label{eq:final_non_kinetic_contributions_comp}
    \mathcal{J}_{\rm nloc} = \left(
         \frac{1}{\rho}\frac{\partial f^{\rm eq}}{\partial \bm{u}}
         - \frac{2}{(D+\delta)\rho}\bm{u}\frac{\partial f^{\rm eq}}{\partial RT} \right)\cdot\bm{F}_{\rm nloc} + \frac{1}{\rho(D+\delta)}\frac{\partial f^{\rm eq}}{\partial RT} Q_{\rm nloc}
    \end{equation}
where the force $\bm{F}_{\rm nloc}$ reads,
    \begin{equation}
    	\bm{F}_{\rm nloc}= -\int\int\int \bm{\nabla}V\left(\lvert \bm{r}-\bm{r}_1\rvert\right)f_{2}(\bm{r},\bm{v}, \bm{r}_1,\bm{v}_1,t) d\bm{v}_1d\bm{r}_1d\bm{v},
    	\label{eq:force_general_comp}
    \end{equation}
where 
    \begin{equation}
        f_{2}(\bm{r},\bm{v}, \bm{r}_1,\bm{v}_1,t)=\int\int f_{2}(\bm{r},\bm{v}, \mathcal{I}, \bm{r}_1,\bm{v}_1,\mathcal{I}_1, t)d\mathcal{I}_1d\mathcal{I},
    \end{equation}
and
    \begin{equation}
    	Q_{\rm nloc}= -2\int\int\int \bm{\nabla}V\left(\lvert \bm{r}-\bm{r}_1\rvert\right)\cdot\bm{v}f_{2}(\bm{r},\bm{v}, \bm{r}_1,\bm{v}_1,t) d\bm{v}_1d\bm{r}_1d\bm{v}.
    	\label{eq:energy_general_comp}
    \end{equation}
Collecting the BGK approximation together with the nonlocal contribution, a generic model for the BBGKY equation may be written,
    \begin{equation}\label{eq:final_kinetic_model_comp}
	    \partial_t f + \bm{v}\cdot\bm{\nabla} f = -\frac{1}{\tau}\left(f - f^{\rm eq}\right) + \mathcal{F} + \mathcal{Q},
    \end{equation}
with:
    \begin{equation}
        \mathcal{F} = \left(
         \frac{1}{\rho}\frac{\partial f^{\rm eq}}{\partial \bm{u}} 
         - \frac{2}{\rho(D+\delta)}\bm{u}\frac{\partial f^{\rm eq}}{\partial RT} \right)\cdot\bm{F}_{\rm nloc}
    \end{equation}
and:
    \begin{equation}
        \mathcal{Q} = \frac{1}{\rho(D+\delta)}\frac{\partial f^{\rm eq}}{\partial RT} Q_{\rm nloc}.
    \end{equation}
The hard-sphere potential contribution is readily shown to lead to:
    \begin{equation}
        \bm{F}_{\rm nloc} = \int \bm{v} \mathcal{J}_{\rm E}^{(1)} d\bm{v} = -\bm{\nabla}b\rho^2\chi RT,
    \end{equation}
and
    \begin{equation}
        Q_{\rm nloc} = -2\bm{\nabla}b\rho^2\chi R T\cdot\bm{u},
    \end{equation}
while the meanfield Vlasov long-range interaction leads to:
    \begin{equation}
        \mathcal{J}_{\rm V} = -\bm{\nabla}\left[2 a \rho(\bm{r}) +\kappa \bm{\nabla}^2 \rho(\bm{r})\right]\cdot\frac{\partial}{\partial\bm{v}}f(\bm{r},\bm{v}).
    \end{equation}
Upon application of the projector operator and addition of the Enskog contribution one gets:
    \begin{equation}
        \bm{F}_{\rm nloc} = -\bm{\nabla}b\rho^2\chi RT + \rho \bm{\nabla}\left[2 a \rho +\kappa \bm{\nabla}^2 \rho\right],
    \end{equation}
and
    \begin{equation}
        Q_{\rm nloc} = -2\bm{\nabla}b\rho^2\chi R T\cdot\bm{u} + 2\rho\bm{u}\cdot\bm{\nabla}\left[2 a \rho +\kappa \bm{\nabla}^2 \rho\right].
    \end{equation}
Upon application of a multi-scale perturbation analysis the generic model can be shown to recover the following system of equation up to NS level:
    \begin{align}
        &  \partial_t\rho+\bm{\nabla}\cdot\rho \bm{u} + \mathcal{O}(\epsilon^3) = 0,\\
        &  \partial_t\rho \bm{u}+\bm{\nabla} \cdot\rho\bm{u}\otimes \bm{u} + \bm{\nabla}\cdot \rho RT\left(1+b\chi\rho-\frac{a\rho}{RT}\right)\bm{I} 
        - \rho \kappa \bm{\nabla} \bm{\nabla}^2 \rho 
         + \bm{\nabla}\cdot\bm{T}_{\rm NS} + \mathcal{O}(\epsilon^3)= 0\\
        & \partial_t \mathcal{E} + \bm{\nabla}\cdot\mathcal{E}\bm{u} 
        + \bm{\nabla}\cdot \rho RT(1+b\chi\rho) \bm{u} - \bm{u}\cdot\bm{\nabla} a\rho^2 - \kappa\rho \bm{u}\cdot\bm{\nabla} \bm{\nabla}^2 \rho 
        + \bm{\nabla}\cdot \bm{u}\cdot\bm{T}_{\rm NS} - \bm{\nabla} \cdot\lambda \bm{\nabla} T + \mathcal{O}(\epsilon^3) = 0.
    \end{align}
Defining the total energy $\mathcal{E}_t$ as the sum of the internal energy $\mathcal{E}$ and potential energy from non-local interaction potentials $\mathcal{E}_{\rm v}$ the following balance equation for total energy is obtained:
    \begin{multline}
        \partial_t \mathcal{E}_t + \bm{\nabla}\cdot\mathcal{E}_t\bm{u} + \bm{\nabla}\cdot\rho RT\left(1+b\chi\rho-\frac{a\rho}{RT} - \frac{\kappa}{2RT}\bm{\nabla}^2 \rho\right)\bm{u} - \left(\kappa\rho\bm{\nabla}\bm{\nabla}\rho + \frac{\kappa}{2}\rho\bm{\nabla}^2\rho\right):\bm{\nabla}\bm{u} \\ + \bm{T}_{\rm NS}:\bm{\nabla} \bm{u} - \bm{\nabla} \cdot\lambda \bm{\nabla} T + \mathcal{O}(\epsilon^3) = 0.
    \end{multline}
In principle, as for the iso-thermal version introduced in \cite{hosseini_towards_2021}, different equations of state, interactions beyond the meanfield approximation and choices of pressure partition can be realized with this generic framework. For instance, introducing the full thermodynamic pressure into the equilibrium distribution function one readily recovers the models in \cite{martys_energy_1999,martys_aclassical_2001,reyhanian_thermokinetic_2020}.
\paragraph{The Enkog-Vlasov-based model of He \& Doolen}
One of the first attempts at proposing a kinetically motivated LBM model for non-ideal fluids was documented in \cite{he_thermodynamic_2002}. Following the original Enskog-Vlasov formalism the collision term is replaced with local and non-local contributions. Additionally the local contributions is replaced with a BGK-type approximation leading to the following discrete evolution equation:
\begin{equation}
    \bar{f}_i(\bm{r}+\bm{c}_i\delta t, t+\delta t) - \bar{f}_i(\bm{r}, t) = -\frac{\chi}{\bar{\tau}}\left( \bar{f}_i(\bm{r}, t) - \bar{f}_i^{\rm eq}(\bm{r}, t)\right) + \delta t(1-\frac{\delta t}{2\bar{\tau}})\mathcal{F}_i \bar{f}_i^{\rm eq}(\bm{r}, t),
\end{equation}
with,
\begin{multline}
    \mathcal{F}_i = \frac{(\bm{c}_i - \bm{u})\cdot(\bm{F} - \bm{\nabla}V)}{RT} \\- b\rho\chi\left[ (\bm{c}_i - \bm{u})\cdot\left( \bm{\nabla}\ln \rho^2 \chi T + \frac{3}{5}\left(\frac{{(\bm{c}_i - \bm{u})}^2}{2RT} - \frac{5}{2}\right)\bm{\nabla}\ln T\right) + \frac{2}{5}\left( \frac{(\bm{c}_i-\bm{u})\otimes(\bm{c}_i-\bm{u})}{RT}:\bm{\nabla}\bm{u} + \left( \frac{{(\bm{c}_i - \bm{u})}^2}{2RT} -\frac{5}{2} \right)\bm{\nabla}\cdot\bm{u}\right) \right].
\end{multline}
The discrete equilibrium distribution function was obtained as a second-order polynomial expansion. While in \cite{he_thermodynamic_2002} the authors only reported the theory and derivation of the model, it was later reprised in \cite{huang_mesoscopic_2021} and used for 1-D and 2-D simulations using a MRT realization.
\subsection{Multiple components}
Another area of active research to extend the range of application of non-ideal LBMs is the introduction of multiple components. While most of the application-oriented litterature in that area relies on the multi-component realization of the Shan-Chen model \cite{shan_multicomponent_1995}, a few different attempts at other models have also been reported. Here we will briefly discuss some of the multi-component models developed within that context.
\subsubsection{Ternary free energy model of W\"ohrwag et al.}
\begin{figure}[h!]
	\centering
	\includegraphics[width=6cm,keepaspectratio]{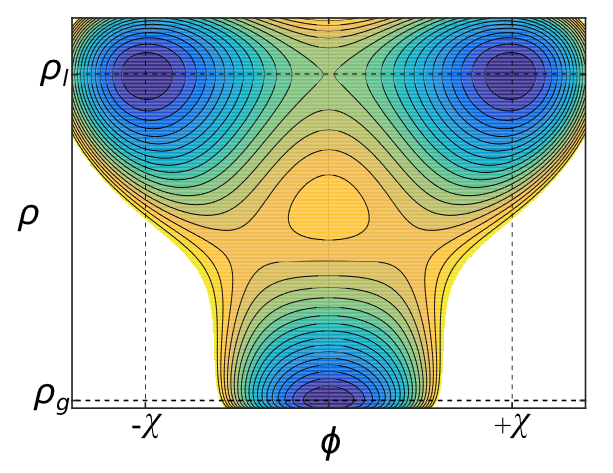}
	\caption{Contour plot of the bulk free energy density of Eq.~\ref{equ:BulkEnergy} as a function of two order parameters, $\rho$ and $\phi$. This plot is reproduced from \cite{wohrwag_ternary_2018}.}
	\label{Fig:ternary_free_energy}
\end{figure}

\begin{figure}[h!]
	\centering
	\includegraphics[width=8cm,keepaspectratio]{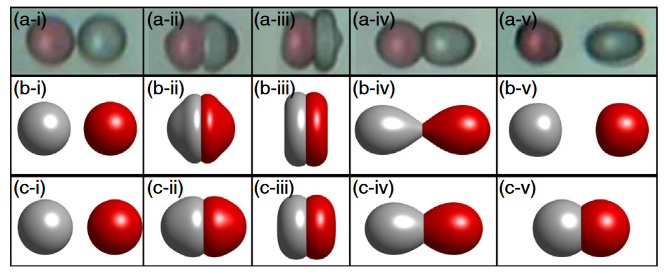}
	\caption{Collision between two immiscible droplets (water and diesel oil). (a) experiments from \cite{chen_collision_2006}. (b),(c) simulation results. This figure is reproduced from \cite{wohrwag_ternary_2018}.}
	\label{Fig:immiscible_collision}
\end{figure}

In an attempt to model ternary systems W\"ohrwag et al. \cite{wohrwag_ternary_2018} introduced a free energy functional allowing three distinct minima, corresponding to one gas and two liquid components. The thermodynamics of the system is governed by two order parameters, namely density, $\rho$, and a phase field, $\phi$. As for the van der Waals fluid the total free energy density consists of two parts, bulk $\mathcal{A}$ and interfacial; The bulk free energy is defined as:
\begin{equation}
\label{equ:BulkEnergy}
\mathcal{A} = \frac{\lambda_1}{2} (\mathcal{A}_{\rm EoS}(\rho) - \mathcal{A}_0) +  \frac{\lambda_2}{2} C_{l1}^2(1-C_{l1})^2 +\frac{\lambda_3}{2} C_{l2}^2(1-C_{l2})^2,
\end{equation}
where $\mathcal{A}_{\rm EoS}(\rho)$ can be derived from integrating 
the equation of state, $P=\rho(\partial \mathcal{A}_{\rm EoS} / \partial \rho)-\mathcal{A}_{\rm EoS}$, with coexisting liquid-gas densities at $\rho_l$ and $\rho_g$. The two last terms in Eq. ~\ref{equ:BulkEnergy} have the form of double well potentials with $C_{l1}$ and $C_{l2}$ the relative concentrations of the two liquid components. Each term has two minima at $C_\#=0$ (component absent) and $C_\#=1$ (present). The relative concentration of the gas phase is defined as $C_g= (\rho-\rho_l)/(\rho_g-\rho_l)$, which is $0$ 
for $\rho = \rho_l$ and $1$ for $\rho = \rho_g$. The relative concentrations are related to the density and phase field through
$C_{l1} = \frac{1}{2} \left[1+\phi/\chi- (\rho-\rho_l)/(\rho_g-\rho_l)\right]$ and $C_{l2} = \frac{1}{2} \left[1-\phi/\chi- (\rho-\rho_l)/(\rho_g-\rho_l)\right]$, with $\chi$ a constant scaling parameter for $\phi$. This form of the bulk free energy leads to three minima located at $(\rho_g,0)$, $(\rho_l,+\chi)$ and $(\rho_l,-\chi)$. The bulk free energy map is shown in Fig.~\ref{Fig:ternary_free_energy}.
For the interfacial free energy density, the authors used
\begin{equation}
 \label{equ:InterfaceEnergy}
 \mathcal{A}_{\rm Inter} = \frac{\kappa_1}{2} (\bm{\nabla} \rho)^2 
 + \frac{\kappa_2}{2}(\bm{\nabla} C_{l1})^2 
 +\frac{\kappa_3}{2} (\bm{\nabla} C_{l2})^2,
\end{equation}
which ca be written as a function of the order parameters $\rho$ and $\phi$,
\begin{equation}
 \mathcal{A}_{\rm Inter} = \left[ \frac{\kappa_1}{2} +  \frac{\kappa_2+\kappa_3}{8(\rho_g-\rho_l)^2} \right] (\bm{\nabla} \rho)^2 + \frac{\kappa_2 +\kappa_3}{8\chi^2}(\bm{\nabla} \phi)^2
 +\frac{\kappa_3 - \kappa_2}{4\chi(\rho_g-\rho_l)} (\bm{\nabla} \rho \cdot \bm{\nabla} \phi),
\end{equation}
which as expected includes three contributions, i.e. density gradient, phase field gradient and mixed density-phase field gradient. This new free energy functional only affects the NS equations through the total pressure tensor:
\begin{equation}
    \bm{\nabla}\cdot\bm{P}_{\rm tot} = \rho\bm{\nabla}\mu_{\rho} + \phi\bm{\nabla}\mu_{\phi},
\end{equation}
with the chemical potentials defined as:
\begin{equation}
    \mu_{k} = \frac{\delta }{\delta k}\int \mathcal{A}+\mathcal{A}_{\rm Inter} dV \lvert_{T,k'},
\end{equation}
with $k\in\{\rho,\phi\}$ and $k'\in\{\phi,\rho\}$. The corresponding system of balance equations was solved using two distribution functions; One for the density and momentum fields and one for the balance of the phase field, i.e. Cahn-Hiliard equation. Using this model the authors successfully modeled collision of drops of immiscible liquids with density ratios of the order of $10^3$. An example is shown in Fig.~\ref{Fig:immiscible_collision}.
\subsubsection{Mixtures with multi-component van der Waals equation of state}
In \cite{ridl_lattice_2018}, Ridl \& Wagner proposed a multi-component extension to the van der Waals second-gradient fluid. Starting with the single-component free energy:
\begin{equation}
    \mathcal{A}_{\rm vdW} = \rho R T \ln\left(\frac{\rho}{1-b\rho}\right) - a\rho^2 + \frac{\kappa}{2} {\lvert\bm{\nabla}\rho\lvert}^2,
\end{equation}
they proposed an N-component van der Waals free energy defined as~\cite{ridl_lattice_2018}:
\begin{equation}
    \mathcal{A}_{\rm vdW-MC} = \sum_{k=1}^{N} \rho_k R T \ln\left(\frac{\rho_k}{1-\sum_{k'=1}^N b_{k'}\rho_{k'}}\right) - \sum_{k'=1}^{N} \left(a_{kk'}\rho_k\rho_{k'} - \frac{\kappa_{kk'}}{2} \bm{\nabla}\rho_k\cdot\bm{\nabla}\rho_{k'}\right),
\end{equation}
with the total bulk thermodynamic pressure obtained as:
\begin{equation}
    P = \sum_{k=1}^N \frac{\rho_k R T}{1-\sum_{k'=1}^N b_{k'}\rho_{k'}} - \sum_{k'=1}^N a_{kk'}\rho_k \rho_{k'}.
\end{equation}
To model the dynamics of the corresponding system they proposed a set of $N$ discrete distribution functions, one for each component, evolving as~\cite{ridl_lattice_2018,li_symmetric_2007}:
\begin{equation}
    f_{ik}(\bm{r}+\bm{c}_i\delta t, t+\delta t) - f_{ik}(\bm{r}, t) = \frac{\delta t}{\bar{\tau}}\left(f_{ik}^{\rm eq}(\bm{r}, t) - f_{ik}(\bm{r}, t)\right) - \mathcal{F}_{ik},
\end{equation}
where the BGK collision operator accounts for the considered component's ideal contributions while the term $\mathcal{F}_{ik}$ takes into consideration non-ideal effects and interaction with other components. The densities and velocities of each individual component are obtained as zeroth- and first-order moments of the corresponding distribution functions while the mixture density and velocity are obtained as the sum of the density of all components and the mass-averaged velocities, respectively. The term $\mathcal{F}_{ik}$ contributes at first order,
\begin{equation}
    \sum_i \bm{c}_i \mathcal{F}_{ik} = \bm{F}_k,
\end{equation}
where the force $\bm{F}_k$ has two contributions: one from the gradient of chemical potential $\bm{F}_{\mu k}$ and one friction between different components $\bm{F}_{f k}$ modeling cross-diffusion effects. The chemical potential gradient contribution is readily obtained as difference between the single-component ideal pressure and multi-component non-ideal one while $\bm{F}_{f k}$ is defined as
\begin{equation}
    \bm{F}_{f k} = -\sum_{k'=1}^N \lambda_{kk'} \frac{\rho_k \rho_{k'}}{\rho_k + \rho_{k'}}\left(\bm{u}_{k'} - \bm{u}_k\right),
\end{equation}
where $\lambda_{kk'}$ are related to binary diffusion coefficients. It is readily demonstrated that under the assumption of symmetry of the diffusion coefficients tensor, i.e. $\lambda_{kk'}-\lambda_{k'k}$ the contribution of cross-diffusion to total momentum is zero:
\begin{equation}
    \sum_{k=1}^{N} \bm{F}_{f k} = 0.
\end{equation}
The authors showed that in the hydrodynamic limit under the assumption of $\lambda_{kk'}=\lambda$, i.e. all components have the same Lewis number, one recovers the N-component Cahn-Hiliard equations. It should also be noted that similar models for two-component non-ideal mixture were also proposed in \cite{guo_finite-difference-based_2005,guo_discrete_2003} where the models were shown to recover non-ideal effects and the Fick diffusion, equivalent to the Maxwell-Stefan system for binary isothermal mixtures.
\subsubsection{The multi-component pseudo-potential method}
While initially developed for non-ideal fluids, extension of the model to multiple components were introduced in \cite{shan_multicomponent_1995}. For a system with $N$ components, the interaction force on component $k$ is given by:
\begin{equation}
    \bm{F}_k = -\psi_k(\bm{r}) \sum_{k'=1}^{N} G_{kk'} \psi_{k'}(\bm{r}+\bm{c}_i\delta t) \bm{c}_i,
\end{equation}
where $G_{kk'}$ sets the interaction strength between components $k$ and $k'$ and forms a symmetrical tensor, i.e. $G_{kk'}=G_{k'k}$. Interestingly enough, upon application of a Taylor expansion the force is:
\begin{equation}
    \bm{F}_k = -\psi_k\sum_{k'=1}^{N} G_{kk'} \bm{\nabla}\psi_{k'} + \frac{G_{kk'}}{3} \bm{\nabla}\Delta \psi_{k'} + \mathcal{O}(\bm{\nabla}^5).
\end{equation}
As for the single-component non-ideal model, it leads to fixed surface tension coefficients and terms that are function of the pseudo-potential $\psi_k$ instead of density $\rho_k$. Introduction of independent capillary coefficients can be readily introduced following the approaches used for the single-component version of the model, i.e. multiple-range interactions.
\pagebreak
\section{Conclusion}
Application of the LBM to non-ideal fluids, and multi-phase flows in general, has been the focus of intense work from the early days of that method with first publications appearing in the early 90s. This rapid extension and growth in popularity is, in part, due to the simple transition from ideal to non-ideal fluids with interfaces in the context of LBM, illustrated best by the simplicity of the original pseudo-potential model.\\
Contrary to diffuse-interface models like the Allen-Cahn equation, the non-ideal LBM models come with interfacial properties, e.g. surface tension, interface thickness, Tolman length etc, linked to bulk physical properties of the considered fluid and follow the van der Waals thermodynamics of interfaces. The interface properties are therefore, in principle, dictated by the thermodynamics of the considered fluid. This was illustrated via the example of a ${\rm N}_2$ liquid/vapor interface showing a thickness of the order of $10^{-7}$~m, see Fig.~\ref{Fig:density_profile_N2}, limiting the grid-size and preventing simulations at larger realistic scales.
As a result they can be perceived as models able to cover a wider range of parameters as compared to classical sharp and/or diffuse interface methods: In the limit of thick interfaces where its width is comparable to the flow characteristic size they recover the van der Waals fluid thermodynamics (or a modified form of it in the case of the pseudo-potential method) and the Korteweg stress tensor at the macroscopic level and can be derived from the Enskog-Vlasov approximation at the level of the kinetic theory. In the limit of thin interfaces, the extreme computational cost imposed by the \emph{physical} thickness of the interface can be considerably reduced by relying on the principle of corresponding states.\\
Non-ideal LBM-based solvers are also subject to a number of issues such as stability resulting in limited range of accessible non-dimensional viscosities, thermodynamic inconsistency leading mismatch in coexistences densities at smaller temperatures, spurious currents appearing at interfaces as a result of the non-isotropy of the discrete system etc. A wide variety of solution to eliminate or reduced the effect of these issues have been proposed allowing simulation with larger density ratios, Weber and Reynolds number to be conducted. With the most recent development around these approaches they have become quite competitive and accurate tools for simulations in the lower Weber number regimes in areas such as microfluidics, drop interaction with solid surfaces etc. For much larger Weber numbers and applications such as multiphase jets with primary and secondary break-up further improvement is needed, especially to extend the range of accessible surface tensions to lower values at given interface thicknesses and acceptable levels of spurious currents. Given the kinetic roots of some of the models in that class of solvers, extension of the models to higher order physics beyond the meanfield approximation would also be a possible area of developmenet in the feature.

\section*{Acknowledgement}
This work was supported by European Research Council (ERC) Advanced Grant no. 834763-PonD (S.A.H and I.K.). Computational resources at the Swiss National Super Computing Center CSCS were provided under grant no. s1066.
\section*{Declaration of interests}
The authors report no conflict of interest.
\appendix
\section{Hermite expansion\label{App:Hermite}}
\noindent {\bf Single variable Hermite polynomials.}
The single variable Hermite polynomial $\mathcal{H}_n$ of order $n$ of a variable $x$ is defined as:
\begin{equation}
    \mathcal{H}_n(x)=\frac{{(-1)}^n}{w(x)}\frac{d^n}{dx^n}w(x),
\end{equation}
where the normalized function $w(x)$ is defined as:
\begin{equation}
    w(x)=\frac{1}{\sqrt{2\pi}}e^{-\frac{x^2}{2}}.
\end{equation}
Based on this definition, the first few polynomials can be computed as:
	\begin{subequations}
		\begin{align}
		\mathcal{H}_0 &= 1, \\
		\mathcal{H}_1 &= x, \\
		\mathcal{H}_2 &= x^2 - 1, \\
		\mathcal{H}_3 &= x^3 - 3x, \\
		\mathcal{H}_4 &= x^4 - 6x^2 + 3, \\
		\mathcal{H}_5 &= x^5 - 10x^3 + 15x, \\
		\mathcal{H}_6 &= x^6 - 15x^4 + 45x^2 - 15.
		\end{align}
	\end{subequations}
These polynomials are mutually orthogonal with respect to the weight function, $w(x)$, i.e.:
\begin{equation}
    \int^{+\infty}_{-\infty} \mathcal{H}_m(x)w(x)\mathcal{H}_n(x)dx=n!\delta_{mn},
\end{equation}
where $\delta_{mn}$ is the Kronecker delta function. Furthermore, they form a complete orthogonal basis of the Hilbert space of functions $f(x)$ satisfying:
\begin{align}
    \int^{+\infty}_{-\infty} {\lvert f(x)  \rvert}^2 w(x)dx<\infty.
\end{align}
As such, one can express the function $f(x)$ as:
\begin{equation}
    f(x) = \sum_{n=0}^{\infty}\frac{1}{n!}\bm{a}_n\mathcal{H}_n(x),
\end{equation}
where $\bm{a}_n$ is the order $n$ Hermite coefficient. Multiplying both sides by $\mathcal{H}_m(x)w(x)$ and integrating over $x$:
\begin{equation}
    \int^{+\infty}_{-\infty} \mathcal{H}_m(x)w(x)f(x)dx=  \sum_{n=0}^{\infty}\frac{1}{n!}\bm{a}_n \int^{+\infty}_{-\infty} \mathcal{H}_m(x)w(x)\mathcal{H}_n dx,
\end{equation}
and using the mutual orthogonality of Hermite polynomials, we get an expression for the Hermite coefficients as:
\begin{equation} \label{Eq:hermite_coefficient_single_var}
    \bm{a}_m = \int^{+\infty}_{-\infty} \mathcal{H}_m(x)w(x)f(x)dx.
\end{equation}
Alternatively, one can also expand the function $f(x)$ as:
\begin{equation}
    f(x) = w(x)\sum_{n=0}^{\infty}\frac{1}{n!}\bm{a}_n\mathcal{H}_n(x),
\end{equation}
resulting in the following expression for the coefficient $\bm{a}^{(m)}$:
\begin{equation} \label{Eq:hermite_coefficient_single_var_alt_form}
    \bm{a}_m = \int^{+\infty}_{-\infty} \mathcal{H}_m(x)f(x)dx.
\end{equation}
To better illustrate this, let us consider the example of the following function:
\begin{equation}\label{eq:sample_distribution}
    f(x) = \rho \frac{1}{\sqrt{2\pi\theta}} e ^{-\frac{{(x-u)}^2}{2\theta}}.
\end{equation}
This function can be shown to be square-integrable with respect to the previously-defined weight function. As such the corresponding Hermite coefficients can be computed through Eq.~\ref{Eq:hermite_coefficient_single_var_alt_form}:
\begin{equation}
    \bm{a}_m = \frac{1}{\sqrt{2\pi\theta}}\int^{+\infty}_{-\infty} \mathcal{H}_m(x) e ^{-\frac{{(x-u)}^2}{2\theta}}dx,
\end{equation}
which using the change of variable $\eta=(x-u)/\sqrt{\theta}$ can be re-written as:
\begin{equation}
    \bm{a}_m = \frac{1}{\sqrt{2\pi}}\int^{+\infty}_{-\infty} \mathcal{H}_m(\sqrt{\theta}\eta+u) e^{-\frac{{\eta}^2}{2}}d\eta.
\end{equation}
The different order coefficients can be easily evaluated using the following integral:
\begin{equation}
    \int_{-\infty}^{+\infty}x^{k}e^{-ax^2}dx = \begin{cases} 0 & k=2k^{'}+1 \\ \frac{(2k^{'}-1)!}{{(2a)}^{k^{'}}}\sqrt{\frac{\pi}{a}} & k=2k^{'} \end{cases},
\end{equation}
leading to the following expansion:
\begin{equation}
    f(x)=\sum_{n=0}^{\infty}\frac{\rho w(x)}{n!} \bm{a}_n\mathcal{H}_n(x).
\end{equation}
The first few terms are given in Table ~\ref{Table:hermite_coefficients_gauss_function}.
	\begin{table}[btp!]
		\centering
		\begin{tabular}{||c|c|c||} 
			\hline
			$n$ & $\bm{a}_n$ & $\mathcal{H}_n(x)$\\ 
			\hline \hline
			0 & 1 & 1 \\
			1 & $u$ & $x$ \\
			2 & $u^2 + (\theta-1)$ & $x^2-1$ \\
			3 & $ u^3 + 3u(\theta-1)$ & $x^3-3x$ \\
			4 & $u^4 + 6 u^2(\theta-1) + 3 {(\theta-1)}^2$ & $x^4 - 6x^2 + 3$ \\
			5 & $u^5 + 10 u^3(\theta-1) + 15 u{(\theta-1)}^2$ & $x^5 - 10x^3 + 15x$ \\
			6 & $u^6 + 15 u^4(\theta-1) + 45 u^2{(\theta-1)}^2 + 15{(\theta-1)}^3$ & $x^6 - 15x^4 + 45x^2 - 15$ \\
			7 & $u^7 + 21 u^5(\theta-1) + 105 u^3{(\theta-1)}^2 + 105u{(\theta-1)}^3$ & $x^7 - 21x^5 + 105x^3-105x$ \\
			8 & $\begin{array}{c}u^8 + 28 u^6(\theta-1) + 210 u^4{(\theta-1)}^2 + 420u^2{(\theta-1)}^3 \\ + 105{(\theta-1)}^4\end{array}$ & $\begin{array}{c}x^8 - 28x^6 + 210x^4-420x^2\\+105\end{array}$ \\
			9 & $\begin{array}{c}u^9 + 36 u^7(\theta-1) + 378 u^5{(\theta-1)}^2 + 1260u^3{(\theta-1)}^3 \\ + 945u{(\theta-1)}^4\end{array}$ & $\begin{array}{c}x^9 - 36x^7 + 378x^5-1260x^3\\+945x\end{array}$ \\
			\hline
		\end{tabular}
		\caption{Hermite polynomials and coefficients for the Gaussian distribution function}
		\label{Table:hermite_coefficients_gauss_function}
	\end{table}
\\ \noindent {\bf Multivariate Hermite polynomials.}
In a $D$-dimensional space the Hermite polynomial of order $n$ is defined as:
\begin{equation}
    \bm{\mathcal{H}_n}\left(\bm{x}\right) = \frac{{(-1)}^n}{w\left(\bm{x}\right)}\bm{\nabla}^n w\left(\bm{x}\right),
\end{equation}
where $\bm{\nabla}^n$ is the $n^\text{th}$ order derivative resulting in a tensor of rank $n$ and $w\left(\bm{x}\right)$ is the weight function defined as:
\begin{equation}
    w\left(\bm{x}\right) = \frac{1}{{2\pi^{D/2}}}e^{-\frac{\bm{x}^2}{2}}.
\end{equation}
Orthogonality of the multivariate Hermite polynomials results in:
\begin{equation}
    \int_{-\infty}^{+\infty}w\left(\bm{x}\right)\bm{\mathcal{H}_i}\left(\bm{x}\right):\bm{\mathcal{H}_j}\left(\bm{x}\right)d\bm{x} = \begin{cases} 0 & i\neq j \\ n!\bm{\delta_{ij}} & i=j \end{cases},
\end{equation}
where $\bm{i}$ and $\bm{j}$ are abbreviations for the set of indices \{$i_1,i_2,\dots,i_n$\} and \{$j_1,j_2,\dots,j_n$\} respectively, and $\bm{\delta_{ij}}$ is equal to unity if $\bm{i}$ is a permutation of $\bm{j}$ and zero otherwise. In a 3-D space the first few Hermite polynomials are computed as:
	\begin{subequations}
		\begin{align}
		\mathcal{H}_0 &= 1, \\
		\mathcal{H}_i &= x_i, \\
		\mathcal{H}_{ij} &= x_i x_j - \delta_{ij}, \\
		\mathcal{H}_{ijk} &= x_i x_j x_k - \left(\delta_{ij} x_k + \delta_{ik} x_j + \delta_{jk} x_i\right), \\
		\mathcal{H}_{ijkl} &= x_i x_j x_k x_l - \left( \delta_{ij} x_k x_l + \delta_{ik} x_j x_l + \delta_{il} x_j x_k + \delta_{jk} x_i x_l + \delta_{jl} x_i x_k + \delta_{kl} x_i x_j\right) \nonumber\\ &+ \left( \delta_{ij}\delta_{kl} + \delta_{ik}\delta_{jl} + \delta_{il}\delta_{jk} \right),\\
		\mathcal{H}_{ijklm} &= x_i x_j x_k x_l x_m - \left( \delta_{lm} x_i x_j x_k + \delta_{km} x_i x_j x_l + \delta_{kl} x_i x_j x_m + \delta_{jm} x_i x_k x_l \right. \nonumber\\ & \left. + \delta_{jl} x_i x_k x_m + \delta_{kl} x_i x_l x_m + \delta_{im} x_j x_k x_l + \delta_{il} x_j x_k x_m + \delta_{ik} x_j x_l x_m + \delta_{ij} x_k x_l x_m \right) \nonumber\\ &+ x_m\left( \delta_{ij}\delta_{kl} + \delta_{ik}\delta_{jl} + \delta_{il}\delta_{jk} \right) + x_l\left( \delta_{ij}\delta_{km} + \delta_{ik}\delta_{jm} + \delta_{im}\delta_{jk} \right) \nonumber\\ & + x_k\left( \delta_{ij}\delta_{lm} + \delta_{il}\delta_{jm} + \delta_{im}\delta_{jl} \right) + x_j\left( \delta_{ik}\delta_{lm} + \delta_{il}\delta_{km} + \delta_{im}\delta_{kl} \right) \nonumber\\ & + x_i\left( \delta_{jk}\delta_{lm} + \delta_{jl}\delta_{km} + \delta_{jm}\delta_{kl} \right),\\
		\mathcal{H}_{ijklmn} &= x_i x_j x_k x_l x_m x_n - \left(x_i x_j x_k x_l \delta_{mn} + x_i x_j x_k x_m \delta_{ln} + x_i x_j x_k x_n \delta_{lm} \right. \nonumber\\ & \left. + x_i x_j x_l x_m \delta_{kn} + x_i x_j x_l x_n \delta_{km} + x_i x_j x_m x_n \delta_{lk} + x_i x_k x_l x_m \delta_{jn} \right. \nonumber\\ & \left. x_i x_k x_l x_n \delta_{jm} + x_i x_k x_m x_n \delta_{jl} + x_i x_l x_m x_n \delta_{jk} + x_j x_k x_l x_m \delta_{in} \right. \nonumber\\ & \left. + x_j x_k x_l x_n \delta_{im} + x_j x_l x_m x_n \delta_{ik} + x_j x_k x_l x_n \delta_{im} + x_j x_k x_m x_n \delta_{il} + x_k x_l x_m x_n \delta_{ij}\right)\nonumber\\ & + x_i x_j\left(\delta_{kl}\delta_{mn} + \delta_{km}\delta_{ln} + \delta_{kn}\delta_{lm} \right) +  x_i x_k \left(\delta_{jl}\delta_{mn} + \delta_{jm}\delta_{ln} + \delta_{jn}\delta_{ml}\right) \nonumber\\ & + x_i x_l \left(\delta_{jk}\delta_{mn} + \delta_{jm}\delta_{kn} + \delta_{jn}\delta_{mk}\right) + x_i x_m \left(\delta_{jk}\delta_{ln} + \delta_{jl}\delta_{kn} + \delta_{jn}\delta_{lk}\right) \nonumber\\ & + x_i x_n \left(\delta_{jk}\delta_{lm} + \delta_{jl}\delta_{km} + \delta_{jm}\delta_{lk}\right) + x_j x_k \left(\delta_{in}\delta_{lm} + \delta_{il}\delta_{nm} + \delta_{im}\delta_{ln}\right) \nonumber\\ & + x_j x_l \left(\delta_{in}\delta_{km} + \delta_{ik}\delta_{nm} + \delta_{im}\delta_{kn}\right) + x_j x_m \left(\delta_{in}\delta_{kl} + \delta_{ik}\delta_{nl} + \delta_{il}\delta_{kn}\right) \nonumber\\ & + x_j x_n \left(\delta_{il}\delta_{km} + \delta_{ik}\delta_{lm} + \delta_{im}\delta_{kl}\right) + x_k x_l \left(\delta_{mn}\delta_{ij} + \delta_{mi}\delta_{nj} + \delta_{mj}\delta_{in}\right) \nonumber\\ & + x_k x_m \left(\delta_{ij}\delta_{ln} + \delta_{il}\delta_{jn} + \delta_{in}\delta_{lj}\right) + x_k x_n \left(\delta_{ml}\delta_{ij} + \delta_{mi}\delta_{lj} + \delta_{mj}\delta_{il}\right) \nonumber\\ & + x_l x_m \left(\delta_{ij}\delta_{kn} + \delta_{ik}\delta_{jn} + \delta_{in}\delta_{kj}\right) + x_l x_n \left(\delta_{mk}\delta_{ij} + \delta_{mi}\delta_{kj} + \delta_{mj}\delta_{ik}\right) \nonumber\\ & + x_n x_m \left(\delta_{ij}\delta_{kl} + \delta_{ik}\delta_{jl} + \delta_{il}\delta_{kj}\right) + \delta_{ij}\left(\delta_{kl}\delta_{mn} + \delta_{km}\delta_{ln} + \delta_{kn}\delta_{ml}\right) \nonumber\\ & + \delta_{ik}\left(\delta_{jl}\delta_{mn} + \delta_{jm}\delta_{ln} + \delta_{jn}\delta_{ml}\right) + \delta_{il}\left(\delta_{kj}\delta_{mn} + \delta_{km}\delta_{jn} + \delta_{kn}\delta_{mj}\right) \nonumber\\ & + \delta_{im}\left(\delta_{jl}\delta_{ln} + \delta_{jk}\delta_{ln} + \delta_{jn}\delta_{kl}\right) + \delta_{in}\left(\delta_{kj}\delta_{ml} + \delta_{km}\delta_{jl} + \delta_{kl}\delta_{mj}\right).
		\end{align}
	\end{subequations}
As for the single variable case, for a square-integrable function $f\left(\bm{x}\right)$, it can be expressed as:
\begin{equation}
    f\left(\bm{x}\right) = w\left(\bm{x}\right)\sum_{n=0}^{\infty} \frac{1}{n!}\bm{a}_n:\bm{\mathcal{H}}_n\left(\bm{x}\right),
\end{equation}
where the Hermite coefficients $\bm{a_n}$ are defined as:
\begin{equation}
    \bm{a}_n = \int_{-\infty}^{+\infty} f\left(\bm{x}\right)\bm{\mathcal{H}}_n\left(\bm{x}\right)d\bm{x},
\end{equation}
resulting in the following coefficients for the multi-variate version of the distribution function of Eq.~\ref{eq:sample_distribution}:
	\begin{subequations}
		\begin{align}
		\bm{a}_0 &= \rho, \\
		\bm{a}_i &= \rho u_i, \\
		\bm{a}_{ij} &= \rho u_i u_j + \rho\left(\theta-1\right)\delta_{ij}, \\
		\bm{a}_{ijk} &= \rho u_i u_j u_k + \rho\left(\theta-1\right)\left(\delta_{ij} u_k + \delta_{ik} u_j + \delta_{jk} u_i\right), \\
		\bm{a}_{ijkl} &= \rho u_i u_j u_k u_l + \rho\left(\theta-1\right)\left( \delta_{ij} u_k u_l + \delta_{ik} u_j u_l + \delta_{il} u_j u_k + \delta_{jk} u_i u_l + \delta_{jl} u_i u_k + \delta_{kl} u_i u_j\right) \nonumber\\ & + \rho{\left(\theta-1\right)}^2\left( \delta_{ij}\delta_{kl} + \delta_{ik}\delta_{jl} + \delta_{il}\delta_{jk} \right),\\
		\bm{a}_{ijklm} &= \rho u_i u_j u_k u_l u_m + \rho\left(\theta-1\right) \left( \delta_{lm} u_i u_j u_k + \delta_{km} u_i u_j u_l + \delta_{kl} u_i u_j u_m + \delta_{jm} u_i u_k u_l \right. \nonumber\\ & \left. + \delta_{jl} u_i u_k u_m + \delta_{kl} u_i u_l u_m + \delta_{im} u_j u_k u_l + \delta_{il} u_j u_k u_m + \delta_{ik} u_j u_l u_m + \delta_{ij} u_k u_l u_m \right) \nonumber\\ & + \rho{\left(\theta-1\right)}^2 \left[ u_m\left( \delta_{ij}\delta_{kl} + \delta_{ik}\delta_{jl} + \delta_{il}\delta_{jk} \right) + u_l\left( \delta_{ij}\delta_{km} + \delta_{ik}\delta_{jm} + \delta_{im}\delta_{jk} \right) \right. \nonumber\\  & \left. + u_k\left( \delta_{ij}\delta_{lm} + \delta_{il}\delta_{jm} + \delta_{im}\delta_{jl} \right) + u_j\left( \delta_{ik}\delta_{lm} + \delta_{il}\delta_{km} + \delta_{im}\delta_{kl} \right) \right. \nonumber\\ & \left. + u_i\left( \delta_{jk}\delta_{lm} + \delta_{jl}\delta_{km} + \delta_{jm}\delta_{kl} \right)\right],\\
		\bm{a}_{ijklmn} &= \rho u_i u_j u_k u_l u_m u_n + \rho\left(\theta-1\right)\left(u_i u_j u_k u_l \delta_{mn} + u_i u_j u_k u_m \delta_{ln} + u_i u_j u_k u_n \delta_{lm} \right. \nonumber\\ & \left. + u_i u_j u_l u_m \delta_{kn} + u_i u_j u_l u_n \delta_{km} + u_i u_j u_m u_n \delta_{lk} + u_i u_k u_l u_m \delta_{jn} \right. \nonumber\\ & \left. u_i u_k u_l u_n \delta_{jm} + u_i u_k u_m u_n \delta_{jl} + u_i u_l u_m u_n \delta_{jk} + u_j u_k u_l u_m \delta_{in} \right. \nonumber\\ & \left. + u_j u_k u_l u_n \delta_{im} + u_j u_l u_m u_n \delta_{ik} + u_j u_k u_l u_n \delta_{im} + u_j u_k u_m u_n \delta_{il} + u_k u_l u_m u_n \delta_{ij}\right)\nonumber\\ & + \rho{\left(\theta-1\right)}^2 \left[u_i u_j\left(\delta_{kl}\delta_{mn} + \delta_{km}\delta_{ln} + \delta_{kn}\delta_{lm} \right) +  u_i u_k \left(\delta_{jl}\delta_{mn} + \delta_{jm}\delta_{ln} + \delta_{jn}\delta_{ml}\right) \right. \nonumber\\ & \left. + u_i u_l \left(\delta_{jk}\delta_{mn} + \delta_{jm}\delta_{kn} + \delta_{jn}\delta_{mk}\right) + u_i u_m \left(\delta_{jk}\delta_{ln} + \delta_{jl}\delta_{kn} + \delta_{jn}\delta_{lk}\right) \right. \nonumber\\ & \left. + u_i u_n \left(\delta_{jk}\delta_{lm} + \delta_{jl}\delta_{km} + \delta_{jm}\delta_{lk}\right) + u_j u_k \left(\delta_{in}\delta_{lm} + \delta_{il}\delta_{nm} + \delta_{im}\delta_{ln}\right) \right. \nonumber\\ & \left. + u_j u_l \left(\delta_{in}\delta_{km} + \delta_{ik}\delta_{nm} + \delta_{im}\delta_{kn}\right) + u_j u_m \left(\delta_{in}\delta_{kl} + \delta_{ik}\delta_{nl} + \delta_{il}\delta_{kn}\right) \right. \nonumber\\ & \left. + u_j u_n \left(\delta_{il}\delta_{km} + \delta_{ik}\delta_{lm} + \delta_{im}\delta_{kl}\right) + u_k u_l \left(\delta_{mn}\delta_{ij} + \delta_{mi}\delta_{nj} + \delta_{mj}\delta_{in}\right) \right.\nonumber\\ & \left.+ u_k u_m \left(\delta_{ij}\delta_{ln} + \delta_{il}\delta_{jn} + \delta_{in}\delta_{lj}\right) + u_k u_n \left(\delta_{ml}\delta_{ij} + \delta_{mi}\delta_{lj} + \delta_{mj}\delta_{il}\right) \right.\nonumber\\ & \left.+ u_l u_m \left(\delta_{ij}\delta_{kn} + \delta_{ik}\delta_{jn} + \delta_{in}\delta_{kj}\right) + u_l u_n \left(\delta_{mk}\delta_{ij} + \delta_{mi}\delta_{kj} + \delta_{mj}\delta_{ik}\right) \right.\nonumber\\ & \left.+ u_n u_m \left(\delta_{ij}\delta_{kl} + \delta_{ik}\delta_{jl} + \delta_{il}\delta_{kj}\right)\right] + \rho{\left(\theta-1\right)}^3\left[\delta_{ij}\left(\delta_{kl}\delta_{mn} + \delta_{km}\delta_{ln} + \delta_{kn}\delta_{ml}\right) \right.\nonumber\\ & \left.+ \delta_{ik}\left(\delta_{jl}\delta_{mn} + \delta_{jm}\delta_{ln} + \delta_{jn}\delta_{ml}\right) + \delta_{il}\left(\delta_{kj}\delta_{mn} + \delta_{km}\delta_{jn} + \delta_{kn}\delta_{mj}\right) \right.\nonumber\\ & \left.+ \delta_{im}\left(\delta_{jl}\delta_{ln} + \delta_{jk}\delta_{ln} + \delta_{jn}\delta_{kl}\right) + \delta_{in}\left(\delta_{kj}\delta_{ml} + \delta_{km}\delta_{jl} + \delta_{kl}\delta_{mj}\right)\right].
		\end{align}
	\end{subequations}

\section{Elements of the von Neumann formalism\label{section:von_neumann}}
Starting with a given set of coupled continuous/discretized PDEs, bound by periodic boundary conditions, defined as:
\begin{equation}\label{eq:3-2-1-1}
\mathcal{L}\left(f_i, \bm{r}, t\right) = 0,
\end{equation}
where $\mathcal{L}$ is the time evolution operator, the equations have to be linearized in order to use the VN method. To achieve this for the LB system of equations one can expand (first-order Taylor-McLaurin expansion) the distribution function around a reference state $f_i\left(\bar{\rho},\bar{u}\right)$:
\begin{equation}
f_i \approx \bar{f}_i + f^{'}_i,
\end{equation}
\begin{equation}
\delta_t\Omega_i (f_i)\approx \delta_t\Omega_i\rvert_{\bar{f}_i} + J_{ij}f^{'}_{j},
\end{equation}
where Einstein's notation (summation) over $j$ is used, and for the sake of clarity, $\bar{f}_i=f_i\left(\bar{\rho},\bar{u}\right)$. Obviously, relying on a first-order expansion around the distribution function this expansion is only valid in the linear regime ({i.e.} small perturbations around the reference state). In addition, $J_{ij}$ is the Jacobian of the collision operator evaluated about $\bar{f}_j$, i.e, $J_{ij}=\partial_{f_{j}}\delta_t\Omega_i\rvert_{\bar{f}_{j}}.$ Placing back these expressions into the discrete LB time-evolution equation:
\begin{equation}
f_i^{'}\left( \bm{r}+\bm{c}_i\delta t,t+\delta t\right) - f_i^{'}\left( \bm{r},t\right) = J_{ij} f_i^{'}\left( \bm{r},t\right) \\ - \underbrace{\left( \bar{f}_i\left( \bm{r}+\bm{c}_i\delta t,t+\delta t\right) - \bar{f}_{i}\left( \bm{r},t\right) -\delta t\Omega_i \rvert_{\bar{f}_{i}}\right)}_{=0},
\end{equation}
and taking out the last terms on the RHS one gets:
\begin{equation}
f_{i}^{'}\left( \bm{r}+\bm{c}_i \delta t,t+\delta t\right) = \left(\delta_{ij}+J_{ij}\right) f_{j}^{'}\left( \bm{r},t\right),
\label{eq:PerturbedDiscreteLBE}
\end{equation}
where $\delta_{ij}$ is the Kronecker delta function.
Using the SRT collision operator for instance, we can then re-write the linearized time-evolution equation as:
\begin{equation}
f_i^{'}\left( \bm{r}+\bm{c}_i\delta t,t+\delta t\right) = \left[ \left(  1-\frac{\delta t}{\bar{\tau}}\right)\delta_{ij}  +\frac{\delta t}{\bar{\tau}}J_{ij}^{\rm eq}\right] f_{j}^{'}\left( \bm{r},t\right),
\end{equation}
with $J_{ij}^{\rm eq}=\partial_{j}f^{\rm eq}_\alpha\rvert_{\bar{f}_{j}}$ and $\bar{f}_{j}=f^{\rm eq}_{j}(\bar{\rho},\bar{\bm{u}})$. To compute the Jacobian matrix of the EDF, knowing that $\partial_{f_{j}}f_{k}=\delta_{jk}$, the following expressions can be used:
\begin{equation}
\partial_{f_{j}} \bm{a}^{\rm eq}_{0} = \partial_{f_{j}}(\rho) = \sum_{k} \delta_{jk} = 1,
\end{equation}
\begin{equation}
\partial_{f_{j}} \bm{a}^{\rm eq}_{1} = \partial_{f_{j}}(\rho \bm{u}) = \sum_{k} \bm{c}_{k} \delta_{jk} = \bm{c}_{j}.
\end{equation}
Once re-written as a function of the conserved Hermite coefficients, computing the Jacobians of higher-order components of the Hermite expansion is straightforward. Let us consider the second-order Hermite coefficient for example:
\begin{equation}
\partial_{f_j}\bm{a}^{\rm eq}_{2} = \partial_{f_j}\frac{\bm{a}^{\rm eq}_1\otimes\bm{a}^{\rm eq}_1}{\bm{a}^{\rm eq}_0} = -\frac{\bm{a}^{\rm eq}_{1}\otimes\bm{a}^{\rm eq}_{1}} {(\bm{a}^{\rm eq}_{0})^2} \\ + \frac{\bm{a}^{\rm eq}_{1}\otimes \bm{c}_{j} + {\left(\bm{a}^{\rm eq}_{1}\otimes \bm{c}_{j}\right)}^\dagger }{\bm{a}^{\rm eq}_{0}}.
\end{equation}
Eventually, for the second-order EDF the Jacobian reads:
\begin{equation}
J_{ij}^{\rm eq} = w_{i}\Bigg( \bm{\mathcal{H}}_{0} + \bm{\mathcal{H}}_{1}(\bm{c}_i):\partial_{f_{j}}\bm{a}^{\rm eq}_{1}  +\bm{\mathcal{H}}_{2}(\bm{c}_i):\frac{\partial_{f_{j}}\bm{a}^{\rm eq}_{2}}{2}\Bigg).
\end{equation}
The last step of the VN analysis is to assume that perturbations $f'_{i}$ are monochromatic plane waves :
$$
f'_{i} = F_{i}\exp{[\sqrt{-1}(\bm{k}\cdot\bm{r}-\omega_i t)]},
$$
where $F_i$ is the wave amplitude, $\sqrt{-1}$ is the imaginary unit, $\vert\vert\bm{k}\vert\vert=k$ is the wave-number, and $\omega$ is the complex time frequency of the wave. $k$ is related to the wave-length of $f'_{i}$, whereas $\Im(\omega)$ and $\Re(\omega)$ are related to its attenuation and propagation speed. By injecting these perturbations into Eq.~\ref{eq:PerturbedDiscreteLBE} one obtains the following eigenvalue problem of size $Q$:
\begin{equation}
\bm{M F} = \exp{(-\sqrt{-1}\omega_i)} \bm{F},
\label{eq:EigenvaluePb}
\end{equation}
where $\bm{F}$ is the eigenvector composed of all amplitudes. It is related to the eigenvalue $\exp{(-\sqrt{-1}\omega)}$. $\bm{M}$ is the matrix associated to Eq.~\ref{eq:PerturbedDiscreteLBE}. Here this matrix can be expressed as :
\begin{equation}
\bm{M} =\bm{E}\left[ \bm{\delta} +\bm{J}\right], 
\end{equation}
with
\begin{equation}
E_{ij} = \exp[-i(\bm{c}_i\cdot\bm{k})]\delta_{ij}.
\end{equation}
It is important to notice that the matrix $\bm{M}$ and the eigenvalue problem~\ref{eq:EigenvaluePb} depend on the mean flow ($\bar{\rho},\bar{\bm{u}}$), the wave-number ($k_x$ and $k_y$ in 2-D) and the relaxation coefficient $\bar{\tau}$, or equivalently the kinematic viscosity $\nu$. This means that for each set of these parameters the eigenvalue problem needs to be solved to obtain the corresponding values of $\Re{(\omega)}$ and $\Im{(\omega)}$. Doing so, the spectral properties (dispersion and dissipation) can be obtained for any given collision model.

\section{Hydrodynamic limit of the Enskog--Vlasov--BGK kinetic model\label{ap:CE_cont}}
\noindent {\bf Chapman--Enskog analysis}
Expanding the distribution function as:
\begin{equation}
	f = f^{(0)} + \epsilon f^{(1)} + \epsilon^2 f^{(2)} + {O}(\epsilon^3),
\end{equation}
introducing it back into \eqref{eq:overall_eq} and separating terms with different orders in $\epsilon$, at order zero one recovers:
\begin{equation}
	f^{(0)} = f^{\rm eq}.
\end{equation}
This latter implies the solvability conditions,
\begin{align}
	& \int f^{(k)} d\bm{v} = 0,\ \forall k\neq 0,\\
	&\int \bm{v} f^{(k)} d\bm{v} = 0,\ \forall k\neq 0.
\end{align}
At order $\epsilon$:
\begin{equation}
	\partial_t^{(1)}f^{(0)}+\bm{v}\cdot\bm{\nabla} f^{(0)} = - \frac{1}{\tau} f^{(1)} - \frac{1}{\rho}\frac{\partial f^{\rm eq}}{\partial \bm{u}}\cdot\bm{F}^{(1)},
\end{equation}
which, upon integration in $\bm{v}$, leads to
\begin{align}
	&  \partial_t^{(1)}\rho+\bm{\nabla}\cdot\rho \bm{u} = 0,\\
	&  \partial_t^{(1)}\rho \bm{u}+\bm{\nabla}\rho \bm{u}\otimes \bm{u} + \bm{\nabla}\cdot P_0\bm{I} + \bm{F}^{(1)} = 0.
\end{align}
At order $\epsilon^2$:
\begin{equation}
	\partial_t^{(2)}f^{(0)} + \partial_t^{(1)}f^{(1)}+ \bm{v}\cdot\bm{\nabla}f^{(1)} = -\frac{1}{\tau} f^{(2)},
\end{equation}
which leads to the following equation for mass conservation:
\begin{equation}
	\partial_t^{(2)}\rho = 0,
\end{equation}
while for momentum:
\begin{equation}
	\partial_t^{(2)}\rho \bm{u} + \bm{\nabla}\cdot\left[\int \bm{v}\otimes\bm{v} f^{(1)}d\bm{v}\right] = 0.
\end{equation}
The last term on the left hand side can be evaluated using the previous order in $\epsilon$ as:
\begin{multline}
	\int \bm{v}\otimes\bm{v} f^{(1)}d\bm{v} = -\tau\left[\partial_t^{(1)}\int \bm{v}\otimes\bm{v} f^{(0)}d\bm{v} + \bm{\nabla}\cdot\int \bm{v}\otimes\bm{v}\otimes\bm{v} f^{(0)}d\bm{v} \right. \\ \left. +\frac{1}{\rho} \int \bm{v}\otimes\bm{v} \frac{\partial f^{\rm eq}}{\partial \bm{u}}\cdot\bm{F}^{(1)} d\bm{v}\right],
\end{multline}
where:
\begin{equation}
	\partial_t^{(1)}\int \bm{v}\otimes\bm{v} f^{(0)}d\bm{v} = -\bm{\nabla}\cdot \rho \bm{u}\otimes\bm{u}\otimes\bm{u} - \left[\bm{u}\otimes(\bm{F}^{(1)}+\bm{\nabla}P_0) +  (\bm{F}^{(1)}+\bm{\nabla}P_0)\otimes\bm{u}\right] + \partial_t^{(1)} P_0\bm{I},
\end{equation}
and:
\begin{equation}
	\bm{\nabla}\cdot\int \bm{v}\otimes\bm{v}\otimes\bm{v} f^{(1)}d\bm{v} = \bm{\nabla}\cdot \rho \bm{u}\otimes\bm{u}\otimes\bm{u} +
	\left[\bm{\nabla}P_0\bm{u} + {\bm{\nabla} P_0\bm{u}}^{\dagger}\right] + \bm{I}\bm{\nabla}\cdot P_0 \bm{u},
\end{equation}
\begin{equation}
	\frac{1}{\rho} \int \bm{v}\otimes \bm{v} \frac{\partial f^{\rm eq}}{\partial \bm{u}}\cdot\bm{F}^{(1)} d\bm{v}= \bm{u} \otimes\bm{F}^{(1)} + {\bm{F}^{(1)}\otimes\bm{u}},
\end{equation}
which leads to:
\begin{equation}
	\int \bm{v}\otimes\bm{v} f^{(1)}d\bm{v} = -\tau\left[ P_0\left(\bm{\nabla}\bm{u}+{\bm{\nabla}\bm{u}}^{\dagger}\right)+ \left(\partial_t^{(1)}P_0 + \bm{\nabla}\cdot P_0 \bm{u}\right)\bm{I}\right],
\end{equation}
where the last two terms can be re-written as:
\begin{align}
	\partial_t^{(1)}P_0 + \bm{\nabla}\cdot P_0 \bm{u}
	= &\frac{\partial P_0}{\partial \rho}(\partial_t^{(1)}\rho + \bm{\nabla}\cdot\rho \bm{u}) + \left(P_0 -\rho\frac{\partial P_0}{\partial \rho}\right)\bm{\nabla}\cdot \bm{u}\nonumber\\
	= & P_0\left(1 - \frac{\partial \ln P_0}{\partial \ln \rho}\right)\bm{\nabla}\cdot\bm{u},
\end{align}
in turn recovering the Navier-Stokes-level momentum equation:
\begin{equation}
	\partial_t^{(2)}\rho \bm{u} - \bm{\nabla}\cdot\mu\left(\bm{\nabla}\bm{u} + {\bm{\nabla}\bm{u}}^{\dagger} - \frac{2}{3}\bm{\nabla}\cdot\bm{u}\bm{I}\right) - \bm{\nabla}\cdot\left(\eta \bm{\nabla}\cdot\bm{u}\bm{I}\right) = 0,
\end{equation}
where:
\begin{align}
	&\mu=\tau P_0,\\
	&\eta=\tau P_0\left(\frac{5}{3} - \frac{\partial \ln P_0}{\partial\ln\rho}\right).
\end{align}
\section{Chapman--Enskog analysis of the lattice Boltzmann method for non-ideal fluids\label{ap:CE}}
Using a Taylor expansion around $(\bm{r},t)$, 
\begin{equation}
	f_i\left(\bm{r}+\bm{c}_i\delta t,t+\delta t\right) - f_i\left(\bm{r},t\right) = \left[\delta t \mathcal{D}_t + \frac{\delta t^2}{2} \mathcal{D}^2_t\right] f\left(\bm{r},t\right) + {O}(\delta t^3)
\end{equation}
the discrete time-evolution equation is re-written as:
\begin{equation}
	\delta t \mathcal{D}_t f_i + \frac{{\delta t}^2}{2}{\mathcal{D}_t}^2 f_i + {O}({\delta t}^3)= \frac{\delta t}{\bar{\tau}}\left(f_i^{\rm eq} - f_i\right) + \left( f^*_i - f_i^{\rm eq}\right),
\end{equation}
where we have only retained terms up to order two. Then introducing characteristic flow size $\mathcal{L}$ and velocity $\mathcal{U}$ the equation is made non-dimensional as:
\begin{equation}
	\left(\frac{\delta r}{\mathcal{L}}\right) \mathcal{D}_t' f_i + \frac{1}{2}{\left(\frac{\delta r}{\mathcal{L}}\right)}^2{\mathcal{D}_t'}^2 f_i = \frac{\delta t}{\bar{\tau}}\left(f_i^{\rm eq} - f_i\right) + \left( f^*_i(\bm{u}'+\frac{\delta u}{ \mathcal{U}}\delta \bm{u}') - f_i^{\rm eq}(\bm{u}')\right),
\end{equation}
where primed variables denote non-dimensional form and
\begin{equation}
	\mathcal{D}_t' = \frac{\mathcal{U}}{c}\left(\partial_t' + \bm{c}_i'\cdot\bm{\nabla}'\right),
\end{equation}
where $c=\delta r/\delta t$. Assuming acoustic, i.e. $\frac{\mathcal{U}}{c}\sim 1$ and hydrodynamic, i.e. $\frac{\delta r}{\mathcal{L}}\sim\frac{\delta u}{\mathcal{U}}\sim\varepsilon$, scaling and dropping the primes for the sake of readability:
\begin{equation}
	\varepsilon \mathcal{D}_t f_i + \frac{1}{2}\varepsilon^2{\mathcal{D}_t}^2 f_i + {O}(\varepsilon^3)= \frac{\delta t}{\bar{\tau}}\left(f_i^{\rm eq} - f_i\right) + \left( f^*_i(\bm{u}+\varepsilon\delta \bm{u}) - f_i^{\rm eq}(\bm{u})\right).
\end{equation}
Then introducing multi-scale expansions:
\begin{eqnarray}
	f_i &=& f_i^{(0)} + \varepsilon f_i^{(1)} + \varepsilon^2 f_i^{(2)} + O(\varepsilon^3),\\
	{f^*_i} &=& {f^*_i}^{(0)} + \varepsilon {f^*_i}^{(1)} + \varepsilon^2 {f^*_i}^{(2)} + O(\varepsilon^3),
\end{eqnarray}
the following equations are recovered at scales $\varepsilon$ and $\varepsilon^2$:
\begin{subequations}
	\begin{align}
		\varepsilon &: \mathcal{D}_{t}^{(1)} f_i^{(0)} = -\frac{\delta t}{\bar{\tau}} f_i^{(1)} + {f^*}_i^{(1)},\\
		\varepsilon^2 &: \partial_t^{(2)}f_i^{(0)} + \mathcal{D}_{t}^{(1)} \left(1-\frac{\delta t}{2\bar{\tau}}\right)f_i^{(1)} = -\frac{\delta t}{\bar{\tau}} f_i^{(2)} + {f^*}_i^{(2)} - \frac{1}{2}\mathcal{D}_{t}^{(1)}{f^*}_i^{(1)},
	\end{align}
	\label{Eq:CE_Eq_orders}
\end{subequations}
with $f_i^{(0)}={f_i^*}^{(0)}=f_i^{\rm eq}$. The moments of the non-local contributions (including both non-ideal contributions to the thermodynamic pressure, surface tension and the correction for the diagonals of the third-order moments tensor) are:
\begin{subequations}
	\begin{align}
		\sum_i {f^*}_i^{(k)} &= 0, \forall k>0,\\
		\sum_i \bm{c}_{i} {f^*_i}^{(1)} &= \bm{F},\\
		\sum_i \bm{c}_{i}\otimes\bm{c}_{i} {f^*_i}^{(1)} & = (\bm{u}\otimes\bm{F} + {\bm{F}\otimes\bm{u}}) + \Phi\\
		\sum_i \bm{c}_{i}\otimes\bm{c}_{i} {f^*_i}^{(2)} & = \frac{1}{\rho}\bm{F}\otimes\bm{F}.
	\end{align}
	\label{Eq:CE_solvability}
\end{subequations}
Taking the moments of the Chapman-Enskog-expanded equation at order $\varepsilon$:
\begin{eqnarray}
	\partial_t^{(1)}\rho + \bm{\nabla}\cdot\rho \bm{u} &=& 0,\label{eq:approach2_continuity1_app}\\
	\partial_t^{(1)}\rho \bm{u} + \bm{\nabla}\cdot\rho \bm{u}\otimes\bm{u} + \bm{\nabla}\cdot P_0\bm{I} + \bm{F} &=& 0,\label{eq:approach2_NS1}
\end{eqnarray}
while at order $\varepsilon^2$ the continuity equation is:
\begin{equation}
	\partial_t^{(2)}\rho + \bm{\nabla}\cdot\frac{\bm{F}}{2} = 0.\label{eq:approach2_continuity2}
\end{equation}
Summing up Eqs.~\ref{eq:approach2_continuity1_app} and \ref{eq:approach2_continuity2} we recover the continuity equation as:
\begin{equation}
	\partial_t \rho + \bm{\nabla}\cdot\rho \bm{U} = 0,
\end{equation}
where $\bm{U} = \bm{u} + \frac{\delta t}{2\rho}\bm{F}$. For the momentum equations we have:
\begin{multline}\label{eq:eps2_mom1_1}
	\partial_t^{(2)}\rho \bm{u} + \frac{1}{2} \partial_t^{(1)} \bm{F} + \frac{1}{2}\bm{\nabla}\cdot\left(\bm{u}\otimes\bm{F} + {\bm{F}\otimes\bm{u}} \right)
	+ \bm{\nabla}\cdot\left(\frac{1}{2}-\frac{\bar{\tau}}{\delta t}\right)\left[\partial_t^{(1)}\Pi_{2}^{(0)}+\bm{\nabla}\cdot\Pi_{3}^{(0)}\right]
	\\ - \bm{\nabla}\cdot\left(\frac{1}{2}-\frac{\bar{\tau}}{\delta t}\right) \left(\bm{u}\otimes\bm{F} + {\bm{F}\otimes\bm{u}}\right) + \bm{\nabla}\cdot\frac{\bar{\tau}}{\delta t}\Phi = 0,
\end{multline}
where $\Pi_{2}^{(0)}$ and $\Pi_{3}^{(0)}$ are the second- and third-order moments of $f_i^{(0)}$ defined as:
\begin{eqnarray}
	\Pi_{2}^{(0)} &=& \rho \bm{u}\otimes\bm{u} + P_0\bm{I},\\
	\Pi_{3}^{(0)} &=& \Pi_{3}^{\rm MB} - \rho\bm{u}\otimes \bm{u} \otimes\bm{u}\circ\bm{J} - 3(P_0 - \rho \varsigma^2)\bm{J}
\end{eqnarray}
where $\Pi_{\alpha\beta\gamma}^{\rm MB}=\rho u_\alpha u_\beta u_\gamma + P_0{\rm perm}(u_\alpha \delta_{\beta\gamma})$ is the third-order moment of the Maxwell-Boltzmann distribution, and for the sake of simplicity we have introduced the diagonal rank three tensor $\bm{J}$, with $J_{\alpha\beta\gamma}=\delta_{\alpha\beta}\delta_{\alpha\gamma}\delta_{\beta\gamma}$ and $\circ$ is the Hadamard product.
The contributions in the fourth term on the left hand side can be expanded as:
\begin{align}
	\partial_t^{(1)} \Pi_{2}^{(0)}=&
	\partial_t^{(1)}\rho \bm{u}\otimes\bm{u} + \partial_t^{(1)} P_0\bm{I}\nonumber\\
	= & \bm{u}\otimes\partial_t^{(1)}\rho \bm{u} + {(\partial_t^{(1)}\rho \bm{u})\otimes\bm{u}} - \bm{u}\otimes\bm{u} \partial_t^{(1)} \rho + \partial_t^{(1)} P_0 \bm{I}\nonumber\\
	=& -\bm{\nabla}\cdot\rho \bm{u}\otimes\bm{u}\otimes\bm{u} 
	-\left[\bm{u}\otimes\left(\bm{\nabla} P_0 - \bm{F}\right) + {\left(\bm{\nabla} P_0 - \bm{F}\right)\otimes\bm{u}}
	\right] + \partial_t^{(1)}P_0 \bm{I}
\end{align}
and:
\begin{equation}
	\bm{\nabla}\cdot\Pi_{3}^{(0)} = \bm{\nabla}\cdot\rho \bm{u}\otimes\bm{u}\otimes\bm{u} 
	+ \left(\bm{\nabla} P_0 \bm{u} + {\bm{\nabla} P_0 \bm{u}}^{\dagger}\right) + (\bm{\nabla}\cdot P_0\bm{u})\bm{I}
	- \bm{\nabla}\cdot\left[\rho\bm{u}\otimes \bm{u} \otimes\bm{u}\circ\bm{J} + 3(P_0 - \rho \varsigma^2)\bm{J}\right],
\end{equation}
resulting in:
\begin{multline}
	\partial_t^{(1)} \Pi_{2}^{(0)} + \bm{\nabla}\cdot\Pi_{3}^{(0)} = P_0\left(\bm{\nabla}\bm{u} + {\bm{\nabla}\bm{u}}^{\dagger} \right) + \left(\bm{u}\otimes\bm{F} + {\bm{u}\otimes\bm{F}}^{\dagger}\right)\\
	+ \left(\bm{\nabla}\cdot P_0 \bm{u} + \partial_t^{(1)}P_0\right)\bm{I} - \bm{\nabla}\cdot\left[\rho\bm{u}\otimes \bm{u} \otimes\bm{u}\circ\bm{J} + 3(P_0 - \rho \varsigma^2)\bm{J}\right].
\end{multline}
Plugging this last equation back into Eq.\ \eqref{eq:eps2_mom1_1}:
\begin{multline}
	\partial_t^{(2)}\rho \bm{u} + \partial_t^{(1)}\frac{\bm{F}}{2} + \frac{1}{2}\bm{\nabla}\cdot(\bm{u}\otimes\bm{F} + {\bm{F}\otimes\bm{u}}) + \bm{\nabla}\cdot\left(\frac{1}{2}-\frac{\bar{\tau}}{\delta t}\right)P_0\left(\bm{\nabla}\bm{u} + {\bm{\nabla}\bm{u}}^{\dagger}\right)\\
	+ \bm{\nabla}\left(\frac{1}{2}-\frac{\bar{\tau}}{\delta t}\right) \left(\partial_t^{(1)}P_0+\bm{\nabla}\cdot P_0 \bm{u}\right) \\+ \bm{\nabla}\cdot\left[\left(\frac{1}{2}-\frac{\bar{\tau}}{\delta t}\right)\bm{\nabla}\cdot\left(\rho\bm{u}\otimes \bm{u} \otimes\bm{u}\circ\bm{J} + 3(P_0 - \rho \varsigma^2)\bm{J}\right) + \frac{\bar{\tau}}{\delta t}\Phi\right] = 0.
\end{multline}
where the last term cancels out by setting:
\begin{equation}
	\Phi = \left(1-\frac{\delta t}{2\bar{\tau}}\right) \bm{\nabla}\cdot\left(\rho\bm{u}\otimes \bm{u} \otimes\bm{u}\circ\bm{J} + 3(P_0 - \rho \varsigma^2)\bm{J}\right),
\end{equation}
and the fourth and fifth terms reduce to the viscous stress tensor by defining $\mu/P_0 = \left(\frac{\bar{\tau}}{\delta t} - \frac{1}{2}\right)$ and:
\begin{equation}
	P_0\left(\frac{2+D}{D} - \frac{\partial\ln P_0}{\partial\ln\rho}\right)\left(\frac{\bar{\tau}}{\delta t} - \frac{1}{2}\right) = \eta.
\end{equation}
Furthermore, using $\bm{U} = \bm{u} + \frac{\delta t}{2\rho}\bm{F}$ and:
\begin{equation}
	\rho \bm{U}\otimes\bm{U} = \rho \bm{u}\otimes\bm{u} + \frac{\delta t}{2}(\bm{u}\otimes\bm{F} + {\bm{F} \otimes\bm{u}}) + \frac{\delta t^2\bm{F}\otimes\bm{F}}{4\rho},
\end{equation}
in combination with the Euler-level equation, and keeping in mind that errors of the form $\bm{\nabla}\cdot\frac{\delta t^2\bm{F}\otimes\bm{F}}{4\rho}$ in the convective term and $\delta t\bm{\nabla}\mu\left(\bm{\nabla}\frac{\bm{F}}{\rho}+{\bm{\nabla}\frac{\bm{F}}{\rho}}^{\dagger}\right)$ in the viscous stress are of order $\varepsilon^3$ one recovers:
\begin{align}
	\partial_t\rho \bm{U} + \bm{\nabla}\cdot \rho \bm{U}\otimes\bm{U} - \bm{\nabla}\cdot\mu\left( \bm{\nabla}\bm{U} + {\bm{\nabla}\bm{U}}^{\dagger} - \frac{2}{D}\bm{\nabla}\cdot\bm{U} \bm{I}\right) - \bm{\nabla}\cdot\left(\eta \bm{\nabla}\cdot\bm{U}\right) + {O}(\varepsilon^3) = 0.
\end{align}
\bibliographystyle{elsarticle-num}
\bibliography{main}
\end{document}